\documentclass[letter,11pt]{article}
\pdfoutput=1
\usepackage{jheppub}
\usepackage{hyperref}
\usepackage{graphicx}
\usepackage{amsmath}
\usepackage{amssymb}
\usepackage{xspace}
\usepackage[small]{subfigure}
\usepackage{slashed}
\usepackage[normalem]{ulem}
\usepackage[dvipsnames]{xcolor}
\usepackage[utf8]{inputenc}
\usepackage[T1]{fontenc}
\usepackage{mathtools}
\usepackage{stmaryrd}
\usepackage{enumerate}

\renewcommand{\sec}[1]{Sec.~\ref{sec:#1}}
\newcommand{\tab}[1]{Tab.~\ref{tab:#1}}
\newcommand{\secs}[2]{Secs.~\ref{sec:#1} and \ref{sec:#2}}

\newcommand{\app}[1]{App.~\ref{app:#1}}
\newcommand{\fig}[1]{Fig.~\ref{fig:#1}}
\newcommand{\figs}[2]{Figs.~\ref{fig:#1} and \ref{fig:#2}}

\newcommand{\Refcite}[1]{Ref.~\cite{#1}}
\newcommand{\Refscite}[1]{Refs.~\cite{#1}}

\newcommand{\eq}[1]{Eq.~(\ref{eq:#1})}
\newcommand{\eqs}[2]{Eqs.~(\ref{eq:#1}) and (\ref{eq:#2})}

\newcommand{\eps}{\epsilon}

\newcommand{\bn}{{\bar{n}}}

\newcommand{\nn}{\nonumber}

\newcommand{\zcut}{z_{\rm cut}}
\newcommand{\tzcut}{\tilde z_{\rm cut}}
\newcommand{\qcut}{Q_{\rm cut}}
\newcommand{\Pythiaeight}{\textsc{Pythia}8\xspace}
\newcommand{\Pythia}{\textsc{Pythia}\xspace}
\newcommand{\Vincia}{\textsc{Vincia}\xspace}
\newcommand{\Herwig}{\textsc{Herwig}\xspace}
\newcommand{\Fastjet}{\textsc{FastJet}\xspace}

\newcommand{\bea}{\begin{eqnarray}}
\newcommand{\eea}{\end{eqnarray}}
\newcommand{\be}{\begin{equation}}
\newcommand{\ee}{\end{equation}}

\def\ln{\textrm{ln}}

\def\nn{\nonumber}

\def\bndry{\varocircle}
\def\figeight{\circ\!\!\circ}

\def\lamcs{{\Lambda\textrm{CS}}}
\allowdisplaybreaks[2]

\preprint{\begin{flushright}
MIT--CTP 5086
\\ UWThPh-2019-15
\end{flushright}}

\title{Nonperturbative Corrections to Soft Drop Jet Mass}

\author[a,b]{Andr\'e H.~Hoang,}
\affiliation[a]{University of Vienna, Faculty of Physics, Boltzmanngasse 5, A-1090 Vienna, Austria}
\affiliation[b]{Erwin Schr\"odinger International Institute for Mathematics and Physics, University of Vienna, Boltzmanngasse 9, A-1090 Vienna, Austria}
\author[c]{Sonny Mantry,}
\affiliation[c]{Department of Physics, University of North Georgia, Dahlonega, GA 30597, U.S.A.}

\author[a]{Aditya Pathak,}
\affiliation[d]{Center for Theoretical Physics, Massachusetts Institute of Technology, Cambridge, MA~02139, U.S.A.}

\author[d]{Iain W.~Stewart}

\emailAdd{andre.hoang@univie.ac.at}
\emailAdd{mantry147@gmail.com}
\emailAdd{aditya.pathak@univie.ac.at}
\emailAdd{iains@mit.edu}

\abstract{We provide a quantum field theory based description of the nonperturbative effects from hadronization for soft drop groomed jet mass distributions using the soft-collinear effective theory and the coherent branching formalism.
There are two distinct regions of jet mass $m_J$ where grooming modifies hadronization effects. In a region with intermediate $m_J$ an operator expansion can be used, and the leading power corrections are given by three universal nonperturbative parameters that are independent of all kinematic variables and grooming parameters, and only depend on whether the parton initiating the jet is a quark or gluon. The leading power corrections in this region cannot be described by a standard normalized shape function.
These power corrections depend on the kinematics of the subjet that stops soft drop through short distance coefficients, which encode a perturbatively calculable dependence on the jet transverse momentum, jet rapidity, and on the soft drop grooming parameters $z_{\rm cut}$ and $\beta$.
Determining this dependence requires a resummation of large logarithms, which we carry out at LL order. For smaller $m_J$ there is a nonperturbative region described by a one-dimensional shape function that is unusual because it is not normalized to unity, and has a non-trivial dependence on $\beta$.
}


\begin{document}
\maketitle

\pagebreak
\section{Introduction}
\label{sec:Intro}

Measurements of jet observables in QCD provide a key tool to test perturbative, resummed, nonperturbative, and Monte Carlo descriptions of QCD dynamics and also are used to probe the presence of new physics. A typical observable receives contributions from perturbative momentum regions, where the description requires fixed order calculations often supplemented with resummations of large logarithms, as well as from the nonperturbative momentum region, related to hadronization. Remarkable progress has been achieved from high precision perturbative calculations, where examples include event shapes in $e^+e^-$ collisions at next-to-next-to-next-to-leading-log order $+ {\cal O}(\alpha_s^3)$~\cite{Gehrmann-DeRidder:2007nzq,GehrmannDeRidder:2007hr,Weinzierl:2008iv,Weinzierl:2009ms,Becher:2008cf,Chien:2010kc,Abbate:2010xh,Hoang:2014wka,Hoang:2015hka}, and Higgs production with a jet veto~\cite{Berger:2010xi,Tackmann:2012bt,Banfi:2012yh,Becher:2012qa,Banfi:2012jm,Liu:2012sz,Stewart:2013faa,Becher:2013xia,Dawson:2016ysj} with a resummation of logarithms at next-to-next-to-leading-log order.

Nonperturbative hadronization corrections can also often be described rigorously from QCD with the help of factorization theorems, for instance by examining operators built out of nonperturbative modes~\cite{Lee:2006fn} within the soft-collinear effective field theory (SCET)~\cite{Bauer:2000ew,Bauer:2000yr,Bauer:2001yt,Bauer:2001ct,Bauer:2002nz}. This program has been successfully carried out for $e^+e^-$ event shapes~\cite{Lee:2006fn,Abbate:2010xh,Mateu:2012nk,Hoang:2014wka}, thrust for DIS with a jet~\cite{,Kang:2013nha}, or the jet-mass in $pp$ collisions~\cite{Jouttenus:2013hs,Stewart:2014nna}. For other methods for examining power corrections using nonperturbative models and other analytic techniques, see for example~\cite{Dokshitzer:1997iz,Salam:2001bd,Dasgupta:2003iq,Dasgupta:2007wa}. The corresponding nonperturbative parameters frequently involve light-like Wilson lines making them hard to evaluate using Lattice QCD. However, their functional dependence and universality can still be determined, and the hadronization effects can then be described by fitting one or more additional parameters in a way consistent with field theory. Often the most important information about hadronization can be encoded in a single parameter, which is the first moment of an underlying nonperturbative shape function.

Another method of accounting for hadronization corrections is to rely on hadronization models that are implemented within Monte Carlo parton shower event generators like Pythia~\cite{Sjostrand:2007gs} and Herwig~\cite{Bahr:2008pv}. The parameters in these models are fixed by tuning them to certain standard observables, and then used to predict hadronization effects in other observables. An advantage of this method is that hadronization effects can be predicted for any observable. However, it is often hard to estimate the accuracy of these models since, unlike the factorization based methods, they are based on an extrapolation rather than systematic expansions. Another problem with the Monte Carlo method is that hadronization parameters are tuned to data using a perturbative accuracy that is often limited to (roughly) next-to-leading logarithmic (NLL) order~\cite{Dasgupta:2018nvj,Hoang:2018zrp} or less (though there are a few notable exceptions that include~\cite{Alioli:2012fc,Hamilton:2013fea,Hoeche:2014aia,Karlberg:2014qua,Hoche:2014dla,Hamilton:2015nsa,Alioli:2013hqa,Alioli:2015toa}). When higher order perturbative precision is available, the use of these Monte Carlo based hadronization estimates becomes problematic since the tuning partially absorbs perturbative corrections beyond NLL, potentially leading to double counting which is hard to control.

Together with these advances in understanding perturbative and nonperturbative dynamics of jets, there has also been a surge of interest in jet substructure techniques~\cite{Butterworth:2008iy,Ellis:2009me,Krohn:2009th,Larkoski:2014wba,Larkoski:2017jix}.
This includes in particular the use of jet grooming to remove soft radiation from jets, with the goal of reducing the effects from hadronization, underlying event, and pileup. Theoretically, the most widely studied jet groomer is the soft drop algorithm~\cite{Larkoski:2014wba} (which includes as a special case the earlier modified mass drop algorithm~\cite{Butterworth:2008iy,Dasgupta:2013ihk}). Soft drop groomed observables have been recently measured by ATLAS~\cite{Aaboud:2017qwh} and CMS~\cite{Sirunyan:2018xdh}. Perturbative methods have been developed to carry out calculations for groomed jets~\cite{Walsh:2011fz,Dasgupta:2013via,Dasgupta:2013ihk,Frye:2016aiz,Marzani:2017mva}, and soft-drop factorization theorems have been derived for $D_2$~\cite{Larkoski:2017iuy}, the 2-point energy correlator, $e^{(\alpha)}_2$~\cite{Frye:2016aiz}, and the jet-mass and angularities for inclusive jets~\cite{Kang:2018jwa,Kang:2018vgn}.
Resummation of groomed event shapes at an $e^+e^-$ collider such as soft drop thrust, hemisphere jet mass, and narrow invariant jet mass were studied in \Refcite{Baron:2018nfz}, and related fixed order corrections at next-to-next-to-leading order (NNLO) accuracy were calculated in \Refcite{Kardos:2018kth}. Resummation of the soft drop jet mass for top jets was studied in \Refcite{Hoang:2017kmk} and $e_2^{(\alpha)}$ for bottom quarks in \Refcite{Lee:2019lge}. So far, either Monte Carlo hadronization models, models based on scaling from single gluon emission, or naive analytical shape functions have been used to estimate hadronization corrections for these observables, and no attempt has been made at obtaining a rigorous operator based description of hadronization corrections after jet grooming.
Scaling results based on the kinematics of single gluon emission in soft drop were considered in Refs.~\cite{Dasgupta:2013ihk,Marzani:2017kqd}, and shape function models for soft drop hadronization corrections have been employed for jet angularities~\cite{Frye:2016aiz}, $D_2$~\cite{Larkoski:2017iuy,Larkoski:2017cqq}, and heavy quark induced jets~\cite{Hoang:2017kmk,Lee:2019lge}.

In this paper we develop a factorization based description of hadronization corrections for jet observables after soft drop grooming. We focus for concreteness on the groomed jet mass $m_J$ for massless jets, though our approach can also be applied to other observables that are not sensitive to soft recoil effects, such as angularties away from the broadening limit for $a < 1$. Soft drop decouples the dependence of the hadronization from aspects of the event that are not associated with the groomed jet being studied, thus our results apply equally well for $e^+e^-$ and $pp$ collisions. We consider two distinct regions for the jet mass, each having a distinct description for their leading hadronization corrections:
\begin{align} \label{eq:NPregions}
& \text{soft drop operator expansion (SDOE) region:}
& & \frac{Q \Lambda_{\rm QCD}}{m_J^2}
\Bigl( \frac{m_J^2}{Q Q_{\rm cut}}\Bigr)^{\frac{1}{2+\beta}}
\ll 1
\,, \nn \\
& \text{soft drop nonperturbative (SDNP) region:}
& & m_J^2 \lesssim Q \Lambda_{\rm QCD} \
\Bigl( \frac{\Lambda_{\rm QCD}}{Q_{\rm cut}}\Bigr)^{\frac{1}{1+\beta}}
\,.
\end{align}
These relations will be discussed in detail in \sec{NPmodes}.
Here $\Lambda_{\rm QCD} \sim 0.5\,{\rm GeV}$ is the typical nonperturbative scale for QCD, and $Q=2 E_J=2 p_T \cosh(\eta_J)$ is a hard scale given by twice the jet energy, which is related to the jet $p_T$ and rapidity $\eta_J$. We assume that $Q\gg m_J$ and note that perturbative resummation is important for both of these regions. Finally, we have defined the smaller soft drop induced scale
\begin{align} \label{eq:Qcut}
Q_{\rm cut} \equiv 2^\beta \tilde z_{\rm cut} Q \,,
\end{align}
which depends on the soft drop parameters $z_{\rm cut}$ and $\beta$ (see \sec{SDdefn} below for the definitions of $\tilde z_{\rm cut}$, $z_{\rm cut}$, and $\beta$).

In the SDOE region, we will demonstrate that the leading hadronization effects are given by two nonperturbative matrix elements
\begin{align}
\Omega_{1\kappa}^{\figeight} \quad \textrm{and} \quad \Upsilon_{1}^\kappa(\beta) \, ,
\end{align}
defined by a field theory based operator expansion, where $\kappa = q,g$ distinguishes quark versus gluon initiated jets. Furthermore, we will show that the $\Upsilon_{1}^\kappa(\beta)$ parameter has a linear dependence on $\beta$ in the SDOE region:
\begin{align}
\label{eq:upslinearity}
\Upsilon_1^\kappa(\beta) = \Upsilon_{1,0}^\kappa + \beta \, \Upsilon_{1,1}^\kappa \,.
\end{align}
This yields in total three universal hadronic parameters, $\Omega_{1\kappa}^{\figeight}$, $\Upsilon_{1,0}^\kappa$, and $\Upsilon_{1,1}^\kappa$ which depend on a universal geometry, that we describe below, and on $\Lambda_{\rm QCD}$.
The matrix elements $\Omega_{1\kappa}^{\figeight}$ and $\Upsilon_1^\kappa(\beta)$ are multiplied by perturbative Wilson coefficients that depend on $\zcut$ and $\beta$, and the jet kinematic variables.
The resulting power corrections depend on the kinematics of the subjet that stops soft drop.
We set up a formalism for determining these coefficients and calculate them with leading logarithmic (LL) resummation accounting for running coupling effects.
The expansions used to derive this form for the power corrections imply that we cannot connect them to the power corrections appearing for the ungroomed jet mass by taking $\beta \to \infty$.

In the SDNP region we find that the leading hadronization effects are given by a non-trivial shape function $F_\kappa^{\otimes}(k,\beta)$. Unlike other examples of shape functions derived in the literature, $F^{\otimes}_\kappa(k,\beta)$ is not normalized to unity when integrated over all $k$. The function $F_\kappa^{\otimes}(k,\beta)$ also depends on the color charge $\kappa$ of the hard particle that initiates the jet, i.e. it depends on whether one considers a quark or a gluon initiated jet. However, it does not depend on $p_T$, $\eta_J$, $z_{\rm cut}$, the jet-radius $R$ or other quantities related to the hard process.

The outline for the further sections is as follows: In \sec{SoftDropObs} we review the soft drop grooming algorithm and setup the leading power soft drop factorization theorem in a convenient manner for our analysis. We then describe the interface between the partonic cross section and the nonperturbative corrections. In \sec{NPmodes} we describe the relevant EFT modes that are responsible for the leading power corrections in the SDOE and SDNP regions. We derive the factorization of the measurement and the matrix elements in the SDOE region in \sec{EFTinOPEregion}, which leads to definitions for the power corrections $\Omega_1^{\figeight}$ and $\Upsilon_1(\beta)$. In \sec{coherentbranching} we calculate the perturbative Wilson coefficients of these power corrections, which are needed to describe the hadronization corrections in the SDOE region. In \sec{NPregion} we analyze the SDNP region using tools of EFT and derive the properties of the shape function that describes power corrections in this region. A comparison with previous work is presented in \sec{comparison}. Section~\ref{sec:montecarlo} presents a parton shower Monte Carlo (MC) event generator study where we confront our field theory based description of the hadronization corrections in the SDOE region with MC results at parton and hadron level. In particular, we test the agreement of MCs with our predictions for universality by fitting the power corrections in the SDOE region to results from MC hadronization models. We conclude in \sec{conclusion}.

\section{Review of Soft Drop and Partonic Factorization}
\label{sec:SoftDropObs}

\subsection{Soft Drop Algorithm and Jet Mass}
\label{sec:SDdefn}

\begin{figure}[t!]
\centering
\includegraphics[width=0.98\columnwidth]{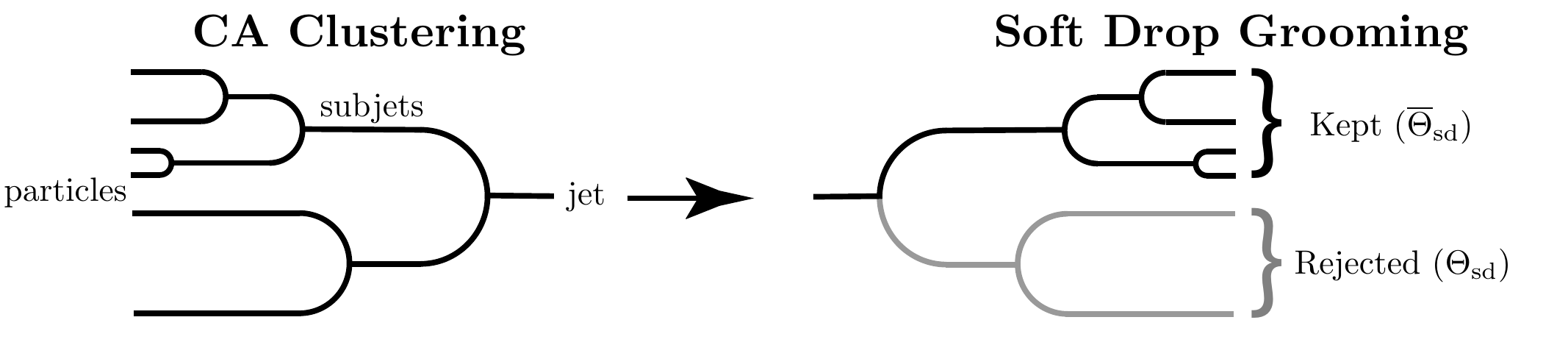}
\caption{Schematic of CA clustering and soft drop grooming algorithm. On the right the branches that fail to satisfy the soft drop criteria, shown in gray, are discarded.}
\label{fig:grooming}
\end{figure}

The soft drop algorithm~\cite{Larkoski:2014wba} considers a jet of radius $R$, reclusters the particles into a angular ordered cluster tree of subjets using the Cambridge-Aachen (CA) algorithm~\cite{Dokshitzer:1997in,Wobisch:1998wt}, and then removes peripheral soft radiation by sequentially comparing subjets $i,j$ in the tree. The grooming stops when a soft drop condition specified by fixed parameters $z_{\rm cut}$ and $\beta$ is satisfied by a pair of subjets. For $pp$ collisions the condition is
\begin{align} \label{eq:SD}
\frac{ {\rm min} [p_{Ti}, p_{Tj} ] }{ (p_{Ti}+p_{Tj}) } > z_{\rm cut} \Bigl( \frac{R_{ij}}{R_0} \Bigr)^\beta \,,
\end{align}
where $R_{ij}$ is the angular distance in the rapidity-$\phi$ plane, $R_{ij}^2= 2\big(\cosh(\eta_i-\eta_j)-\cos(\phi_i-\phi_j)\big)$ or $R_{ij}^2 = \sqrt{(\eta_i - \eta_j)^2 + (\phi_i - \phi_j)^2}$ (definitions that are equivalent in the boosted limit, and the latter being the one implemented in the soft drop algorithm). In general $R_0$ is a parameter that is part of the definition of the soft drop algorithm which is often chosen to be the jet radius. In the actual implementation of the soft drop algorithm one, however, defines $R_{ij}$ in terms of a Euclidean distance in $(\eta,\phi)$ plane, such that $R_{ij}^2 = \sqrt{(\eta_i - \eta_j)^2 + (\phi_i - \phi_j)^2}$. The two definitions are equivalent in the boosted limit.
For $e^+e^-$ collisions the condition is
\begin{align} \label{eq:SDee}
\frac{ {\rm min} [E_{i}, E_{j} ] }{ (E_{i}+E_{j}) }
> z_{\rm cut} \biggl( \sqrt{2}\, \frac{\sin(\theta_{ij}/2)}{\sin(R_0^{ee}/2)} \biggr)^\beta \,.
\end{align}
This is illustrated in \fig{grooming} where $\overline{\Theta}_{\rm sd}=1-{\Theta}_{\rm sd}$ represents the pass/fail test being applied by the soft drop groomer.
Once Eq.~(\ref{eq:SD}) or Eq.~(\ref{eq:SDee}) is satisfied all subsequent constituents in the tree are kept, thus setting a new jet radius $R_g<R$ for the groomed jet.

In the limit $R_{ij}\ll 1$, with jet constituents close to the jet axis, we can rewrite \eq{SD} in terms of the energies $E_i =p_{Ti} \cosh\eta_i$ and polar angles $\theta_{ij}\ll 1$, so that the $pp$ formula becomes
\begin{align} \label{eq:SD2}
\frac{ {\rm min} [E_{i}, E_{j} ] }{ (E_{i}+E_{j}) }
> \tilde z_{\rm cut} \, \theta_{ij}^\beta
\,,
\end{align}
where here we introduced the shorthand notation
\begin{align}
\label{eq:ppzcut}
\tilde z_{\rm cut} = z_{\rm cut} \, \frac{\cosh^\beta\eta_J}{R_0^\beta}
\qquad\quad \text{($pp$ case)}\,.
\end{align}
To obtain this result we used $R_{ij}^2 \simeq \cosh\eta_{i}\cosh\eta_j\,\theta_{ij}^2$ and $\cosh\eta_i = \cosh\eta_j +{\cal O}(\theta_{ij}) \simeq \cosh(\eta_J)$ to write the extra factors in terms of the jet's rapidity $\eta_J$. For $e^+e^-$, the result in the $\theta_{ij}\ll 1$ limit takes the same form in \eq{SD2}, but with
\begin{align}
\label{eq:eezcut}
\tilde z_{\rm cut} = z_{\rm cut} \bigg(\sqrt{2}\sin\Big(\frac{R_0^{ee}}{2}\Big)\bigg)^{-\beta}
\qquad\quad \text{($e^+e^-$ case)} \, .
\end{align}
With our definitions of $\tzcut$ in \eqs{ppzcut}{eezcut}, the formula in \eq{SD2} applies both for $pp$ and $e^+e^-$ when $\theta_{ij}\ll 1$. For our analysis we will always assume
\begin{align}
\tilde z_{\rm cut}\ll 1 \,.
\end{align}
\begin{figure}[t!]
\centering
\includegraphics[width=0.75\textwidth]{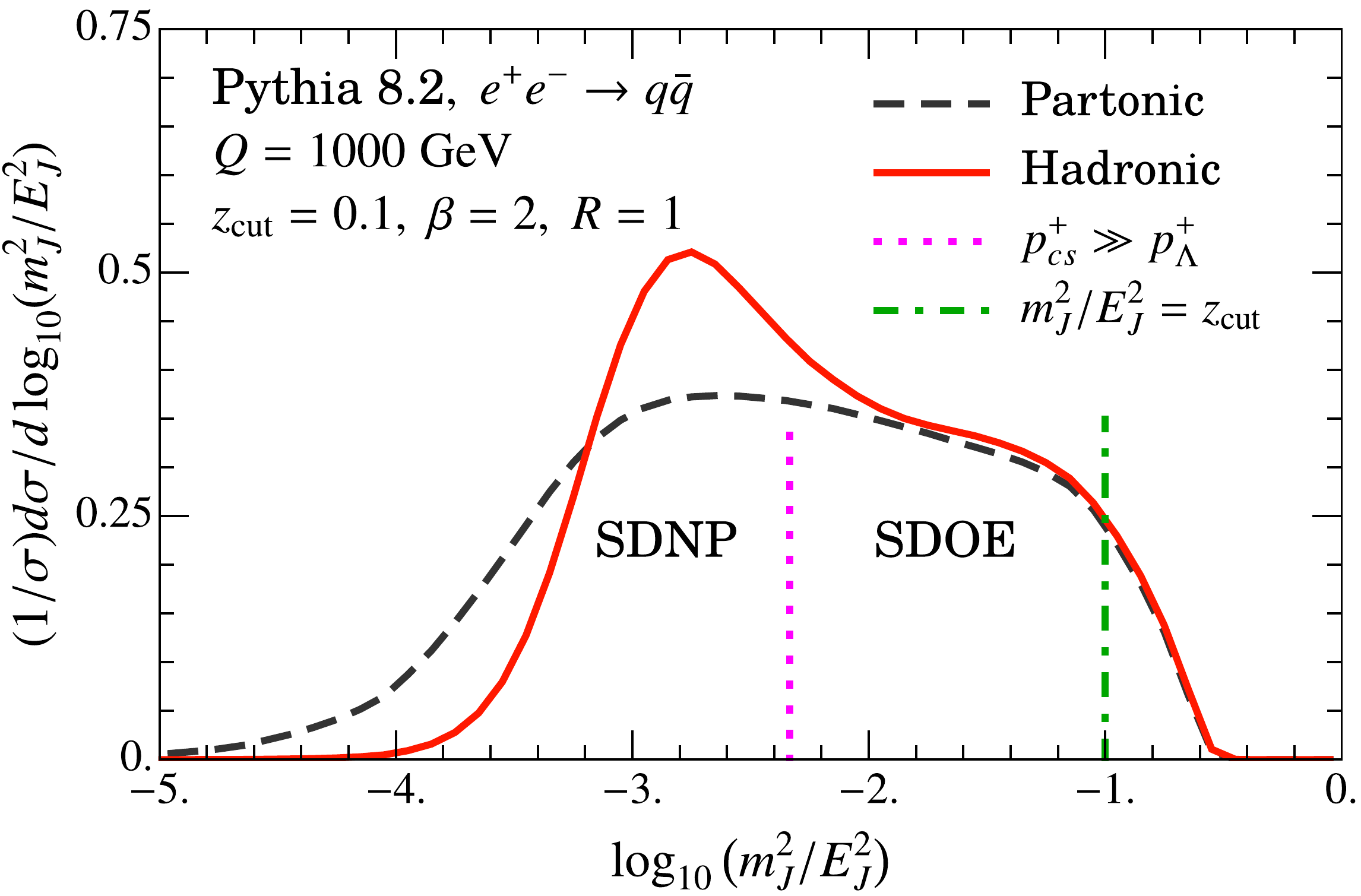}\\
\vspace{-0.2cm}
\caption{Pythia prediction for $m_J^2/E_J^2$ distribution indicating three different regions of the spectrum.
Here we take $R_0^{ee} = \frac{\pi}{2}$ and the inequality ``$\gg$'' in the ``$p_{cs}^+ \gg p_\Lambda^+$'' constraint for the SDOE region is replaced by a factor of 5.}
\label{fig:logrhoplot}
\vspace{-0.2cm}
\end{figure}
With soft drop grooming the jet mass is defined by starting with the constituents of the jet of radius $R$ and summing only over the constituents ${\cal J}_{sd}$ that remain after soft-drop has been applied,
\begin{align} \label{eq:mJ}
m_J^2 = \biggl(\: \sum_{i\in {\cal J}_{sd}} p_i^\mu \biggr)^2 \,.
\end{align}

In \fig{logrhoplot} we show a \Pythiaeight Monte Carlo prediction for this groomed jet mass spectrum with $z_{\rm cut}=0.1$ and $\beta=2$.
We distinguish three relevant regions of the spectrum: the soft drop nonperturbative region (SDNP) to the far left (left of the magenta dashed line), the soft drop operator expansion region (SDOE) in the middle (between the dashed lines), and the ungroomed resummation region on the far right where soft drop turns off but the log resummation for the ungroomed case is still active (right of the green dashed line). The distinction between the SDNP and SDOE regions is determined by \eq{NPregions}, while the distinction between the SDOE and ungroomed resummation region is given by
\begin{align} \label{eq:sdlimit}
\hspace{0.2cm}
& \text{soft drop operator expansion (SDOE):}
\nn \\
&\quad m_J^2 \ll z_{\rm cut}\, p_T^2R^2\Bigl(\frac{R}{R_0}\Bigr)^\beta \ (pp)
& & \textrm{or}
\qquad m_J^2 \ll Q\qcut\tan^{\beta+1}\Bigl(\frac{R}{2}\Bigr) \ \ (ee)
\,, \hspace{3cm} \nn \\
& \text{ungroomed resummation region:}
\\
&\quad m_J^2 \gtrsim z_{\rm cut} \,p_T^2R^2\Bigl(\frac{R}{R_0}\Bigr)^\beta \ (pp)
& & \textrm{or}
\qquad m_J^2 \gtrsim Q \qcut\tan^{\beta+1}\Bigl(\frac{R}{2}\Bigr) \ \ (ee)
\,. \nn
\end{align}
For the hemisphere case in $e^+e^-$ one has $R=\pi/2$.
For $R=1$, in the $e^+e^-$ case this boundary roughly corresponds to $m_J^2/E_J^2 = z_{\rm cut}$ which we use in \fig{logrhoplot}.
In all three of these regions the resummation of large logarithms is important, though the precise nature of this resummation is different.
There is an additional transition between resummation and fixed-order regions which occurs on the very far right of \fig{logrhoplot}, near where $d\sigma/dm_J$ goes to zero (not indicated in the plot).

\subsection{Partonic Factorization for Light Quark and Gluon Jets} \label{sec:massless}

In this section we review the pioneering partonic massless soft drop factorization theorem derived in~\cite{Frye:2016okc,Frye:2016aiz}, working in the same limit
\begin{align}
\label{eq:partfactconstr}
4m_J^2 /Q^2 \ll z_{\rm cut} \ll 1 \,.
\end{align}
In Secs.~\ref{sec:NPmodes}--\ref{sec:NPregion} we will extend this factorization based description to hadron level by including the leading final state hadronization effects. This will be achieved by incorporating field theoretically derived nonperturbative matrix elements for the SDOE region, and by setting up a novel shape function for the SDNP region.

Consider the groomed jet mass measurement from jets initiated by light quarks or gluons. A partonic factorization formula for the soft dropped jet mass in the limit in \eq{partfactconstr}, was derived in Ref.~\cite{Frye:2016aiz}:
\begin{align} \label{eq:SDmasslessFrye}
\frac{d\hat\sigma}{dm_J^2}
&= \sum_{\kappa=q,g} \tilde N_{\kappa}\,
\tilde J_\kappa(m_J^2/Q^2) \otimes \tilde S_C^\kappa(z_{\rm cut}m_J/Q) \,.
\end{align}
Here $\otimes$ denotes a convolution, $\kappa$ sums over quark and gluon jets, the $\tilde N_{\kappa}$ are normalization factors, and $\tilde J_\kappa$ and $\tilde S_C^\kappa$ are dimensionless jet and collinear-soft functions.
The $\tilde N_{\kappa}$ determine what fraction of the jets are produced by gluons or quarks given an underlying hard process. For $pp$ it incorporates the initial state parton distribution functions (PDFs). It also contains the global soft function, the hard function, as well as any other functions describing other aspects of the event. The jet mass spectrum is determined by the inclusive jet function $\tilde J_\kappa$ that encodes the distribution of collinear radiation in the jet and a collinear-soft function $\tilde S_C$ that describes the influence of soft radiation retained by soft drop. $\tilde S_C$ depends on $Q$ and the soft drop parameters $z_{\rm cut}$ and $\beta$. The dependence on the variables $Q$ and $z_{\rm cut}$ is remarkably simple since they appear only in a single combination with $m_J$~\cite{Frye:2016aiz}.
At (modified) LL order, \eq{SDmasslessFrye} agrees with the results derived in the original soft drop paper~\cite{Larkoski:2014wba} and at LL for $\beta = 0$ with the results for modified Mass Drop Tagger derived in \Refcite{Dasgupta:2013ihk} using the coherent branching formalism.
Results for $\beta\ge 0$ have also been derived in~\Refscite{Marzani:2017mva,Marzani:2017kqd}.

We choose to write the partonic factorization theorem for the jet mass spectrum in a more explicit form as
\begin{align} \label{eq:masslessFact0}
\frac{d \hat \sigma}{dm_J^2 d \Phi_J}
&= \sum_{\kappa={q,g}} N_\kappa(\Phi_J, R, \zcut, \beta,\mu)
\, \qcut^{\frac{1}{1+\beta}}
\int\!\! d\ell^+ \: J_\kappa \big(m_J^2 - Q\ell^+ ,\, \mu \big)
\: S_c^\kappa \Big[ \ell^+ \qcut^{\frac{1}{1+\beta}},\beta,\mu\Big]
\,.
\end{align}
Here we collectively denote the jet kinematic variables by $\Phi_J=\{p_T,\eta_J\}$ for $pp$ and $\Phi_J=\{E_J\}$ for $ee$, and have displayed the form of the convolution between the jet and collinear-soft functions. This expression involves the soft drop modified hard scale $\qcut$ defined in \eq{Qcut}.
In our notation the $J_\kappa$ and $S_c^\kappa$ functions have non-zero mass dimensions and hence differ from those in \eq{SDmasslessFrye}.

In \eq{masslessFact0} we use the standard SCET jet function $J_\kappa$ for quarks and gluons which has mass dimension $-2$~\cite{Bauer:2001yt}, rather than the $\tilde J_\kappa$ in \eq{SDmasslessFrye} that was a function of a dimensionless variable.
The jet function $J_\kappa$ encodes collinear modes that have the momentum scaling given by
\begin{align}
p_C^\mu \sim \Bigl( \frac{m_J^2}{Q}, Q, m_J \Bigr) \,,
\end{align}
where we show the light-cone momentum components
\begin{align}
\label{eq:lightcone}
(p^+,p^-,p_\perp)\equiv (n_J\cdot p,\bar n_J\cdot p,p_\perp) \,,
\end{align}
relative to the jet axis $\hat n_J$, with $n_J=(1,\hat n_J)$ and $\bar n_J=(1,-\hat n_J)$. In terms of these coordinates the angle $\theta$ relative to the jet axis is given by $\sin(\theta) = |\vec p_\perp|/p^z$ which for $\theta \ll 1$ gives
\begin{align} \label{eq:theta}
\frac{\theta}{2} \simeq \frac{|\vec p_\perp|}{p^-} \,.
\end{align}
Hence the collinear particles spread out relative to the jet axis with a typical angle $\theta_c \sim 2 m_J/Q$. The scaling for the global soft modes that contribute to $N_\kappa$ is
\begin{align}
\label{eq:pGS}
p_{gs}^\mu \sim Q_{\rm cut} (1,1,1)\quad (e^+e^-) \,,
\qquad
\qquad
p_{gs}^\mu \sim p_T z_{\rm cut} (1,1,1)\quad (pp) \,.
\end{align}
This soft scaling differs from that of the ultrasoft mode, $p_{us}^\mu \sim m_J^2/Q (1,1,1)$, which is relevant in the ungroomed case.

Finally, rather than using a dimensionless collinear-soft function $\tilde S_C^\kappa$ as in \eq{SDmasslessFrye}, we use a collinear-soft function with dimension $(-2-\beta)/(1+\beta)$:
\begin{align} \label{eq:SCdefn}
& S_c^\kappa\Bigl[\ell^+\, Q_{\rm cut}^{\frac{1}{1+\beta}}, \beta,\mu \Bigr] \equiv \frac{Q_{\rm cut}^{\frac{-1}{1+\beta}}}{n_\kappa} \:
{\rm tr} \bigl\langle 0 \big| \bar T X_{n\kappa}^\dagger V_{n\kappa} \delta \bigl(\ell^+- \overline\Theta_{\rm sd}\, \hat p^+\bigr) T V_{\kappa n}^\dagger X_{n\kappa}
\big| 0 \bigr\rangle_\mu
\,,
\end{align}
where the subscript $\mu$ on the RHS indicates that this matrix element is renormalized in the $\overline{\rm MS}$ scheme.
The normalization convention we adopt for this definition ensures that $S_c^\kappa$ is truly a function of only the three variables shown, which makes manifest the non-trivial connection between $\ell^+$, $Q$ and $z_{\rm cut}$ derived in Ref.~\cite{Frye:2016aiz}, which imply that it is only a function of $\ell^+ \qcut^{\frac{1}{1+\beta}}$. Note that in the $pp$ case this combination is independent of the jet rapidity $\eta_J$ since the $\cosh\eta_J$ factors in $\qcut^{\frac{1}{1+\beta}}/Q $ cancel:
\begin{align}
\ell^+ \qcut^{\frac{1}{1+\beta}}
\sim \frac{m_J^2}{Q} \qcut^{\frac{1}{1+\beta}}
= \frac{m_J^2}{p_T} \big( p_T z_{\rm cut} R_0^{-\beta} \big)^{\frac{1}{1+\beta}}
\,. \qquad
\end{align}
Note that the prefactor of $ \qcut^{\frac{-1}{1+\beta}}$ in \eq{SCdefn} is compensated by the prefactor in the factorization formula in \eq{masslessFact0}. In \eq{SCdefn} $n_q=N_c$ and $n_g=N_c^2-1$ normalize the color trace, where $N_c=3$ is the number of colors.
Also, $\overline\Theta_{\rm sd}$ is the soft drop measurement function that selects the collinear-soft particles that pass soft drop. We discuss the measurement operator $\overline\Theta_{\rm sd}\,\hat p^+$ in \eq{SCdefn} in \sec{measurement}. Up to the overall normalization factor, the RHS of \eq{SCdefn} is identical to the perturbative collinear-soft function defined in Ref.~\cite{Frye:2016aiz} (setting their $\alpha=2$). The terms $V_{n\kappa}=V_{n\kappa}[\bn\cdot A_{cs}]$ and $X_{n\kappa}=X_{n\kappa}[n\cdot A_{cs}]$ in \eq{SCdefn} are collinear-soft Wilson lines in the fundamental ($\kappa=q$) or adjoint ($\kappa=g$) representations. They are easily derived following the procedure in {\rm SCET}$_+$ (see Refs.~\cite{Bauer:2011uc, Procura:2014cba, Larkoski:2015zka, Pietrulewicz:2016nwo}).
This function $S_c^\kappa$ encodes the dynamics of the collinear-soft modes, whose momentum components scale as
\begin{align} \label{eq:pCS}
p_{cs}^\mu &\sim \frac{m_J^2}{Q\zeta_{cs}}
\Bigl( \zeta_{cs} ,\frac{1}{\zeta_{cs}} , 1 \Bigr)
\,,
\qquad \zeta_{cs} \equiv \Bigl( \frac{m_J^2}{Q \qcut} \Bigr)^{\frac{1}{2+\beta}}
\,.
\end{align}
Here $\zeta_{cs}\ll 1$ follows because $m_J^2 \ll Q \qcut$ holds true in the SDOE region as seen from \eq{sdlimit}, for $R \lesssim 1$.
The scaling in \eq{pCS} is determined by demanding that $p_{cs}^+\sim m_J^2/Q$ so that this mode contributes to the jet mass measurement, and that it saturates the soft drop condition, hence satisfying $p^-_{cs} \sim Q \tilde z_{\rm cut} \theta_{cs}^\beta = Q_{\rm cut} (\theta_{cs}/2)^\beta$. From \eq{theta} we have
\begin{align}
\frac{\theta_{cs}}{2} \sim \zeta_{cs} \gg
\frac{\theta_c}{2} \sim \frac{m_J}{Q}
\,,
\end{align}
so that the collinear-soft modes probe the edge of the groomed jet, while the collinear modes are well inside.
At one-loop the result for the collinear-soft function in the $\overline{\rm MS}$ scheme is
\begin{align}
\label{eq:sc1loop}
S_c^\kappa\Bigl[\ell^+\, Q_{\rm cut}^{\frac{1}{1+\beta}}, \beta,\mu \Bigr]
&= \delta\Bigl( \ell^+ Q_{\rm cut}^{\frac{1}{1+\beta}}\Bigr)
+ \frac{C_\kappa\alpha_s(\mu)}{\pi} \Biggl\{ \!
\frac{-2(1+\beta)}{(2\!+\!\beta)\mu^{\frac{2+\beta}{1+\beta}}}
\,{\cal L}_1\!\biggl[ \frac{\ell^+ Q_{\rm cut}^{\frac{1}{1+\beta}}}{\mu^{\frac{2+\beta}{1+\beta}}} \biggr]
+ \frac{\pi^2}{24} \frac{2\!+\!\beta}{1\!+\!\beta} \,
\delta\Bigl( \ell^+ Q_{\rm cut}^{\frac{1}{1+\beta}}\Bigr)
\!\Biggr\}
,
\end{align}
where ${\cal L}_1(x)= \big[\Theta(x) \frac{\ln x}{x} \big]_+$ is the standard plus function which integrates to zero on $x\in [0,1]$, $C_q=C_F$, and $C_g=C_A$. Here we see explicitly that the collinear-soft function is only a function of the three variables shown in the arguments in the left hand side of \eq{sc1loop}.
The momentum space result in \eq{sc1loop} agrees with the Laplace space result in Eq.(E.4) of Ref.~\cite{Frye:2016aiz}.

The form of the convolution shown in \eq{masslessFact0} follows from the fact that the jet mass, when decomposed into contributions from the collinear and collinear-soft modes, is
\begin{align}
\label{eq:mJexpn}
m_J^2 = (p_n +p_{cs})^2 = p_n^2 + Q\, p_{cs}^+ + \ldots \,,
\end{align}
where the ellipses are higher order in the power counting. Here $p_n^2$ is the argument of the jet function $J_\kappa$, and $p_{cs}^+ = \ell^+$ is the variable that appears in $S_c^\kappa$. The fact that the sum of collinear and collinear-soft momenta gives the total jet mass leads to the convolution. Including the resummation of large logarithms, the factorization theorem in \eq{masslessFact0} becomes
\begin{align} \label{eq:masslessFact1}
\frac{d \hat \sigma}{dm_J^2 d \Phi_J} &= \sum_{\kappa={q,g}}
N_\kappa(\Phi_J, R, \zcut, \beta,\mu_h,\mu_{gs})\,
U_{S_G}(\qcut,\mu_{gs},\mu_{cs})\,
\qcut^{\frac{1}{1+\beta}} \int\!\! d\ell^+ ds \:
J_\kappa \big(m_J^2 - s ,\, \mu_{J} \big)
\nn \\[-5pt]
&\qquad
\times
U_J(s-Q\ell^+,\mu_J,\mu_{cs})
\: S_c^\kappa \Big[ \ell^+ \qcut^{\frac{1}{1+\beta}},\beta,\mu_{cs}\Big]
\,, \nn \\
&\equiv \sum_{\kappa={q,g}} N_\kappa(\Phi_J, R, \zcut, \beta,\mu_h,\mu_{gs})
\, \frac{d \hat \sigma_\kappa}{dm_J^2}(\mu_{gs})
\,,
\end{align}
where $U_{S_G}$ and $U_J$ are RG evolution kernels. Note that the hard scale $\mu_h$ indicates an upper limit for an evolution that takes place inside $N_\kappa$, and since this factor also includes the boundary condition at $\mu_h$ it is formally $\mu_h$ independent. In the third line of \eq{masslessFact1} we have defined a notation for the partonic cross sections $d \hat \sigma_\kappa/dm_J^2$ for the individual $\kappa=q,g$ channnels which we will use later on. This formula does not account for final state hadronization effects. The canonical global soft, jet, and collinear-soft scales appearing in \eq{masslessFact1} are
\begin{align}
\mu_h &= p_T R \quad (pp) \quad \textrm{or} \ \
\mu_h= Q\tan\Bigl(\frac{R}{2}\Bigr)\ \ (ee) \,,
\nn\\
\mu_{gs} &= p_T R\, z_{\rm cut} \Big(\frac{R}{R_0}\Big)^\beta \ \ (pp)
\quad \textrm{or} \quad
\mu_{gs}=\qcut \tan^{1+\beta}\Bigl(\frac{R}{2}\Bigr) \ \ (ee) \,,
\nn\\
\mu_J & = m_J \,, \qquad\qquad
\mu_{cs} = \big(\qcut\big)^{\frac{1}{2+\beta}}
\Bigl( \frac{m_J^2}{Q}\Bigr)^{\frac{1+\beta}{2+\beta}} \,.
\end{align}
The renormalization group evolution between these scales sums the associated large logarithms. Note that we will always have $\mu_J \gg \mu_{cs}$, since $\mu_{cs}/\mu_J =(m_J/Q)^{\frac{\beta}{2+\beta}} (\qcut/Q)^{\frac{1}{2+\beta}}\ll 1$, and also $\mu_{gs}\gg \mu_{cs}$, since $\mu_{cs}/\mu_{gs} \sim (m_J^2/Q\qcut)^{\frac{1+\beta}{2+\beta}}\ll 1$. On the other hand there is no universal hierarchy between the scales $\mu_{gs}$ and $\mu_J$. The functions $N_\kappa$ involve both the hard scale $\mu_h\sim Q\sim p_T$ and the global-soft scale $\mu_{gs}\sim \qcut$, and additional logarithms of $z_{\rm cut}$ from the hierarchy $Q\gg \qcut$ can be summed inside $N_\kappa$ if desired. The integrations in \eq{masslessFact1} can be evaluated analytically with standard techniques and give predictions at LL, NLL, etc, for the soft drop groomed spectrum. Results up to NNLL were obtained in Ref.~\cite{Frye:2016aiz} for $\beta=0$ and $\beta=1$.

\section{Nonperturbative Modes for Soft Drop}
\label{sec:NPmodes}

To determine the leading hadronization corrections we first determine for our observable the dominant nonperturbative modes $\Lambda$ with momenta $(p_\Lambda^+,p_\Lambda^-,p_\Lambda^\perp)$. This is done separately for the operator expansion (SDOE) and nonperturbative (SDNP) regions, see \eqs{NPregions}{sdlimit}.

\begin{figure}[t!]
\hspace{-1cm}
\includegraphics[width=0.54\textwidth]{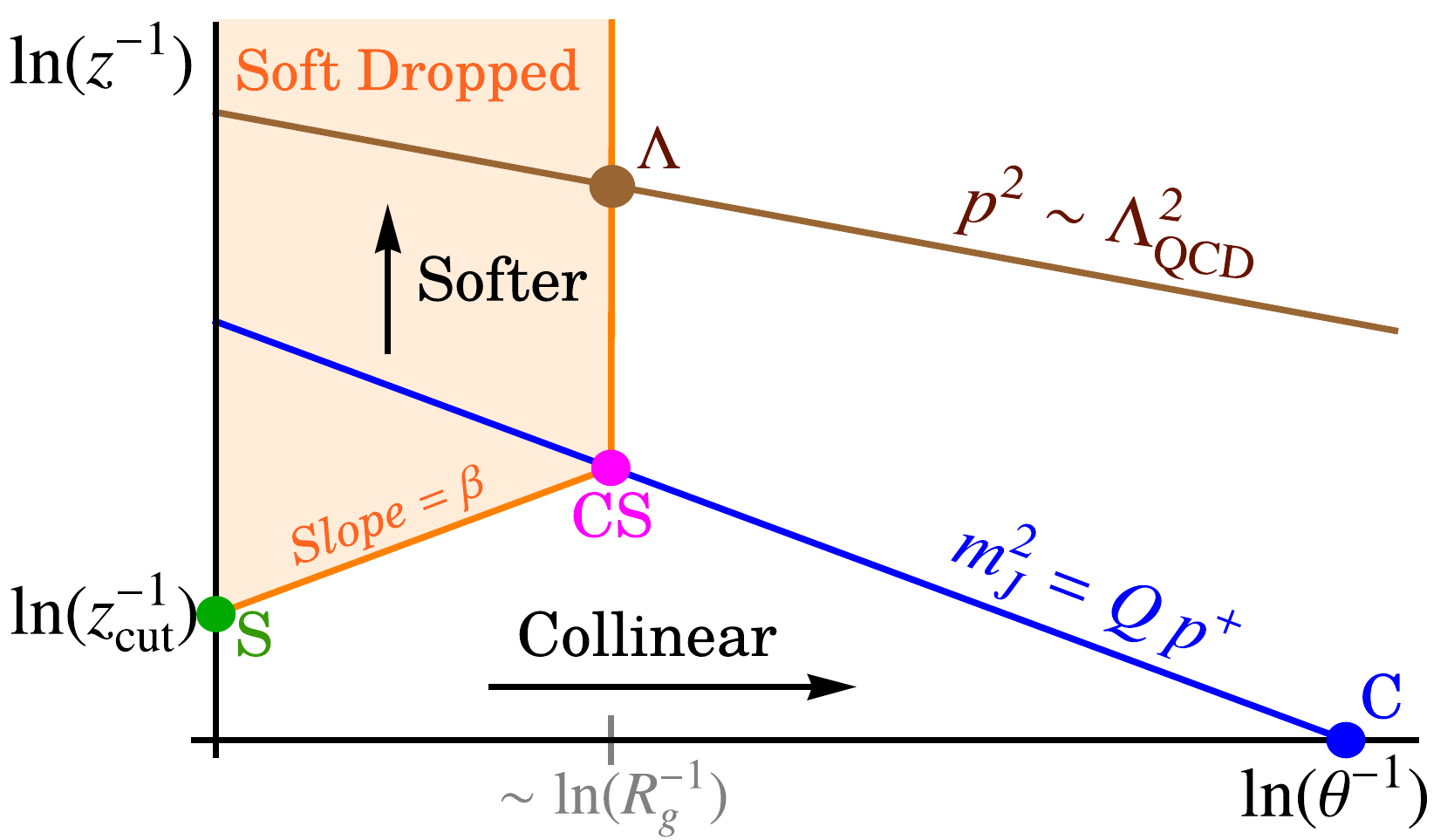}
\includegraphics[width=0.54\textwidth]{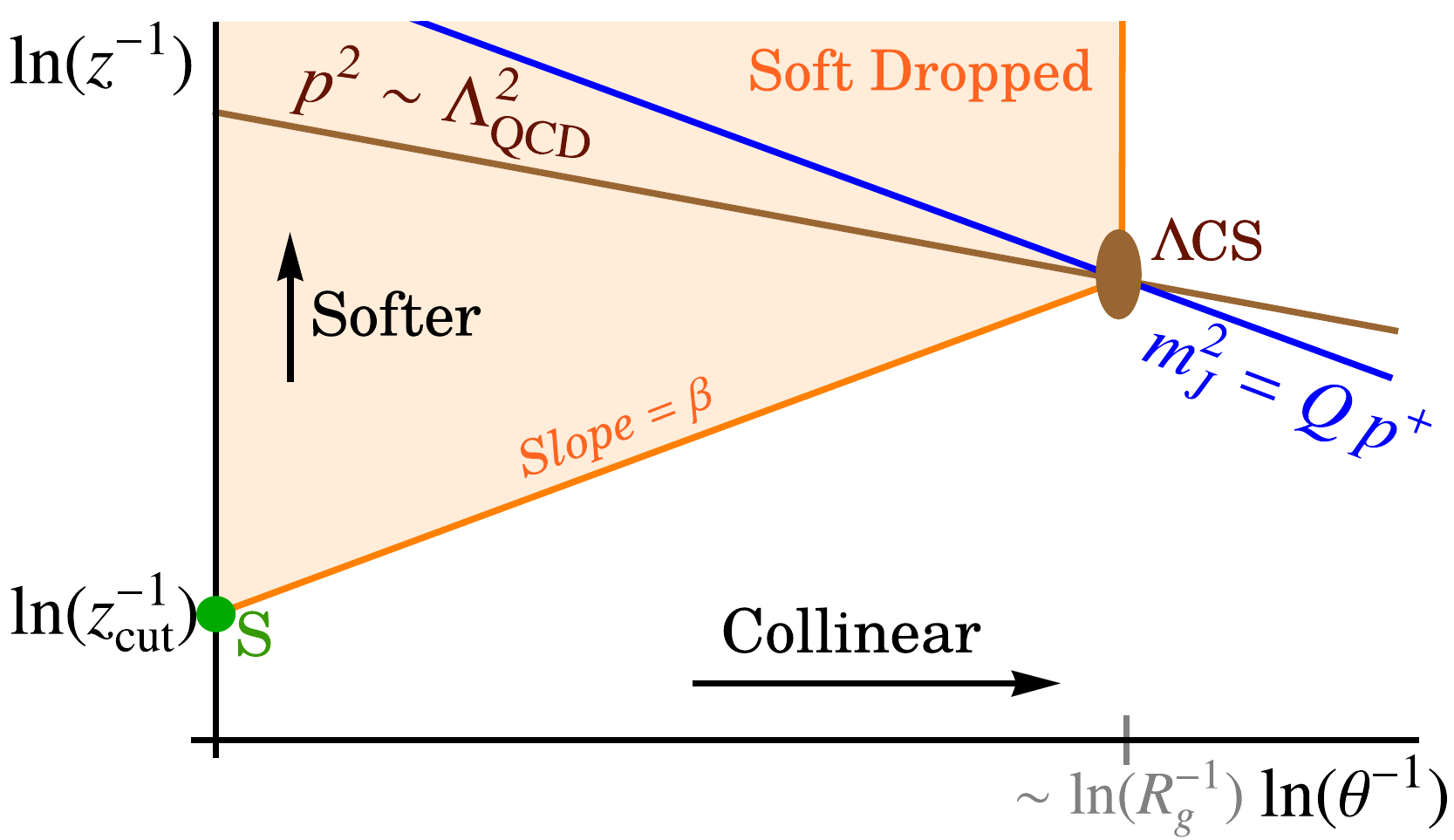}
\\[-10pt]
\phantom{x}\hspace{-0.3cm} a) \hspace{8.2cm} b) \\[-15pt]
\caption{Modes appearing in the soft drop factorization formula for jets initiated from massless quarks or gluons. Panel a) shows a $m_J$ value in the operator expansion region, while panel b) shows a $m_J$ in the nonperturbative region. Here $z$ is the energy fraction relative to the total jet energy, and $\theta$ is the polar angle relative to the jet-axis.
The hard modes $H$ at $z\sim 1$ and $\theta \sim 1$ are not shown.}
\label{fig:modesmassless}
\vspace{-0.2cm}
\end{figure}

In \fig{modesmassless}a we show all the relevant perturbative and nonperturbative SCET modes for $m_J$ values in the SDOE region. In \fig{modesmassless}b we show the relevant modes when $m_J$ is in the SDNP region. These figures shows only scaling relations, so that the momentum scaling of modes at different locations are separated by a $\ll$ inequality. Any particles with momenta satisfying a $\sim$ relation appear at the same point, with the scaling of the mode at that point. Here C denotes the collinear modes appearing in $J_\kappa$, which sit on the blue measurement curve labeled $m_J^2=Q p^+$ at small $\theta_c\sim 2m_J/Q$.
The slanted orange line for $z\simeq \tilde z_{\rm cut} \theta^\beta$ bounds the momentum region removed by soft drop and is labeled by ``$\text{slope} =\beta$''. The magenta point labeled CS denotes the collinear-soft modes which determine $S_c^\kappa$, whose scaling was given in \eq{pCS}, and is determined by the intersection of the slanted orange line and the blue curve. Finally, $S$ denotes the global soft modes that are groomed away. Their presence is required for renormalization group (RG) consistency, as part of the calculation of the no-emission probability, and are included in $N_{\kappa}$. Note that ultra-soft modes, sensitive to other parts of the event, sit at the intersection of the blue curve and y-axis, and are removed by soft drop.

Figure~\ref{fig:modesmassless}a for the SDOE region is identical to the mode picture in~\cite{Frye:2016aiz} with two exceptions that did not matter there but are important for our analysis. The first is that we have
added the brown line, $p^2\sim \Lambda_{\rm QCD}^2$, which denotes where the dominant modes responsible for hadronization effects are located.
The second is that the shaded orange region, which denotes the region removed by soft drop, is truncated in the $\theta$ direction by the vertical orange line at the angle $\theta\simeq R_g$ where the iterative grooming stops, see~\cite{Larkoski:2014wba}.
At leading power in the SDOE region, soft drop is always stopped by comparing a perturbative CS subjet with the subjet containing the collinear particles. Thus the vertical line occurs at the location of the CS mode which has a parametrically larger angle than the collinear mode. To understand this, note that the CS modes saturate the soft drop condition, i.e. they sit at the largest possible angle from the jet axis and have large enough energies to enable them to pass soft drop. So when pairs of subjets are tested as we traverse the CA clustering tree backwards, there will be a subjet in the tree that carries all the C particles, and another with one or more CS particles. As long as at least one CS subjet is kept by soft drop, then the stopping subjet will have collinear-soft scaling.

In the SDOE region a perturbative CS subjet will always stop soft drop, yielding $R_g\sim \theta_{cs}$.
This then determines the dominant nonperturbative mode for the SDOE region, labeled by $\Lambda$ in \fig{modesmassless}a, which has the same parametric angle as the CS mode. Its $(p^+,p^-,p^\perp)$ momentum components therefore have the scaling
\begin{align} \label{eq:pL}
p_\Lambda^\mu
&\sim \Lambda_{\rm QCD} \Bigl( \zeta_{cs} ,\frac{1}{\zeta_{cs}} , 1 \Bigr)
\,.
\end{align}
These $\Lambda$ modes yield the most important nonperturbative contribution to the jet mass since, as seen from \eq{mJexpn}, in the collinear-soft limit the contribution to the jet mass is given by $Q p^+$ and the $\Lambda$ modes have the largest $p^+$ momentum component among all modes on the brown line that are not completely removed by soft drop.
To see this we note that for massless modes $p^+\sim Q z (\theta/2)^2\sim \Lambda_{\rm QCD} (\theta/2)$, where the second $\sim$ follows from $p^2\sim\Lambda_{\rm QCD}^2$ when $\theta\ll 1$. Thus the nonperturbative mode with the largest $p^+$ is the one with the largest $\theta$ that is not removed by soft drop, given by $\theta_{cs}/2\sim \zeta_{cs}$.
Comparing $p^+$ momenta in \eqs{pCS}{pL} we see that for $p_{cs}^+\gg p_\Lambda^+$ we have $(Q\Lambda_{\rm QCD}/m_J^2) (m_J^2/QQ_{\rm cut})^{\frac{1}{2+\beta}}\ll 1$, which is precisely the equation for the SDOE region in \eq{NPregions}. In this region the hadronization contributions from $\Lambda$ are power corrections.
We note that the dominant power corrections appear only from corrections to the collinear-soft sector. The nonperturbative corrections in the collinear sector are subleading since nonperturbative corrections to the jet function are suppressed by $\Lambda_{\rm QCD}/m_J\ll 1$.

\begin{figure}[t!]
\centering
\includegraphics[width=0.5\textwidth]{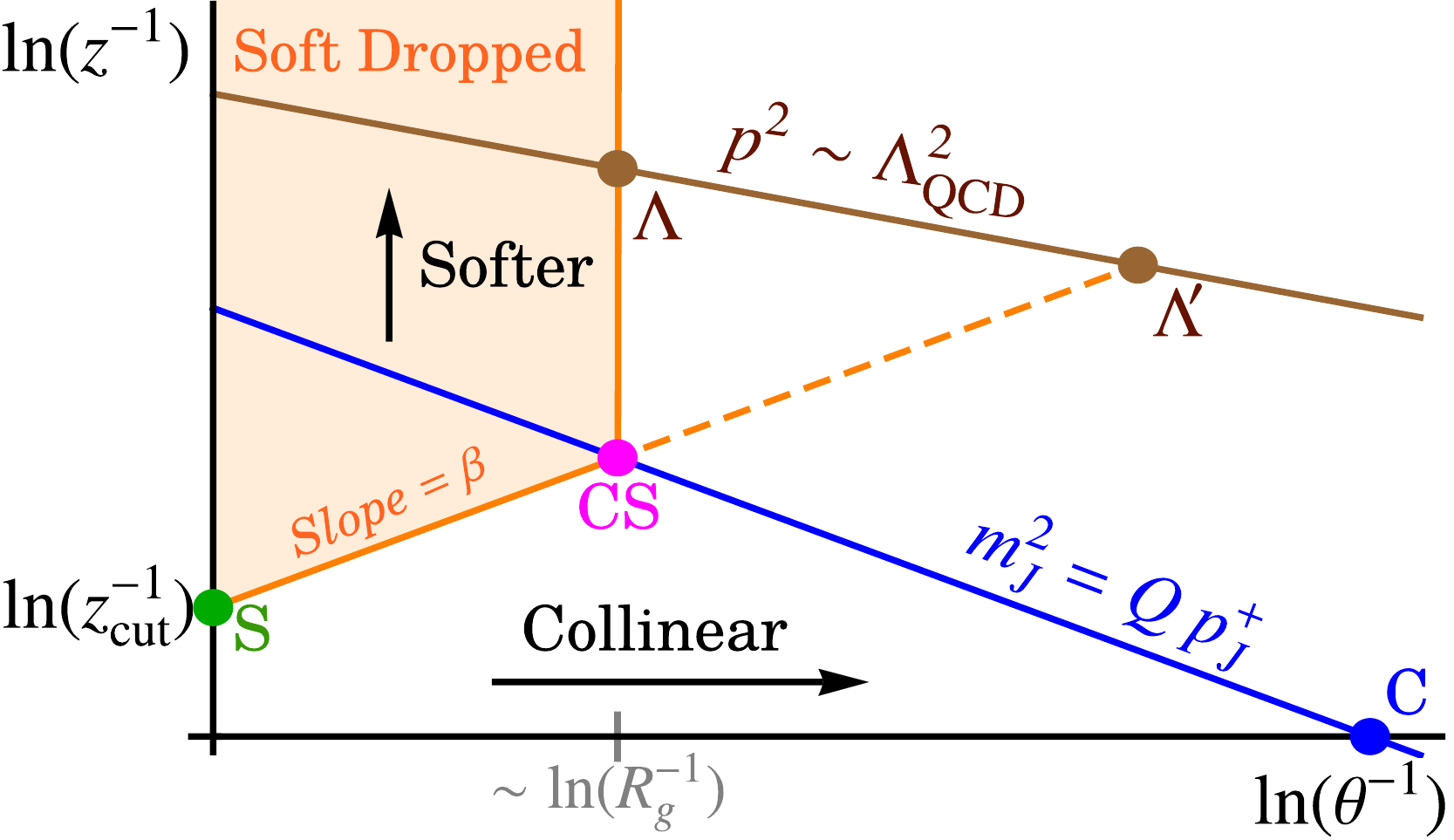}\\
\vspace{-0.2cm}
\caption{Comparison of the leading nonperturbative $\Lambda$ mode and a subleading $\Lambda^\prime$ mode in the SDOE region.}
\label{fig:modesLambdaprime}
\vspace{-0.2cm}
\end{figure}

We could also consider the case where there is no perturbative CS particle retained by soft drop in the SDOE region, so that all CS subjets are eliminated. In this case soft drop will be stopped by a nonperturbative mode whose momentum would sit at a location determined by extending the slanted orange line until it intersects the brown curve, indicated by the $\Lambda^\prime$ mode in \fig{modesLambdaprime}. However, in the SDOE region the probability for such an event is exponentially suppressed by the ratio of Sudakov exponentials that arise in the cases with or without a perturbative CS mode, describing the respective no radiation probability. This is because in the scenario without a CS mode there is a significantly larger region of phase space without an emission until the brown line is reached where a NP particle with $\Lambda^\prime$ scaling is always found that stops soft drop.

This suppression can be examined more explicitly by considering the expression for the LL cumulant of the soft drop jet mass distribution discussed below in \sec{partonic},
\begin{align} \label{eq:cumulant0}
\hat \Sigma(m_J^2)
= \exp\bigg[-{\cal R}_q\Big(\frac{4m_J^2}{R^2Q^2},RQ,\tilde z_{\rm cut}R^\beta,\beta\Big)\bigg] \, ,
\end{align}
where the radiator ${\cal R}_q$ is defined in \eq{radiator} and corresponds with an area in \fig{modesLambdaprime} where emissions did not occur. For a given jet mass $m_J^2$ in the SDOE region, the probability that there is no perturbative gluon emission, so that soft drop is stopped by a $\Lambda^\prime$ mode, can be obtained by comparing to the radiator evaluated at the smaller value $m_J^2 = Q\Lambda_{\rm QCD}\big({\Lambda_{\rm QCD}}/{\qcut}\big)^{\frac{1}{1+\beta}}$ as in \eq{NPregions} [noting that $\Lambda^\prime$ mode corresponds to the CS mode in this case]. This yields the probability for having no perturbative CS emission as
\begin{align}
\label{eq:probNoCS}
\textrm{Prob[no pert.\,CS]} \simeq \frac{\Sigma\bigg(Q \Lambda_{\rm QCD}\Big(\frac{\Lambda_{\rm QCD}}{\qcut}\Big)^{\frac{1}{1+\beta}} \bigg)}{\Sigma(m_J^2)} \, ,
\end{align}
which we see is exponentially suppressed for jet masses in the SDOE region satisfying \eq{NPregions}. Taking the example shown in \fig{logrhoplot} for $Q = 1000$ GeV, $\zcut = 0.1$, $\beta = 2$, setting $\Lambda_{\rm QCD} = 0.3$ GeV, and choosing $m_J^2$ to lie in the SDOE region with $\log_{10}(m_{J,\rm SDOE}^2/E_J^2) \sim -2$, we find the probability in \eq{probNoCS} is Prob[no pert.\,CS] $\simeq 12\%$, which is compatible with the size of other power corrections. As we will see later, this is also compatible with the uncertainty of our leading log treatment for the perturbative coefficients of the leading power corrections. The $\Lambda^\prime$ in \fig{modesLambdaprime} is thus a subleading mode for power corrections in the SDOE region.

In contrast, for $p_{cs}^+\sim p_\Lambda^+$ the CS and $\Lambda$ modes become parametrically close, merging into a single mode, which is labeled by $\Lambda$CS in \fig{modesmassless}b. Using \eqs{pCS}{pL} we find that this parametric relation corresponds to jet masses $m_J^2 \sim Q\Lambda_{\rm QCD} (\Lambda_{\rm QCD}/Q_{\rm cut})^{\frac{1}{1+\beta}}$. This scaling relation agrees with Ref.~\cite{Frye:2016aiz}. Jet masses with these or smaller values\footnote{ Note that there is another transition for even smaller jet masses, $m_J^2\sim \Lambda_{\rm QCD}^2$, which corresponds to the resonance region where the jet is reduced to having the invariant mass of a single hadron. In this region the brown line in \fig{modesmassless} runs through the location of the blue collinear modes, which themselves become nonperturbative. For our analysis we always consider bins of jet mass whose size is parametrically much larger than this. } correspond to the SDNP region, quoted above in \eq{NPregions}\footnote{Note that if we rewrote the parametric inequality for the SDOE region as $ Q\Lambda_{\rm QCD} (\Lambda_{\rm QCD}/Q_{\rm cut})^{\frac{1}{1+\beta}}\ll m_J^2 $ it would be satisfied for somewhat smaller $m_J$ values than the relation in \eq{NPregions}, due to the different meaning for the $\ll$ symbol. For example, replacing the $\ll$ by a factor of 3, we see that \eq{NPregions} leads to a larger lower bound for jet masses in the SDOE region $m_J^2 = 3^{\frac{2+\beta}{1+\beta}} \,Q\Lambda_{\rm QCD} (\Lambda_{\rm QCD}/Q_{\rm cut})^{\frac{1}{1+\beta}}$ than $m_J^2 = 3 \,Q\Lambda_{\rm QCD} (\Lambda_{\rm QCD}/Q_{\rm cut})^{\frac{1}{1+\beta}}$. This difference is particularly relevant for $\beta=0$. Thus we quoted \eq{NPregions} without making manipulations that take powers of both sides.}. Here it is a $\Lambda$CS mode in SCET that stops the soft drop groomer. In this region we have $\theta_{cs}/2 \sim \zeta_{np}$ and the scaling
\begin{align} \label{eq:LamCS}
p_{\Lambda{\rm CS}}^\mu \sim
\Lambda_{\rm QCD} \Bigl( \zeta_{np} ,\frac{1}{\zeta_{np}} , 1 \Bigr)
\,,\qquad\quad
\zeta_{np} = \, \Bigl(\frac{\Lambda_{\rm QCD}}{\qcut}\Bigr)^{\frac{1}{1+\beta}}
\sim \Bigl( \frac{m_J^2}{Q \qcut} \Bigr)^{\frac{1}{2+\beta}}
\,,
\end{align}
where $\zeta_{np}$ is still parametrically $\ll 1$. Since the $\Lambda$CS modes sit on the blue line in \fig{modesmassless}b, there are leading nonperturbative corrections to the jet mass spectrum in this SDNP region.

The above nonperturbative modes are the new ingredients needed for our analysis of the SDOE and SDNP regions.
These modes with $p^2 \sim \Lambda_{\rm QCD}^2$ will determine
the nonperturbative matrix elements (SDOE) or functions (SDNP) that contribute to the jet mass cross section.
Note that the need to consider the Sudakov exponentials for the analysis of power corrections in the SDOE region is novel, and differs from SCET analyses in other contexts, such as ungroomed event shapes. This implies that at least leading logarithmic (LL) resummation will need to be considered in order to properly incorporate the dominant hadronization corrections in the SDOE region. The manner in which the nonperturbative corrections appear can also depend on the order in resummed perturbation theory considered. For our analysis of power corrections we will make use of leading-logarithmic resummation, leaving results at higher order to future work.

Using the SCET based techniques of Refs.~\cite{Hoang:2007vb,Ligeti:2008ac,Abbate:2010xh,Mateu:2012nk}, which here involves factorizing perturbative and nonperturbative contributions in the collinear-soft region, we can derive a factorized form for these power corrections. This involves factorizing the measurement operator and matrix elements, as well as considering the role of CA clustering, which we consider in Secs.~\ref{sec:measurement}-\ref{sec:EFTmatching} in order to obtain the final result for the SDOE region in \sec{factorization}. The extension to the SDNP region is considered in \sec{NPregion}.

\section{Nonperturbative Corrections in the Operator Expansion Region}
\label{sec:EFTinOPEregion}

In this section we reconsider the soft drop jet mass factorization formula in order to include the leading nonperturbative effects related to final state hadronization in the SDOE region.
The expansions for the SDOE region are based on the ratios from comparing \eqs{pCS}{pL},
\begin{align}\label{eq:minusexpnGroomed}
\textrm{SDOE expansions:} \qquad &
\frac{p^-_\Lambda}{p^-_{cs}} \> \sim
\frac{p^\perp_\Lambda}{p^\perp_{cs}} \> \sim \> \frac{p^+_\Lambda}{p^+_{cs}} \> \sim \> \frac{Q \Lambda_{\rm QCD}}{m_J^2}\Bigl( \frac{m_J^2}{Q Q_{\rm cut}}\Bigr)^{\frac{1}{2+\beta}}
\ll 1
\, .
\end{align}

\subsection{Expansion of the Measurement Operator}
\label{sec:measurement}

The measurement operator for the groomed jet mass cross section involves CA clustering, subsequent grooming, and the measurement of the observable on the remaining set of particles as illustrated in \fig{grooming}. In this section we extend the analysis of the partonic measurement operator to include leading power corrections. Before we embark on the calculation we first set up some useful notation. Consider first the case of plain jet mass (without grooming). The differential cross section can be written as
\begin{align}
\frac{d \hat \sigma}{d m_J^2d\Phi_J} = \sum_{\kappa={q,g}}\sum_{X,X'} H^\kappa_{\cal IJ}\langle 0 | {\cal O^\kappa_J}\, \hat \delta | X X' \rangle \langle X X' | {\cal O^\kappa_I}^\dagger | 0 \rangle \, ,
\end{align}
where $H^\kappa_{\cal IJ}$ encodes the dependence on the hard production process and initial state, ${\cal O^\kappa}$ represents the SCET operator which includes fields for $\kappa =q$ or $g$ which initiate the jet, $X$ are the final state particles in the jet region of interest, $X'$ are other final state particles,
and ${\cal I, J}$ are shorthand for any color and spin indices.
Since soft drop decouples the jet mass distribution from the behavior of particles outside the jet region, we can focus on $X$ for our analysis.
The measurement operator in the soft or collinear-soft limit can be written as
\begin{align}
\hat \delta= \delta \bigl (m_J^2 -Q \, \hat p^+\bigr) \, ,
\end{align}
where the operator $\hat p^+$ measures the $p^+$ momentum of final state radiation. The action of $\hat p^+$ on a $n$-particle final state is simply the sum of all the individual $+$-momentum components:
\begin{align}
\hat p^+ | \{i_1, \, i_2, \ldots i_n \} \rangle = (p_{i_1}^+ + p_{i_2}^+ + \ldots + p_{i_n}^+) | \{i_1, \, i_2, \ldots i_n \} \rangle \, .
\end{align}

In the case of the groomed jet mass the situation is more complicated since the particles cannot be simply selected from a simple geometrical region due to CA clustering and the soft drop test. Hence, the $\hat p^\mu$ operator above is not sufficient to define the groomed jet mass measurement operator consistently. To ameliorate the problem, we define a ``soft drop momentum operator'', $\hat p_{\rm sd}^\mu(X; \tilde z_{\rm cut}, \beta)$, that takes into account the CA clustering and grooming for a given multiparticle reference state $|X\rangle$. Its action on a single particle state $| i\rangle $ that may or may not be in $|X\rangle$ is defined as follows
\begin{align}
\hat p_{\rm sd}^\mu(X; \tilde z_{\rm cut}, \beta) | i\rangle&= \bigl [ \,\hat p^\mu\, \overline \Theta_{\rm sd} \bigl(\hat p^\mu, \{p_{j}^\mu ; \, j\in X\}; \tilde z_{\rm cut}, \beta\bigr) \bigr]\, | i\rangle \\
&\equiv\, p_i^\mu\, \overline \Theta_{\rm sd} \bigl( p_i^\mu, \{p_{j}^\mu; \, j\in X \}; \tilde z_{\rm cut}, \beta\bigr) | i\rangle \nn \, .
\end{align}
The operator $\overline \Theta_{\rm sd} \bigl( p_i^\mu, \{p_{j}^\mu ;\, j\in X\}; \tilde z_{\rm cut}, \beta\bigr)$ is defined to be 1 if the particle $|i\rangle$ is not groomed away and 0 otherwise. For simplicity we suppress the dependence on the soft drop parameters $\tilde z_{\rm cut}$ and $\beta$ in the argument of $\overline \Theta_{\rm sd}$ below. The usefulness of $\hat p_{\rm sd}^\mu(X)$ becomes apparent when its action on a multiparticle state is considered:
\begin{align}
\label{eq:hatpsdmulti}
\hat p_{\rm sd}^\mu(X) | \{i_1, \, i_2, \ldots i_n \} \rangle = \bigg( \sum_{\alpha = 1}^n \overline \Theta_{\rm sd} \bigl (p_{i_\alpha}^\mu, \{p_j^\mu;\, j\in X\} \bigr)\, p_{i_\alpha}^\mu \bigg) | \{i_1, \, i_2, \ldots i_n \} \rangle \, .
\end{align}
If the particles $ \{i_1, \, i_2, \ldots i_n \} $ are contained in $X$ then each particle $|i_\alpha\rangle$ is individually tested for passing soft drop amongst the particles in $|X \rangle $.
Thus $\hat p_{\rm sd}^\mu(X)$ yields a measurement of the groomed jet momentum. In this notation, the measurement operator for the groomed jet mass for the final state $|X \rangle $ of collinear-soft particles is simply
\begin{align}
\label{eq:measurement0}
\hat \delta_{\rm sd} = \delta \Bigl( (m_J^2)_{cs} - Q\, \hat p_{\rm sd}^+(X) \Bigr)\, .
\end{align}
The action of $\hat \delta_{\rm sd} $ on $| X\rangle $ will precisely collect the $p^+$ momenta of only those particles that remain after grooming.

Having established some notation, we are now prepared to consider leading hadronization corrections in the SDOE region. Upon including nonperturbative (NP) particles, represented by a multiparticle state $|X_\Lambda\rangle $, together with perturbative particles $|X\rangle$, \eq{measurement0} now reads
\begin{align}
\label{eq:measurement1}
\hat \delta_{\rm sd} = \delta \Bigl( (m_J^2)_{cs} - Q\, \hat p_{\rm sd}^+(X, X_{\Lambda}) \Bigr)\, .
\end{align}
The NP contribution to the jet mass is then given by
\begin{align}
\label{eq:measurement2}
Q\, \hat p_{\rm sd}^+(X, X_{\Lambda}) | X_\Lambda \rangle = Q \, p_{\Lambda {\rm sd} }^+ | X_\Lambda \rangle \, ,
\end{align}
where $p_{\Lambda {\rm sd} }^+$ is the sum of all $p^+$ momenta of the nonperturbative particles kept after grooming.
Here we have made use of the fact that the combined state $ |X X_\Lambda \rangle$ factorizes into $|X\rangle |X_\Lambda\rangle$ in the SDOE region. This follows because the $\Lambda$ and CS modes have the same boost but hierarchically different momentum components, and hence factorize in their respective Lagrangians. Furthermore, the momentum operator now uses the full hadronic state as its reference state:
\begin{align}
\label{eq:measurement3}
\hat p_{\rm sd}^\mu(X, X_\Lambda) &=\bigl [\hat p^\mu \, \overline \Theta_{\rm sd} \bigl(\hat p^\mu, \{p_j^\mu;\,j\in X \cup X_\Lambda \}\bigr)\bigr]\, .
\end{align}
We identify two types of hadronization corrections:
\begin{enumerate}
\item A ``shift'' correction: contribution to the observable from the NP radiation kept in the groomed jet, given by $Q \, p_{\Lambda {\rm sd} }^+ $ in \eq{measurement2},
\item A ``boundary'' correction: modification of the soft drop test for a perturbative subjet in presence of NP radiation, as seen from the $X_\Lambda$ dependence of $\overline \Theta_{\rm sd}$ in \eq{measurement3}.
\end{enumerate}
We will show below that both of these power corrections modify the shape of the spectrum and cannot simply be included via a shape function. In general both of these corrections are tied to the clustering history of other perturbative particles in the jet, hence complicating the nonperturbative factorization. However, there are some key simplifications one can make in the SDOE region up to LL accuracy, which we address in the following.

\begin{figure}[t!]
\centering
\includegraphics[width=0.98\columnwidth]{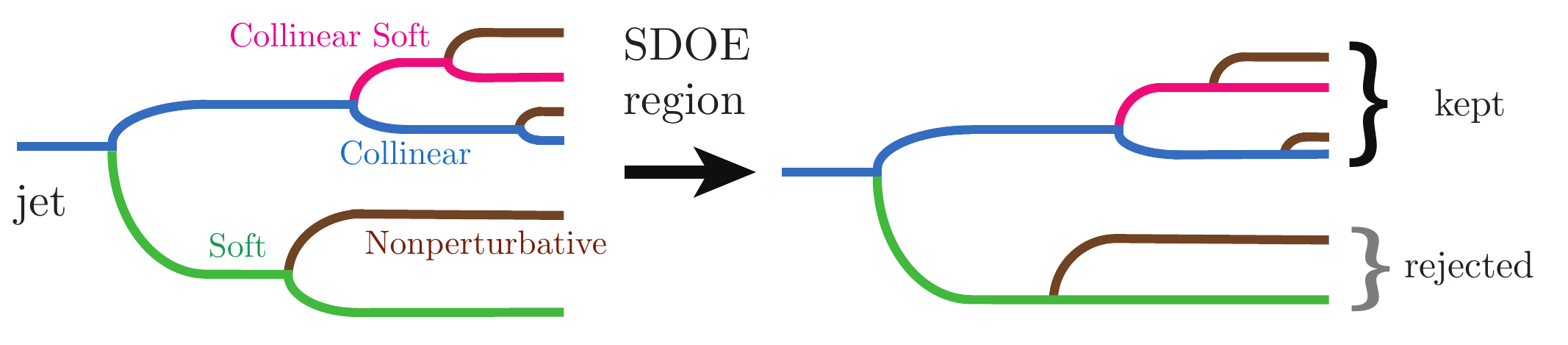}
\caption{Schematic of a CA clustered tree and its simplification in the SDOE region. The vertical separation between the branches corresponds to the angular distances and the CA clustering proceeds from right to left. The colors correspond to the modes displayed in \fig{modesmassless}. The angular locations of perturbative subjets in the SDOE region at leading power remain unperturbed as nonperturbative particles are added.}
\label{fig:grooming-sdoe}
\end{figure}
To help visualize the problem we show in \fig{grooming-sdoe} the same schematic as \fig{grooming}, but with the momentum scaling of the branches made explicit. The labels and the colors correspond to the EFT modes shown in \fig{modesmassless}. The perturbative branches are effectively immersed in a bath of nonperturbative particles distributed at all angles, corresponding to the brown line in \fig{modesmassless}. The scaling of the combined subjet depends on the dominant mode of the pair, with the collinear modes having the highest energy. Hence, the collinear subjet undergoes the smallest relative change in the subjet momentum during clustering.
The soft drop grooming will be stopped by a comparison involving branches that have collinear and CS scaling.
As discussed above in \sec{NPmodes}, in the SDOE region there is always a perturbative CS subjet that stops soft drop.

In general, the angular location of a subjet changes at each stage of clustering as the subjets are combined, as shown in the left figure in \fig{grooming-sdoe}. At leading power in the SDOE region the shift to the momentum of perturbative subjets on adding NP particles is small and can be ignored. At the first subleading power where the hadronization effects enter, the NP particles that determine the shift term are the ones that belong to the same leading power collinear or CS subjets. In calculating the shift term we thus ignore the effects of NP particles on the clustering and soft drop comparison of perturbative subjets, as shown in the right schematic in \fig{grooming-sdoe}, and hence can use $\hat p^+_{\rm sd}(X,X_\Lambda) \simeq \hat p^+_{\rm sd}(X)$ for the shift term. In contrast, the boundary term separately captures the effect of the leading NP modification to the subjet geometric regions, which modifies the amount of perturbative $p^+$ momentum kept in the groomed jet.

\subsubsection{Expansion for the Shift Term}
\begin{figure}[t!]
\centering
\includegraphics[width=0.5\textwidth]{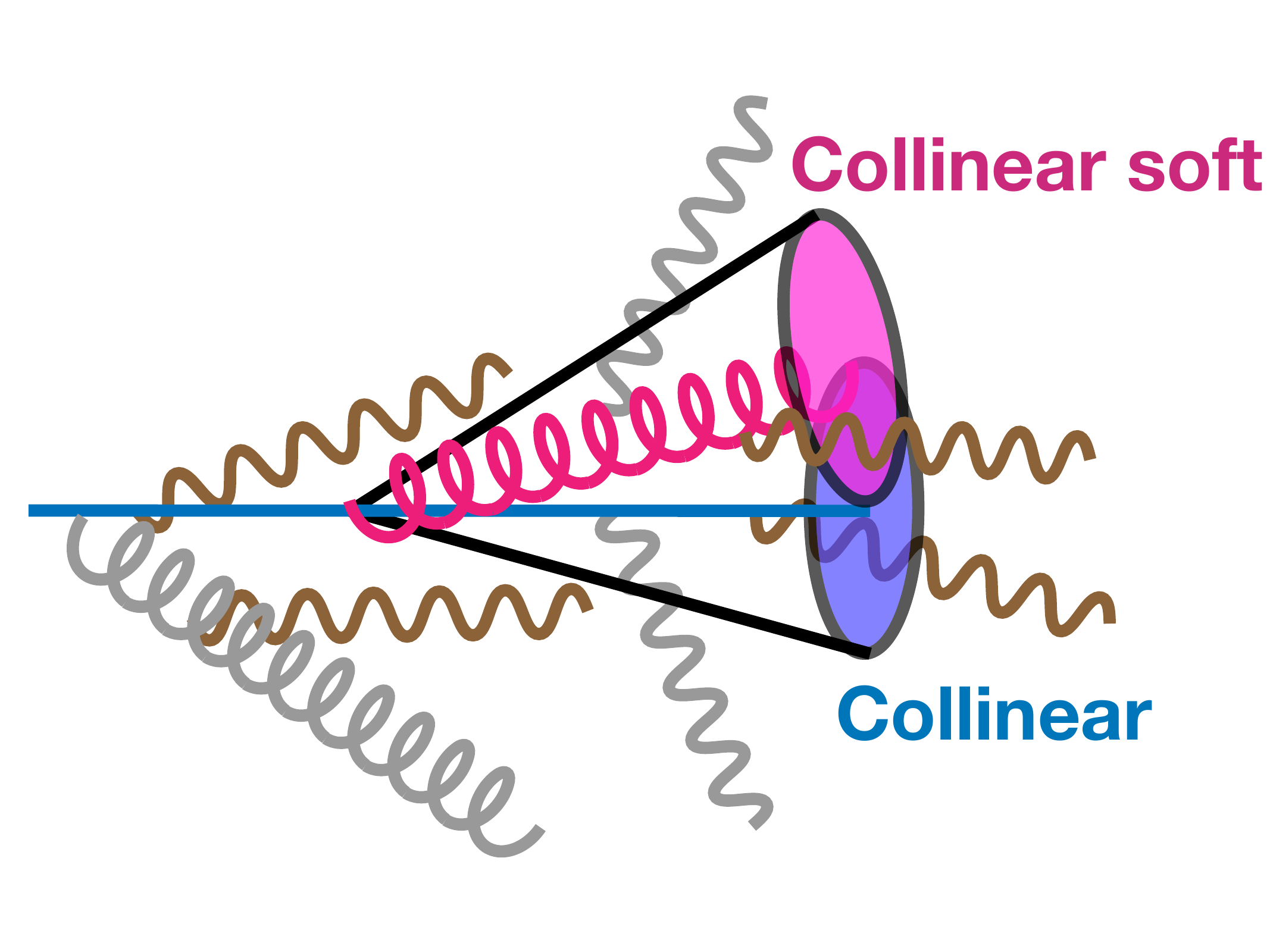}\\
\caption{The catchment area (blue and pink shaded regions) of nonperturbative particles (wavy lines) set by the perturbative collinear and collinear-soft subjets with angular ordering of perturbative subjets. The particles shown in gray are groomed away. NP particles are not assumed to obey angular ordering.
\label{fig:catchmentarea}}
\vspace{-0.2cm}
\end{figure}
At LL one can make a further approximation of assuming strong angular ordering of the perturbative emissions.
This dramatically simplifies the complexity of the CA clustering in the measurement operator. Strong angular ordering implies that all the perturbative emissions subsequent to the one that stops soft drop lie at much smaller angles. We illustrate in \fig{catchmentarea} the region of momentum space that forms the catchment area of the kept NP particles at LL (blue and pink shaded regions). The perturbative emissions that occur after the emission that stops soft drop will also lie within this region. Each cone is centered on one of the subjets that stops soft drop, and the conic sections correspond to the regions where the nonperturbative radiation is collected by each of these subjets. To extend this formalism to NLL would require considering modifications of the catchment area in \fig{catchmentarea} due to inner resolved perturbative subjets (which are not necessarily strongly ordered), which we leave to future work.

\begin{figure}[t!]
\centering
\includegraphics[width=0.9\textwidth]{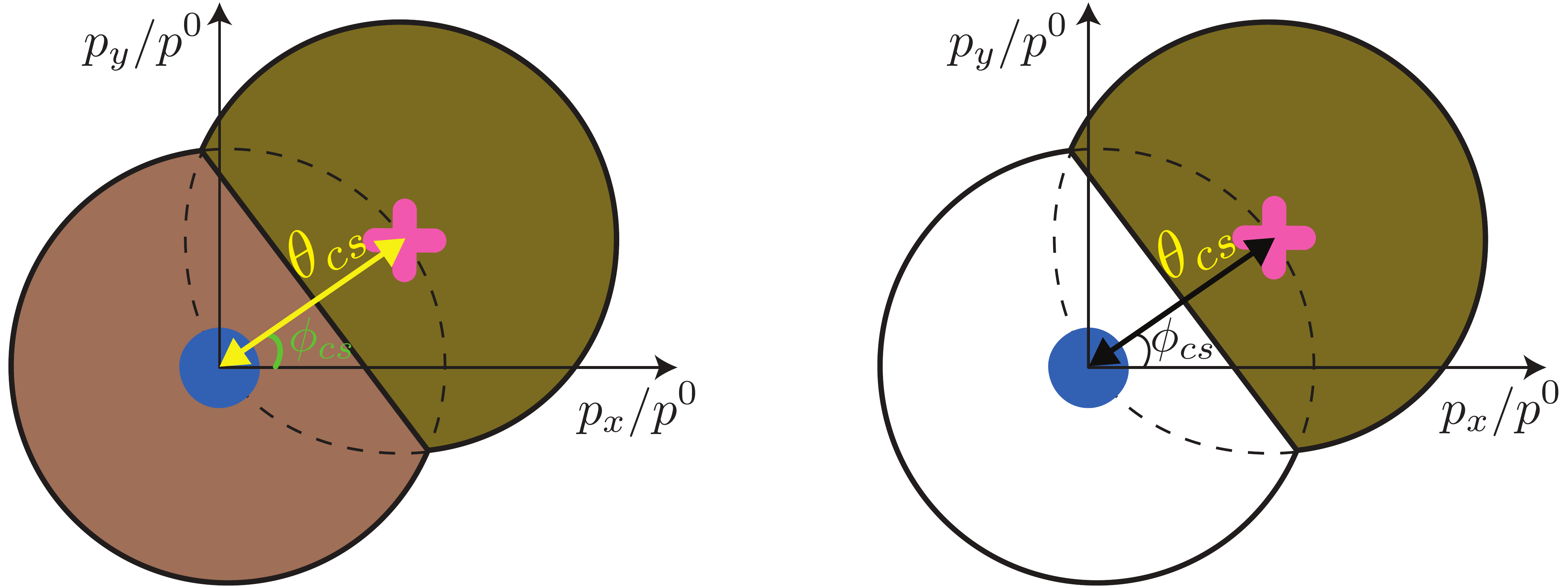}\\
[-20pt]
\phantom{x} \hspace{-4.8cm}a) \hspace{7.8cm} b) \\[-5pt]
\caption{The catchment area of nonperturbative modes relevant for a) the shift term and b) the boundary term up to LL, pictured from above looking down the jet axis (taken along the $z$-direction). These modes are clustered with either the collinear subjet located on the jet axis (blue dot), or the stopping collinear-soft subjet (pink cross) as indicated by the shaded brown regions. The overlapping circles both have radius $\theta_{cs}$.
\label{fig:NPfig8}}
\vspace{-0.2cm}
\end{figure}

Looking down the jet axis, in the small angle approximation, the size and the alignment of the region is determined by $\{\theta_{cs},\phi_{cs}\}$, the polar and azimuthal location of the CS subjet measured relative to the jet axis, as shown in \fig{NPfig8}a. Since the collinear subjet carries the majority of the jet energy, we assume that the jet axis is aligned with that of the collinear subjet. As a result the contribution of the NP particles to the observable via the shift term will only come from the catchment area of the collinear and collinear-soft subjet. A similar geometry but for a different application of pile-up and underlying event subtraction in the case of CA clustering was also explored in \Refcite{Cacciari:2008gn}.

Thus, at leading log we can make \eq{measurement2} manifest with a simpler geometrical constraint for the shift term in the SDOE region:
\begin{align} \label{eq:shiftapprox}
Q \, \hat p_{\rm sd}^+(X)\, | X_\Lambda\rangle \stackrel{\rm LL}{\simeq} Q \, \hat p^+_{\figeight} (\theta_{cs}, \phi_{cs})\, | X_\Lambda \rangle =Q\, p_{\Lambda{\rm sd}}^+\,| X_\Lambda\rangle \, ,
\end{align}
where $Q\, p_{\Lambda{\rm sd}}^+$ is the contribution of the nonperturbative particles to the groomed jet mass.
Note the use here of state $|X\rangle$ in the operator $\hat p_{\rm sd}^+(X)$ as opposed to $\hat p_{\rm sd}^+(X,X_\Lambda)$ in \eq{shiftapprox}, since the difference is a subleading power correction as explained above. The operator $\hat p^+_{\figeight} (\theta_{cs}, \phi_{cs})$ gives the $p^+$ momentum of all the particles clustered with the collinear or CS subjet, and is defined as:
\begin{align}
\label{eq:hatpfig8}
\hat p_{\figeight}^\mu (\theta_{cs}, \phi_{cs}) \equiv \bigl [ \, \hat p^\mu \,\overline \Theta_{\rm NP}^{\, \figeight}(\hat p^\mu, \, \theta_{cs}, \phi_{cs}) \,\bigr] \, ,
\end{align}
where the operator $ \overline \Theta^{\,\figeight}_{\rm NP}$ is defined to be 1 when a NP subjet in $X_\Lambda$ is clustered with either the collinear or CS subjets, as given by the shaded region in the $p_x$-$p_y$ plane shown in \fig{NPfig8}a. The operator $\hat p^\mu_{\figeight}$ acts on a nonperturbative multiparticle state the same way as $\hat p_{\rm sd}^\mu$ does in \eq{hatpsdmulti}. The condition $\theta_{cs}\ll 1$ in the SDOE region implies that the two circles simply have radius $\theta_{cs}$, yielding a compact expression for $ \overline \Theta^{\, \figeight}_{\rm NP}$:
\begin{align}
\label{eq:Thetafig8}
\overline \Theta_{\rm NP}^{\, \figeight}( p_{\Lambda }^\mu, \, \theta_{cs}, \phi_{cs}) &= \Theta\bigg(| \Delta \phi| - \frac{\pi}{3} \bigg)\Theta\bigg(1 -\frac{ \theta_{\Lambda}}{\theta_{cs}} \bigg)+ \Theta\bigg(\frac{\pi}{3} - | \Delta \phi| \bigg)\Theta \bigg(2 \cos (\Delta \phi) - \frac{\theta_{\Lambda}}{\theta_{cs}}\bigg) \, \nn \\
&\equiv\overline \Theta_{\rm NP}^{\,\figeight}(\theta_{\Lambda}, \, \theta_{cs},\, \Delta \phi)
\,,
\end{align}
where $\Delta \phi = \phi_{\Lambda} - \phi_{cs}$ is the relative azimuthal angle in the plane perpendicular to the jet axis, and $\theta_{\Lambda}$ the polar angle relative to the jet axis. In the second line, we identified the only arguments that the operator depends on, namely the angular locations of the CS and the NP subjets in momentum space.

\subsubsection{Expansion for the Boundary Term}
\label{sec:BndrytermExpn}

We now turn to the boundary term. We first rewrite \eq{measurement3} as
\begin{align}
\hat p_{\rm sd}^+(X, X_{\Lambda}) = \hat p_{\rm sd}^+(X) + \Delta \hat p^+_{\rm sd}(X, X_{\Lambda}) \, ,
\end{align}
with the boundary power correction being given by
\begin{align}
\label{eq:measurement4}
\Delta \hat p_{\rm sd}^+(X, X_{\Lambda}) &=\bigl [ \hat p^+\, \Delta \overline \Theta_{\rm sd}(X, X_\Lambda)\, \bigr] \\
&\equiv\bigg [ \, \hat p^+ \, \Big (\overline \Theta_{\rm sd} \bigl(\hat p^\mu, \{p_{j}^\mu;\,j\in X \cup X_\Lambda \}\big) - \overline \Theta_{\rm sd} \bigl(\hat p^\mu, \{p_{j}^\mu;\, j\in X \}\bigr)\Big )\bigg] \, . \nn
\end{align}
Unlike the shift term in \eq{shiftapprox}, that only contributed through the catchment area defined by the final collinear and CS subjets, the effect of the boundary term is to modify the soft drop condition by $\Delta \overline \Theta_{\rm sd}(X, X_\Lambda)$ for every step of comparison.
In the collinear-soft limit the soft drop test for a soft subjet with perturbative momentum $p^\mu$, accounting for the additional momentum $q^\mu$ from hadronization effects, reads
\begin{align}
\label{eq:sdthetapq}
\overline \Theta_{\rm sd}^{\, p+q} = \Theta \bigg ( \frac{p^- + q^-}{Q} - \tilde z_{\rm cut} \, \Big(\frac{2 \,|\vec p_{\perp} + \vec q_{\perp}| }{p^- + q^-} \Big)^\beta \bigg) \, .
\end{align}
We can expand the expression above in the limit where all the components of $q^\mu$ are parametrically smaller than those of $p^\mu$, while the angles are of the same order $\theta_p \sim \theta_{q}\,$:
\begin{align}
\label{eq:DeltaSD}
\overline \Theta_{\rm sd}^{\, p+q} - \overline \Theta_{\rm sd}^{\, p} = \delta \big ( z_{p} - \tilde z_{\rm cut}\, \theta_{p}^\beta \big)\, \frac{ q^- }{Q}\Bigg[ 1+ \beta \, \bigg (1 - \, \frac{\theta_{q}}{\theta_{p}} \, \cos(\Delta \phi) \bigg) \Bigg] + \ldots \, ,
\end{align}
with $\overline \Theta_{\rm sd}^{\, p}$ denoting the soft drop condition applied on $p$ alone. Here $z_{p} = p^-/Q$, $\theta_p = 2 | \vec p_{\perp}|/p^-$, $\theta_{q} = 2 | \vec q_{\perp}|/q^-$, and $\Delta \phi = \phi_{q} - \phi_{p}$.
The boundary correction is also influenced by momentum that is removed from the subjet due to hadronization, in which case the correction is
\begin{align}
\label{eq:DeltaSD2}
\overline \Theta_{\rm sd}^{\, p-q} - \overline \Theta_{\rm sd}^{\, p} = -\delta \big ( z_{p} - \tilde z_{\rm cut}\, \theta_{p}^\beta \big)\, \frac{ q^- }{Q}\Bigg[ 1+ \beta \, \bigg (1 - \, \frac{\theta_{q}}{\theta_{p}} \, \cos(\Delta \phi) \bigg) \Bigg] + \ldots \, ,
\end{align}
which is just the negative of that in \eq{DeltaSD}. \eqs{DeltaSD}{DeltaSD2} describe scenarios where hadronization causes NP momentum to enter and leave the soft subjet respectively. For the analysis of boundary correction we can ignore the power suppressed nonperturbative corrections to the collinear subjet momentum.

The correction in \eq{measurement4} can affect the collinear-soft function $S^\kappa_c$ as well as the normalization factor $N_\kappa$ in \eq{masslessFact0} that accounts for the global soft modes. However, the CA clustering and the scaling of the NP modes in SCET implies that only the boundary correction to the collinear-soft modes need to be considered for the leading order NP power corrections. In order for a NP mode to modify the soft drop condition for a perturbative global soft mode at LL (or NLL), it must sit at the same parametric angle (due to CA clustering), which yields $p_{\Lambda(us)}^\mu\sim \Lambda_{\rm QCD}(1,1,1)$. These corrections in the global soft region consist entirely of subjets that fail to pass the soft drop condition. However,
$p_\Lambda^\mu/p_{cs}^\mu \gg p_{\Lambda(us)}^\nu/p_{gs}^\nu$ for all components in the light cone basis, so the power corrections to global soft are always further suppressed, as can be seen from the momentum scalings in Eqs.~(\ref{eq:pGS}), (\ref{eq:pCS}) and (\ref{eq:pL}).
There are also additional modifications in the collinear-soft region from subjets that fail soft drop, which first become non-trivial at ${\cal O}(\alpha_s^2)$. These subjets do contribute to the leading power NP corrections, but are beyond LL accuracy, as we discuss in more detail in \app{scfail}. Hence for the remainder of this section we will only focus on the correction to the soft drop test for the collinear-soft subjet that stops soft drop.

The geometric region at LL that corresponds to the catchment area of the CS subjet is shown in \fig{NPfig8}b. The projection operator that selects the NP emissions in the CS subjet is given by
\begin{align}
\label{eq:Thetabndry}
\overline \Theta_{\rm NP}^{\, \bndry} (p^\mu_{\Lambda}, \, \theta_{cs},\, \phi_{cs}) &= \Theta\bigg(\frac{\pi}{3} - | \Delta \phi | \bigg)\Theta \bigg(\frac{\theta_{\Lambda}}{\theta_{cs}} - \frac{1}{2 \cos (\Delta \phi)}\bigg)\Theta \bigg(2 \cos (\Delta \phi) - \frac{\theta_{\Lambda}}{\theta_{cs}}\bigg) \, , \nn \\
&\equiv \overline \Theta_{\rm NP}^{\, \bndry} (\theta_{\Lambda}, \, \theta_{cs},\,\Delta \phi)
\,,
\end{align}
with $\Delta \phi = \phi_{\Lambda} - \phi_{cs} \, $. This is shown as the brown shaded region in \fig{NPfig8}b that represents the region where the NP particles are clustered with the collinear-soft subjet (and not with the collinear subjet). We also define $\Theta_{\rm NP}^{\, \bndry} \equiv 1 - \overline \Theta_{\rm NP}^{\, \bndry}$, which describes the complimentary region.

For the collinear-soft subjet that stops soft drop, the results in \eqs{DeltaSD}{DeltaSD2} can be combined to give the leading result for the
eigenvalue of the operator $\Delta \overline \Theta_{\rm sd}(X, X_\Lambda)$, as
\begin{align}
\label{eq:measurement6}
\Delta \overline \Theta^{\, cs}_{\rm sd} = \delta \big ( z_{cs} - \tilde z_{\rm cut}\theta_{cs}^\beta \big)\,\frac{ q^{ \bndry}_{cs}(\theta_{cs}, \phi_{cs},\beta)}{Q} \,,
\end{align}
where $q^{\bndry}_{cs}(\theta_{cs}, \phi_{cs},\beta)$ is the non-perturbative momentum entering into or leaving from the collinear-soft subjet (the shaded region in \fig{NPfig8}b) that is at the angles $\theta_{cs}$ and $\phi_{cs}$. It can be defined by the eigenvalue equation $\hat{p}_{\bndry}(\theta_{cs}, \phi_{cs},\beta)\, | X_\Lambda \rangle = q^{\bndry}_{cs}(\theta_{cs}, \phi_{cs},\beta) \, | X_\Lambda \rangle$, where
the relevant operator for the boundary term is given by
\begin{align}
\label{eq:hatpbndry}
\hat{ p} _{\bndry}(\theta_{cs}, \phi_{cs},\beta) &\equiv \bigg[ \, \hat{ \tilde p}(\theta_{cs}, \phi_{cs},\beta)\,
\Big( \overline \Theta_{\rm NP}^{\, \bndry} (\hat p^\mu, \theta_{cs}, \phi_{cs})
- \Theta_{\rm NP}^{\, \bndry} (\hat p^\mu, \theta_{cs}, \phi_{cs}) \Big)
\,\bigg] \, ,
\nn\\
\text{with}\quad \hat{ \tilde p}(\theta_{cs},& \phi_{cs},\beta) \equiv \hat p^- + \beta \, \Big( \hat p^-- \frac{2\hat p_\perp}{\theta_{cs}} \, \cos(\hat \phi - \phi_{cs})\Big) \,.
\end{align}
Here the $\overline \Theta_{\rm NP}^{\, \bndry}$ term yields the corrections corresponding with \eq{DeltaSD}, while the $\Theta_{\rm NP}^{\, \bndry}$ term yields those for \eq{DeltaSD2}.
The results from non-perturbative matrix elements in the operator expansion enable us to encode effects where a simple replacement of partonic momenta by hadronic momenta variables does not suffice.
In \app{bndry} we present further justification for \eq{hatpbndry} via a simple illustrative example.

Note that $\hat{ \tilde p}(\theta_{cs}, \phi_{cs},\beta)$ in \eq{hatpbndry} is linear in $\beta$. This will lead to the linear dependence of the boundary power correction on $\beta$ that was noted above in \eq{upslinearity}.
For $\beta = 0$, the boundary power corrections are particularly simple, where from \eqs{DeltaSD}{DeltaSD2} we can see that they entirely result from an expansion in the minus momentum of the nonperturbative and collinear-soft modes. This has no analogue in the case of the plain (i.e. ungroomed) jet mass since that measurement solely involves the plus component. The groomed jet mass requires a minimum $p^- = Q \zcut $ for the collinear-soft mode to pass, which is susceptible to power corrections that are of the same order as the shift corrections to the jet mass measurement. For $\beta > 0$ the boundary power correction has additional angular dependence at the same order, as seen from \eq{sdthetapq} or \eq{hatpbndry}.
We also note from examining \eqs{minusexpnGroomed}{hatpbndry} that we cannot take $\beta$ too large if we want the SDOE expansion in \eqs{DeltaSD}{DeltaSD2} to remain valid. This implies a constraint on the $\beta$ values:
\begin{align}
\label{eq:betaconstr}
\beta\ \frac{q^-}{Q} \sim\beta\ \frac{Q \Lambda_{\rm QCD}}{m_J^2}\Bigl( \frac{m_J^2}{Q Q_{\rm cut}}\Bigr)^{\frac{1}{2+\beta}} \ll 1 \,,
\end{align}
so that the limit $\beta\to \infty$ is not compatible with the expansions in the SDOE region.
In the numerical analysis in \secs{coherentbranching}{montecarlo} (also in \fig{logrhoplot} above), we replace the inequality defining the SDOE region in \eq{NPregions} by 1/5.
Hence, to avoid violating \eq{betaconstr} we will impose $\beta \le 2$ for our numerical analysis.

\subsubsection{Rescaling}

Summarizing the results from the previous sections, the two nonperturbative power corrections to the jet mass measurement in the SDOE regions from a set of NP particles $\{q^\mu_i\}$ using the LL approximation can be expressed as
\begin{align}
\label{eq:sdoefull}
(m_J^2)_{cs} &\stackrel{\rm SDOE}{\simeq} Q \, p_{cs}^+ + Q \sum_i q^+_{i\figeight}+Q \,p_{cs}^+\, \delta \big ( z_{cs} - \tilde z_{\rm cut}\theta_{cs}^\beta \big) \sum_i
\frac{q_{i\bndry}}{Q}
\,,
\end{align}
where
\begin{align}
\label{eq:sdoefull2}
q^+_{i\figeight} |q_i^\mu \rangle &\equiv \hat p_{\figeight}^\mu (\theta_{cs}, \phi_{cs})|q_i^\mu \rangle =
q_i^+\, \overline \Theta_{\rm NP}^{\, \figeight}( \theta_{q_i}, \, \theta_{cs}, \Delta \phi_i) |q_i^\mu \rangle \,,\\
q_{i\bndry} |q_i^\mu \rangle&\equiv\nn\hat p_{\bndry} (\theta_{cs}, \phi_{cs}, \beta)|q_i^\mu \rangle \nn\\
&= \bigg [q_i^- + \beta \, \Big( q_i^-- \frac{2q_{i\perp}}{\theta_{cs}} \, \cos(\Delta \phi_i)\Big)\bigg]
\Big( \overline \Theta_{\rm NP}^{\, \bndry} (\theta_{q_i}, \theta_{cs}, \Delta \phi_i) - \Theta_{\rm NP}^{\, \bndry} (\theta_{q_i}, \theta_{cs}, \Delta \phi_i)\Big)
|q_i^\mu \rangle
\nn
\end{align}
Here, $(m_J^2)_{cs}$ refers to the contribution to the jet mass from the collinear-soft sector and the NP particles $\{q_i\}$, and $\Delta \phi_i = \phi_{q_i} - \phi_{cs}$. The first term is the leading contribution to the measurement on the perturbative collinear-soft mode and the next two terms result from the shift and boundary corrections respectively. The measurement operators $\hat p_{\figeight}^\mu (\theta_{cs}, \phi_{cs}) $ and $\hat p_{\bndry} (\theta_{cs}, \phi_{cs}, \beta)$ in \eq{sdoefull2} are a simplification over the full soft drop operator $\hat p_{\rm sd}^\mu(X, X_\Lambda)$ since the same constraint now applies to all the NP subjets without involving additional nontrivial modifications due to CA clustering. The catchment area only depends on the kinematics of the perturbative radiation, that, however, varies at each point in the jet mass spectrum - a novel feature for the groomed jet mass. This is related to the observation in \sec{NPmodes} that the $\Lambda$ mode has the same parametric angle as the CS mode. Hence, the dominant NP radiation lies on and inside the boundary of the catchment area defined by the overlapping cones in \fig{NPfig8} where $\theta_{\Lambda} \sim \theta_{cs}$, and moves with the location of CS mode along the spectrum. The next step towards deriving nonperturbative factorization is then to factorize this perturbative dependence of the power corrections from the purely nonperturbative contribution to the observable.

\begin{figure}[t!]
\centering
\includegraphics[width=0.9\textwidth]{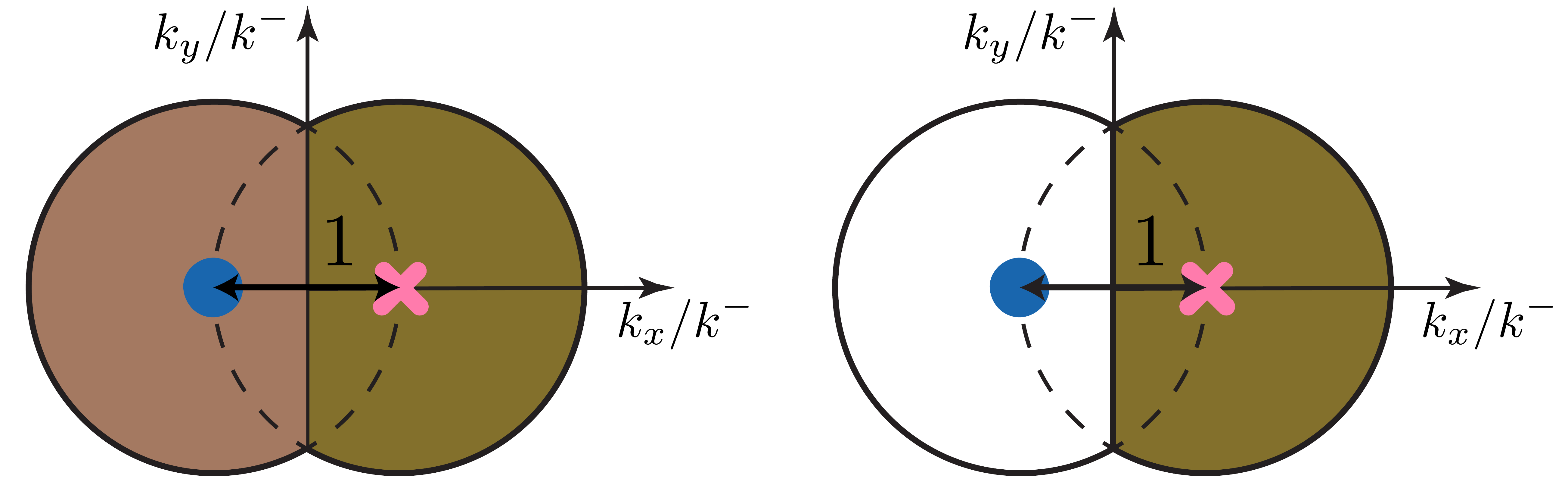}\\
[-20pt]
\phantom{x} \hspace{-4.8cm}a) \hspace{7.8cm} b) \\[-5pt]
\caption{The catchment area of nonperturbative modes kept relevant for a) shift and b) boundary terms at LL, in the respective boosted frame, as pictured from above looking down the jet axis (taken along the $z$-direction). These modes are clustered with either the collinear subjet located on the jet axis (blue dot), or the stopping collinear-soft subjet (pink cross) as indicated by the shaded brown regions. The overlapping circles both have radius 1 for the given choice of axes.
\label{fig:NPfig8rescaled}}
\vspace{-0.2cm}
\end{figure}

We observe from \eqs{Thetafig8}{Thetabndry} that only the ratio of polar angles $\theta_\Lambda/\theta_{cs}$ and relative azimuthal angles $\phi_\Lambda - \phi_{cs}$ appear in the projection operators. We thus make the following change of variables from the momenta $q_i \sim p_\Lambda\sim \Lambda_{\rm QCD}\Bigl(\frac{\theta_{cs}}{2},\frac{2}{\theta_{cs}},1\Bigr)$ in \eq{sdoefull2} to momenta $k^\mu_i$ defined by:
\begin{align}
\label{eq:rescaling}
q_i^+ = \frac{\theta_{cs}}{2} \, k_i^+ = \sqrt{\frac{p_{cs}^+}{p_{cs}^-}}\, k_i^+ , \qquad q_i^- = \frac{2}{\theta_{cs}} \, k_i^- = \sqrt{\frac{p_{cs}^-}{p_{cs}^+}}\, k_i^-\, , \qquad q_{i\perp} = k_{i\perp} \, , \qquad \phi_{q_i} = \phi_{k_i} + \phi_{cs} \, .
\end{align}
This implies
\begin{align}
\theta_{q_i} = \frac{2 \, q_{i\perp}}{q_i^-} = \theta_{cs} \frac{k_{i\perp}}{k_i^-} \, , \qquad \Delta \phi_i= \phi_{k_i} \,, \qquad k_i^\mu \sim \Lambda_{\rm QCD}\big(1,1,1\big) .
\end{align}
In terms of rescaled momentum $k_i^\mu$ in \eq{sdoefull2} the projection operators defined in \eqs{Thetafig8}{Thetabndry} read
\begin{align}
\label{eq:projectionboosted}
\overline \Theta_{\rm NP}^{\, \figeight}( \theta_{q_i}, \, \theta_{cs}, \Delta \phi_i)
&= \overline \Theta_{\rm NP}^{\, \figeight} \Big(\frac{k_{i\perp}}{k_i^-}, 1 , \phi_k \Big)
\\
&\equiv \Theta\bigg(| \phi_{k_i}| - \frac{\pi}{3} \bigg)\Theta\bigg(1 -\frac{ k_{i\perp}}{k_i^-}\bigg)+ \Theta\bigg(\frac{\pi}{3} - | \phi_k| \bigg)\Theta \bigg(2 \cos ( \phi_{k_i}) - \frac{ k_{i\perp}}{k_i^-}\bigg) \, , \nn \\
\label{eq:bndryboosted}
\overline \Theta_{\rm NP}^{\, \bndry} (\theta_{q_i}, \, \theta_{cs},\, \Delta \phi_i)
&= \overline \Theta_{\rm NP}^{\, \bndry} \Big(\frac{k_{i\perp}}{k_i^-}, 1, \phi_{k_i} \Big)
\\
&\equiv \Theta\bigg(\frac{\pi}{3} - | \phi_{k_i}| \bigg)\Theta \bigg(\frac{ k_{i\perp}}{k_i^-} - \frac{1}{2 \cos ( \phi_{k_i})}\bigg)\Theta \bigg(2 \cos ( \phi_{k_i}) - \frac{k_{i\perp}}{k_i^-}\bigg) \,, \nn
\end{align}
In \fig{NPfig8rescaled} we show the catchment area of the nonperturbative particles in the $k_x$-$k_y$ plane.
In the rescaled $\overline\Theta_{\rm NP}^{\, \figeight}$ and $\overline\Theta_{\rm NP}^{\, \bndry}$ the second argument $1$ corresponds to the unit radius appearing in these figures.
Note that in contrast to \fig{NPfig8} the axes are now $k_{x,y}/k^-$.

We can think of the rescaling in \eq{rescaling} as boosting the nonperturbative momenta along the jet axis, and rotating by $\phi_{cs}$ in the plane perpendicular to the jet axis, which we can approximate to be aligned with the collinear subjet. To see this explicitly we first define the Lorentz operator $\hat \Lambda (\gamma,\phi)$ for a boost $\gamma$ along the jet direction and a rotation by $\phi$:
\begin{align}
\hat \Lambda (\gamma,\phi)\, |\big(p^+, p^-, p_\perp \big) \rangle &= | \Lambda^{\mu}_\nu(\gamma,\phi)\, p^\nu \rangle =\bigg |\Big(\gamma\, p^+, \, \frac{1}{\gamma} \, p^-, \, R_\phi p_\perp \Big) \bigg \rangle \,,
\end{align}
where $R_\phi$ is a $2\times 2$ rotation matrix in the transverse plane. Hence
\begin{align} \label{eq:boostk}
\hat \Lambda^{-1} \Big(\frac{\theta_{cs}}{2},\phi_{cs}\Big) |\big(q_i^+, q_i^-, q_{i\perp} \big) \rangle = | (k_i^+, k_i^-,k_{i\perp}) \rangle \, .
\end{align}
with $\hat \Lambda^{-1}$ being the corresponding inverse Lorentz transformation. Then the operators in \eqs{hatpfig8}{hatpbndry} for the shift and boundary power corrections assume a simple form
\begin{align}
\label{eq:hatpboosted1}
\hat p_{\figeight}^+ (\theta_{cs}, \phi_{cs}) &= \frac{ \theta_{cs}}{2} \, \bigg [ \hat \Lambda\Big (\frac{\theta_{cs}}{2},\phi_{cs}\Big) \, \hat p^+ \,\overline \Theta_{\rm NP}^{\, \figeight}\Big(\frac{\hat p_{\perp }}{\hat p^-}, 1, \hat \phi\Big) \, \hat \Lambda^{-1} \Big(\frac{\theta_{cs}}{2},\phi_{cs}\Big)\bigg] \, , \\
\label{eq:hatpboosted2}
\hat{ p}_{\bndry}(\theta_{cs}, \phi_{cs},\beta) &= \frac{2}{\theta_{cs}} \bigg[ \,\hat \Lambda\Big (\frac{\theta_{cs}}{2},\phi_{cs}\Big) \, \Big(\, \hat p^- + \beta \, \big( \hat p^-- \hat p_\perp\, \cos(\hat \phi)\big) \,\Big)\,
\nn \\
& \qquad \times \bigg (\overline \Theta_{\rm NP}^{\, \bndry} \Big(\frac{\hat p_{\perp }}{\hat p^-},1, \hat \phi\Big) - \Theta_{\rm NP}^{\, \bndry} \Big(\frac{\hat p_{\perp }}{\hat p^-}, 1, \hat \phi\Big) \bigg) \, \hat \Lambda^{-1} \Big(\frac{\theta_{cs}}{2},\phi_{cs}\Big)\,\bigg]
\, ,
\end{align}
The measurement in the square brackets is performed on the state $|X_\Lambda\rangle$ as seen in \eq{shiftapprox}, and thus yields momenta $k_i^\mu \sim \Lambda_{\rm QCD}$ in both the cases, whereas the simple angular factors outside are purely perturbative. Thus we see that, despite their $\pm$ superscripts, the new variables $k_i^+$ and $k_i^-$ are invariant under physical boosts along the jet axis, which follows from their definitions in \eq{rescaling} and \eq{boostk}, since any boost to $q_i^\pm$ is compensated by that to $\theta_{cs}/2 = \sqrt{p_{cs}^+/p_{cs}^-}$.

Thus we observe that by performing the measurement on the nonperturbative subjets in an appropriately boosted frame we are able to completely factorize the perturbative and the nonperturbative dependence of the power corrections induced through the angles of the subjet. We note that the small angle approximation is crucial for this derivation. As already mentioned above near \eq{betaconstr}, in the limit $\beta \to \infty$ the modes that stop soft drop have $\theta \sim 1$, and we are no longer able to factor out the perturbative dependence of the measurement. Hence, our results for the soft drop power corrections do not have any simple connection to power corrections for the plain jet mass spectrum.

\subsection{CA Clustering within the Nonperturbative Sector}
\label{sec:CAforNP}

We remind the reader that our method of treating hadronization uses nonperturbative modes with virtuality $p_\Lambda^2 \sim \Lambda_{\rm QCD}^2$ that in conjunction with the perturbative modes account for the full hadronic cross section. Both the NP and perturbative modes are described by different fields in the SCET Lagrangian, with their own individual contributions to matrix elements. Concerning \eqs{hatpfig8}{hatpbndry}, we note that a key feature of the operators $\hat p_{\figeight}^\mu (\theta_{cs}, \phi_{cs}) $ and $\hat p_{\bndry} (\theta_{cs}, \phi_{cs}, \beta)$, used to calculate the power corrections, is that they implement a single-particle and purely geometrical constraint on the NP emissions state $|X_\Lambda\rangle$ based on the location of the perturbative subjets and their catchment areas. As a consequence, in the SDOE region we were able to decouple the effects of CA clustering between the perturbative and nonperturbative sectors. However, the CA clustering is still important within the nonperturbative sector: the angular locations of the NP branches can change significantly if other NP branches with similar momentum scaling get paired with it. In this section we address this issue by clarifying what we mean precisely by ``NP subjets'' for the purpose of defining our NP source function.

\begin{figure}[t!]
\includegraphics[width=0.45\textwidth]{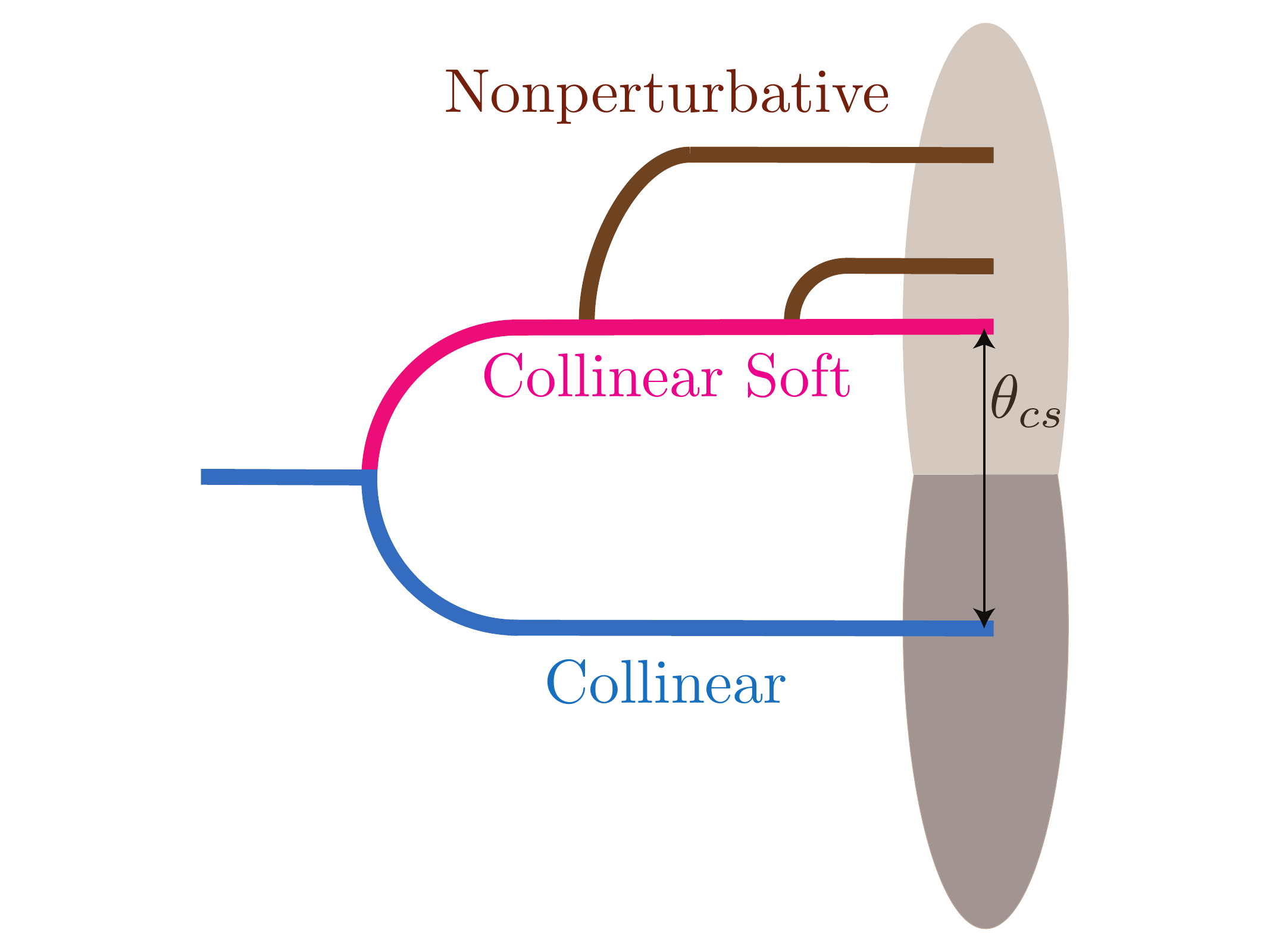}
\hspace{0.1\textwidth}
\includegraphics[width=0.45\textwidth]{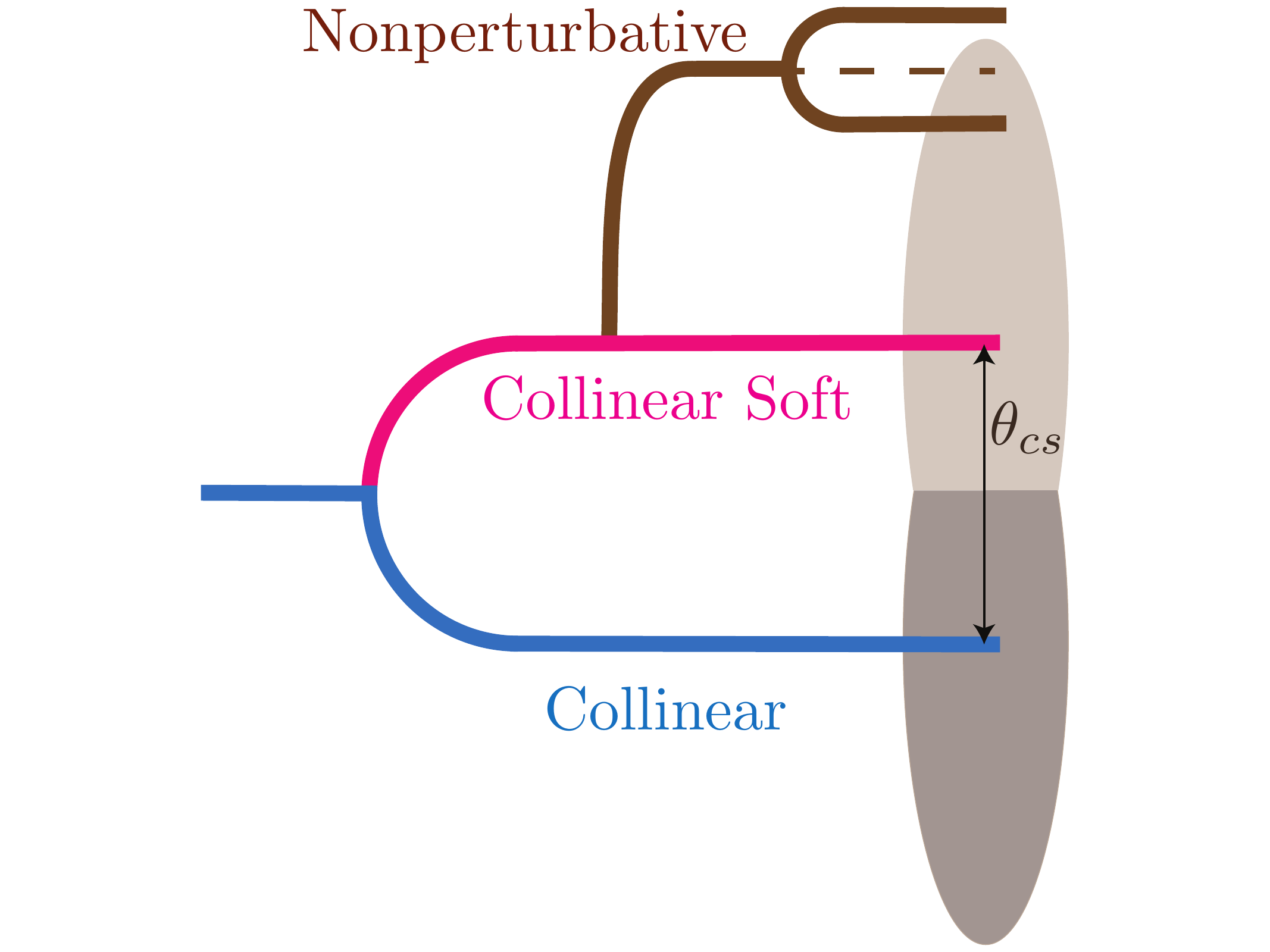}
\\[-10pt]
\phantom{x} \hspace{0.4cm}a) \hspace{9.3cm} b) \\[-15pt]
\caption{Two example scenarios where (a) the NP subjets get clustered at different stages and (b) where the NP subjets are clustered together first before they get paired with another perturbative branch. In (b) although one the NP particles lies outside the cone of radius $\theta_{cs}$ centered at the CS subjet, the combined NP branch gets eventually clustered. This requires $\overline \Theta_{\rm NP}^{\, \figeight}$ to act on NP subjets rather than individual particles.}
\label{fig:npclustering}
\vspace{-0.2cm}
\end{figure}

As an example we consider two scenarios with two NP particles and two perturbative branches as shown in \fig{npclustering}: in scenario (a) both of the NP particles get clustered with the perturbative tree at different stages, and in scenario (b) they get clustered together first and the combined branch is then paired with a perturbative branch (here the CS branch). In the scenario a) both the nonperturbative particles can be combined with the CS subjet only if each of them falls in the catchment area shown in \fig{NPfig8}, and hence individually satisfy $ \overline \Theta^{\figeight}_{\rm NP}=1$. After the first NP particle closer to the CS subjet is clustered, the angular location of the resulting subjet direction is roughly the same, and hence the same geometrical constraint applies for the second NP particle clustered later. In scenario (b), however, one of the NP particles may not lie in the region of overlapping cones, because only the combination of them needs to. Hence, in order to make our operators in \eqs{hatpfig8}{hatpbndry} account for such cases it is mandatory to make the meaning of the state $|X_\Lambda \rangle$ more precise.

Given a set of particles to be reclustered for grooming, the EFT provides a natural Lorentz invariant distinction between perturbative and NP particles without introducing a hard momentum cutoff. Physically, the momentum distribution of non-perturbative particles peaks at smaller momenta in the SDOE region.
We demand that the operators $\hat p_{\figeight}^\mu (\theta_{cs}, \phi_{cs}) $ and $\hat p_{\bndry}^\mu (\theta_{cs}, \phi_{cs}, \beta)$ should be applied to ``NP subjets'' instead of being tested on individual NP particles, where these NP subjets are
obtained by CA clustering of all the NP particles treating perturbative particles as ``beam'' directions.
These NP subjets are defined in the following manner:
\begin{enumerate}
\item All NP particles are called NP subjets.
\item The pair of NP subjets with smallest relative angular distance $\Delta \theta$ is grouped into a new NP subjet, if the angular separation of each of the two NP subjets to any perturbative particle is larger than $\Delta \theta$. The grouping of the NP subjets continues until the latter condition fails.
\end{enumerate}
The set of NP subjets that result from this grouping defines the ``multi-particle'' state $|X_\Lambda\rangle $ that is tested according to the geometrical constraint set by the collinear and CS subjets. We also note that the NP subjets then themselves have energy $\sim \Lambda_{\rm QCD}$. With this refinement of the meaning of $|X_\Lambda\rangle$ \eqs{hatpfig8}{hatpbndry} now account for all the clustering cases with arbitrary number of particles.

Note that the steps outlined above yield the same CA clustered tree as one would obtain via the usual CA procedure that starts with clustering the closest pair regardless of their energy. We will make use of this procedure in the Monte Carlo studies presented below in \sec{montecarlo} to demonstrate the validity of our kinematic approximations in the SDOE region.

\subsection{Factorization for Matrix Elements}
\label{sec:npsource}

Having simplified the form of the measurement operator we now consider the nonpertubative factorization for the corresponding matrix elements. In this section we shed light on the properties of the power corrections for groomed event shapes, via fixed order calculations working at LL in the perturbative emissions and using Feynman gauge, in order to demonstrate the factorization of the perturbative and nonperturbative parts of the matrix element.
The part of these perturbative calculations involving nonperturbative modes simply serve as a proxy to probe the corresponding matrix elements.

In standard event shapes, without jet grooming, the nonperturbative effects are often sourced by Wilson lines that know only about the direction of the energetic collinear parton~\cite{Belitsky:2001ij,Lee:2006fn,Abbate:2010xh,Mateu:2012nk}. This is due to the fact that the leading power correction is governed by soft modes that cannot resolve the details of collinear splittings of quarks and gluons that constitute the internal perturbative structure of the jet. For $e^+e^-\to $ dijets~\cite{Lee:2006fn}, or jet production in $pp$ with small jet radius $R$~\cite{Stewart:2014nna}, the matrix element of nonperturbative radiation becomes invariant under boosts along the collinear direction. This is a desirable property given the simplifications we obtained for the groomed jet measurement operators in \eqs{hatpboosted1}{hatpboosted2} on boosting the NP sector to an appropriate reference frame along the jet direction. In our analysis below, the abelian graphs without gluon splitting exhibit a factorization of nonperturbative and perturbative matrix elements (without making a boost since they are boost invariant).
In contrast, the non-abelian graphs, where the NP gluon is emitted from the collinear-soft gluon, yield an expression that is apparently not factorized into perturbative and nonperturbative matrix elements.
However, the change of variables in \eq{rescaling}, corresponding to a boost and a rotation of the NP sector alone that depend on the angles of the collinear-soft subjet, does in fact yields a factorization of the perturbative and nonperturbative matrix elements in the new frame. This transformation yields precisely the same perturbative Wilson coefficients as the abelian diagrams, showing that the transformation in \eq{rescaling} is essential not only for factorization of the measurement but also for the matrix element.

To determine the perturbative coefficients multiplying the non-perturbative matrix elements in the operator expansion we follow the logic of \Refcite{Mateu:2012nk}, where a source NP gluon replaces the NP mode. As discussed above, the dominant effect of NP modes is induced only via the collinear-soft function $S_c$, and thus we do not consider nonperturbative effects in the global soft function or other functions. We consider a case of a quark or a gluon initiated jet in the single emission picture with a perturbative gluon $p^\mu$, that has the collinear-soft scaling, and a nonperturbative gluon $q^\mu$. With this set up we demonstrate how the nonperturbative power corrections can be factorized from the perturbative matrix element. The corresponding Feynman diagrams are shown in \figs{np-abelian}{np-nonabelian}.
We expand the interactions to the leading non-trivial power, which leads to eikonal couplings to the energetic source lines. We do not consider cut vacuum polarization graphs for quarks or gluons as they yield subleading nonperturbative corrections.
The nonperturbative gluons are brown, whereas the perturbative gluons are magenta.
With the dashed line representing plus momentum measurement with the soft drop test, $\delta (\ell^+ - \hat p_{\rm sd}^+)$, these graphs precisely correspond to the ${\cal O}(\alpha_s)$ perturbative corrections with an additional nonperturbative gluon.

\addtocontents{toc}{\protect\setcounter{tocdepth}{1}}
\subsubsection{Abelian Graphs} \label{sec:abelian}
\addtocontents{toc}{\protect\setcounter{tocdepth}{2}}

\begin{figure}[t!]
\centering
\includegraphics[width=\columnwidth]{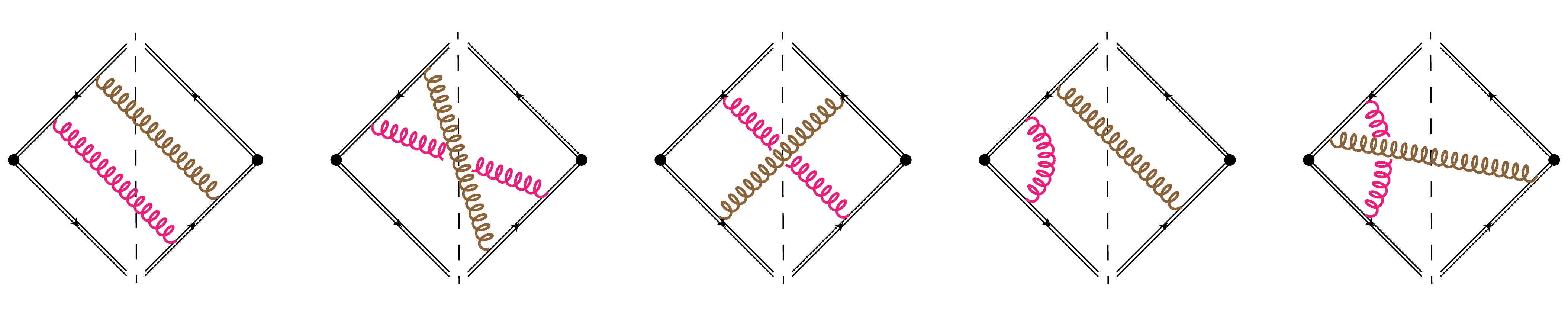}
\caption{Examples of abelian diagrams with one perturbative (magenta) CS gluon and a nonperturbative (brown) gluon. Dashed line represents measurement of the plus momentum along with the soft drop operator. Graphs with a virtual NP gluon are not considered.}
\label{fig:np-abelian}
\end{figure}

We first consider the abelian graphs shown in \fig{np-abelian}. The result for the one loop collinear-soft function in Feynman gauge including the effect of the NP gluon reads
\begin{align}
\label{eq:scabelian}
S_{c}^{\,\kappa, \rm had}\Bigl( &\ell^+,\qcut,\beta, \mu\Bigr)
= S_{c}^{\,\kappa,\rm pert}\Bigl(\ell^+\qcut^{\frac{1}{1+\beta}},\beta, \mu \Bigr)
\\
&\hspace{-1cm}
+\! \int\!\! \frac{d^d q}{(2\pi)^d}\,
\bigg[ \frac{(4 C_\kappa-2C_A) g^2 \, \tilde \mu^{2\epsilon}\, \tilde{ \cal C}(q) }{ q^+ \,q^-} \bigg]
\bigg \{ \Delta \tilde S_{c}^{\, \kappa,\figeight} \Bigl(\ell^+, q^\mu,\qcut,\beta, \mu\Bigr)
+\Delta \tilde S_c^{\, \kappa,\bndry}\Bigl(\ell^+, q^\mu,\qcut,\beta, \mu\Bigr)\bigg\}
, \nn
\end{align}
where the first term is simply the perturbative ${\cal O}(\alpha_s)$ collinear-soft function quoted above in \eq{sc1loop}, and the remaining terms are the power corrections with an integral over the momentum $q$ of the non-perturbative source gluon. Here $\tilde \mu^{2\epsilon} = (\mu^2 e^{\gamma_E}/(4\pi))^\eps$ and $C_\kappa = C_F$ or $C_A$ for quark and gluon initiated jets respectively.
The details of the derivation are presented in \app{Eikonal}.
At first order in the power expansion, only diagrams where the NP gluon attaches after the perturbative gluon (next to the cut) contribute. Thus in \fig{np-abelian} the 1st graph does not contribute, but the the 2nd and 3rd graphs do.
The power corrections for the shift and the boundary terms involve phase space integrals over the perturbative gluon
\begin{align}
\label{eq:scfig8}
\Delta\tilde S_c^{\,\kappa,\figeight}\Bigl(\ell^+, q^\mu, \qcut, \beta, \mu\Bigr)
&= \qcut^{\frac{-1}{1+\beta}}\, \frac{\alpha_s C_\kappa}{\pi}
\frac{(\mu^2 e^{\gamma_E})^\eps}{\Gamma(1-\eps)}
\int_0^\infty\!\! \frac{d p^+ \,dp^-}{(p^+\, p^-)^{1+\eps}}
\int_0^{2\pi}\!\frac{d\phi_p}{2\pi} \,
\Theta \bigg ( \frac{p^- }{Q} - \tilde z_{\rm cut} \, \theta_p^\beta \bigg)
\nn\\
&\quad \times \overline \Theta_{\rm NP}^{\,\figeight}(\theta_{q}, \, \theta_{p},\, \Delta \phi) \ \Big[ \delta\big(\ell^+ - p^+ - \, q^+\big) - \delta(\ell^+ - p^+)\Big]
\, , \nn \\
\Delta\tilde S_c^{\,\kappa, \bndry} \Bigl(\ell^+, q^\mu, \qcut, \beta, \mu\Bigr)
&= \qcut^{\frac{-1}{1+\beta}}\, \frac{\alpha_s C_\kappa}{\pi}
\frac{(\mu^2 e^{\gamma_E})^\eps}{\Gamma(1-\eps)}
\int_0^\infty\!\! \frac{d p^+ \,dp^-}{(p^+\, p^-)^{1+\eps}}
\int_0^{2\pi}\!\frac{d\phi_p}{2\pi} \,
\delta \bigg ( \frac{p^- }{Q} - \tilde z_{\rm cut} \, \theta_p^\beta \bigg)
\nn\\
&\ \times \Big( \overline \Theta_{\rm NP}^{\,\bndry} (\theta_{q}, \, \theta_p,\, \Delta \phi) - \Theta_{\rm NP}^{\,\bndry} (\theta_{q}, \, \theta_p,\, \Delta \phi) \Big)\,
\nn \\
&\ \times \frac{q^- }{Q} \, \bigg [\, 1 +\beta \,\Big(1- \frac{\theta_{q}}{\theta_p} \, \cos( \Delta \phi) \Big)\bigg] \Big[ \delta\big(\ell^+ - p^+ \big) - \delta(\ell^+)\Big]
\, .
\end{align}
Here $\theta_p=2\sqrt{p^+/p^-}$, $\theta_q=q_\perp/|\vec q|$, $\Delta\phi=\phi_q-\phi_p$ and the $\overline\Theta_{\rm NP}^{\,\figeight}$ and $\overline\Theta_{\rm NP}^{\,\bndry}$ were defined in \sec{measurement}.
The results in \eq{scfig8} refer to the remaining pieces after subtracting the perturbative $S_{c}^{\,\kappa,\rm pert}$ from the full expression $S_{c}^{\,\kappa, \rm had}$. In \eq{scabelian} we have further factored out the matrix element for the NP gluon, such that the term in the square brackets serves as a proxy for a nonperturbative source function:
\begin{align}
\label{eq:Fabelian}
\tilde F^{\, \rm ab.}_\kappa (q^\mu)
\equiv \frac{(4 C_\kappa-2C_A) g^2\,\tilde \mu^{2\epsilon}\, \tilde{ \cal C}(q) }{ q^+ \,q^-} \, ,
\end{align}
with the `ab.' superscript emphasizing that this is derived from the abelian graphs, and the subscript $\kappa$ that it is dependent on the jet initiating parton. Here $\tilde {\cal C}(q)$ is defined in \eq{cut}. (At the end of our analysis the source functions will be replaced by a full non-perturbative function rather than some perturbative approximation.)
In \eq{scabelian} we did not add diagrams with virtual NP gluons because they only affect the overall normalization. We comment further on the normalization of the nonperturbative source function below.
The two graphs in \fig{np-abelian} where only a NP gluon crosses the cut do not contribute in the SDOE region, as discussed further in \app{Eikonal}.

We observe that $\tilde F (q^\mu)$ and the measure $d^d q$ are individually invariant under boosts along the jet direction. On performing the boost and the rotation defined in \eq{rescaling} taking $\theta_{cs}= \theta_p$ and $\phi_{cs} = \phi_p$ we find
\begin{align}
\theta_q = \frac{\, q_\perp}{q^-} = \theta_p \frac{k_\perp}{k^-} \, , \qquad
\tilde F^{\, \rm ab.}_\kappa (q^\mu) = \tilde F^{\, \rm ab.}_\kappa (k^\mu) \, , \qquad d^d q = d^d k \, .
\end{align}
As a result of which \eq{scabelian} becomes
\begin{align}
\label{eq:scabeliank}
S_c^{\,\kappa,\rm had}(\ell^+,\qcut,\beta, \mu)
&= S_c^{\,\kappa,\rm pert}\Bigl(\ell^+ \qcut^{\frac{1}{1+\beta}},\beta, \mu\Bigr)
- \Omega^{\, \rm ab.}_{1\kappa} \, \frac{d}{d \ell^+}\,\Delta S_c^{\,\kappa,\figeight} (\ell^+,\qcut,\beta, \mu)
\nn\\
&\ + \frac{\Upsilon^{\, \rm ab.}_{1\kappa}(\beta)}{Q} \, \Delta S_c^{\,\kappa, \bndry} (\ell^+,\qcut,\beta, \mu) \, ,
\end{align}
where the power corrections $\Omega_1^{\figeight}$ and $\Upsilon_1$ involve measurements on the NP radiation in the boosted frame:
\begin{align}
\label{eq:O1Ab}
\Omega^{\, \rm ab.}_{1\kappa} &\equiv \int \frac{d^d k}{(2\pi)^d}\, k^+ \,\overline \Theta_{\rm NP}^{\, \figeight} \Big(\frac{k_\perp}{k^-}, \, 1, \, \phi_k \Big) \, \tilde F_{\kappa}^{\, \rm ab.} (k^\mu)
\,, \\
\label{eq:Ups1Ab}
\Upsilon^{\, \rm ab.}_{1\kappa}(\beta) &\equiv \int \frac{d^d k}{(2\pi)^d}\, \Big (k^- + \beta\, \big( k^- - k_\perp \, \cos \phi_k\big) \Big)
\nn \\
&\qquad \times \bigg[\overline \Theta_{\rm NP}^{\, \bndry} \Big(\frac{k_\perp}{k^-}, \, 1, \, \phi_k \Big) - \Theta_{\rm NP}^{\, \bndry} \Big(\frac{k_\perp}{k^-}, \, 1, \, \phi_k \Big)\bigg] \,\, \tilde F_{\kappa}^{\, \rm ab.} (k^\mu) \,,
\end{align}
with the projection operators given by \eq{projectionboosted}. Since $k^\pm$ are boost invariant, so are the full definitions in \eqs{O1Ab}{Ups1Ab}. We note that the shift term is $\beta$ and $\tilde z_{\rm cut}$ independent, whereas the boundary term has a linear dependence on $\beta$. The power corrections in \eq{scabelian} are multiplied by perturbative Wilson coefficients given by
\begin{align}\label{eq:scfig8bndryk}
\Delta S_c^{\,\kappa,\figeight} (\ell^+,\qcut,\beta, \mu)
&= \qcut^{\frac{-1}{1+\beta}}\,\frac{\alpha_s C_\kappa}{\pi}
\frac{(\mu^2 e^{\gamma_E})^\eps}{\Gamma(1-\eps)}
\int_0^\infty\!\! \frac{d p^+ \,dp^-}{(p^+\, p^-)^{1+\eps}} \, \frac{\theta_p}{2} \Theta \Bigl( \frac{p^- }{Q} - \tilde z_{\rm cut} \, \theta_p^\beta \Bigr) \, \delta(\ell^+ - p^+)\, ,\nn
\\
\Delta S_c^{\,\kappa, \bndry} (\ell^+,\qcut,\beta, \mu)
&= \qcut^{\frac{-1}{1+\beta}} \,\frac{\alpha_s C_\kappa}{\pi}
\frac{(\mu^2 e^{\gamma_E})^\eps}{\Gamma(1-\eps)}
\int_0^\infty \!\!\frac{d p^+ \,dp^-}{(p^+\, p^-)^{1+\eps}} \, \frac{2}{\theta_p} \delta \Bigl( \frac{p^- }{Q} - \tilde z_{\rm cut} \, \theta_p^\beta \Bigr) \nn \\
&\qquad \times \Big[ \delta\big(\ell^+ - p^+ \big) - \delta(\ell^+)\Big] \, .
\end{align}

Thus we see from \eq{scabeliank} the perturbative and the nonperturbative contributions have been successfully decoupled. Here we see a direct application of the result in \eqs{hatpboosted1}{hatpboosted2} - the measurement in the boosted frame yields the nonperturbative moments in \eqs{O1Ab}{Ups1Ab}, and the residual factors of $\theta_{cs}/2$ and $2/\theta_{cs}$ are part of the perturbative Wilson coefficients in \eq{scfig8bndryk}.

\addtocontents{toc}{\protect\setcounter{tocdepth}{1}}
\subsubsection{Non-Abelian Graphs} \label{sec:nonabelian}
\addtocontents{toc}{\protect\setcounter{tocdepth}{2}}

\begin{figure}[t!]
\centering
\includegraphics[width=\columnwidth]{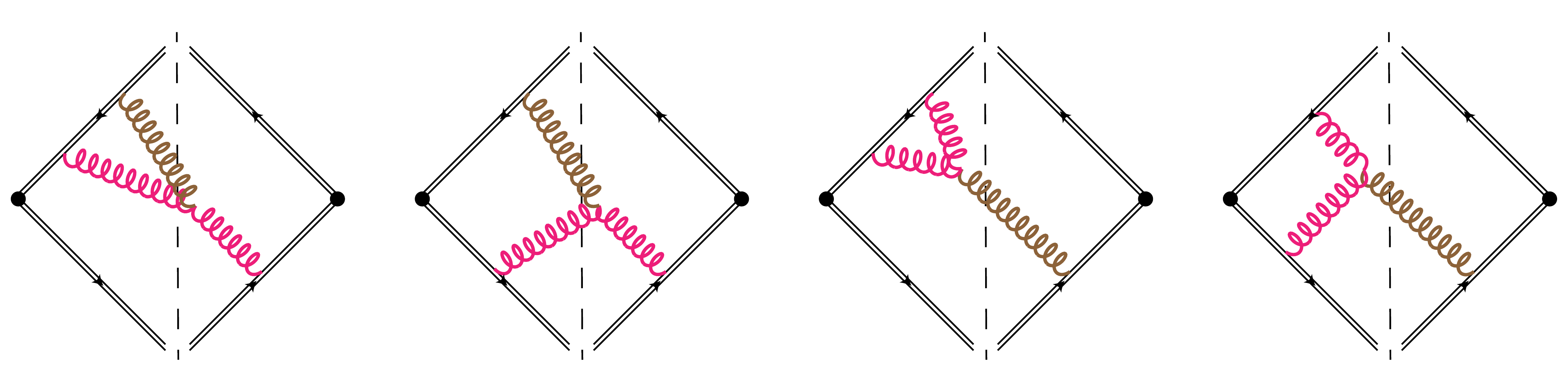}
\caption{Examples of non-abelian diagrams with one perturbative (magenta) CS gluon and a nonperturbative (brown) gluon. Dashed line represents measurement of the plus momentum along with the soft drop operator. Graphs with a virtual NP gluon are not considered.}
\label{fig:np-nonabelian}
\end{figure}

We now turn to the non-abelian contributions shown in \fig{np-nonabelian}. Here the NP gluon is radiated off the perturbative gluon. These diagrams contribute at the same order as the abelian ones, while graphs that are not shown (such as a cut gluon vacuum polarization graph) are higher order in the power expansion. The sum over all the non-abelian graphs is discussed in \app{Eikonal} and the result reads
\begin{align}
\label{eq:scnonabelian}
S_{c}^{\,\kappa,\rm had, \, n.a.} &(\ell^+,\qcut,\beta, \mu)= \qcut^{\frac{-1}{1+\beta}}\, \frac{\alpha_s C_\kappa}{\pi}\frac{(\mu^2 e^{\gamma_E})^\eps}{\Gamma(1-\eps)} \int_0^\infty \frac{d p^+ \,dp^-}{(p^+\, p^-)^{1+\eps}} \, \int \frac{d^d q}{(2\pi)^d}\, \frac{2\, g^2\,C_A\,\tilde \mu^{2\epsilon}\, \tilde{ \cal C}(q) }{ q^+ \,q^-} \\
&\qquad \times\, \big[ {\cal M}^{p+q} - {\cal M}^{q} \big] \frac{q^+ p^- + p^+ q^-}{p^+ q^- + q^+ p^- - 2 \sqrt{p^+ p^-} | \vec q_\perp| \cos (\Delta \phi)} \, ,\nn
\end{align}
where the measurement functions ${\cal M}^{p+q}$ and ${\cal M}^q$ are given in \eqs{MpqSDOE}{MqSDOE}.

Unlike the abelian graphs it appears that we cannot simply carry out the $p^+$ and $p^-$ integrations to express $S_c^{\rm had, \, n.a.}$ in a factorized form analogous to \eq{scabelian}. Since the last factor in the second line of \eq{scnonabelian} non-trivially couples the nonperturbative and the perturbative momentum dependence. This term also invalidates a definition of $F$ in terms of $n$-$\bar{n}$ Wilson lines. However, if carry out the rescaling according to \eq{rescaling},
\begin{align}
\label{eq:rescaling2}
q^+ = \frac{\theta_p}{2} \, k^+ = \sqrt{\frac{p^+}{p^-}} \, k^+ \, , \qquad q^- = \frac{2}{\theta_p} \, k^- = \sqrt{\frac{p^-}{p^+}}\,k^-\, , \qquad q_{\perp} = k_\perp \, ,\qquad \phi_q = \phi_k + \phi_p \, ,
\end{align}
we find that the nonperturbative and the perturbative factors in in the last term of \eq{scnonabelian} completely decouple:
\begin{align} \label{eq:NAfactor}
\frac{q^+ p^- + p^+ q^-}{p^+ q^- + q^+ p^- - 2 \sqrt{p^+ p^-} | \vec q_\perp| \cos (\Delta \phi)}
&= \frac{k^+ + k^-}{k^+ + k^- - 2 \,| \vec k_\perp| \cos (\phi_k)} \,.
\end{align}
Note that the factor in \eq{NAfactor} is boost invariant.
Using the expansions for ${\cal M}^{p+q}$ and ${\cal M}^q$ the leading non-perturbative power corrections in the SDOE region \eq{scnonabelian} are then
\begin{align}
\label{eq:scnonabeliank}
S_c^{\,\kappa,\rm had,\,n.a.}(\ell^+,\qcut,\beta, \mu)
&= S_c^{\,\kappa,\rm pert}\Bigl(\ell^+ \qcut^{\frac{1}{1+\beta}},\beta, \mu\Bigr)
- \Omega^{\, \rm n.a.}_{1} \, \frac{d}{d \ell^+}\,\Delta S_c^{\,\kappa,\figeight} (\ell^+,\qcut,\beta, \mu)
\nn\\
&\ + \frac{\Upsilon^{\, \rm n.a.}_{1}(\beta)}{Q} \, \Delta S_c^{\,\kappa,\bndry} (\ell^+,\qcut,\beta, \mu) \, ,
\end{align}
where the perturbative coefficients $\Delta S_c^{\,\figeight}$ and $\Delta S_c^{\, \bndry}$ are exactly the same as in the abelian case, see \eq{scfig8bndryk}, and the nonperturbative moments are given by
\begin{align}
\label{eq:O1NA}
\Omega^{\, \rm n.a.}_1 &\equiv \int \frac{d^d k}{(2\pi)^d}\, k^+ \,\overline \Theta_{\rm NP}^{\, \figeight} \Big(\frac{k_\perp}{k^-}, \, 1, \, \phi_k \Big) \, \tilde F^{\, \rm n.a.} (k^\mu) \,, \\
\label{eq:Ups1NA}
\Upsilon^{\, \rm n.a.}_1(\beta) &\equiv \int \frac{d^d k}{(2\pi)^d}\,\Big (k^- + \beta\, \big( k^- - k_\perp \, \cos \phi_k\big) \Big)\nn\\
&\qquad \times \bigg[\overline \Theta_{\rm NP}^{\, \bndry} \Big(\frac{k_\perp}{k^-}, \, 1, \, \phi_k \Big) - \Theta_{\rm NP}^{\, \bndry} \Big(\frac{k_\perp}{k^-}, \, 1, \, \phi_k \Big)\bigg] \,\, \tilde F^{\, \rm n.a.} (k^\mu) \,.
\end{align}
Here $\tilde F^{\, \rm n.a.} (k^\mu)$ is a proxy for the nonperturbative source function for the non-abelian graphs:
\begin{align}
\label{eq:Fnonabelian}
\tilde F^{\, \rm n.a.} (k^\mu) \equiv \frac{2C_A\,g^2\,\tilde \mu^{2\epsilon}\, \tilde{ \cal C}(k) }{ k^+ \,k^-} \, \frac{k^+ + k^-}{k^+ + k^- - 2 \,| \vec k_\perp| \cos (\phi_k)} \, .
\end{align}
This expression demonstrates that the power corrections we are considering here can not be expressed in terms of a non-perturbative matrix element of a Wilson line operator. From the analysis of the non-abelian graphs we see that the rescaling in \eq{rescaling} is crucial to achieve a separation of the nonperturbative matrix elements and perturbative Wilson coefficients.

The factorization property of the source function continues to hold in the presence of additional perturbative gluons at LL. This occurs because the nonperturbative gluon momentum is relevant only in one denominator, which is either from a gluon propagator or Wilson line regardless of the number of perturbative emissions.
In \app{2pert} we perform an explicit check that this property holds with two perturbative gluons.
We note that in our one-loop fixed order analysis we only considered the boundary power correction to the subjet stopping soft drop. In case of multiple perturbative emissions there are subjets with collinear-soft scaling that fail soft drop and their contribution to the normalization of the cross section can receive a boundary power correction as well. However, we show in \app{scfail} that such corrections only enter beyond LL order.

Overall we can combine the abelian and non-abelian results from \eqs{scabeliank}{scnonabeliank} to obtain:
\begin{align}
\label{eq:scfullk}
S_c^{\,\kappa,\rm had}(\ell^+,\qcut,\beta, \mu)
&= S_c^{\,\kappa,\rm pert}\Bigl(\ell^+ \qcut^{\frac{1}{1+\beta}},\beta, \mu\Bigr)
- \Omega_{1\kappa}^{\figeight} \, \frac{d}{d \ell^+}\,\Delta S_c^{\,\kappa,\figeight} (\ell^+,\qcut,\beta, \mu)
\nn\\
&\qquad + \frac{\Upsilon^\kappa_1(\beta)}{Q} \, \Delta S_c^{\,\kappa,\bndry} (\ell^+,\qcut,\beta, \mu) \, ,
\end{align}
with $\Omega_{1\kappa}^{\figeight} = \Omega_{1\kappa}^{\, \rm ab.} + \Omega_1^{\, \rm n.a.} $ and $\Upsilon_{1}^\kappa(\beta) = \Upsilon_{1\kappa}^{\, \rm ab.}(\beta) + \Upsilon_1^{\, \rm n.a.}(\beta) $. We note that we did not explicitly consider diagrams with virtual nonperturbative gluons in the analysis above, however they do not affect in any way the conclusion of this factorization analysis.
The essential point is that our analysis clarifies that the interface between NP and perturbative modes, and in particular the shift and boundary corrections, are governed by a nonperturbative source function $\tilde F_{\kappa} (k^\mu)$ which, eventually, has to be determined using methods outside perturbation theory.

\subsection{${\cal O}(\alpha_s)$ Matching Coefficients}
\label{sec:EFTmatching}

In this section we evaluate the Wilson coefficients in \eq{scfullk} at the cross section level at ${\cal O}(\alpha_s)$
It is straightforward to evaluate the integrals in \eq{scfig8bndryk} for $\epsilon=0$, and their results read
\begin{align}\label{eq:C1C2oneloop}
\qcut^{\frac{1}{1+\beta}} \Delta S_c^{\,\kappa, \figeight} (\ell^+,\qcut,\beta, \mu)
&= \frac{\alpha_s C_\kappa}{\pi}
\frac{2}{\ell^+} \biggl( \frac{\ell^+}{\qcut}\biggr)^{\frac{1}{2+\beta}}
\,, \\
Q^{-1} \qcut^{\frac{1}{1+\beta}} \Delta S_c^{\,\kappa,\bndry} (\ell^+,\qcut,\beta, \mu)
&= \frac{\alpha_s C_\kappa}{\pi}
\,\frac{2}{2+\beta}\, \frac{1}{(\ell^+)^2} \biggl( \frac{\ell^+}{\qcut}\biggr)^{\frac{1}{2+\beta}}
\,. \nn
\end{align}
We parameterize the hadron level differential cross section with the leading power shift and boundary nonperturbative corrections in the SDOE region in the form
\begin{align}
\label{eq:sighad}
\frac{d \sigma^{\rm had}}{d m_J^2 d\Phi_J} = \sum_{\kappa={q,g}} N_\kappa(\Phi_J,z_{\rm cut},\beta,\mu_{gs}) \bigg[
\frac{d\hat \sigma_\kappa}{d m_J^2}
+ \frac{d \sigma^{\rm shift}_\kappa}{d m_J^2}
+ \frac{d \sigma^{\rm bndry}_\kappa}{d m_J^2} \bigg] \, .
\end{align}
Combining \eq{C1C2oneloop} with tree level ingredients in \eq{masslessFact1} we can derive results for the power corrections at the cross section level with ${\cal O}(\alpha_s)$ soft-collinear matching coefficients:
\begin{align}\label{eq:sigshiftbndry}
\frac{d \sigma^{{\rm shift}}_\kappa}{d m_J^2} \bigg|_{{\cal O}(\alpha_s)}
&= Q \,\Omega_{1\kappa}^{\figeight} \, \frac{d}{d m_J^2} \bigg(
\int_0^\infty d \ell^+ \delta (m_J^2 - Q \ell^+) \>\qcut^{\frac{1}{1+\beta}}\Delta S_c^{\,\kappa, \figeight} (\ell^+,\qcut,\beta, \mu)
\bigg)
\nn \\
& = Q \,\Omega_{1\kappa}^{\figeight} \, \frac{d}{d m_J^2} \Bigg(
2 \frac{\alpha_s C_\kappa}{\pi} \>
\frac{1}{m_J^2} \biggl( \frac{m_J^2}{Q \qcut}\biggr)^{\frac{1}{2+\beta}}
\Bigg)
\nn\\
&= -\frac{Q \,\Omega_{1\kappa}^{\figeight}}{m_J^4}
\Big(\frac{\alpha_s C_\kappa}{\pi} \Big) \bigg[\,\frac{2(1+\beta)}{2+\beta} \,
\biggl( \frac{m_J^2}{Q \qcut}\biggr)^{\frac{1}{2+\beta}}\bigg]
\,, \nn \\
\frac{d \sigma^{{\rm bndry}}_\kappa}{d m_J^2}
\bigg|_{{\cal O}(\alpha_s)}
&=
\frac{\Upsilon^\kappa_1(\beta)}{Q} \,
\int d \ell^+ \delta (m_J^2 - Q \ell^+) \>
\qcut^{\frac{1}{1+\beta}}\Delta S_c^{\,\kappa, \bndry} (\ell^+,\qcut,\beta, \mu)
\nn \\
&= \frac{Q\Upsilon^\kappa_1(\beta) }{m_J^4}
\Big(\frac{\alpha_s C_\kappa}{\pi} \Big)\,
\bigg[\frac{2}{2+\beta} \,
\biggl( \frac{m_J^2}{Q \qcut}\biggr)^{\frac{1}{2+\beta}}
\bigg]
\, .
\end{align}
The form of both results in \eq{sigshiftbndry} makes explicit that they are the same order in the power counting, as argued above. We also note that for the $pp$ case the combinations in \eq{sigshiftbndry} are independent of $\cosh\eta_J$ since
\begin{align}\label{eq:indepcosheta}
Q \biggl( \frac{m_J^2}{Q \qcut}\biggr)^{\frac{1}{2+\beta}}
= p_T
\biggl(
\frac{m_J^2}{p_T^2 z_{\rm cut} R_0^{-\beta} }\biggr)^{\frac{1}{2+\beta}}
\,.
\end{align}

Note that the presence of an explicit $\alpha_s$ factor in the first order SDOE power corrections is explained by the need for a perturbative emission which stops soft drop. Furthermore we emphasize that the coefficients for the shift and boundary power corrections contain large logarithms from higher orders so one should not conclude that the ${\cal O}(\alpha_s)$ terms shown in \eq{sigshiftbndry} suffice to determine these coefficients at the leading log level.
Due to the inherent separation of scales that is present with the collinear-soft, jet, and global soft scales, these results will be dressed by further perturbative emissions which produce a tower of leading logarithms.

\subsection{Results for Power Corrections from the Operator Expansion}
\label{sec:factorization}

Based on the factorization for the power corrections derived in Secs.~\ref{sec:NPmodes}--\ref{sec:EFTmatching} it is natural to write the hadronic cross section as
\begin{align}
\label{eq:sigfull0}
\frac{d \sigma^{\rm had}_\kappa}{d m_J^2}
&= \frac{d \hat \sigma_\kappa}{d m_J^2}
- \frac{Q\Omega_{1\kappa}^{\figeight}}{m_J^2} \
\frac{d \Delta\hat \sigma^{\figeight}_\kappa}{d m_J^2}
+ \frac{Q \Upsilon_{1}^\kappa (\beta)}{m_J^2} \
\frac{d \Delta\hat\sigma^{\bndry}_\kappa}{d m_J^2}
\,,
\end{align}
where the leading power corrections are projections on a non-perturbative source distribution $\tilde F_{\kappa}(k^\mu)$ given by
\begin{align} \label{eq:O1Ups1}
\Omega_{1\kappa}^{\figeight} &\equiv \int \frac{d^d k}{(2\pi)^d}\, k^+ \,\overline \Theta_{\rm NP}^{\, \figeight} \Big(\frac{k_\perp}{k^-}, \, 1, \, \phi_k \Big) \, \tilde F_{\kappa}(k^\mu)
\,, \\
\Upsilon_1^\kappa(\beta) &= \Upsilon_{1,0}^\kappa + \beta \, \Upsilon_{1,1}^\kappa
\,,\nn\\
\Upsilon_{1,0}^\kappa &\equiv \int \frac{d^d k}{(2\pi)^d}\, k^- \, \bigg[\overline \Theta_{\rm NP}^{\, \bndry} \Big(\frac{k_\perp}{k^-}, \, 1, \, \phi_k \Big) - \Theta_{\rm NP}^{\, \bndry} \Big(\frac{k_\perp}{k^-}, \, 1, \, \phi_k \Big)\bigg] \,
\tilde F_\kappa(k^\mu)
\,,\nn \\
\Upsilon_{1,1}^\kappa &\equiv \int \frac{d^d k}{(2\pi)^d}\,\big ( k^- - k_\perp \, \cos\phi_k \big) \, \bigg[\overline \Theta_{\rm NP}^{\, \bndry} \Big(\frac{k_\perp}{k^-}, \, 1, \, \phi_k \Big) - \Theta_{\rm NP}^{\, \bndry} \Big(\frac{k_\perp}{k^-}, \, 1, \, \phi_k \Big)\bigg] \,
\tilde F_\kappa (k^\mu) \,
\,. \nn
\end{align}
Here we have used the fact that $\Upsilon_1(\beta)$ is linear in $\beta$. Thus the leading non-perturbative power corrections in the SDOE region are expressed in terms of three hadronic parameters, $\Omega_{1\kappa}^{\figeight}$, $\Upsilon_{1,0}^\kappa$, and $\Upsilon_{1,1}^\kappa$. These parameters are each ${\cal O}(\Lambda_{\rm QCD})$, depend on whether the jet is initiated by a quark or gluon via $\kappa=q,g$, and are independent of all other variables.\footnote{We ignore possible dependence on the renormalization scale $\mu$ because we do not attempt to sum large logarithms occurring between the $\mu_{cs}$ and $\Lambda_{\rm QCD}$ scales. Single logarithms of this type are known to appear for $e^+e^-$ event shapes~\cite{Mateu:2012nk}.} Since the momentum variables $k^\pm$ are defined as boost invariant along the jet axis, so are these hadronic parameters. We stress that these parameters are only defined for groomed jet mass in the SDOE region, with geometry determined by the $\overline \Theta_{\rm NP}^{\, \figeight}$ and $\overline \Theta_{\rm NP}^{\, \bndry}$ functions, and have no connection to the nonperturbative matrix element(s) that govern the case of plain jet mass.

In \eq{sigfull0} the terms ${d \Delta\hat \sigma^{\figeight}_\kappa}/{d m_J^2}$ and ${d \Delta\hat\sigma^{\bndry}_\kappa}/{d m_J^2}$ are perturbative coefficients containing terms scaling as
\begin{align} \label{eq:sigscaling}
\frac{d \Delta\hat \sigma^{\figeight}_\kappa}{d m_J^2}
\sim \frac{d \Delta\hat\sigma^{\bndry}_\kappa}{d m_J^2}
\sim \biggl( \frac{m_J^2}{Q \qcut}\biggr)^{\frac{1}{2+\beta}}
\frac{\alpha_s}{m_J^2}
\bigg[ \sum_{k=0}^\infty (\alpha_s L^2)^k + \ldots \bigg]
\,,
\end{align}
where $L$ denotes a generic large logarithm in the SDOE region (which will be determined in \sec{coherentbranching}).
The displayed terms are at LL order, while the ellipses denote terms at higher orders in the resumed perturbation theory.

In \sec{coherentbranching} we will show that, in fact, the LL series for each of ${d \Delta\hat \sigma^{\figeight}_\kappa}/{d m_J^2}$ and ${d \Delta\hat\sigma^{\bndry}_\kappa}/{d m_J^2}$ can be related to the LL series in the leading power cross section, ${d \hat \sigma_\kappa}/{d m_J^2}$.
This simplification can be expressed by rewriting the hadronic cross section with leading power corrections in the form:
\begin{align}
\label{eq:sigfullk}
\frac{d \sigma^{\rm had}_\kappa}{d m_J^2}
&= \frac{d \hat \sigma_\kappa}{d m_J^2} - Q\, \Omega_{1\kappa}^{\figeight} \, \frac{d}{d m_J^2} \bigg(C^\kappa_1(m_J^2, Q, \tilde z_{\rm cut}, \beta, R) \, \frac{d \hat \sigma_\kappa}{d m_J^2}\bigg) + \frac{Q\Upsilon_{1}^\kappa (\beta)}{m_J^2} \, C^\kappa_2(m_J^2, Q, \, \tilde z_{\rm cut}, \beta, R) \, \frac{d \hat \sigma_\kappa}{d m_J^2} \,.
\end{align}
Where we have introduced two additional functions $C_1^\kappa(m_J^2, Q, \tilde z_{\rm cut}, \beta, R)$ and $C_2^\kappa(m_J^2, Q, \tilde z_{\rm cut}, \beta, R)$ which we will refer to as the Wilson coefficients for the shift and the boundary corrections respectively. Their arguments reflect the fact that they are not constants along the spectrum, and that they depend the grooming parameters. When there is no cause for confusion we will suppress the $\{Q, \tilde z_{\rm cut}, \beta, R\}$ arguments for simplicity. Hence, the correction to the normalization and the shape of the partonic spectrum is given by
\begin{align}
\label{eq:sigsplit}
\frac{d \sigma^{\rm had}_\kappa}{d m_J^2}
&=\bigg[1 - Q\, \Omega_{1\kappa}^{\figeight}
\frac{d C_1^\kappa(m_J^2)}{d m_J^2} +
\frac{Q\Upsilon_{1}^\kappa(\beta)}{m_J^2} \, C_2^\kappa(m_J^2)\bigg]
\frac{d \hat \sigma_\kappa}{d m_J^2} - Q\, \Omega_{1\kappa}^{\figeight} C_1^\kappa(m_J^2) \frac{d}{d m_J^2} \, \frac{d \hat \sigma_\kappa}{d m_J^2} \, .
\end{align}
Equation~(\ref{eq:sigfullk}), or equivalently \eq{sigsplit}, are our main results for the operator expansion with leading power corrections in the SDOE region.

In the next section we will show that a leading double logarithmic series is absent from $C_1^\kappa$ and $C_2^\kappa$ when defined as in \eq{sigfullk}, so that their parametric forms include terms
\begin{align} \label{eq:C1C2scaling}
C_1^\kappa \sim C_2^\kappa
\sim \biggl( \frac{m_J^2}{Q \qcut}\biggr)^{\frac{1}{2+\beta}}
\frac{1}{L}
\bigg[ 1+ \sum_{k=1}^\infty (\alpha_s L)^k + \ldots \bigg] \,.
\end{align}
Here we work at LL order with a running coupling, and hence only single logarithms from the running of $\alpha_s(\mu)$ are included.
The determination of the full set of terms $\sum_{k=1} (\alpha_s L)^k$ requires a NLL calculation for ${d \Delta\hat \sigma^{\figeight}_\kappa}/{d m_J^2}$ and ${d \Delta\hat\sigma^{\bndry}_\kappa}/{d m_J^2}$ which we leave to future work. We remind the reader that the factors in \eq{C1C2scaling} are such that in the $pp$ case the combination of factors of $Q$ and coefficients $C_{1,2}^\kappa$ yield $\cosh\eta_J$ independent results for these power corrections, see \eq{indepcosheta}.

It can be convenient to express the shift power correction as the moment of a one dimensional distribution:
\begin{align}
\label{eq:moments}
\Omega_{1\kappa}^{\figeight} &= \int_0^\infty d k \>k\, \tilde F_\kappa^{\, \circ \! \! \circ} (k) \,,
\end{align}
where
\begin{align}
\label{eq:shiftshape}
\tilde F_\kappa^{\, \circ \! \! \circ} (k' ) &\equiv \int d^d k \>
\delta ( k' - k^+) \, \tilde F_\kappa (k^\mu) \, \overline \Theta_{\rm NP}^{\, \figeight} \Big(\frac{k_\perp}{k^-}, \, 1, \, \phi_k \Big) \, .
\end{align}
Interestingly, \eq{shiftshape} implies that $\Omega_1^{\figeight}$ in \eq{moments} is effectively obtained from an unnormalized distribution, which differs completely from the case of ungroomed event shapes:
\begin{align}
\int_0^\infty d k \>\tilde F_\kappa^{\,\figeight} (k) = f_0 \equiv \int d^d k \> \tilde F_\kappa (k^\mu)\, \overline \Theta_{\rm NP}^{\, \figeight} \Big(\frac{k_\perp}{k^-}, \, 1, \, \phi_k \Big) \, .
\end{align}
At this order we can reexpress the results in terms of a normalized distribution $F_\kappa^{\, \figeight}(q')$ defined as
\begin{align}
F_\kappa^{\, \figeight}(q') &\equiv \frac{1}{f_0^2} \tilde F_{\kappa}^{\, \figeight} \Big(\frac{q'}{f_0}\Big)= \frac{1}{f_0^2} \int d^d k\>\ \tilde F_\kappa(k)\, \delta \Big(\frac{q'}{f_0} - k^+\Big) \, \overline \Theta_{\rm NP}^{\, \figeight} \Big(\frac{k_\perp}{k^-}, \, 1, \, \phi_k \Big) \, ,
\end{align}
such that
\begin{align}
\label{eq:moments2}
\int_0^\infty d q' \> F_{\kappa}^{\, \figeight} (q') = 1 \,, \qquad \int_0^\infty d q' \> q' F_\kappa^{\,\figeight} (q') = \Omega_{1\kappa}^{\figeight} \, .
\end{align}
Thus using \eqs{sigfullk}{moments2} we arrive at an expression of the hadronic cross section valid for terms up to the leading nonperturbative corrections:
\begin{align}
\label{eq:factnonpert}
\frac{d \sigma^{\rm had}_\kappa}{d m_J^2} &= \int_0^\infty d k \, \frac{d \hat \sigma_\kappa}{d m_J^2}\big(m_J^2 - Q \, k \, C_1^\kappa(m_J^2) \big) \\
&\qquad \times \bigg[1 - Q\,k \, \frac{d C_1^\kappa(m_J^2)}{d m_J^2} + \frac{Q\Upsilon_{1}^\kappa(\beta)}{m_J^2} \, C_2^\kappa(m_J^2)\bigg] \, F_\kappa^{\,\figeight}(k) \, .\nn
\end{align}
Written in this manner we see more explicitly that the $C_1^\kappa$ term in the first line acts like a jet mass dependent shift, while the $C_1^\kappa$ and $C_2^\kappa$ in the second line contribute to a jet mass dependent normalization correction.
This form may be more useful than \eq{sigfullk} as it allows the higher order power corrections to be modeled by higher order moments of $F_\kappa^{\figeight}(k)$.

Combining the nonperturbative factorization in \eq{factnonpert} with the partonic factorization theorem in \eq{masslessFact0} we obtain
\begin{align}
\label{eq:factfull}
&\frac{d \sigma^{\rm had}}{dm_J^2}
= \sum_{\kappa={q,g}} N_\kappa(\Phi_J, R,\tilde z_{\rm cut}, \beta,\mu)\, \qcut^{\frac{1}{1+\beta}}
\int_0^\infty\!\! d\ell^+ \int_0^\infty \! \! d k \: J_\kappa \big(m_J^2 - Q\,\ell^+ ,\, \mu \big) \\
&\qquad \times S_c^\kappa \Big[\big( \ell^+ - C_1^\kappa(m_J^2)\,k \big) \qcut^{\frac{1}{1+\beta}},\beta,\mu\Big] \, \bigg(1 - Q\,k \, \frac{d C_1^\kappa(m_J^2)}{d m_J^2} + \frac{Q\Upsilon_{1}^\kappa(\beta)}{m_J^2} \, C_2^\kappa(m_J^2)\bigg)\, F_\kappa^{\,\figeight}(k) \nn
\, .
\end{align}
Note that \eq{factfull} contains a nontrivial form of convolution between the collinear-soft function $S_c$ and the normalized shape function $F_\kappa^{\,\figeight}$ due to the appearance of the Wilson coefficient $C_1^\kappa(m_J^2)$. This encapsulates the effect of the jet mass dependent NP catchment area. In contrast, for an ungroomed event shape, such as thrust or a hemisphere mass, the dijet region receives leading power NP corrections that represent a constant jet mass independent shift to the spectrum. Furthermore, for the leading power corrections there is also no analog of the $m_J$ dependent normalization corrections.

\section{Resummation for Matching Coefficients for Hadronic Corrections}
\label{sec:coherentbranching}

The goal of this section is to calculate the perturbative Wilson coefficients $C_1^\kappa$ and $C_2^\kappa$ in \eq{sigfullk}, and demonstrate that they do not contain LL double logarithms. We make use of a combination of the coherent branching formalism~\cite{Catani:1992ua} and input about the nature of the expansion from SCET. Since $C_1^\kappa$ and $C_2^\kappa$ are properties of the groomed jet, it suffices to calculate them for quark dijets from an $e^+e^-$ collision, with the extension to gluon jets obtained by a simple replacement. We will also quote the corresponding results for $pp$ collisions.

In what follows, we first review the derivation of the partonic resummation formula for the groomed jet mass using the coherent branching formalism presented in \Refscite{Dasgupta:2013ihk,Larkoski:2014wba}. We then make use of the key results derived from the EFT analysis above to setup a coherent branching calculation of the Wilson coefficients in \eq{sigshiftbndry}.
In coherent branching, the resummation is implemented via a sum over real emissions, where one can, in analogy to a coherent branching parton shower, track the kinematic information of the sequence of emissions, making the calculation quite intuitive.
This novel coherent branching calculation corresponds to carrying out a resummation for an observable (here $m_J^2$) while simultaneously weighting its phase space by a function of another observable (the stopping angle, $\theta_{cs}$).
Schematically, the calculation of $C^\kappa_1$ and $C_2^\kappa$ corresponds to evaluating the following resummed averages of the powers of the soft drop stopping angle:
\begin{align}
\label{eq:C1C2sim}
C_1^\kappa(m_J^2) & \sim
\left \langle \frac{\theta_{cs}(m_J^2)}{2}\right \rangle \, ,
\qquad
C_2^\kappa(m_J^2) \sim
\left\langle\frac{2}{\theta_{cs}(m_J^2)}\, \frac{m_J^2}{Q^2}\, \delta \big ( z_{cs} - \tzcut \theta_{cs}^\beta \big) \right\rangle
\, .
\end{align}

We note that it is not obvious how to define this result from coherent branching alone, since coherent branching does not provide the power expansion or the formulation of fields needed for describing non-perturbative corrections. On the other hand, at LL order in resummed perturbation theory, using the coherent branching formalism is simpler than setting up the necessary (novel) SCET formalism for the resummation of large logarithms.

\subsection{Review of Parton Level Resummation in Coherent Branching}
\label{sec:partonic}

We start with a series of angular-ordered emissions off an energetic parton. These are being clustered in a jet of radius $R$, with the previous emissions off the parton being at wider angles, $\theta_1 > \theta_2 > \ldots > \theta_n$. Subsequently the radiation is groomed. We now assume that the CA clustering proceeds such that, at each step, an emission is paired with the central collinear subjet, and not with another emission. Thus at every stage of unclustering we will recover the emissions in the order they were emitted~\cite{Catani:1992ua,Dokshitzer:1998kz,Banfi:2004yd,Banfi:2014sua}.
It is useful to define
\begin{align}
\tilde \theta_i = \frac{\theta_i}{R} \,,
\end{align}
which satisfies $\tilde\theta_i\le 1$. We also define $z_i$ as the energy fraction of the $i$'th emission with respect to the jet's energy.
At this point we replace the angular ordering with ordering in the variable~\cite{Dasgupta:2013ihk,Marzani:2017mva}:
\begin{align}
\label{eq:rhoi}
\rho_i = z_i\,\tilde \theta_i^2 \,,
\qquad
\rho_1 > \rho_2 > \ldots > \rho_n \,.
\end{align}
Using this chain of ordered emissions we review the known resummation for the parton level jet mass cross section in this section. This provides the basis for the calculation of the resummed expressions for the Wilson coefficients $C_1(m_J^2)$ and $C_2(m_J^2)$ discussed in \sec{resumC1C2}.

In the collinear-soft and soft limit the contribution to the jet mass from the $i$'th emission is
\begin{align}
\Delta m_{Ji}^2 &= Q p_i^+ = \frac{1}{4} \rho_iR^2Q^2
\,,
\end{align}
Thus we see that the ordering in $\rho_i$ is equivalent to the ordering in the contribution to the jet mass of each emission carried away from the collinear parton.
For simplicity we consider a quark jet for our discussion, noting that the result for gluon jets at NLL is simply obtained via a substitution for the color factor $C_F \rightarrow C_A$ and the splitting function $p_{gq}(z) \to p_{gg}(z)$. We will therefore frequently suppress the $\kappa = q$ subscript in the following.
For convenience, we also use a short hand notation for the single emission phase space and matrix element:
\begin{align}
\label{eq:omegai}
\int d^2 \omega_i \equiv \int_0^1 d z_i \int_0^1 \frac{d\tilde\theta_i^2}{\tilde\theta_i^2} \> \frac{\alpha_s (z_i
\tilde\theta_iR Q/2 ) C_F}{\pi} p_{gq} (z_i) \, .
\end{align}
Here $p_{gq}(z)$ is the one-loop collinear splitting function which reads
\begin{align}
\label{eq:pgq}
p_{gq} (z) = \frac{1 + (1-z)^2 }{2z} \, .
\end{align}
We also include a running coupling in \eq{omegai} as a part of the LL resummation. Here the running coupling is evaluated at the scale of the $\perp$ momentum of the emission with respect to the jet axis.
At LL accuracy we can limit ourselves to the case that the emission with the largest $\rho_n$ that passes soft drop also sets the value of measured jet mass $m_J^2$.
Emissions with $\rho_i < \rho_n$ (for $i > n$) that are kept can then be considered unresolved. Thus the ordering in $\rho_i$ ensures that there is a natural upper cutoff for the unresolved emissions. Further, in the LL approximation the $\rho_i$ can be assumed to be strongly ordered, with the inequality `$>$' in \eq{rhoi} replaced by strong inequality `$\gg$'. We also can ignore the perturbative radiation off the emitted gluons.
If the stopping pair that sets the value of the groomed jet mass $m_J$ is found after $n$ unclusterings then the normalized perturbative differential cross section is given by
\begin{align}
\label{eq:partonic}
\frac{1}{\hat \sigma} \frac{d\hat \sigma}{d m_J^2} &= \delta(m_J^2) + \sum_{n = 1}^{\infty} \bigg[ \prod_{i = 1}^{n} \int d^2\! \omega_i \bigg] \,
\Big\{\overline \Theta_{\rm sd}^{\, n}\, \delta\Big(m_J^2 - \frac{1}{4}\rho_n R^2Q^2 \Big) + \Theta_{\rm sd}^{\, n}\,\delta (m_J^2) - \delta (m_J^2) \Big \}
\nn \\
&\qquad\qquad\qquad\ \times
\prod_{j = 1}^{n-1} \big\{ \Theta_{\rm sd}^{\, j} - 1 \big\} \Theta(\rho_{j}-\rho_{j+1}) \, ,
\end{align}
where $\overline \Theta_{\rm sd}^{\, i}$ and $\Theta_{\rm sd}^{\, i}$ are soft drop passing and failing conditions for the $i^{\rm th}$ subjet:
\begin{align}
\overline \Theta_{\rm sd}^{\, i} = \Theta \big (z_i - \tilde z_{\rm cut}( \tilde \theta_iR)^\beta \big) \, , \qquad \Theta_{\rm sd}^{\, i} = 1 - \overline \Theta_{\rm sd}^{\, i} \, .
\end{align}
The $-\delta(m_J^2)$ and $-1$ terms in \eq{partonic} correspond to the virtual contributions in the passing and failing subjets respectively and the terms $ \Theta(\rho_{i}-\rho_{i+1}) $ impose the ordering. The term with $\Theta_{\rm sd}^{\, n}$ represents a unique scenario where all the $n$ emissions in the jet fail soft drop and a zero contribution to the jet mass $m_J^2$ is obtained. The scenario where there is yet another emission after the $n^{\rm th}$ one is accounted for by the $(n+1)^{\rm th}$ term in the sum, and so on. Hence, there is no double counting in the resummation formula in \eq{partonic}.

It is convenient to work with the cumulant of the cross section defined by
\begin{align}
\hat \Sigma (m_J^2) = \int_{0}^{\infty} d m_J^{\prime 2} \, \frac{1}{\hat \sigma} \frac{d\hat \sigma}{dm_J^{\prime 2}}\, \Theta\big (m_J^2 - m_J^{\prime 2}\big ) \, ,
\end{align}
which leads to exponentiation for the emissions:
\begin{align} \label{eq:cumulant}
\hat \Sigma(m_J^2)
&= \Theta(m_J^2) + \Theta (m_J^2) \sum_{n = 1}^{\infty}
\bigg[ \prod_{i = 1}^{n} \int d^2\! \omega_i \bigg] \,
(-\overline \Theta_{\rm sd}^{\, n})\, \Theta \Big(\frac{1}{4}\rho_n R^2Q^2-m_J^2 \Big)
\prod_{i = 1}^{n-1}\,
\bigl(- \overline \Theta_{\rm sd}^{\, i} \bigr)\, \Theta(\rho_{i}-\rho_{i+1}) \,
\nn \\
&= \Theta(m_J^2) \exp\bigg[-{\cal R}_q\Big(\frac{4m_J^2}{R^2Q^2},RQ,\tilde z_{\rm cut}R^\beta,\beta\Big)\bigg] \, ,
\end{align}
where the radiator ${\cal R}_q$ for quark jets in the Sudakov factor reads
\begin{align}
\label{eq:radiator}
{\cal R}_q\big(\rho\big)
&= {\cal R}_q\big(\rho,RQ,\tilde z_{\rm cut} R^\beta,\beta\big) \\
&\equiv \int_{0}^{1}\! \frac{d\tilde \theta^{\prime 2}}{\tilde\theta^{\prime 2}} \int_0^1 \!\!d z^\prime \> \frac{\alpha_s(z^\prime\tilde \theta^\prime R Q/2)C_F }{\pi} \, p_{gq} (z^\prime )
\Theta \Big(z^\prime - \tilde z_{\rm cut} \big(\tilde\theta^\prime R\big)^\beta\Big) \
\Theta\Big(z^\prime \tilde\theta^{\prime\,2} - \rho\Big)\, \nn
\end{align}
Note that the radiator and cumulant only depend on $R$ in combinations with other variables. The cumulant requires $m_J^2 \geq 0$ and we will leave the $ \Theta(m_J^2)$ implicit below. For the gluon jet radiator ${\cal R}_g$ we can replace $C_F p_{gq(z')}\to C_A p_{gg(z')}$ to capture the leading logarithms. Lastly, the exponentiation in \eq{cumulant} resulted from the replacement of the angular ordering of the emissions by $\rho_i$ ordering in \eq{rhoi}, which yields the same constraint on the $n^{\rm th}$ emission as the previous ones.

The result for the differential jet mass distribution in $e^+e^-$ annihilation then reads
\begin{align}
\label{eq:sigpert}
\frac{1}{\hat \sigma} \frac{d\hat \sigma}{d m_J^2}
= \frac{d}{dm_J^2} \hat \Sigma(m_J^2)
&= \frac{1}{m_J^2} \,C^q_0\big (m_J^2,RQ,\tilde z_{\rm cut} R^\beta,\beta\big) \, , \qquad m_J^2 > 0
\end{align}
where we find it convenient to work with the result in terms of $C_0^q$ defined as
\begin{align}
\label{eq:C0}
C_0^q(m_J^2) &= C_0^q\big (m_J^2,RQ,\tilde z_{\rm cut} R^\beta,\beta\big)
\\
&
\equiv \exp\bigg[-{\cal R}_q\Bigl(\frac{4m_J^2}{R^2Q^2}\Bigr)\bigg]
\int_{0}^{1} \frac{d\tilde \theta^2}{\tilde \theta^2}\: \alpha_s\Big(\frac{2\,m_J^2}{\tilde \theta RQ} \Big)\frac{C_F}{\pi} \, \frac{4\,m_J^2}{\tilde \theta^2R^2 Q^2 } \: p_{gq}\Big(\frac{4\,m_J^2}{\tilde \theta^2R^2 Q^2 }\Big)\nn \\
&\qquad \times \Theta \Bigl(\tilde \theta -\frac{2\,m_J}{RQ} \Bigr) \, \Theta\Bigl(\tilde \theta^\star(m_J^2) - \tilde \theta \Bigr), \nn
\end{align}
where $\tilde \theta^\star$ is given by
\begin{align}
\label{eq:thetastar}
\tilde \theta^\star(m_J^2) =
\tilde \theta^\star(m_J^2, RQ, \tilde z_{\rm cut} R^\beta, \beta )
\equiv \frac{2}{R}\,\bigg(\frac{ m_J^2}{ Q \qcut}\bigg)^{\frac{1}{2 + \beta}} = \frac{2}{R} \, \zeta_{cs} \,.
\end{align}
Note that to simplify the notation we have here defined $C_0^q$, ${\cal R}_q$ and $\tilde \theta^\star$ with abbreviated arguments, and we continue to use this notation below.
The $\zeta_{cs}$ is defined in \eq{pCS} and appeared in the definition of CS mode, which captures the scaling of the softest subjet that satisfies the soft drop condition. We see that in \eq{C0} the angle of the stopping subjet lies between the angle of the collinear modes and the collinear-soft modes:
\begin{align}
\label{eq:thetastop}
\tilde \theta_{c}= \frac{\theta_c}{R} = \frac{2m_J}{RQ} \leq \tilde\theta \leq \min\Big\{\tilde \theta^\star(m_J^2) , 1\Big\} \, ,
\end{align}
where the minimum condition accounts for the transition to the ungroomed resummation region when $\tilde \theta^\star(m_J^2) \sim 1$.

For a quark jet from a $pp$ collision with $R\lesssim 1$, the angles are cutoff by $\theta_i\le R/\cosh\eta_J$, so we would define $\tilde\theta_i = \theta_i \cosh\eta_J/R$ to have $0\le \tilde\theta_i\le 1$. Repeating the analysis for this case we obtain
\begin{align}
\label{eq:C0pp}
C^{q(pp)}_0(m_J^2) &= C^{q(pp)}_0\big (m_J^2,Rp_T, z'_{\rm cut} R^\beta,\beta\big)
\\
&
\equiv \exp\bigg[-{\cal R}_q^{pp}\bigl(\frac{m_J^2}{R^2p_T^2},Rp_T, z'_{\rm cut}R^\beta,\beta\bigr)\bigg]
\nn \\
&\qquad \times \int_{0}^{1} \frac{d\tilde \theta^2}{\tilde \theta^2}\: \alpha_s\Big(\frac{m_J^2}{\tilde \theta Rp_T} \Big)\frac{C_F}{\pi} \, \frac{m_J^2}{\tilde \theta^2R^2 p_T^2 } \: p_{gq}\Big(\frac{m_J^2}{\tilde \theta^2R^2 p_T^2 }\Big)\Theta \Bigl(\tilde \theta -\frac{m_J}{Rp_T} \Bigr) \, \Theta\Bigl(\tilde \theta_{pp}^\star(m_J^2) - \tilde \theta \Bigr), \nn
\end{align}
where $\tilde \theta^\star$ is now given by
\begin{align}
\label{eq:thetastarpp}
\tilde \theta_{pp}^\star(m_J^2) =
\tilde \theta^\star(m_J^2, Rp_T, z'_{\rm cut} R^\beta, \beta )
\equiv \frac{2\cosh\eta_J}{R}\,\bigg(\frac{ m_J^2}{ Q \qcut}\bigg)^{\frac{1}{2 + \beta}}
= \frac{2\cosh\eta_J}{R} \, \zeta_{cs} \,.
\end{align}
Eq.~(\ref{eq:C0pp}) also involves the radiator for the $pp$ case which is
\begin{align}
\label{eq:radiatorpp}
{\cal R}_q^{pp}\big(\rho,Rp_T, z'_{\rm cut} R^\beta,\beta\big)
&= \int_{0}^{1}\! \frac{d\tilde \theta^{\prime 2}}{\tilde\theta^{\prime 2}} \int_0^1 \!\!d z^\prime \: \frac{\alpha_s(z^\prime\tilde \theta^\prime R p_T)C_F }{\pi} \, p_{gq} (z^\prime )
\Theta \Big(z^\prime - z'_{\rm cut} \big(\tilde\theta^\prime R\big)^\beta\Big)
\Theta\Big(z^\prime \tilde\theta^{\prime\,2} - \rho\Big) \,,
\end{align}
with
\begin{align}
\label{eq:zcutprime}
z'_{\rm cut} = z_{\rm cut}/R_0^\beta \, .
\end{align}
Note that with the rescaled variables the results for $C^{q(pp)}_0(m_J^2)$, $\tilde \theta^\star(m_J^2)$, and ${\cal R}_q^{pp}$ are all independent of $\cosh\eta_J$. In the case of $\theta^\star(m_J^2)$ the displayed $\cosh\eta_J$ dependence cancels with the implicit dependence contained in the $Q$ and $\qcut$, see \eq{indepcosheta}.

\subsection{Resummed Matching Coefficients $C_1$ and $C_2$ for Subleading Power}
\label{sec:resumC1C2}
We now turn to a calculation of the matching coefficients $C_1$ and $C_2$ of the leading power nonperturbative corrections, which we carry out at LL order.

\addtocontents{toc}{\protect\setcounter{tocdepth}{1}}
\subsubsection{Resummation for the Shift Power Correction}
\addtocontents{toc}{\protect\setcounter{tocdepth}{2}}
\label{sec:shift}

In \sec{measurement} we showed that the shift correction corresponds to the total NP radiation captured by the groomed jet area. From \eq{hatpboosted1} we see that this correction can be obtained by performing a measurement in an appropriately boosted frame and rescaling the result by $\theta_{cs}/2$, the opening angle of the soft drop stopping pair. From the analysis in \sec{npsource} we learned that the nonperturbative and perturbative matrix elements can be factored by boosting the nonperturbative momentum to an appropriate reference frame, see \eq{rescaling}. Using these results the jet mass cumulant including the shift correction is given by
\begin{align}
\label{eq:shift01}
\hat \Sigma(m_J^2) &+ \Sigma^{\rm shift} (m_J^2)
= 1 + \int \frac{d^d k}{(2 \pi)^d}\:\tilde F(k^\mu) \:
\sum_{n =1 }^{\infty} \bigg[ \prod_{i = 1}^{n} \int d^2\! \omega_i \bigg]
\: \\
& \times \Bigg\{ \overline \Theta_{\rm sd}^{\, n}\, \Theta\bigg(m_J^2 - \frac{1}{4} \rho_n R^2Q^{2} - RQ\,\frac{\tilde\theta_n}{2}\, k^+\, \overline \Theta_{\rm NP}^{\, \figeight} \Big(\frac{k_\perp}{k^-}, \, 1, \, \phi_k \Big) \bigg)+ \Theta_{\rm sd}^{\, n}\, \Theta(m_J^2) -\Theta(m_J^2)\Bigg \}\nn\\
& \times\> \prod_{i = 1}^{n-1} \big (- \overline \Theta_{\rm sd}^{\, i} \big)\, \Theta(\rho_{i}-\rho_{i+1})\, , \nn
\end{align}
In \eq{shift01} the measurement on the NP radiation is performed in the boosted frame with respect to the collinear-soft momentum of emission $n$, which corresponds to setting $\theta_{cs}$ to $\theta_n = R \tilde \theta_n$ in \eq{hatpboosted1}. The term $\tilde F(k^\mu)$ is the NP source function. Separating the perturbative and nonperturbative components we get
\begin{align}
\label{eq:shift03}
\Sigma^{\rm shift} (m_J^2 ) &= - RQ\, \Omega_1^{\figeight} \, \frac{d}{d m_J^2} \Sigma^{\figeight} (m_J^2) \, ,
\end{align}
where
\begin{align}
\label{eq:shift04}
\Sigma^{\figeight} (m_J^2) &= \sum_{n =1 }^{\infty} \bigg[ \prod_{i = 1}^{n} \int d^2\! \omega_i \bigg]\> \frac{ \tilde\theta_n}{2} \,(-\overline \Theta_{\rm sd}^{\, n})\, \Theta \Big(\frac{1}{4}\rho_n R^2Q^2-m_J^2 \Big) \, \prod_{i = 1}^{n-1}\> \bigl(- \overline \Theta_{\rm sd}^{\, i} \bigr)\, \Theta(\rho_{i}-\rho_{i+1})\, ,
\end{align}
and just as above in \eqs{O1Ab}{O1NA} we have identified
\begin{align}
\label{eq:omega1}
\Omega_{1}^{\figeight} &\equiv \int \frac{d^d k}{(2\pi)^d}\, k^+ \,\overline \Theta_{\rm NP}^{\, \figeight} \Big(\frac{k_\perp}{k^-}, \, 1, \, \phi_k \Big) \, \tilde F (k^\mu) \, .
\end{align}
The series in \eq{shift04} exponentiates as in \eq{cumulant} yielding
\begin{align}
\label{eq:shift05}
\Sigma^{\figeight} (m_J^2)&=-\int_{0}^1 d z \int_0^1 \frac{d\tilde \theta^2}{\tilde \theta^2}\, \frac{ \tilde\theta}{2} \, \frac{\alpha_s(z\, \tilde \theta\, R Q/2)C_F }{\pi} p_{gq} (z)\, \Theta \Big(z-\tilde z_{\rm cut} \big(\tilde\theta R\big)^\beta \Big) \\
&\qquad\qquad\times \, \Theta \Big( \frac{1}{4}z \tilde \theta^2 R^2 Q^2-m_J^2 \Big)e^{-{\cal R}_q (z \tilde \theta^2)}\, , \nn
\end{align}
Thus the shift correction to the differential cross section is given by
\begin{align}
\label{eq:sigshift}
\frac{1}{\hat \sigma} \frac{d \sigma^{\rm shift}}{d m_J^2}
&= \frac{d \Sigma^{\rm shift}}{d m_J^2}
= - R Q\, \Omega_1^{\figeight} \, \frac{d}{d m_J^2} \frac{d}{d m_J^2} \Sigma^{\figeight} (m_J^2) \, , \nn \\
&=
- RQ\, \Omega_1^{\figeight} \, \frac{d}{d m_J^2} \bigg[\tilde C_1^q \big (m_J^2, RQ, \tilde z_{\rm cut}R^\beta, \beta\big)\, \frac{1}{\hat \sigma}\frac{d\hat \sigma}{d m_J^2} \bigg] \, ,
\end{align}
which have rewritten in the desired form of \eq{sigfullk}. Then from \eqs{shift05}{sigpert} we can identify
\begin{align}
\label{eq:C1}
\tilde C_1^q \big (m_J^2, &RQ, \tilde z_{\rm cut}R^\beta, \beta\big)= \frac{m_J^2}{C_0^q(m_J^2)} \, \frac{d}{d m_J^2} \Sigma^{\figeight} (m_J^2) \\
&= \frac{e^{-{\cal R}_q\bigl(\frac{4 m_J^2}{R^2Q^2}\bigr) } }{C_0^q(m_J^2)}
\int_{\frac{4 m_J^2}{R^2Q^2}}^{1} \frac{d\tilde \theta^2}{\tilde \theta^2}\>\frac{\tilde \theta}{2}\,
\alpha_s\Big(\frac{2\,m_J^2}{\tilde \theta \, RQ} \Big)
\frac{C_F}{\pi} \,\frac{4\,m_J^2}{\tilde \theta^2 R^2Q^2}
\> p_{gq}\Big(\frac{4\,m_J^2}{\tilde \theta^2 R^2Q^2} \Big) \,
\Theta\big(\tilde \theta^\star(m_J^2) - \tilde \theta \big)
\nn \\
&= \frac{ \int_{\tilde\theta_c^2}^{1} \frac{d\tilde \theta^2}{\tilde \theta^2}\>
\frac{\tilde \theta}{2}\,
\alpha_s\Big(\frac{2\,m_J^2}{\tilde \theta \, RQ} \Big)
\frac{C_F}{\pi} \,\frac{4\,m_J^2}{\tilde \theta^2 R^2Q^2}
\> p_{gq}\Big(\frac{4\,m_J^2}{\tilde \theta^2 R^2Q^2} \Big) \,
\Theta\big(\tilde \theta^\star(m_J^2) - \tilde \theta \big) }
{ \int_{\tilde\theta_c^2}^{1}
\frac{d\tilde \theta^2}{\tilde \theta^2}\> \,
\alpha_s\Big(\frac{2\,m_J^2}{\tilde \theta \, RQ} \Big)
\frac{C_F}{\pi} \,\frac{4\,m_J^2}{\tilde \theta^2 R^2Q^2}
\> p_{gq}\Big(\frac{4\,m_J^2}{\tilde \theta^2 R^2Q^2} \Big) \,
\Theta\big(\tilde \theta^\star(m_J^2) - \tilde \theta \big) }
\equiv \frac{ \big\langle \frac{\tilde \theta}{2} \big\rangle(m_J^2) }{ \langle 1 \rangle (m_J^2)}
\,,\nn
\end{align}
where $\tilde \theta_c = 2m_J/(QR)$.
Thus the Wilson coefficient $C_1^q$ in \eq{sigfullk} for the $e^+e^-$ case is given by
\begin{align}
\label{eq:C1R}
C_1^q(m_J^2, Q, \tilde z_{\rm cut}, \beta, R)
= R \, \tilde C_1^q \big (m_J^2, RQ, \tilde z_{\rm cut}R^\beta, \beta\big)
\, .
\end{align}
Note that in the ratio $e^{-{\cal R}_q}/C_0^q$ of the Sudakov exponential and the leading power cross section, all the double logarithmic terms cancel out.

From \eq{C1} we see that $C_1$ is given by the average resummed opening angle of the stopping pair that passes soft drop at a given value of $m_J$. Our result is similar to the average of the groomed jet radius, $R_g$, calculated in \Refcite{Larkoski:2015lea}, except that the result is evaluated at a fixed value of $m_J$ that sets the range of possible values for $\theta_{\rm stop}$ in \eq{thetastop}. We find that a rough approximation is
\begin{align}
\label{eq:C1approx}
& C_1^q (m_J^2, Q, \tilde z_{\rm cut}, \beta, R) \sim 0.66 \, \zeta_{cs} \, .
\end{align}
This demonstrates the validity of the power counting estimate of the opening angle of collinear soft emissions given in \eq{pCS}. Interestingly, the result is approximately $R$ independent in the SDOE region.

We now give the corresponding result for $C_1^q(m_J^2)$ for a $pp$ collider. Here we find
\begin{align}
\tilde C^{q(pp)}_1 \big (m_J^2, & Rp_T, \zcut^\prime R^\beta, \beta\big) = \frac{1}{C^{q(pp)}_0(m_J^2)}\exp\bigg[-{\cal R}_q^{pp}\bigl(\frac{m_J^2}{R^2p_T^2},Rp_T, z'_{\rm cut}R^\beta,\beta\bigr)\bigg] \\
& \times \int_{0}^{1} \frac{d\tilde \theta^2}{\tilde \theta^2}\>\frac{\tilde \theta}{2}\,
\alpha_s\Big(\frac{m_J^2}{\tilde \theta \, Rp_T} \Big)
\frac{C_F}{\pi} \,\frac{m_J^2}{\tilde \theta^2 R^2p_T^2}
\> p_{gq}\Big(\frac{m_J^2}{\tilde \theta^2 R^2p_T^2} \Big) \,\Theta \Bigl(\tilde \theta -\frac{m_J}{Rp_T} \Bigr) \,
\Theta\big(\tilde \theta_{pp}^\star(m_J^2) - \tilde \theta \big)\nn \,,
\end{align}
such that
\begin{align} \label{eq:C1pp}
C_1^{q(pp)}(m_J^2, Q, \tilde z_{\rm cut}, \beta, R)
= \frac{R}{\cosh \eta_J} \, \tilde C_1^{q(pp)} \big (m_J^2, Rp_T, z_{\rm cut}'R^\beta, \beta\big)
\, ,
\end{align}
where $\zcut'$ and $\theta^\star_{pp}(m_J^2)$ are given above in \eqs{zcutprime}{thetastarpp}. Here a rough approximation is
\begin{align}
\label{eq:C1ppapprox}
& C_1^{q(pp)}(m_J^2, Q, \tilde z_{\rm cut}, \beta, R)
\sim \frac{0.66}{2\cosh \eta_J} \biggl( \frac{m_J^2}{p_T^2\, \zcut R_0^{-\beta}} \biggr)^{\frac{1}{2+ \beta}}
\,.
\end{align}

\addtocontents{toc}{\protect\setcounter{tocdepth}{1}}
\subsubsection{Resummation for the Boundary Power Correction}
\addtocontents{toc}{\protect\setcounter{tocdepth}{2}}
\label{sec:boundary}

We now turn to the boundary power correction that originates from the fact that the soft drop comparison for the collinear-soft subjet receives a correction given by \eq{measurement6}. The perturbative cumulant with these corrections then reads
\begin{align}
\label{eq:bndry01}
\hat \Sigma (m_J^2) &+ \Sigma^{\rm bndry} (m_J^2) = 1 + \int \frac{d^d k}{(2 \pi)^d}\> \tilde F(k^\mu) \> \sum_{n =1 }^{\infty}\Bigg[ \prod_{i = 1}^{n} \int d^2 \omega_i \Bigg] \bigg[ -\Theta \Big(\frac{1}{4} \rho_n R^2 Q^2-m_J^2 \Big) \bigg]\, \nn \\
&\qquad \qquad\times \big [\overline \Theta_{\rm sd}^{\, n} + \Delta \overline \Theta_{\rm sd}^{\, n}\big] \,\prod_{i = 1}^{n-1} \> \Theta(\rho_{i}-\rho_{i+1})\, \big[-\overline \Theta_{\rm sd}^{\, i} \big ] \, ,
\end{align}
where $\Delta \overline \Theta_{\rm sd}^n$ is the correction in the soft drop test for the stopping subjet $n$ stated in \eq{measurement6}. Performing the change of variables in \eq{rescaling} we can again factorize the measurement such that
\begin{align}
\label{eq:deltaSD}
\int \frac{d^d k}{(2 \pi)^d}\> \tilde F(k^\mu) \> \Delta \overline \Theta_{\rm sd}^{\, n} = \frac{2}{\tilde \theta_n R}\, \delta \Big ( z_{n} - \tilde z_{\rm cut} \big(\tilde \theta_{n} R\big)^\beta \Big)\,\frac{ \Upsilon_{1}(\beta)}{Q} \, ,
\end{align}
where just as \eqs{Ups1Ab}{Ups1NA} we identify
\begin{align}
\label{eq:upsbeta}
\Upsilon_{1}(\beta) &= \int \frac{d^d k}{(2\pi)^d}\, \Big (k^- - \beta \, \big (k^- - k_\perp \, \cos( \phi_k)\big) \Big) \\
&\qquad \times \bigg(\overline \Theta_{\rm NP}^{\, \bndry} \Big(\frac{k_\perp}{k^-}, \, 1, \, \phi_k \Big) - \Theta_{\rm NP}^{\, \bndry} \Big(\frac{k_\perp}{k^-}, \, 1, \, \phi_k \Big)\bigg) \,\, \tilde F (k^\mu)
\nn\\
&= \Upsilon_{1,0} + \beta \, \Upsilon_{1,1}
\,.\nn
\end{align}
We note again that $\Upsilon_1(\beta)$ is linear in $\beta$. The power correction can be factorized from the perturbative component as:
\begin{align}
\label{eq:bndry02}
\Sigma^{\rm bndry} (m_J^2)
= \frac{\Upsilon_1(\beta)}{RQ} \, \Sigma^{\bndry}(m_J^2) \, ,
\end{align}
where
\begin{align}
\label{eq:bndry03}
\Sigma^{\bndry} (m_J^2) &=\int_{0}^1 d z \int_0^1 \frac{d\tilde \theta^2}{\tilde \theta^2} \, \frac{\alpha_s(z\, \tilde \theta\, R \, Q/2)C_F }{\pi} p_{gq} (z) \,\frac{2}{\tilde\theta}\, \delta \Big ( z - \tilde z_{\rm cut} \big(\tilde\theta R\big)^\beta \Big) \\
&\qquad\qquad\times \Big[ -\Theta \Big( \frac{1}{4}z \tilde \theta^2 R^2 Q^2-m_J^2 \Big) \Big] \, e^{-{\cal R}_q (z \tilde \theta^2)}\, . \nn
\end{align}
Comparing this result to \eq{sigfullk}
\begin{align}
\label{eq:sigbndry}
\frac{1}{\hat \sigma }\frac{d \sigma^{\rm bndry}}{d m_J^2}
= \frac{d \Sigma^{\rm bndry}}{d m_J^2}
= \frac{Q\Upsilon_1(\beta)}{m_J^2}\, C_2 (m_J^2, Q, \, \tilde z_{\rm cut}, \beta, R)\,
\frac{1}{\hat \sigma } \frac{d \hat \sigma}{d m_J^2} \, ,
\end{align}
allows us to identify $C_2$ coefficient for a quark jet in the $e^+e^-$ case as
\begin{align}
\label{eq:C2def}
C^q_2(m_J^2, Q, \, \tilde z_{\rm cut}, \beta, R) = \frac{1}{R} \frac{m_J^2}{Q^2} \,\tilde C^q_2(m_J^2, RQ, \tilde z_{\rm cut} R^\beta, \beta)
\,,
\end{align}
where $\tilde C_2^q$ is given by
\begin{align}
\label{eq:C2}
&\tilde C^q_2(m_J^2, RQ, \tilde z_{\rm cut}R^\beta, \beta) = \frac{m_J^2}{C_0^q(m_J^2)} \frac{d }{d m_J^2} \Sigma^{\bndry} (m_J^2) \\
&= \frac{e^{-{\cal R}_q\bigl(\frac{4 m_J^2}{R^2Q^2}\bigr) } }{C_0^q(m_J^2)} \
\frac{2}{2+\beta} \
\frac{2}{\tilde \theta^\star(m_J^2)} \
\alpha_s \bigg(\frac{1}{2}\,\tilde z_{\rm cut} R^\beta\, \tilde \theta^\star (m_J^2)^{\beta+1} \,RQ \bigg)
\frac{C_F}{\pi} p_{gq} \Big(\tilde z_{\rm cut} \,\big(\tilde \theta^\star(m_J) R \big)^\beta\Big) \nn \\
&\qquad \qquad \times \Theta\Big(1 - \tilde z_{\rm cut} \,\big(\tilde \theta^\star(m_J^2) R \big)^\beta \Big)\, \Theta\big(1 - \tilde \theta^\star(m_J)\big) \nn \\
&\equiv \frac{\langle 2/\tilde \theta\ \delta \big(z - \tzcut(R \tilde \theta)^\beta \big) \rangle(m_J^2)}{\langle 1 \rangle(m_J^2)}\nn \, .
\end{align}
Here $\tilde \theta^\star(m_J^2)$ is defined as in \eq{thetastar}. Thus we see that $\tilde C_2^q(m_J^2)$ is the average of $2/\tilde \theta$ evaluated at the boundary of soft drop at a given $m_J^2$.

The result in the $pp$ case can be computed in a similar manner, and is given by
\begin{align} \label{eq:C2pp}
C_2^{q(pp)}(m_J^2, Q, \tilde z_{\rm cut}, \beta, R)
&= \frac{R}{\cosh \eta_J} \frac{m_J^2}{4(Rp_T)^2}\,
\tilde C_2^{q(pp)} \big (m_J^2, Rp_T, z_{\rm cut}'R^\beta, \beta\big) \, ,
\end{align}
where $\tilde C_2^{q(pp) }$ reads
\begin{align}
\tilde C_2^{q(pp) } \big (m_J^2, Rp_T, &\zcut^\prime R^\beta, \beta\big) = \frac{1}{C^{q(pp)}_0(m_J^2)}\exp\bigg[-{\cal R}_q^{pp}\bigl(\frac{m_J^2}{R^2p_T^2},Rp_T, z'_{\rm cut}R^\beta,\beta\bigr)\bigg]
\nn\\
&\times\frac{2}{2+\beta}\ \frac{2}{\tilde \theta_{pp}^\star(m_J^2)}
\alpha_s \bigg(z_{\rm cut}^\prime R^\beta\, \tilde \theta_{pp}^\star (m_J^2)^{\beta+1} \,Rp_T\bigg) \frac{C_F}{\pi} p_{gq} \Big(z^\prime_{\rm cut} \,\big(\tilde \theta^\star(m_J) R \big)^\beta\Big) \nn \\
& \times \Theta\Big(1 - z^\prime_{\rm cut} \,\big(\tilde \theta_{pp}^\star(m_J^2) R \big)^\beta \Big)\, \Theta\big(1 - \tilde \theta^\star_{pp}(m_J)\big) \,,
\end{align}
with $\zcut'$ and $\theta^\star_{pp}(m_J^2)$ are given above in \eqs{zcutprime}{thetastarpp}.

\begin{figure}[t!]
\centering
\includegraphics[width=0.45\textwidth]{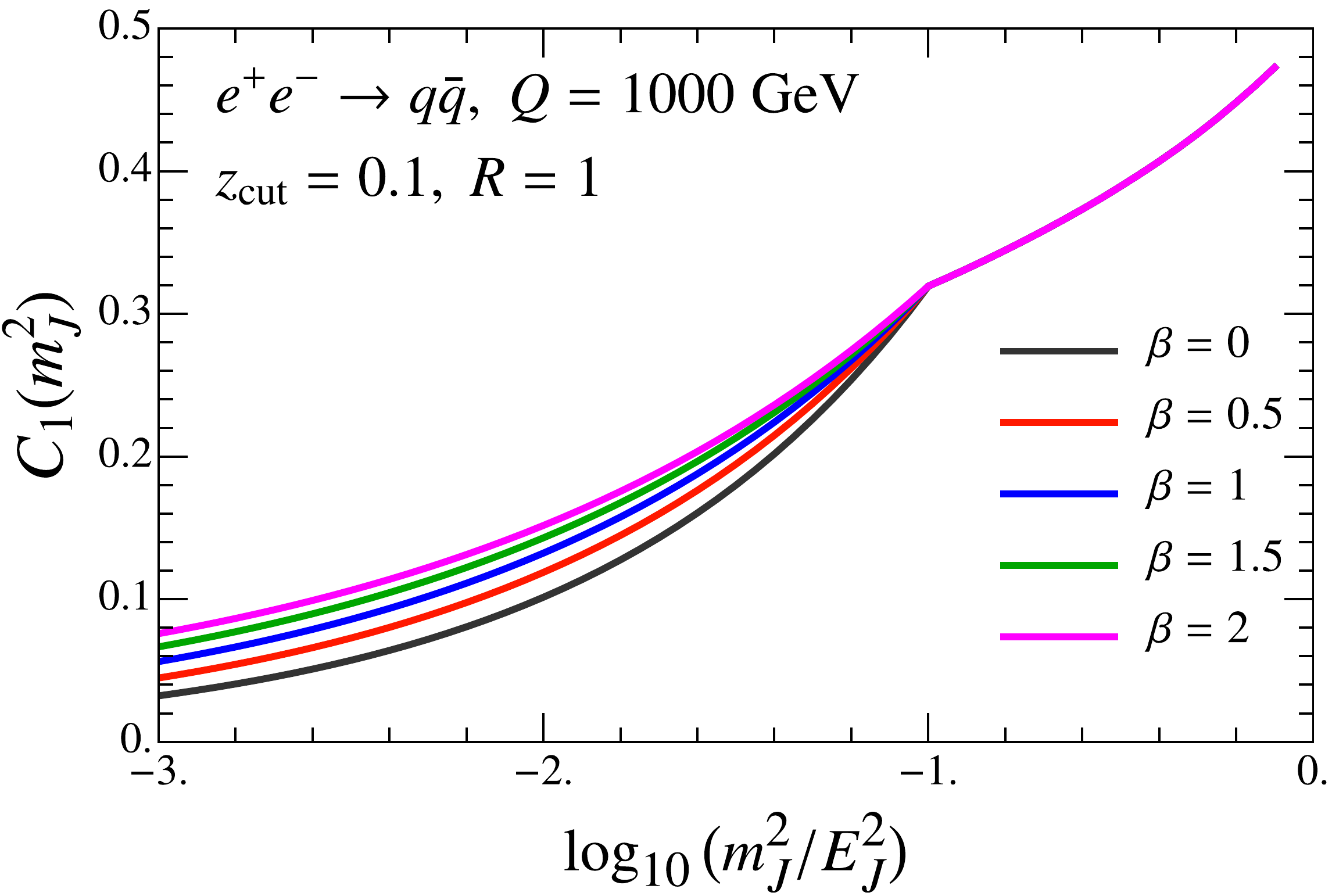}
\includegraphics[width=0.45\textwidth]{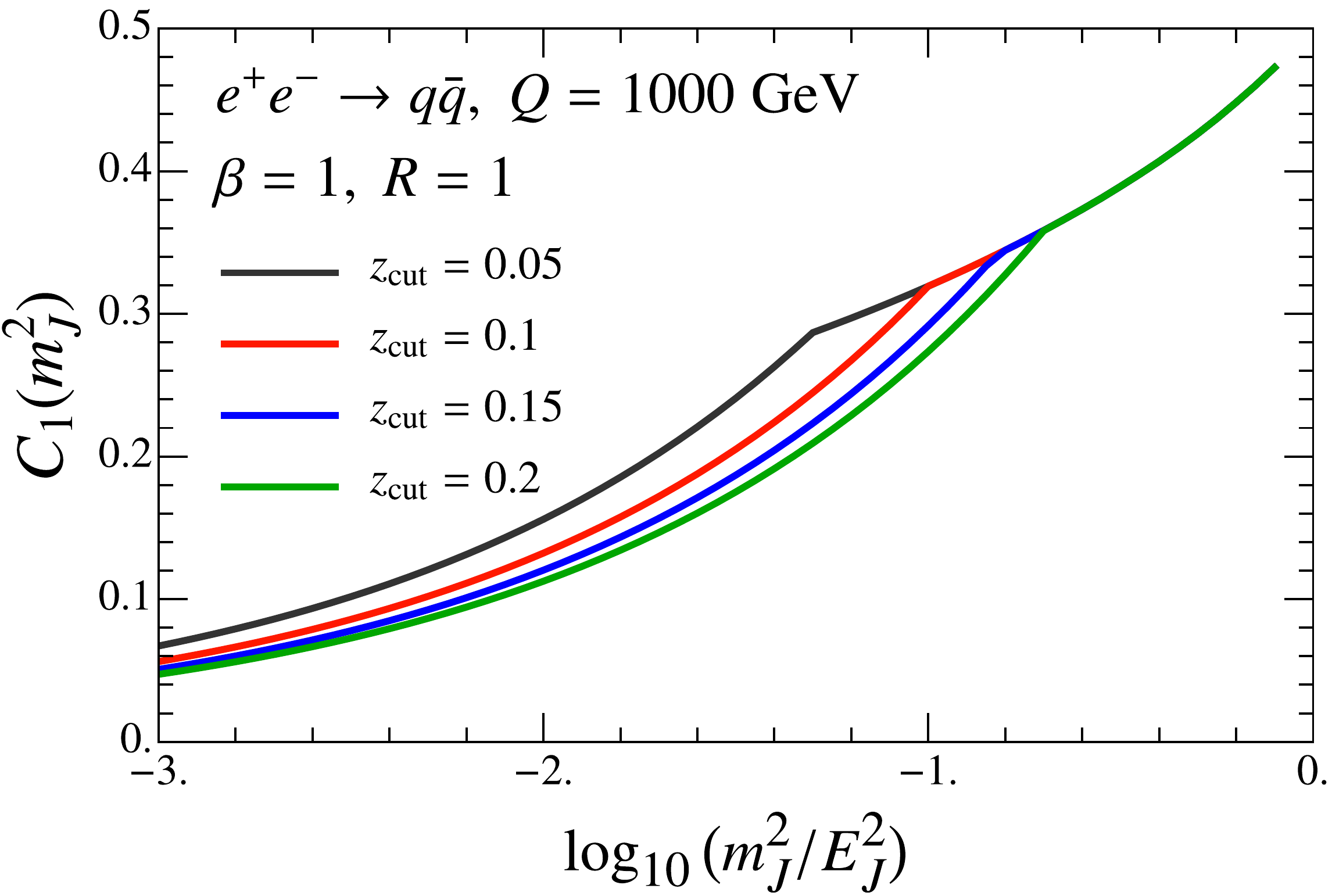}
\\
\includegraphics[width=0.45\textwidth]{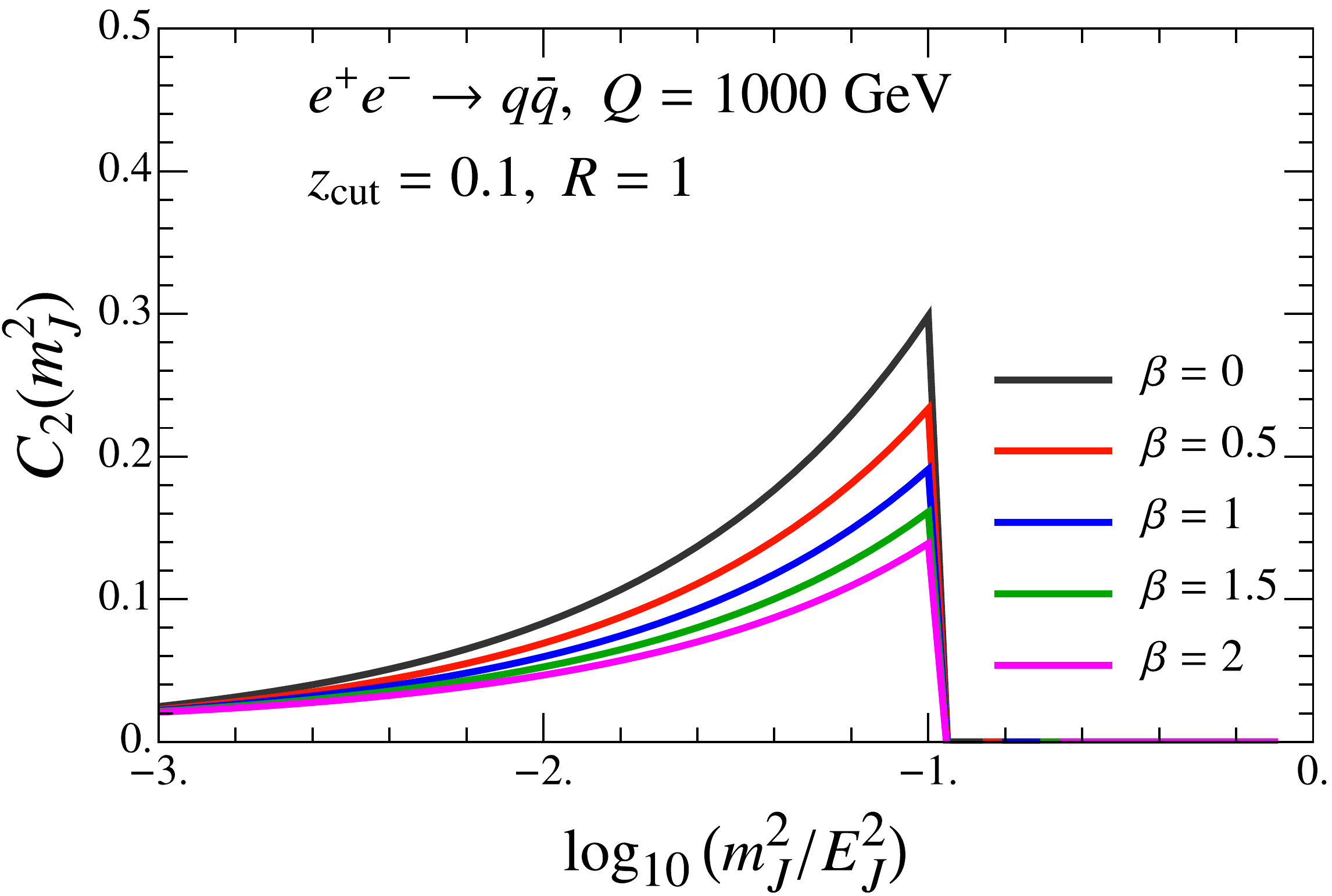}
\includegraphics[width=0.45\textwidth]{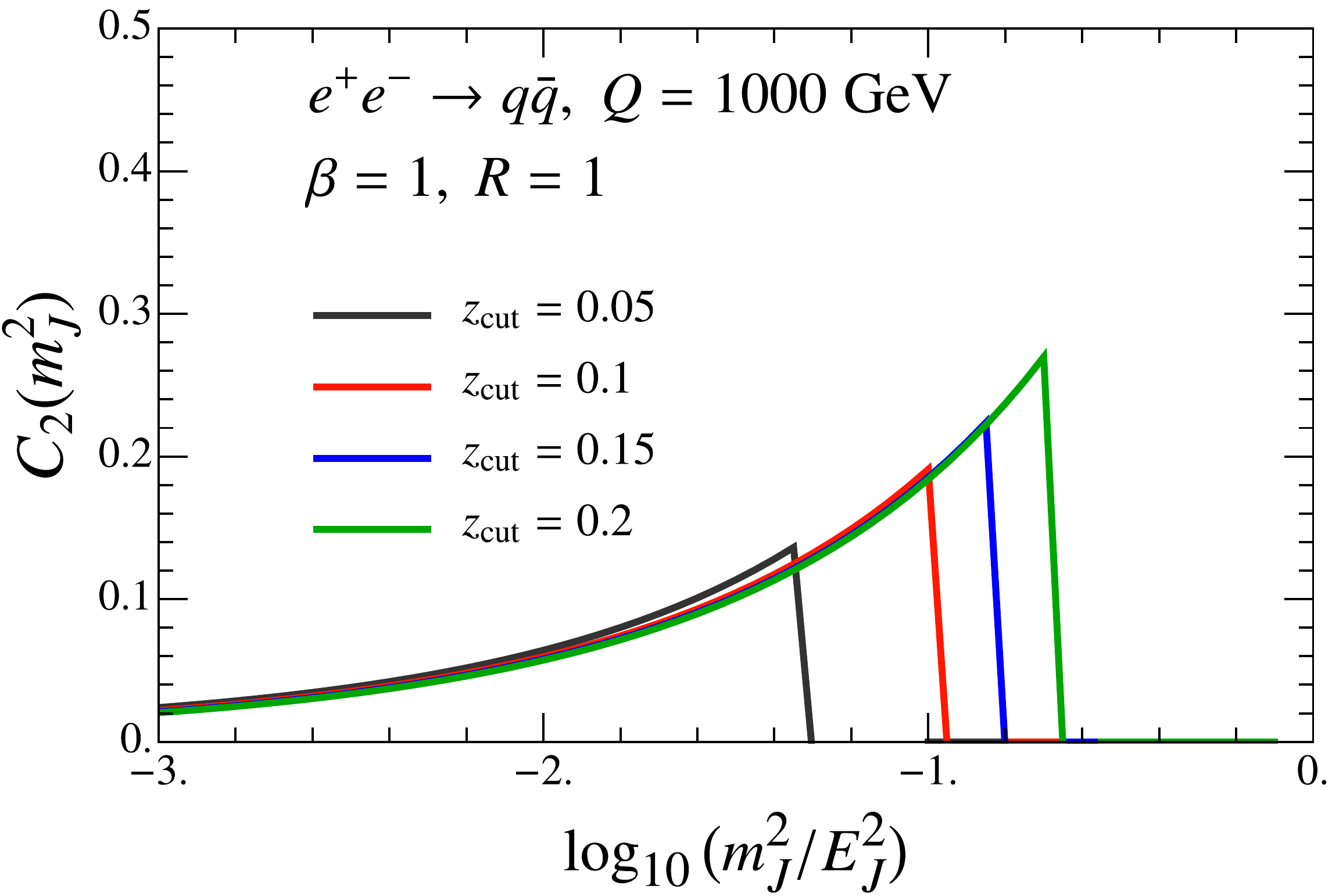}
\caption{Results for $C_1^q(m_J^2)$ and $C_2^q(m_J^2)$ across the jet mass spectrum for different values of $\zcut$ and $\beta$. Here the jet energy, $E_J = Q/2$.
\label{fig:Civaryzcutbeta}}
\vspace{-0.2cm}
\end{figure}

\subsection{Numerical Results for $C_1$ and $C_2$}
\label{sec:c1c2numerical}

Having derived the expressions for $C_1(m_J^2)$ and $C_2(m_J^2)$ we now present some numerical results for quark jets $\kappa=q$. We use 2-loop running for $\alpha_s(\mu)$:
\begin{align}
\frac{1}{\alpha_s(\mu)} = \frac{1}{\alpha_s(\mu_0)} + \frac{\beta_0}{2\pi} \, \ln\Big(\frac{\mu}{\mu_0}\Big) + \frac{\beta_1}{4 \pi \beta_0} \ln \bigg[1 + \frac{\beta_0}{2 \pi}\alpha_s(\mu_0)\,\ln \Big(\frac{\mu}{\mu_0}\Big)\bigg] \, ,
\end{align}
with $\alpha_s(\mu_0=m_Z)=0.118$ and
\begin{align}
\beta_0 = \frac{11}{3} C_A -\frac{2}{3} n_f \, , \qquad \beta_1 = \frac{34}{3} C_A^2 - \frac{10}{3} C_A n_f - 2 C_F n_f\, .
\end{align}
Given that for $Q = $ 500--1000 GeV and the typical values of $z_{\rm cut}$ we adopt both the collinear-soft and jet scales lie well below the top mass, we employ the evolution for $n_f = 5$ dynamical flavors. We choose to freeze the coupling at $\alpha_s(1.5\,{\rm GeV})$ for scales $\mu \le 1.5\,{\rm GeV}$.

In \fig{Civaryzcutbeta} we show the $e^+e^-$ results for $C_1^q(m_J^2)$ and $C_2^q(m_J^2)$ as a function of the jet mass, picking $Q = 1000$ GeV and various values of $\zcut$ and $\beta$. Here the jet energy is set to $E_J=Q/2$.
We note that both coefficients depend on $m_J$ and hence have a nontrivial affect on the shape of the spectrum.
In addition, we find that on holding $m_J^2/Q^2$ fixed, the coefficients have little remaining dependence on $Q$, consistent with the expectation from running coupling effects.
To estimate the uncertainty of our LL calculations, we compare the results for $C_1^q(m_J^2)$ and $C_2^q(m_J^2)$ evaluated with a frozen coupling and $\mu=m_Z$ and with a running coupling. We find that for $C_1^q(m_J^2)$ the variations are small at larger $m_J$ and grow to at most $-5\%$ at the lower limit of the SDOE region, i.e. when $\log_{10}(m_J^2/E_J^2) \sim -3$. For $C_2^q(m_J^2)$ the variation is larger, varying between $0\%$ and $-20\%$ as we go from larger to smaller jet masses. The running coupling effects can of course be included exactly. Note that we expect the uncertainty to be fairly moderate since the Wilson coefficients involve a single logarithmic series. Moreover, since they are related to ratios of two perturbative series, see \eqs{C1}{C2}, there are also cancellations from higher order contributions that affect the overall normalization of the numerator and denominator.
To account for missing higher order corrections beyond our LL analysis, we conservatively take $\pm 20\%$ as an estimate for the uncertainty in our determination of $C_{1,2}$.

Due to the dependence on $\zcut$ and $\beta$ one may obtain the hadronic
matrix elements $\Omega^{\figeight}_{1\kappa}$ and $\Upsilon_{1,i}^\kappa$ from data by considering different sets of $\zcut$ and $\beta$ values, as we show below through Monte Carlo studies in \sec{montecarlo}.
The kinks in the curves in \fig{Civaryzcutbeta} occur at the transition point from the SDOE to the ungroomed resummation region, where $m_J^2/E_J^2 \simeq z_{\rm cut}$.
Above this transition point the description of power corrections with \eq{sigfullk} no longer applies. It is instead replaced by an analogous formula with $C_2=0$, $C_1=1$, and a different definition for the parameter $\Omega_{1\kappa}^{\figeight}\to \Omega_{1\kappa}(R)$, see~\cite{Stewart:2014nna} for its definition.

\begin{figure}[t!]
\includegraphics[width=0.33\textwidth]{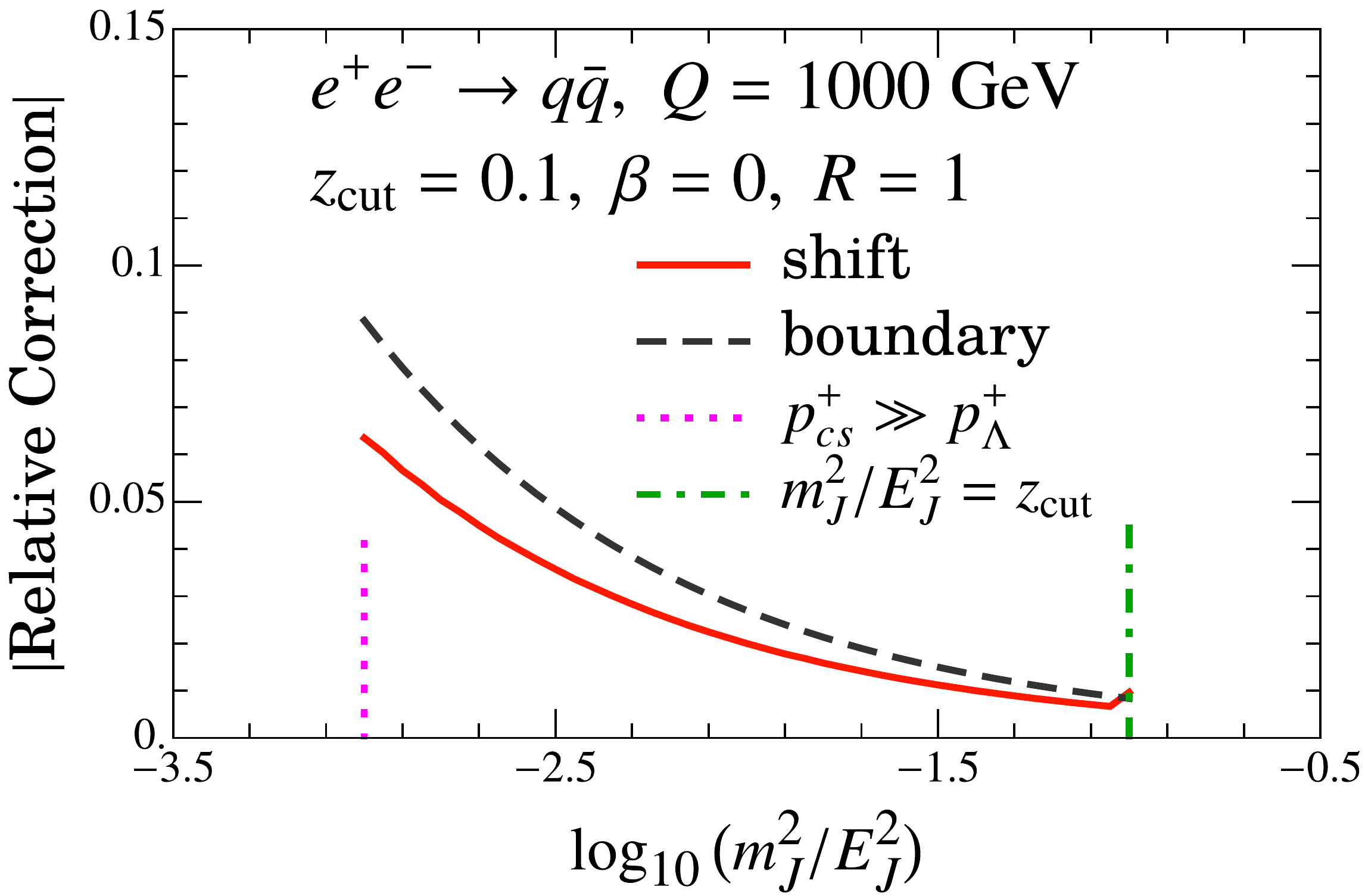}
\includegraphics[width=0.33\textwidth]{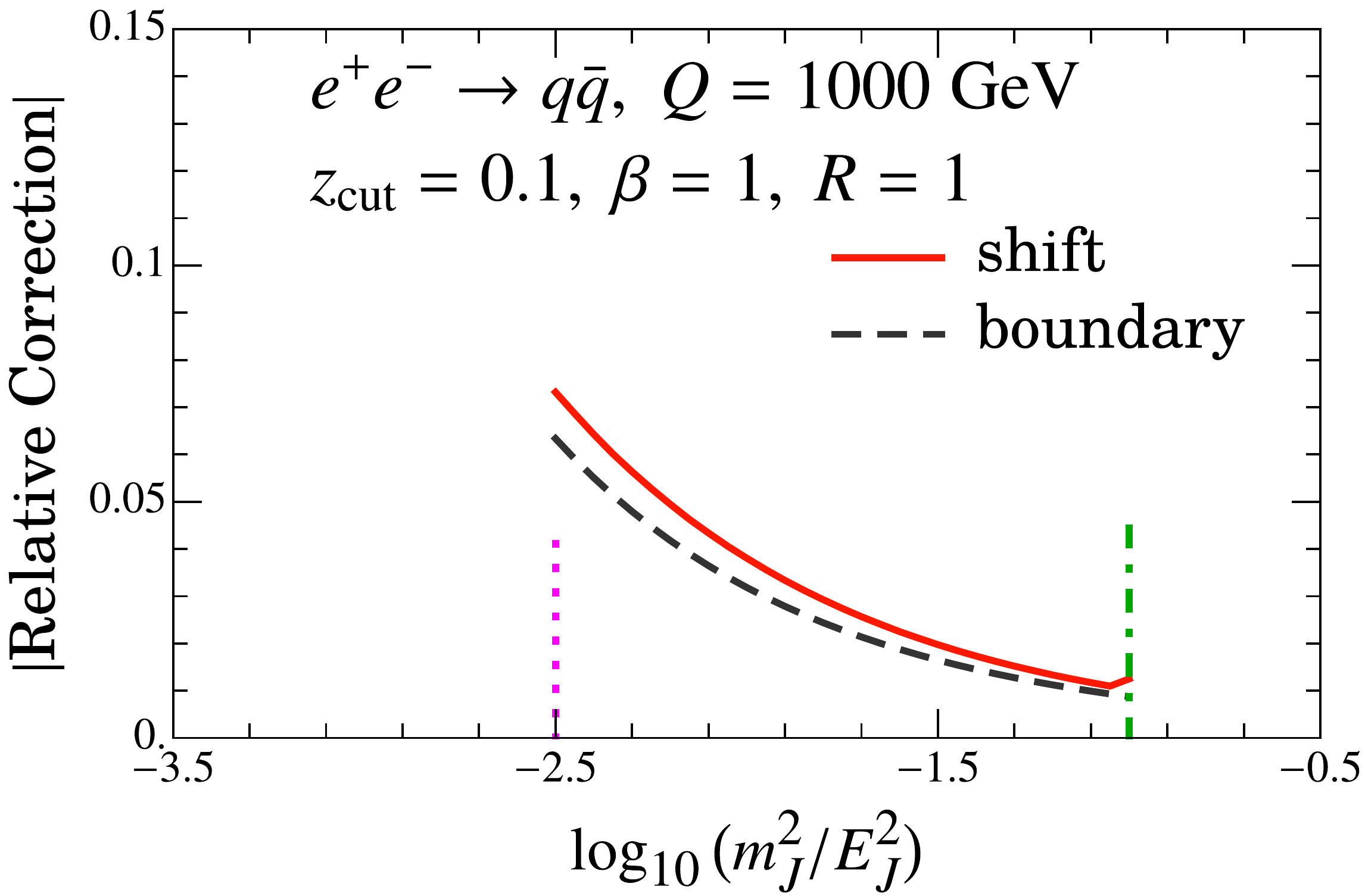}
\includegraphics[width=0.33\textwidth]{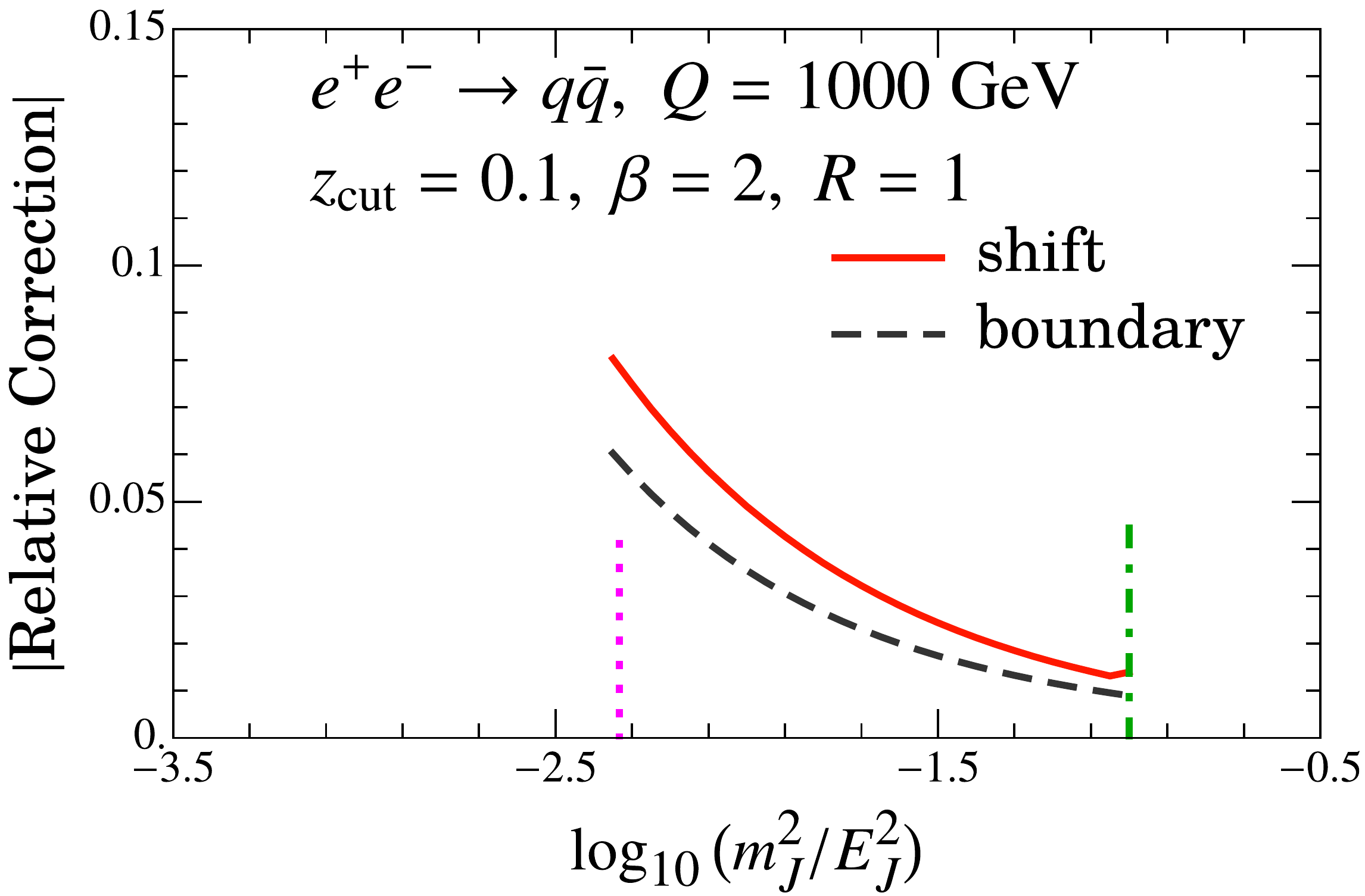}
\\
\includegraphics[width=0.33\textwidth]{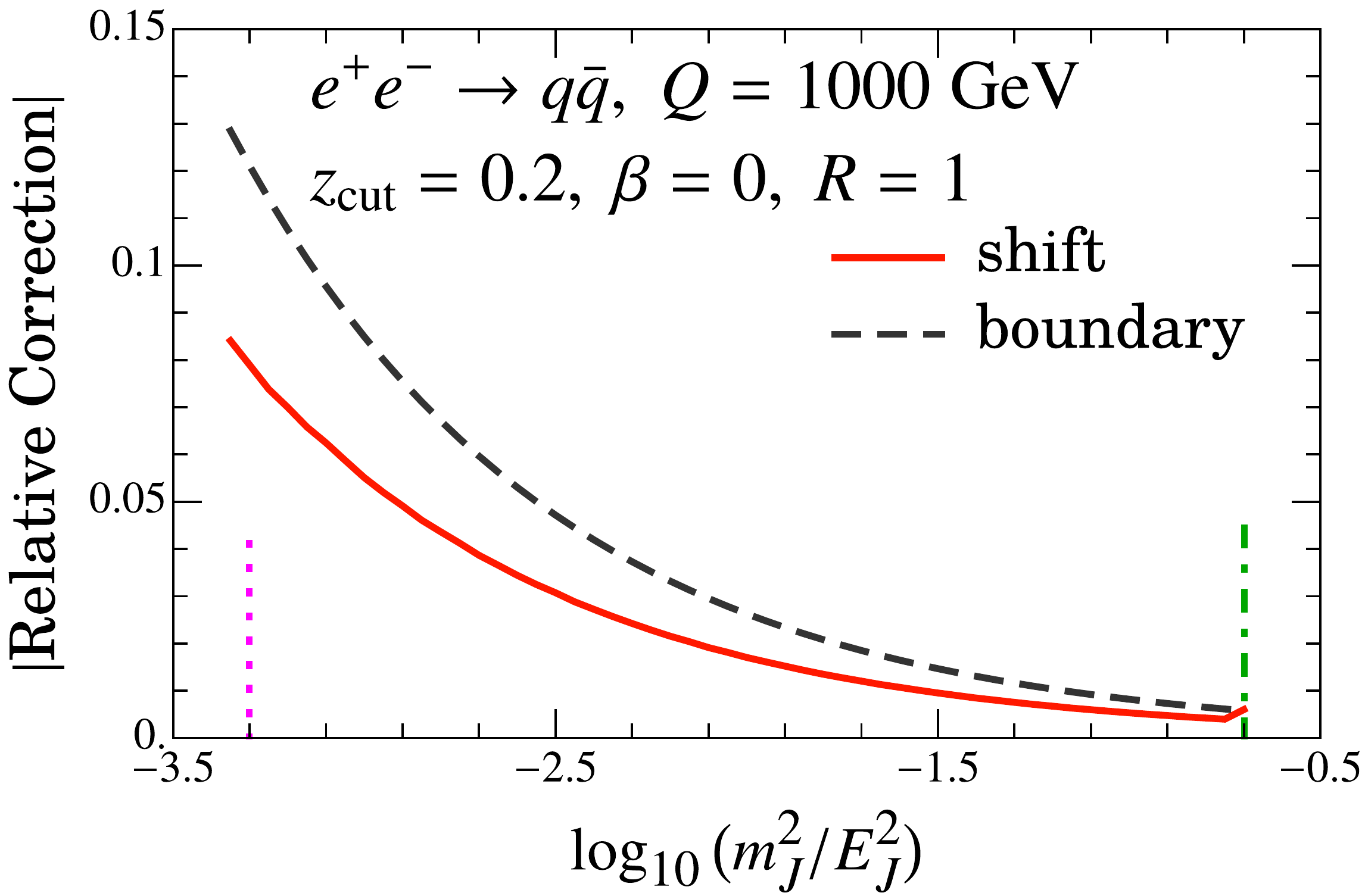}
\includegraphics[width=0.33\textwidth]{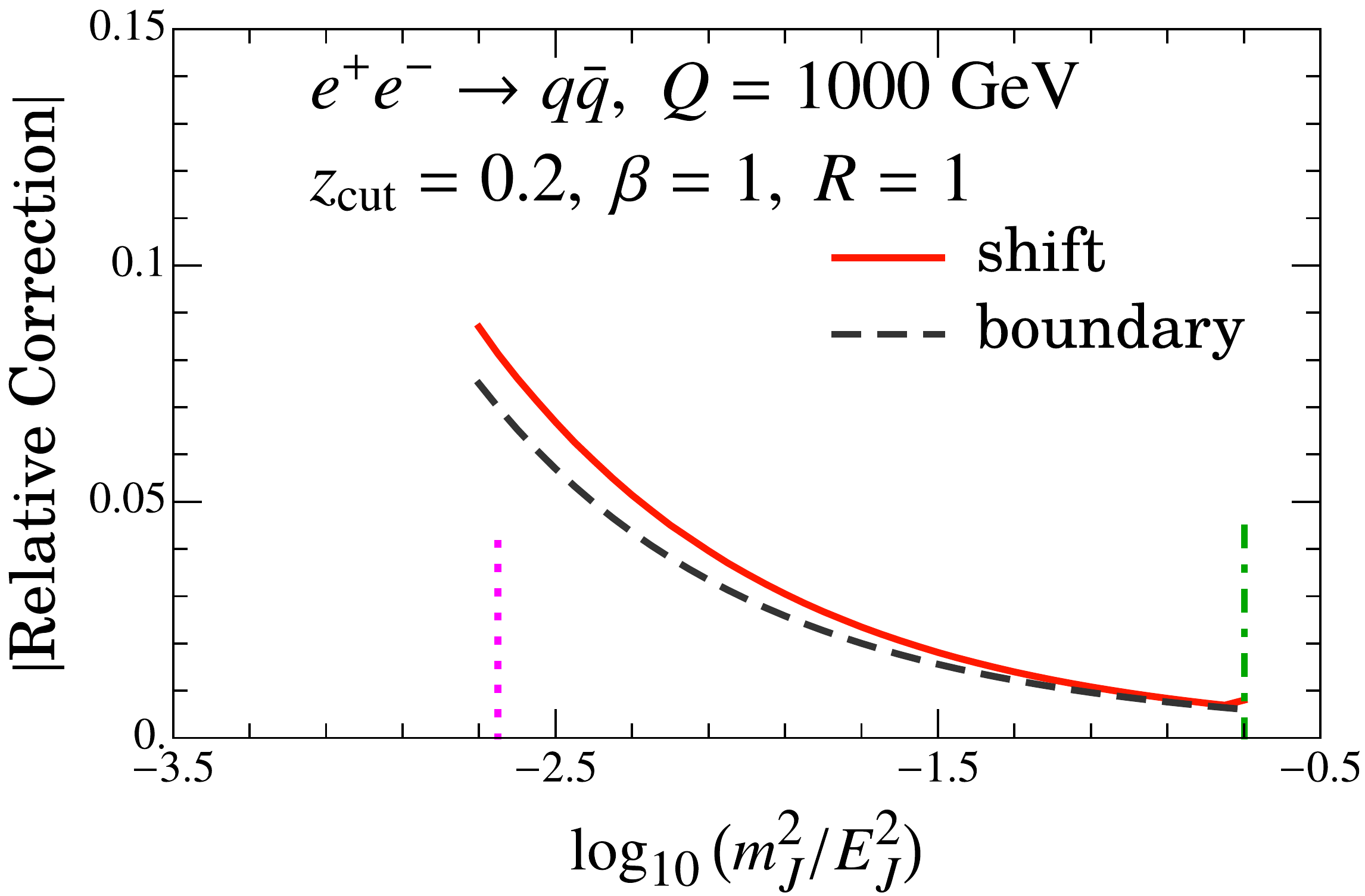}
\includegraphics[width=0.33\textwidth]{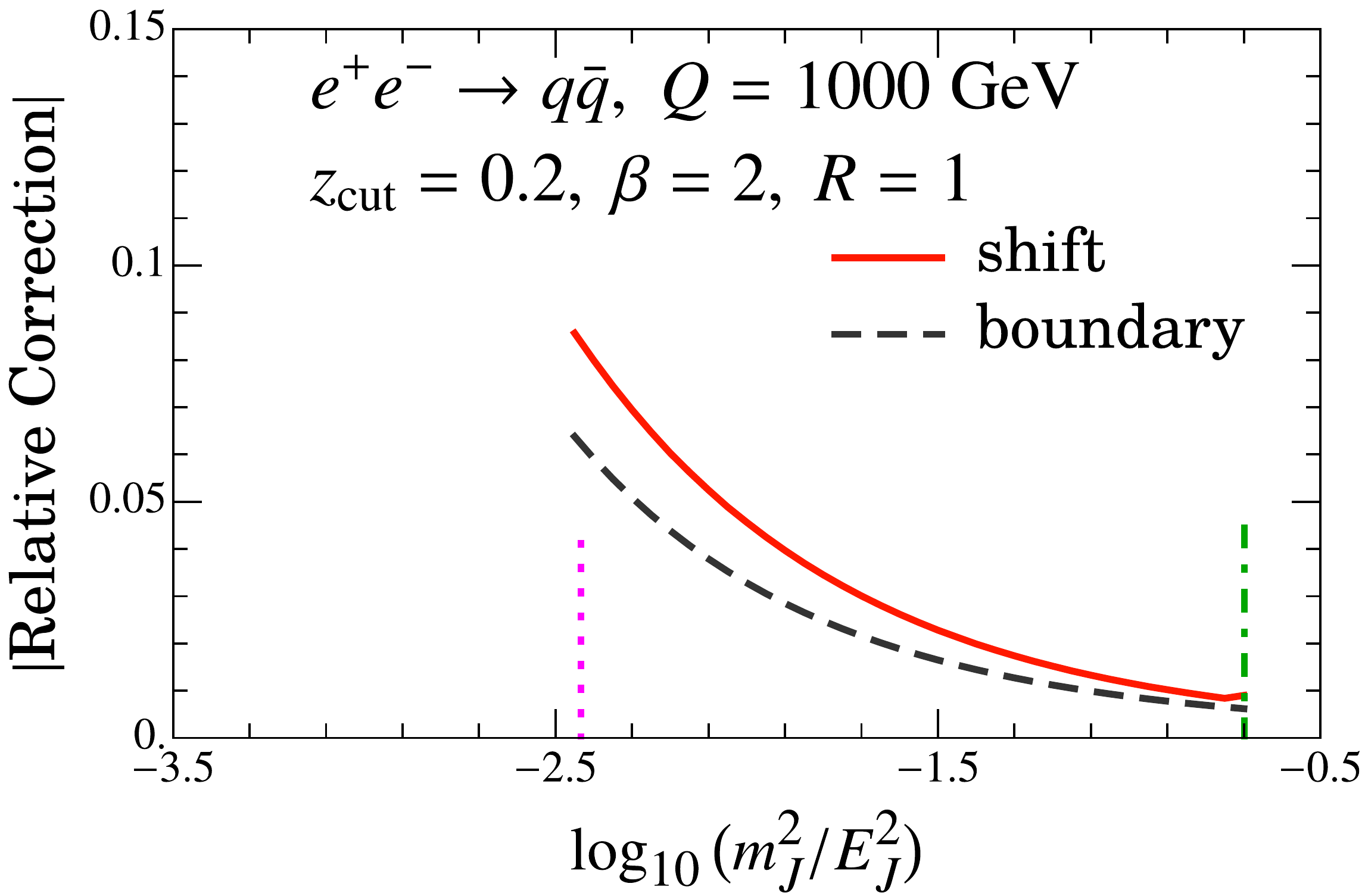}
\caption{Absolute value of the fractional power correction induced by the leading nonperturbative shift and boundary corrections, choosing for illustration $\Omega_{1}^{\, \figeight} =1.0$ GeV, $\Upsilon_{1,0} = 0.7$ GeV, $\Upsilon_{1,1} = 0.4$ GeV.
The vertical lines as in \fig{logrhoplot} indicate the extent of the SDOE region, as labeled in the top left panel.
\label{fig:relcorrection}}
\vspace{-0.2cm}
\end{figure}

In \fig{relcorrection} we show corrections due to the shift and boundary power corrections using \eq{sigfullk}. For the nonperturbative parameters we take the representative values of $\Omega_{1}^{\, \figeight} =1.0$ GeV, $\Upsilon_{1,0} = 0.7$ GeV, $\Upsilon_{1,1} = 0.4$ GeV and consider several values of $\zcut$ and $\beta$. The curves named ``shift'' and ``boundary'' correspond to the two leading power correction terms in \eq{sigfullk} divided by the parton level cross section $d \hat \sigma/dm_J^2$. The ``shift'' term involves derivative of the partonic cross section times $C_1(m_J^2)$ and is proportional to $\Omega_{1\kappa}^{\figeight}$, and the ``boundary'' term is the correction to the normalization due to $C_2(m_J^2)$ times $\Upsilon_{1}^\kappa(\beta) = \Upsilon_{1,0}^\kappa + \beta \Upsilon_{1,1}^\kappa$:
\begin{align}
\text{|shift correction|} &= \bigg | - Q\, \Omega_{1\kappa}^{\figeight} \, \frac{d}{d m_J^2} \bigg(C^\kappa_1(m_J^2, Q, \tilde z_{\rm cut}, \beta, R) \, \frac{d \hat \sigma_\kappa}{d m_J^2}\bigg) \bigg/\frac{d \hat \sigma_\kappa}{d m_J^2} \bigg| \\
\text{|boundary correction|} &= \bigg | \frac{Q\Upsilon_{1}^\kappa (\beta)}{m_J^2} \, C^\kappa_2(m_J^2, Q, \, \tilde z_{\rm cut}, \beta, R) \, \frac{d \hat \sigma_\kappa}{d m_J^2} \bigg/\frac{d \hat \sigma_\kappa}{d m_J^2} \bigg|\nn \, .
\end{align}
The region between the vertical dashed and dashed-dotted lines indicates the appropriate SDOE region where this description is valid. We see that the power corrections amount to about $\sim 1$--$10\%$ in the SDOE region and grow larger as we approach the SDNP region.
These plots display the absolute value of the power corrections, and hence do not illustrate the possibility that the two nonperturbative corrections may have different signs. In general we expect a positive value for the shift parameter $\Omega^{\figeight}_{1\kappa}$. We will see in \sec{MCfits} that MC event generators favor a negative value for $\Upsilon_{1,0}^\kappa$.

\section{Hadronization Effects in the Nonperturbative Region}
\label{sec:NPregion}

Next we consider the nonperturbative (SDNP) region of the groomed jet mass spectrum, as defined in \eq{NPregions}. Here the nonperturbative effects contribute at leading order, and we show that they can be accounted for via a shape function. We demonstrate that this shape function exhibits universality by being independent of the jet kinematics $\Phi_J$ and $z_{\rm cut}$, but does depend on $\beta$ in a non-trivial fashion. Furthermore, it exhibits an unusual feature that it is not normalized to 1.

\subsection{Shape Function in the Nonperturbative Region}
\label{sec:SDNP}
Consider the factorization for the cross section in the SDNP region, as illustrated in \fig{modesmassless}b. In this region we have the same hierarchy of modes as were considered for the perturbative cross section in the SDOE region, with $\{$H,C,S,CS$\}$
mapped to $\{$H,C,S,$\Lambda$CS$\}$. The essential difference is that now the collinear-soft $\Lambda$CS mode is nonperturbative. Since the steps in the derivation of the factorization theorem with SCET that lead to \eq{masslessFact0} do not rely on a perturbative expansion, the factorization formula in \eq{masslessFact0} and its properties are also valid for the SDNP region. However now the collinear-soft function $S_c^\kappa\Big(\ell^+\qcut^{\frac{1}{1+\beta}},\beta,\mu\Big)$ defined in \eq{SCdefn} is nonperturbative since it involves modes that have $p^2\sim \Lambda_{\rm QCD}^2$. Here $\ell^+$ represents the total plus momentum of the collinear-soft sector. In the following we show how to best describe this collinear-soft function, and how this leads to a conceptually and practically viable version of the factorization theorem of \eq{masslessFact0} involving a shape function with prescribed properties.

Note that in the SDNP region there is no analog of a separation into shift and boundary corrections as seen in the SDOE region, since here the NP subjets themselves determine when the soft drop is stopped. So there is a single leading order effect to the jet mass measurement from particles in the SDNP region which enters through $S_c^\kappa$.

In \Refcite{Frye:2016aiz} it was shown that the $Q$ and $\zcut$ dependence in the collinear-soft function $S_c^\kappa$ in \eq{SCdefn} only enters in the combination $\ell^+ \qcut^{\frac{1}{1+\beta}}$. Although the analysis in \Refcite{Frye:2016aiz} relied on considering the phase space integrals for on-shell particles, it is also straightforward to see that it also holds when considering off-shell particles in the nonperturbative region (as is necessary when considering the interpolating operators for hadrons). The proof follows from expressing the momenta $p_i^\mu$ in the lab frame in terms of dimensionless rescaled coordinates $\{k_i^a,k_i^\perp,k_i^b\}$ which in our notation corresponds to the following change of variables
\begin{align}
\label{eq:qcutrescaling}
p_i^+ = (\ell^+)\ k_i^a \,
\qquad
p_{i\perp} = \Bigl(\qcut^{\frac{1}{2+\beta}}\big(\ell^+\big)^{\frac{1+\beta}{2+\beta}} \Bigr) \ k_{i\perp} \,,
\qquad
p_i^- = \Bigl( \qcut^{\frac{2}{2+\beta}}\big(\ell^+\big)^{\frac{\beta}{2+\beta}} \Bigr) \ k_i^b \, .
\end{align}
As discussed in detail in \Refcite{Frye:2016aiz}, the CA clustering and soft drop test on a cluster of particles become independent of $\qcut$ upon making this rescaling.\footnote{The proof in \Refcite{Frye:2016aiz} did not explicitly consider the fact that emissions lying at angles smaller than the passing subjet do not get tested for soft drop. The additional constraint imposed by this is the comparison of angles of these subjets with that of the stopping subjet. This angular comparison is also invariant under the rescaling in \eq{qcutrescaling} and hence is fully compatible with the proof.}
The Wilson lines appearing in the definition in \eq{SCdefn} are also invariant under this rescaling.
Finally, Lorentz invariant combinations of momenta such as
\begin{align}
\label{eq:pipjkakb}
p_i \cdot p_j = \frac{p_i^+ p_j^-+p_j^+ p_i^-}{2} +p_{i\perp}\cdot p_{j\perp} = \bigg(\big(\qcut^{\frac{1}{1+\beta}}\ell^+\big)^{\frac{1+\beta}{2+\beta}}
\bigg)^2 \ \bigg(\frac{k_i^a k_j^b+k_j^a k_i^b}{2} + k_{i\perp}\cdot k_{j\perp}\bigg) \,,
\end{align}
also involve only the combination $\ell^+ \qcut^{\frac{1}{1+\beta}}$.
Applying this change of variables in \eq{SCdefn} gives a rescaling to
the explicit momentum operator $\hat p^+$ and the momentum operators in $\overline \Theta_{\rm sd}(\hat p_i^\mu)$, which can be arranged into the form
\begin{align} \label{eq:Screscaled}
S_c^\kappa\Bigl[\ell^+\, Q_{\rm cut}^{\frac{1}{1+\beta}}, \beta,\mu \Bigr] &= \frac{1}{n_\kappa} \:
{\rm tr} \bigl\langle 0 \big| T X_{n\kappa}^\dagger V_{n\kappa} \delta \Bigl(Q_{\rm cut}^{\frac{1}{1+\beta}}\ell^+ - Q_{\rm cut}^{\frac{1}{1+\beta}}\, \ell^+ \hat k^a \overline \Theta_{\rm sd}(\hat k_i) \Bigr) \bar T V_{\kappa n}^\dagger X_{n\kappa}
\big| 0 \bigr\rangle_\mu
\,,
\end{align}
where the new operators $\hat k_i$ give the rescaled momenta of \eq{qcutrescaling}.
Thus the proof for the dependence on only this combination holds also with having offshellness of ${\cal O}(\Lambda^2_{\rm QCD})$.
Note that due to the ultraviolet structure of the collinear-soft function that there are contributions from $p_\perp^{-2\epsilon}$, which leads to another source of $\ell^+$ dependence in \eq{Screscaled} through distributions involving $(\ell^+ Q_{\rm cut}^{\frac{1}{1+\beta}}/\mu^{\frac{2+\beta}{1+\beta}})$. Therefore one should not simply pull the distribution variable $\ell^+$ in \eq{Screscaled} out of the $\delta$-function.

The function $S_c^\kappa$ depends on three arguments, but we can factor out its dependence on the $\overline{\rm MS}$ UV renormalization scale $\mu$ by following Refs.~\cite{Hoang:2007vb,Ligeti:2008ac}. First we define the Fourier transform
\begin{align}
\tilde S_c^\kappa(y,\beta,\mu)
\equiv \int d\ell^+\qcut^{\frac{1}{1+\beta}} \:
\exp\Bigl( -i\, y\, \ell^+\qcut^{\frac{1}{1+\beta}} \Bigr) \
S_c^\kappa\Bigl[\ell^+\, Q_{\rm cut}^{\frac{1}{1+\beta}}, \beta,\mu \Bigr]
\,,
\end{align}
so that the variable $y$ has mass dimensions $-(2+\beta)/(1+\beta)$ and the function $\tilde S_c^\kappa$ is dimensionless.
In position space the RGE equation for $\tilde S_c^\kappa(y,\beta,\mu)$ is multiplicative. Therefore we can define a new $\mu$ independent nonperturbative shape function $\tilde F_\kappa^\otimes(y,\beta)$ via
\begin{align} \label{eq:SCFy}
\tilde S_c^\kappa(y,\beta,\mu) \equiv
\tilde S_c^{\kappa,{\rm pert}}(y,\beta,\mu) \
\tilde F_\kappa^\otimes(y,\beta) \,,
\end{align}
the perturbative collinear-soft function $\tilde S_c^{\kappa,{\rm pert}}$ is included to the order in the $\alpha_s$ expansion (or in resummed perturbation theory) required for the precision of the prediction. When using $\overline{\rm MS}$ for the computation of $\tilde S_c^{\rm pert}$ this leads to the nonperturbative function $\tilde F_\kappa^\otimes$ being defined in the $\overline{\rm MS}$ scheme. At this point other solutions with reduced infrared-sensitivity, which could allow to control renormalon effects, are also possible following the approach of \Refcite{Hoang:2007vb}. The superscript $\otimes$ in \eq{SCFy} is meant to distinguish it from the source function introduced in \sec{npsource} that describes the $\Lambda$ mode in SDOE region.

The collinear-soft function $\tilde S_c^\kappa$ depends on the jet initiating hard parton $\kappa = q,g$. Furthermore, all possible logarithmic dependence on $y$ is associated with the $\mu$ dependence captured by $\tilde S_c^{\kappa,{\rm pert}}$, implying that it is confinement which restricts the possible form of the shape function $\tilde F_\kappa^\otimes$. In momentum space this implies that all moments of $F_\kappa^\otimes$ exist, so that it falls off at large momentum faster than any power law (such as exponentially). In position space this implies that all derivatives of the function exist at the origin $y=0$. Next we define the momentum space shape function $F_\kappa^\otimes$ using a dimension $1$ momentum space variable $k_{\rm NP}$ via
\begin{align} \label{eq:Fotimesdefn}
F_\kappa^\otimes(k_{\rm NP},\beta)
\equiv \frac{2+\beta}{1+\beta}\: k_{\rm NP}^{\frac{1}{1+\beta}}\!
\int \frac{dy}{2\pi}\: \exp\Bigl( i \,y\, k_{\rm NP}^{\frac{2+\beta}{1+\beta}} \Bigr)
\: \tilde F_\kappa^\otimes(y,\beta)
\,,
\end{align}
so that $F_\kappa^\otimes$ has mass dimension $-1$. The prefactors in \eq{Fotimesdefn} allow switching from the natural momentum space variable $k_{\rm NP}^{\frac{2+\beta}{1+\beta}}$ to $k_{\rm NP}$ and absorb the resulting Jacobian as part of the definition of $F_\kappa^\otimes(k_{\rm NP},\beta)$. With this definition we naturally have the scaling $k_{\rm NP}\sim \Lambda_{\rm QCD}$ and we can assume that the function $F_\kappa^\otimes(k_{\rm NP},\beta)$ typically only has non-trivial support in this momentum range.
With the definition in \eq{Fotimesdefn}, the Fourier transform of \eq{SCFy} now yields a convolution with this momentum space shape function,
\begin{align}
\label{eq:SDNPfact}
S_c^\kappa\Bigl[\ell^+\, Q_{\rm cut}^{\frac{1}{1+\beta}}, \beta,\mu \Bigr] =
\int d k_{\rm NP} \
S_c^{\, \kappa, \rm pert}\Bigl[ \ell^+ \qcut^{\frac{1}{1+\beta}} -k_{\rm NP}^{\frac{2+\beta}{1+\beta}}, \beta, \mu\Bigr] \
F^\otimes_\kappa(k_{\rm NP},\beta)
\,.
\end{align}
This is the final form that we will use for describing the collinear-soft function in the SDNP region. Note that the shape function $F^\otimes_\kappa (k_{\rm NP}, \beta)$ depends only on $\beta$ and the jet initiating parton $\kappa$, but not on $Q$ or $\zcut$, demonstrating this universality. We also note that the condition that only the combination $\ell^+ \qcut^{\frac{1}{1+\beta}}$ appears in $S_c^\kappa$ already implies the form of the convolution in \eq{SDNPfact}.

An important difference to the case of ungroomed event shapes is that the normalization of $F^\otimes_\kappa$ is not constrained to unity. In the ungroomed case the normalization condition follows from the fact that the shape function represents a unitary non-perturbative redistribution of the partonic plus momentum. Moreover, in both the nonperturbative peak region and the operator expansion region, the leading nonperturbative mode is a wide angle soft mode that scales as $\Lambda_{\rm QCD} (1, 1,1)$, enabling a connection to be made between these two regions, constraining the normalization condition and the way how the hadronic parameters enter~\cite{Hoang:2007vb}.
In contrast, in the groomed case, the power corrections in the SDOE region involved both a term with a derivative of the cross section, as well as terms without. This can be seen from \eq{factfull} where in the context of our SDOE analysis we went as far as possible to rewrite the result to a form involving a one-dimensional shape function. The form of terms involving $\Upsilon_1^\kappa(\beta)$ and $dC_1^\kappa/dm_J^2$ makes it clear that the same normalization condition can not be derived in this case, so the shape function $F_\kappa^\otimes$ of \eq{SDNPfact} is not normalized to unity. Furthermore, there is no simple connection between the shape function $F^\otimes_\kappa(k_{\rm NP},\beta)$ and the function $F_\kappa^{\,\figeight}(k)$ of \eq{factfull} since the perturbative $m_J$-dependent Wilson coefficients $C_1$ and $C_2$ must emerge when we transition from the SDNP to the SDOE region. This is because there are two different leading nonperturbative modes, the $\lamcs$ mode for the SDNP region and the $\Lambda$ mode for the SDOE region, so the transition requires a more complicated interpolation between them which we do not consider further here.
In the next section we discuss an approach for building a model for $F^\otimes_\kappa(k_{\rm NP},\beta)$ in order to study this function in greater detail.

\subsection{A Model for the SDNP Region Shape Function}
\label{sec:sdnpmodel}

Here we discuss a model for the nonperturbative function $F_\kappa^\otimes(k_{\rm NP},\beta)$ in the SDNP region which enables a more detailed exploration of its properties.
Unlike the perturbative CS mode in the SDOE region, the relevant nonperturbative mode for this region, $\Lambda$CS, both stops the groomer and is responsible for the nonperturbative effects.
Consider a nonperturbative subjet with momentum $q^\mu$ that is being tested for soft drop:
\begin{align}
\label{eq:SDNPpass}
\overline \Theta_{\rm sd}^q &
= \Theta\bigg(\frac{q^-}{Q} - \tzcut \Big(2 \frac{q_\perp}{q^-}\Big)^\beta \bigg)
= \Theta\bigg(2\Big(\frac{q_\perp}{\qcut}\Big)^{\frac{1}{1+\beta}} - 2 \frac{q_\perp}{q^-}\bigg) = \Theta\big(\theta_\lamcs(q_\perp) - \theta_q \big)\, ,
\end{align}
where we have expressed the angle in terms of $q_\perp$ and $q^-$ since the subjet does not need to obey an onshell condition, but is still boosted.
Here we have identified for a given $q_\perp$ the maximum angle a NP subjet can have to pass soft drop:
\begin{align}
\label{eq:thetalamcs}
\theta_{\lamcs} (q_\perp) \equiv 2 \Big(\frac{q_\perp}{\qcut}\Big)^{\frac{1}{1+\beta}} \,, \qquad q_\perp \sim \Lambda_{\rm QCD} \, .
\end{align}
\begin{figure}[t!]
\centering
\includegraphics[width=0.8\textwidth]{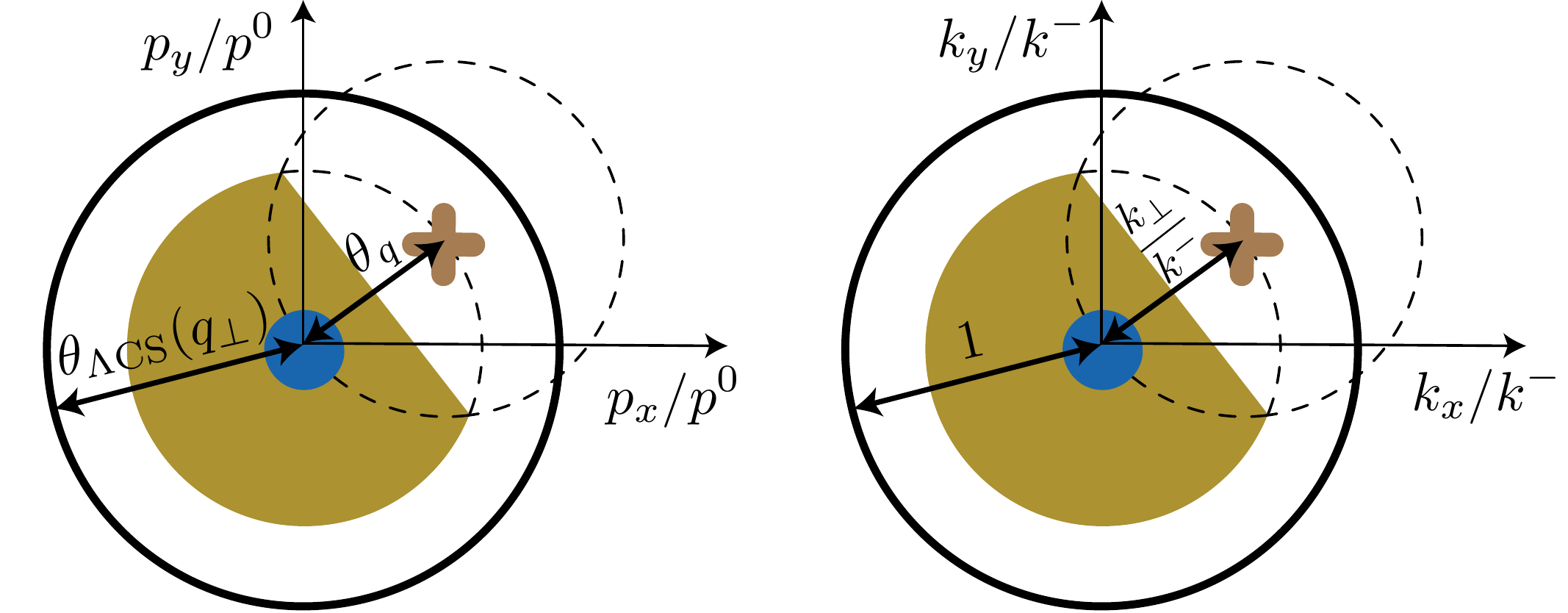}
\\[-20pt]
\phantom{x} \hspace{-3.8cm}a) \hspace{5.8cm} b) \\[-5pt]
\caption{The catchment area of nonperturbative modes in the SDNP region is shown. The brown cross indicates a NP subjet that stops soft drop in the SDNP region and has an angle $\theta_q < \theta_{\Lambda{\rm CS}} (q_\perp)$. Blue dot refers to the collinear parton aligned with the jet axis. The brown shaded region corresponds to the region where any other NP subjets could lie clustered with the collinear subjet. In b) the same geometry is shown in the rescaled coordinates.
\label{fig:NPfig8SDNP}}
\vspace{-0.2cm}
\end{figure}

In the SDNP region, CA clustering of commensurate momenta will lead to significant changes in the direction of the subjet as the particles are clustered. However, we can simplify the analysis by considering subjets of NP particles defined via the procedure described above in \sec{CAforNP}. In the SDNP region the only perturbative subjet relevant is the collinear jet-initiating parton. This implies that once the NP particles have been clustered to yield a set of NP subjets, the subsequent CA clustering will only combine an NP subjet with the collinear subjet.

We show in \fig{NPfig8SDNP}a a NP subjet at angle $\theta_q$ that stops soft drop, which from \eq{SDNPpass} satisfies $\theta_q \leq \theta_\lamcs(q_\perp)$. The brown shaded region is the catchment area for any other NP subjets that contribute to the jet mass which have been clustered with the collinear parton.
To construct our model, we consider two NP subjets with momenta $q^\mu$ and $r^\mu$, where $q^\mu$ stops soft drop. In the SDNP region we then express the collinear-soft function as follows
\begin{align}
\label{eq:ScSDNP}
&S_{c}^{\,\kappa} \Bigl(\ell^+\qcut^{\frac{1}{1+\beta}}, \beta, \mu \Bigr) =
\qcut^{\frac{-1}{1+\beta}} \int d^d q \int d^d r \
F^\otimes_\kappa(q^\mu, r^\mu, \mu)\; \Theta\big(\theta_\lamcs (q_\perp) - \theta_q \big) \\
&\qquad \times \Big \{\overline \Theta_{\rm NP}^\otimes \big(r^\mu, \theta_q, \Delta \phi\big) \delta(\ell^+ -r^+- q^+)
+ \Theta_{\rm NP}^{\figeight} \big(r^\mu, \theta_q, \Delta \phi\big)
\Theta\big(\theta_r - \theta_{\lamcs}(r_\perp)\big)
\delta(\ell^+ -q^+)
\Big\}\,. \nn
\end{align}
The distribution of the two subjets in the SDNP region including its $\mu$ dependence is described by a nonperturbative source function $F^\otimes_\kappa(q^\mu, r^\mu, \mu)$. The function does not depend on the soft drop parameters which are implemented explicitly by the measurement in \eq{ScSDNP}. (For simplicity we do not write the symmetric case when $r^\mu$ stops soft drop and $q^\mu$ is kept or rejected.) Here $\overline \Theta_{\rm NP}^\otimes$ denotes the catchment area for $r^\mu$ related to the collinear subjet when $q^\mu$ stops soft drop at angle $\theta_q$. It can be expressed with the help of the operators defined in \eqs{Thetafig8}{Thetabndry} above, which we introduced in our analysis for shift and boundary terms:
\begin{align}
\label{eq:thetaotimes}
\overline \Theta_{\rm NP}^\otimes(r^\mu, \theta_q, \Delta \phi) &= \overline \Theta_{\rm NP}^{\figeight}(r^\mu, \theta_q, \Delta \phi) \ \Theta_{\rm NP}^{\bndry}(r^\mu, \theta_q, \Delta \phi) \, , \nn \\
&=\Theta\bigg(| \Delta \phi| - \frac{\pi}{3} \bigg)\Theta\bigg(1 -\frac{ \theta_{r}}{\theta_{q}} \bigg)+ \Theta\bigg(\frac{\pi}{3} - | \Delta \phi| \bigg)\Theta \bigg(\frac{1}{2 \cos (\Delta \phi)} - \frac{\theta_{r}}{\theta_{q}}\bigg) \nn \\
&\equiv\overline \Theta_{\rm NP}^\otimes(\theta_r, \theta_q, \Delta \phi) \, ,
\end{align}
where $\Delta \phi = \phi_r - \phi_q$. This is shown as the brown shaded area in \fig{NPfig8SDNP}a. In \eq{ScSDNP} the first term in the second line corresponds to the case where $r^\mu$ is already a part of the collinear subjet when $q^\mu$ is being tested and hence it contributes to the measurement. Since we have assumed that $q^\mu$ stops soft drop, the second term in \eq{ScSDNP} describes the scenario where $r^\mu$ lies outside the combined catchment area of $q^\mu$ and the collinear subjet and fails soft drop.

The expression in \eq{ScSDNP} can be further simplified by expressing all nonperturbative momenta in terms of variables in an appropriately boosted frame with respect to the $q^\mu$ subjet in a fashion similar to \eq{rescaling}. Here the analog is to choose the relevant angle for the boost to be $\theta_\lamcs(q_\perp)$, the maximum angle allowed for the momentum $q^\mu$ of the stopping NP subjet:
\begin{align}
\label{eq:rescalingSDNP}
q^{+} &= \frac{\theta_\lamcs(q_\perp)}{2} \, k^+ \, , \qquad q^{-}= \frac{2}{\theta_\lamcs(q_\perp)} \, k^- \, , \qquad q_{\perp} = k_\perp \, , \nn \\
r^+ &= \frac{\theta_\lamcs(q_\perp)}{2} \, \tilde k^+ \, , \qquad r^- = \frac{2}{\theta_\lamcs(q_\perp)} \, \tilde k^- \, , \qquad r_{\perp} =\tilde k_\perp \, .
\end{align}
Note that since $\theta_\lamcs(q_\perp)$ does not transform under boosts along the jet direction, the variables $k^\mu$ in \eq{rescalingSDNP} are not boost invariant and transform as $q^\mu$ does under boosts (unlike the rescaled variables in the SDOE case in \eq{rescaling}). We will assume that the nonperturbative function $ F^\otimes_\kappa(q^{\mu}, r^\mu, \mu)$ is invariant under boosts along the jet direction, and thus $F^\otimes_\kappa(q^{\mu}, r^\mu, \mu) = F^\otimes_\kappa (k^\mu, \tilde k^\mu, \mu)$. Applying this rescaling we can rewrite \eq{ScSDNP} in the form.
\begin{align}
\label{eq:ScSDNP2}
S_{c}^{\,\kappa} \Bigl(\ell^+ \qcut^{\frac{1}{1+\beta}}, \beta, \mu \Bigr)
&=
\int \!d k^{\prime+}\! \!\int\! d^d k\! \int\! d^d \tilde k \
F^\otimes_\kappa (k^\mu, \tilde k^\mu, \mu)\,
\Theta\Big(1 - \frac{k_\perp}{k^-} \Big) \\
& \
\times \delta\Big(\ell^+ \qcut^{\frac{1}{1+\beta}} -k^{\prime+} \big(k_\perp\big)^{\frac{1}{1+\beta}}
\Big) \bigg \{\overline \Theta_{\rm NP}^\otimes \Big(\frac{\tilde k_\perp}{\tilde k^-}, \frac{k_\perp}{k^-}, \Delta \phi\Big) \
\delta(k^{\prime+} -k^+ -\tilde k^+)
\nn \\
& \ \ \
+\Theta_{\rm NP}^{\figeight} \Big(\frac{\tilde k_\perp}{\tilde k^-}, \frac{k_\perp}{k^-}, \Delta \phi\Big)
\Theta\bigg(\frac{\tilde k_\perp}{\tilde k^-} - \Big(\frac{\tilde k_\perp}{k_\perp}\Big)^{\frac{1}{1+\beta}}\bigg)\
\delta(k^{\prime+} -k^+)
\bigg\}\,. \nn
\end{align}
We thus see that after the change of variable, the dependence on $\qcut$ drops out in the soft drop passing and failing conditions for the two subjets consistent with our discussion in \sec{SDNP}. In \fig{NPfig8SDNP}b we show the catchment area in the rescaled coordinates. We have also introduced an auxiliary variable $k^{\prime+}$ that encodes the total NP $+$ momentum.

We can rewrite \eq{ScSDNP2} including $\mu$ evolution down to a low nonperturbative scale $\mu_\Lambda \sim \Lambda_{\rm QCD}$ via:
\begin{align}
\label{eq:ScSDNP3}
&S_{c}^{\,\kappa} \Bigl(\ell^+ \qcut^{\frac{1}{1+\beta}}, \beta, \mu \Bigr) =
\int \! d \big(k_{\rm NP}^{\frac{2+\beta}{1+\beta}}\big)\! \!\int \!d k^{\prime+}\! \!\int\! d^d k\! \int\! d^d \tilde k \
U^{\kappa}_{S_c}\Bigl(\ell^+\qcut^{\frac{1}{1+\beta}} - k_{\rm NP}^{\frac{2+\beta}{1+\beta}}, \mu, \mu_{\Lambda}\Bigr)
\, F^\otimes_\kappa (k^\mu, \tilde k^\mu, \mu_\Lambda) \nn \\
& \qquad\times
\delta\Big(k_{\rm NP}^{\frac{2+\beta}{1+\beta}} -k^{\prime+} \big(k_\perp\big)^{\frac{1}{1+\beta}}
\Big) \Theta\Big(1 - \frac{k_\perp}{k^-} \Big) \,
\bigg \{\overline \Theta_{\rm NP}^\otimes \Big(\frac{\tilde k_\perp}{\tilde k^-}, \frac{k_\perp}{k^-}, \Delta \phi\Big) \
\,
\delta(k^{\prime+} -k^+ -\tilde k^+)
\nn \\
&\qquad
+\Theta_{\rm NP}^{\figeight} \Big(\frac{\tilde k_\perp}{\tilde k^-}, \frac{k_\perp}{k^-}, \Delta \phi\Big)
\Theta\bigg(\frac{\tilde k_\perp}{\tilde k^-} - \Big(\frac{\tilde k_\perp}{k_\perp}\Big)^{\frac{1}{1+\beta}}\bigg)\
\delta(k^{\prime+} -k^+)
\bigg\}\,,
\end{align}
where we introduced $k_{\rm NP}^{\frac{2+\beta}{1+\beta}}$ as a convenient integration variable, and note that the $\delta$ function sets
$k_{\rm NP}^{\frac{2+\beta}{1+\beta}} = k^{\prime+} \big(k_\perp\big)^{\frac{1}{1+\beta}}$.
At NLL the perturbative collinear-soft function only consists of a RG evolution factor since the boundary condition at tree level is a $\delta$-function, so
\begin{align}
\hat S_c^{\, \kappa,\rm NLL} \Big(\ell^+ \qcut^{\frac{1}{1+\beta}}, \beta ,\mu\Big) = U^{ \kappa}_{S_c}\Bigl(\ell^+\qcut^{\frac{1}{1+\beta}}, \mu, \mu_{\Lambda}\Bigr) \, .
\end{align}
As a result of this \eq{ScSDNP3} becomes:
\begin{align}
\label{eq:SDNPfactmodel}
S_{c}^{\,\kappa} \Bigl(\ell^+ \qcut^{\frac{1}{1+\beta}}, \beta, \mu \Bigr) =
\int d k_{\rm NP} \
\hat S_c^{\, \kappa,\rm NLL} \Big(\ell^+ \qcut^{\frac{1}{1+\beta}} -k_{\rm NP}^{\frac{2+\beta}{1+\beta}}, \beta ,\mu\Big) \, F^\otimes_\kappa (k_{\rm NP}, \beta)\,,
\end{align}
where
\begin{align}
\label{eq:Fotimesdef}
F^\otimes_\kappa (k_{\rm NP}, \beta) &\equiv
\frac{2+\beta}{1+\beta} \int\! d^d k\! \int\! d^d \tilde k \
\Big(\frac{k_{\rm NP}}{k_\perp}\Big)^{\frac{1}{1+\beta}} \,
F^\otimes_\kappa (k^\mu, \tilde k^\mu,\mu_\Lambda)\,
\Theta\Big(1 - \frac{k_\perp}{k^-} \Big) \\
& \times \bigg \{\overline \Theta_{\rm NP}^\otimes \Big(\frac{\tilde k_\perp}{\tilde k^-}, \frac{k_\perp}{k^-}, \Delta \phi\Big)\ \delta\bigg(k_{\rm NP} \Big(\frac{k_{\rm NP} }{k_\perp}\Big)^{\frac{1}{1+\beta}} -k^+ -\tilde k^+\bigg)
\nn \\
&
\ \ +\Theta_{\rm NP}^{\figeight} \Big(\frac{\tilde k_\perp}{\tilde k^-}, \frac{k_\perp}{k^-}, \Delta \phi\Big)
\Theta\bigg(\frac{\tilde k_\perp}{\tilde k^-} - \Big(\frac{\tilde k_\perp}{k_\perp}\Big)^{\frac{1}{1+\beta}}\bigg)\
\delta\bigg(k_{\rm NP} \Big(\frac{k_{\rm NP} }{k_\perp}\Big)^{\frac{1}{1+\beta}} -k^+ \bigg)
\bigg\}\,. \nn
\end{align}
The result in \eq{SDNPfactmodel} is the analog of what we derived above in \eq{SDNPfact} from general considerations. The result in \eq{Fotimesdef} provides an explicit model for $F^\otimes_\kappa (k_{\rm NP}, \beta)$ in terms of a two-variable source function $F^\otimes_\kappa (k^\mu, \tilde k^\mu,\mu_\Lambda)$ that is independent of the soft drop parameters. This expression clearly illustrates the non-trivial $\beta$ dependence of the non-perturbative shape function. Note that this model also includes the same overall factors that appeared in the prefactor of \eq{Fotimesdefn}.

Using the properties of the $\Theta$ functions, we can also investigate the normalization of $F^\otimes_\kappa (k_{\rm NP}, \beta)$:
\begin{align}
\label{eq:Fotimesnorm}
\int_0^\infty d k_{\rm NP}F^\otimes_\kappa (k_{\rm NP}, \beta) & =
\int\! d^d k\! \int\! d^d \tilde k \
F^\otimes_\kappa (k^\mu, \tilde k^\mu,\mu_\Lambda)\,
\Theta\Big(1 - \frac{k_\perp}{k^-} \Big) \\
& \qquad \times \bigg \{\overline \Theta_{\rm NP}^\otimes \Big(\frac{\tilde k_\perp}{\tilde k^-}, \frac{k_\perp}{k^-}, \Delta \phi\Big)
+\Theta_{\rm NP}^{\figeight} \Big(\frac{\tilde k_\perp}{\tilde k^-}, \frac{k_\perp}{k^-}, \Delta \phi\Big)
\Theta\bigg(\frac{\tilde k_\perp}{\tilde k^-} - \Big(\frac{\tilde k_\perp}{k_\perp}\Big)^{\frac{1}{1+\beta}}\bigg)
\bigg\}\, \nn \\
&< \int\! d^d k\! \int\! d^d \tilde k \
F^\otimes_\kappa (k^\mu, \tilde k^\mu,\mu_\Lambda)\,
\Theta\Big(1 - \frac{k_\perp}{k^-} \Big) < \int\! d^d k\! \int\! d^d \tilde k \
F^\otimes_\kappa (k^\mu, \tilde k^\mu,\mu_\Lambda)\, .\nn
\end{align}
Thus we see explicitly that the presence of the soft drop grooming induced $\Theta$-function projection operators decreases the normalization. A practical assumption for this model would be then to take $F^\otimes_\kappa (k^\mu, \tilde k^\mu,\mu_\Lambda)$ as normalized, which would then imply $\int_0^\infty d k_{\rm NP}F^\otimes_\kappa (k_{\rm NP}, \beta) \equiv f_0^{\rm NP} <1$.

\section{Comparison with Previous Work}
\label{sec:comparison}

Nonperturbative corrections to groomed jet mass spectra have been studied previously as an extension of perturbative calculations, both with Monte Carlo studies~\cite{Dasgupta:2013ihk,Larkoski:2014wba,Frye:2016aiz,Marzani:2017mva,Marzani:2017kqd}, analytic models~\cite{Dasgupta:2013ihk,Marzani:2017kqd} and shape function models~\cite{Frye:2016aiz}. In general, hadronization corrections are found to be reduced by the jet grooming~\cite{Larkoski:2014wba}. Here we will make a comparison of our results with the analytic estimates for hadronization corrections from \Refscite{Dasgupta:2013ihk,Marzani:2017kqd,Frye:2016aiz}.

For $pp$ collisions \Refcite{Dasgupta:2013ihk} considered the modified mass-drop tagger, which closely corresponds to soft drop with $\beta = 0$. Analogs of both the shift and boundary corrections were considered in the region where $m_J^2 > \Lambda_{\rm QCD}^2/\zcut$, which corresponds to the SDOE region in \eq{NPregions} with $\beta = 0$. These estimates were extended to $\beta > 0$ in \Refcite{Marzani:2017kqd} where hadronization corrections were considered due to a shift in $m_J$ and a reduction in $p_T$ of soft subjets from nonperturbative particles,
\begin{align}
\label{eq:npMSS0}
\delta m_J^2 = C_i \Lambda_{\rm hadr} p_T R_{\rm eff} \, ,\qquad \delta p_T = -C_A \frac{\Lambda_{\rm hadr}}{R_{\rm eff}} \, .
\end{align}
Here $R_{\rm eff} = m_J/(p_T\sqrt{z(1-z)})$ is the effective radius of the jet formed by a single perturbative splitting, and $\Lambda_{\rm hadr} \sim \Lambda_{\rm QCD}$ is a nonperturbative parameter common to both the corrections. \eq{npMSS0} was then used to calculate the leading power corrections from these two types of hadronization effects to the jet mass cross section by averaging over $z$ with the soft drop passing constraint:
\begin{align}
\label{eq:npMSS}
\frac{d\sigma_\kappa}{d m_J}\bigg|_{\rm hadr}^{m_J\,\rm shift}
&= \frac{d\hat \sigma_\kappa}{d m_J} \times \Big(1+\frac{C_\kappa \Lambda_{\rm hadr}}{m_J}\frac{z_{\rm sd}^{-1/2}-\Delta_\kappa}{\ln z_{\rm sd}^{-1} + B_i}\Big)
\, ,\\
\frac{d\sigma_\kappa}{d m_J}\bigg|_{\rm hadr}^{p_T\,\rm shift}
&= \frac{d\hat \sigma_\kappa}{d m_J} \times \Big(1- \frac{C_A \Lambda_{\rm hadr}}{m_J}\frac{z_{\rm sd}^{-1/2}}{\ln z_{\rm sd}^{-1} + B_\kappa}\Big)
\, .\nn
\end{align}
Here $d \hat \sigma/dm_J$ is the partonic cross section and an expansion in $z_{\rm sd} = \zcut^{\frac{2}{2+\beta}}\Bigl(\frac{m_J}{Rp_T}\Bigr)^{\frac{2\beta}{2+\beta}}\ll 1$ was performed. The constants $\Delta_q ={3\pi}/{8}$, $\Delta_g = {(15C_A - 6 n_f T_F)\pi}/{(32C_A)}$, $B_q = -{3}/{4}$, $B_g = -{11}/{12} + {n_f T_F}/{(3C_A)}$ are the first subleading terms from this expansion. The logarithm can also be expressed as $\ln z_{\rm sd}^{-1} = \ln\bigl[\tilde \theta^\star(m_J^2)/\tilde \theta_c(m_J^2)\bigr]$, and hence is the same as the first logarithm that appears in an $\alpha_s$ expansion of $C_0(m_J^2)$ in \eq{C0}.
The two corrections in \eq{npMSS} have the same scaling for $z_{\rm sd}\ll 1$. Taking $d\hat\sigma/dm_J\sim (\alpha_s \ln z_{\rm sd}^{-1} ) /m_J$, the lowest order terms in $\alpha_s$ have the scaling
\begin{align} \label{eq:npMSS2}
\frac{d\sigma}{d m_J}\bigg|_{\rm hadr} - \frac{d\hat \sigma}{d m_J}
\sim \alpha_s \frac{\Lambda_{\rm QCD}}{Rp_T} \Big(\frac{R^2p_T^2}{m_J^2}\Big)^{\frac{1 +\beta }{2+\beta}} \Big(\frac{1}{\zcut}\Big)^{\frac{1}{2+\beta}}
\, .
\end{align}
Comparing \eq{npMSS2} with our \eqs{sigfull0}{sigscaling} we find that our results agree as far as the scaling for the term is concerned.

When compared in more detail there are, however, significant differences between the model of \Refscite{Dasgupta:2013ihk,Marzani:2017kqd} shown in \eq{npMSS} and our final $pp$ results from Eqs.~(\ref{eq:sigfullk}, \ref{eq:C1pp}, \ref{eq:C2pp}), which we now discuss. One difference is that our shift correction in \eq{sigfullk} involves a derivative of the differential jet mass cross section, implying that the shift corrections in \eqs{sigfullk}{npMSS} do not agree at LL order, and hence differ for example in their $m_J$ and $\beta$ dependence. This is not unexpected as \Refscite{Dasgupta:2013ihk,Marzani:2017kqd} did not carry out a LL analysis. Comparing our boundary correction with the $p_T$-shift term, we see that both results are proportional to the leading power cross section at this order, so perturbatively the only difference is due to running coupling effects kept in our $C_2^{pp}$, compared to the fixed coupling that was used to obtain the expressions in \eq{npMSS} (this difference is also present for the shift term). The results also differ because ours depend on three hadronic parameters for each of the quark and gluon induced jets $(\Omega_{1,\kappa}^{\figeight},\Upsilon_{1,0}^\kappa,\Upsilon_{1,1}^\kappa)$. In contrast only one hadronic parameter $\Lambda_{\rm hadr}$ is employed in the two corrections in \eq{npMSS} for both the $m_J$ and $p_T$ shift correction terms with color prefactors $C_\kappa$ that do not appear in this manner for us (since in general both abelian and non-abelian attachments of non-perturbative gluons are present as sources). Furthermore, since the source function $\tilde F_\kappa$ in \eq{O1Ups1} has been derived after performing a rescaling in the NP sector following \eq{rescaling}, it cannot be simply related to models for nonperturbative functions that have appeared in the previous literature. Since different projection operators appear in the definition of $\Omega_{1\kappa}^{\figeight}$ and $\Upsilon_{1}^\kappa(\beta)$ in \eq{O1Ups1} there is no relation that connects these two quantities in our treatment. The $\Upsilon_{1,1}^\kappa$ also signifies additional linear $\beta$ dependence in our boundary correction that is not present in the $p_T$-shift correction of \eq{npMSS}, since \Refscite{Dasgupta:2013ihk,Marzani:2017kqd} did not account for the additional $\beta$ dependence from expanding the soft drop condition in their treatment of \eqs{DeltaSD}{DeltaSD2}. Finally, in our treatment the signs of our $\Upsilon_{1,i}^\kappa$ are apriori not determined as the softer subjet can in principle both gain or lose momentum from hadronization.

\begin{figure}[t!]
\hspace{-0.7cm}
\includegraphics[width=0.34\textwidth]{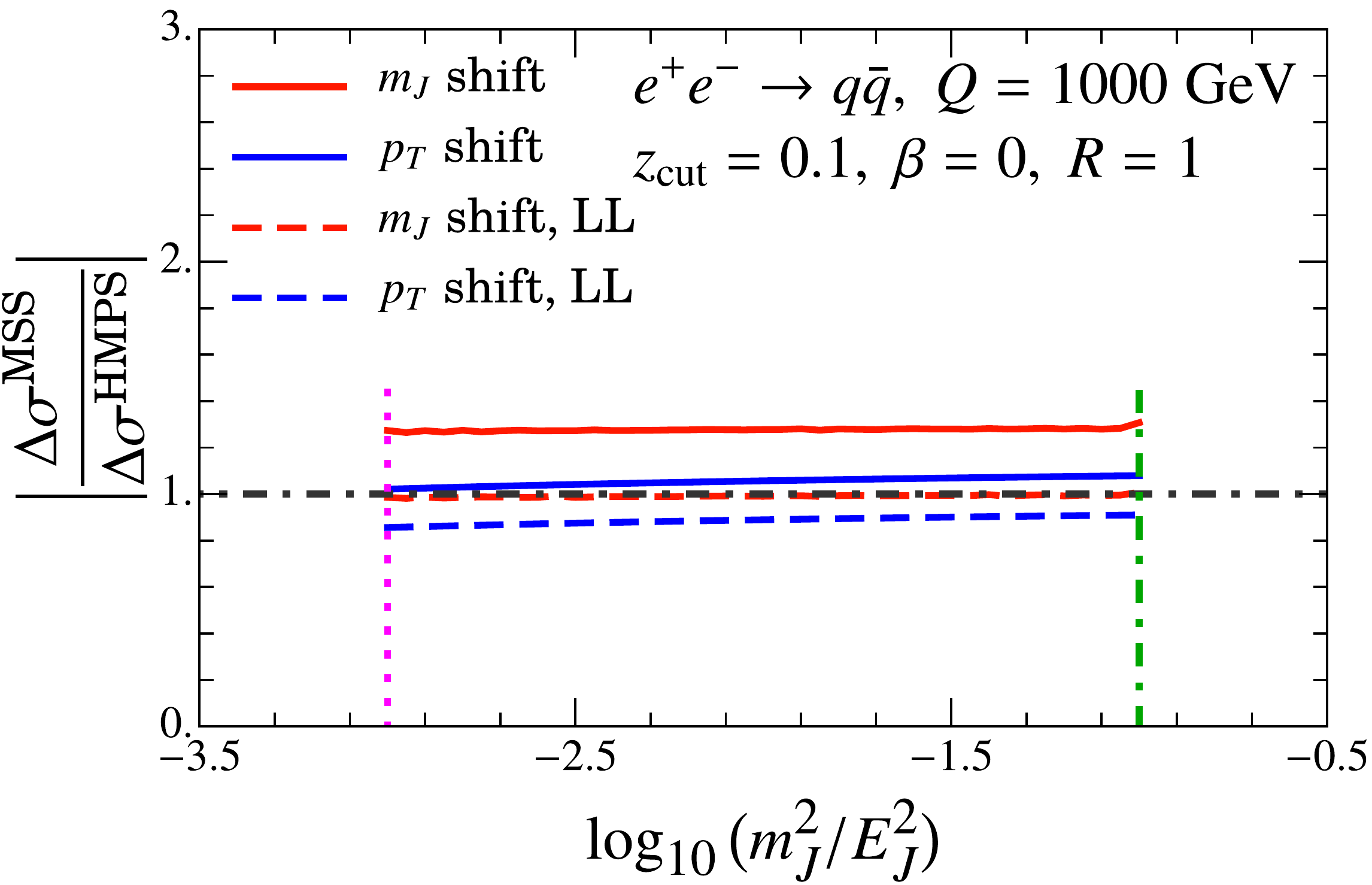}
\includegraphics[width=0.34\textwidth]{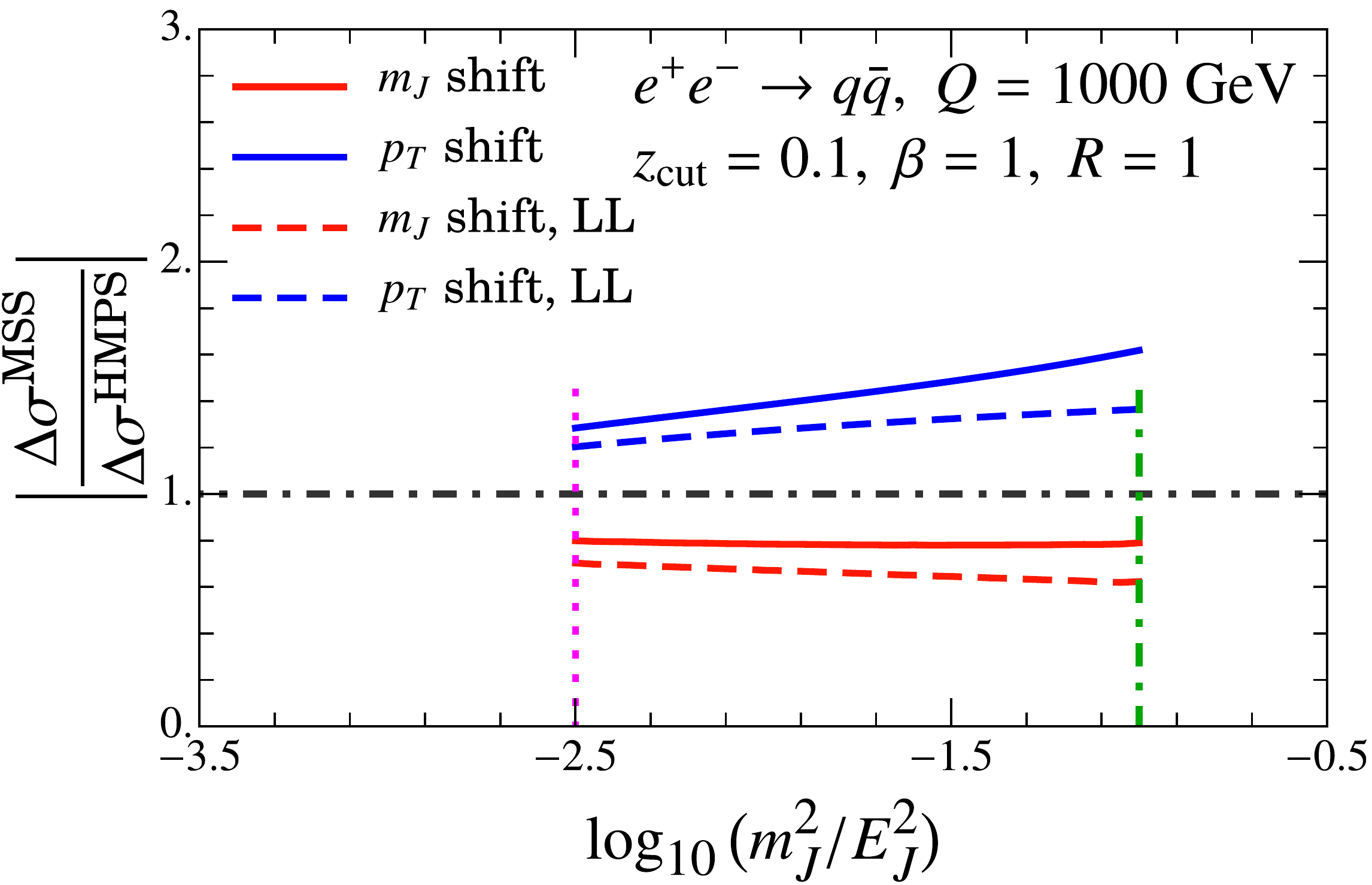}
\includegraphics[width=0.34\textwidth]{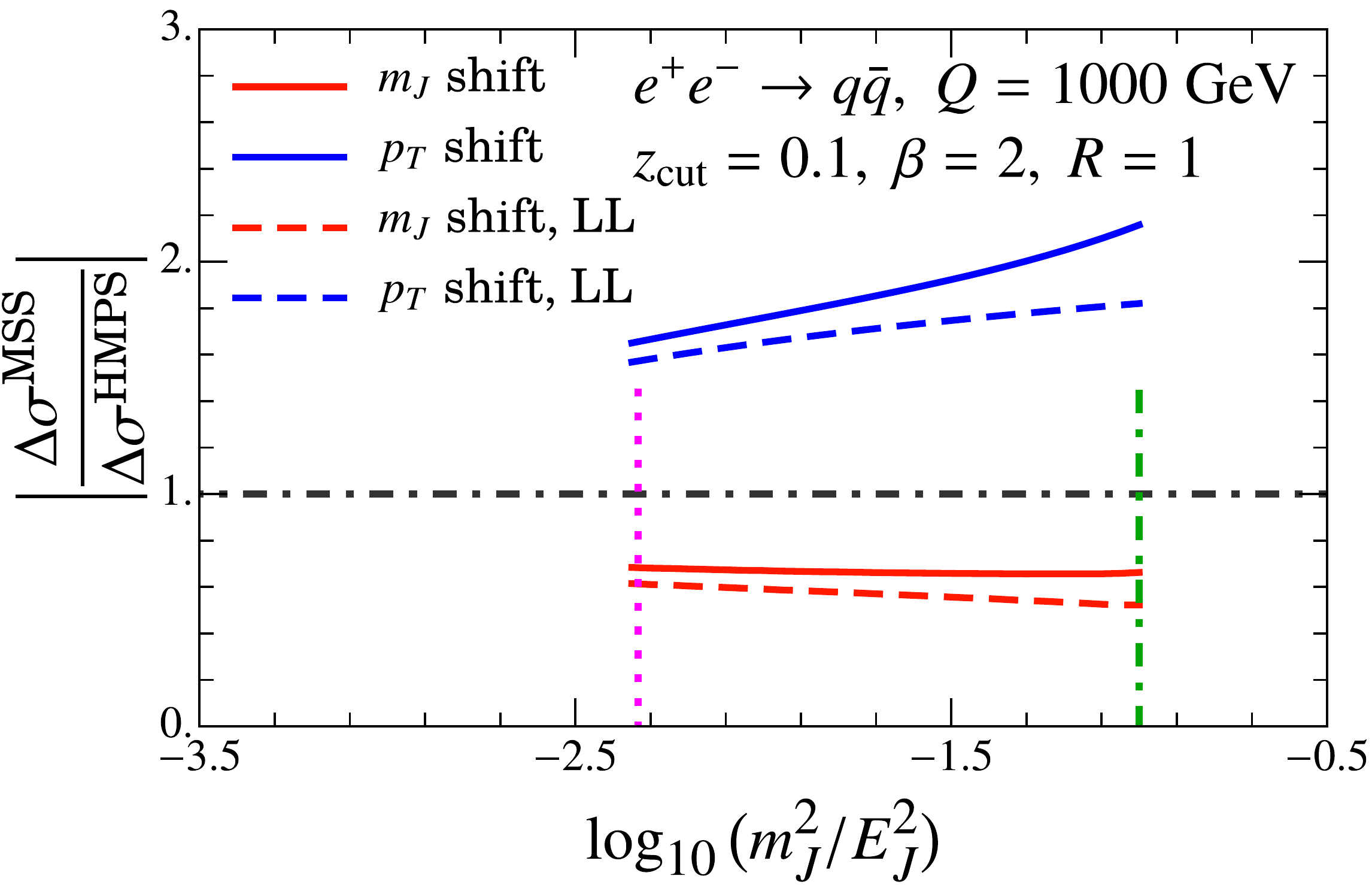}
\caption{Comparison of absolute value of the power corrections from our $m_J$ shift and $p_T$ shift terms (labeled as ``HMPS'') with the results from \Refcite{Marzani:2017kqd} (labeled as ``MSS'').\label{fig:relcorrectionMSS} }
\vspace{-0.2cm}
\end{figure}
For a numerical comparison of the $m_J$- and $\beta$-dependence of our shift and boundary corrections with the corresponding results of \Refcite{Marzani:2017kqd} given in \eq{npMSS} we consider the quark jet case, and set $\Omega_{1q}^{\figeight}= C_F \Lambda_{\rm had} = 1\,{\rm GeV}$, $\Upsilon_{1,0}^q = C_A \Lambda_{\rm had} = 1\,{\rm GeV}$, and $\Upsilon_{1,1}^q= 0\,{\rm GeV}$.
We then consider the following ratios:
\begin{align}
\label{eq:MSSvsHMPS}
\frac{\Delta \sigma^{m_J\,\textrm{shift, MSS}}}{\Delta \sigma^{\textrm{shift, HMPS}}} &\equiv \bigg( \dfrac{d\sigma^{\rm MSS}}{d m_J}\bigg|_{\rm hadr}^{m_J\,\rm shift}-\dfrac{d\hat\sigma}{d m_J} \bigg )\bigg/ \dfrac{d \sigma^{\rm shift}}{d m_J} \, , \\
\frac{\Delta \sigma^{p_T\,\textrm{shift, MSS}}}{\Delta \sigma^{\textrm{bndry, HMPS}}} &\equiv \bigg( \dfrac{d\sigma^{\rm MSS}}{d m_J}\bigg|_{\rm hadr}^{p_T\,\rm shift}-\dfrac{d\hat\sigma}{d m_J} \bigg )\bigg/ \dfrac{d \sigma^{\rm bndry}}{d m_J} \, ,
\end{align}
where the numerators (labeled ``MSS'') are predictions from \Refcite{Marzani:2017kqd} in \eq{npMSS}, and the denominators (labeled ``HMPS'') are our predictions for shift and the boundary terms in \eqs{sigshift}{sigbndry}.
We plot these ratios in \fig{relcorrectionMSS}, where the three panels correspond to three choices of $\beta$. Differences can be as large as 30\% for $\beta=0$, and they become even larger for larger $\beta$. For the dashed curves we keep only LL terms from the splitting function, $p_{gq}(z)=1/z$, and other $z$ dependent terms, and use a fixed scale for the coupling, so that higher order terms that are treated differently in the two formulas are uniformly dropped. For the solid curves we use the full ${\cal O}(\alpha_s)$ splitting function $p_{gq}(z)$ given in \eq{pgq}. The differences between dashed and solid curves is at the 10--20\% level.
The difference in the $m_J$ dependence appears to be rather moderate for the shift term, but is more significant for the boundary term except when $\beta=0$. Note that the presence of a non-zero hadronic parameter $\Upsilon_{1,1}^q$ will induce an even larger differences.

In Ref.~\cite{Frye:2016aiz} hadronization corrections to soft dropped jets were investigated in the SDNP region with a simple convolution with a normalized model function $F_{\rm shape}$ to obtain an alternate estimate for these corrections, based on the analogy with event shapes in $e^+e^-$ collisions. Their model corresponds to using
\begin{align} \label{eq:FryeNP}
\int\!\! dk^+\: S_c^\kappa\biggl[ \biggl( \ell^+ - \Bigl(\frac{k^+}{\qcut}\Bigr)^{\frac{1}{1+\beta}} k^+ \biggr) \qcut^{\frac{1}{1+\beta}} ,\beta,\mu\biggr] F_{\rm model}(k^+)\,
,
\end{align}
in place of the last line in our \eq{factfull}. In~\cite{Frye:2016aiz} the approximate agreement of this implementation with Monte Carlo was taken as evidence that there might indeed exist a normalized shape function for describing hadronization in groomed jet observables.
Given that we have derived a shape function description of the power corrections for the jet mass in \eq{SDNPfact} in the SDNP region we can directly compare with \eq{FryeNP}.
In particular we note that \eq{FryeNP} has the same form as our result in the SDNP region, given in \eq{SDNPfact} (though the integration variable is not strictly speaking a $+$-momentum). The most important difference, however, is that the norm of our shape function $F^\otimes_\kappa(k_{\rm NP},\beta)$ is not constrained, and is instead determined by an additional nonperturbative parameter. Our $F^\otimes_\kappa(k_{\rm NP},\beta)$ is independent of $z_{\rm cut}$, but depends on $\beta$ and the jet initiating parton $\kappa=q,g$.

The scaling analysis of Ref.~\cite{Frye:2016aiz} effectively corresponds to considering nonperturbative modes $\Lambda'$ with the scaling
\begin{align} \label{eq:Lpmode}
p_{\Lambda'}^\mu &\sim \Lambda_{\rm QCD} \Bigl(\zeta_0',\frac{1}{\zeta_0'},1\Bigr)
\,,
\text{ with }\
\zeta_0' \equiv \Bigl(\frac{\Lambda_{\rm QCD}}{\qcut}\Bigr)^{\frac{1}{1+\beta}}
.
\end{align}
\begin{figure}[t!]
\includegraphics[width=0.33\textwidth]{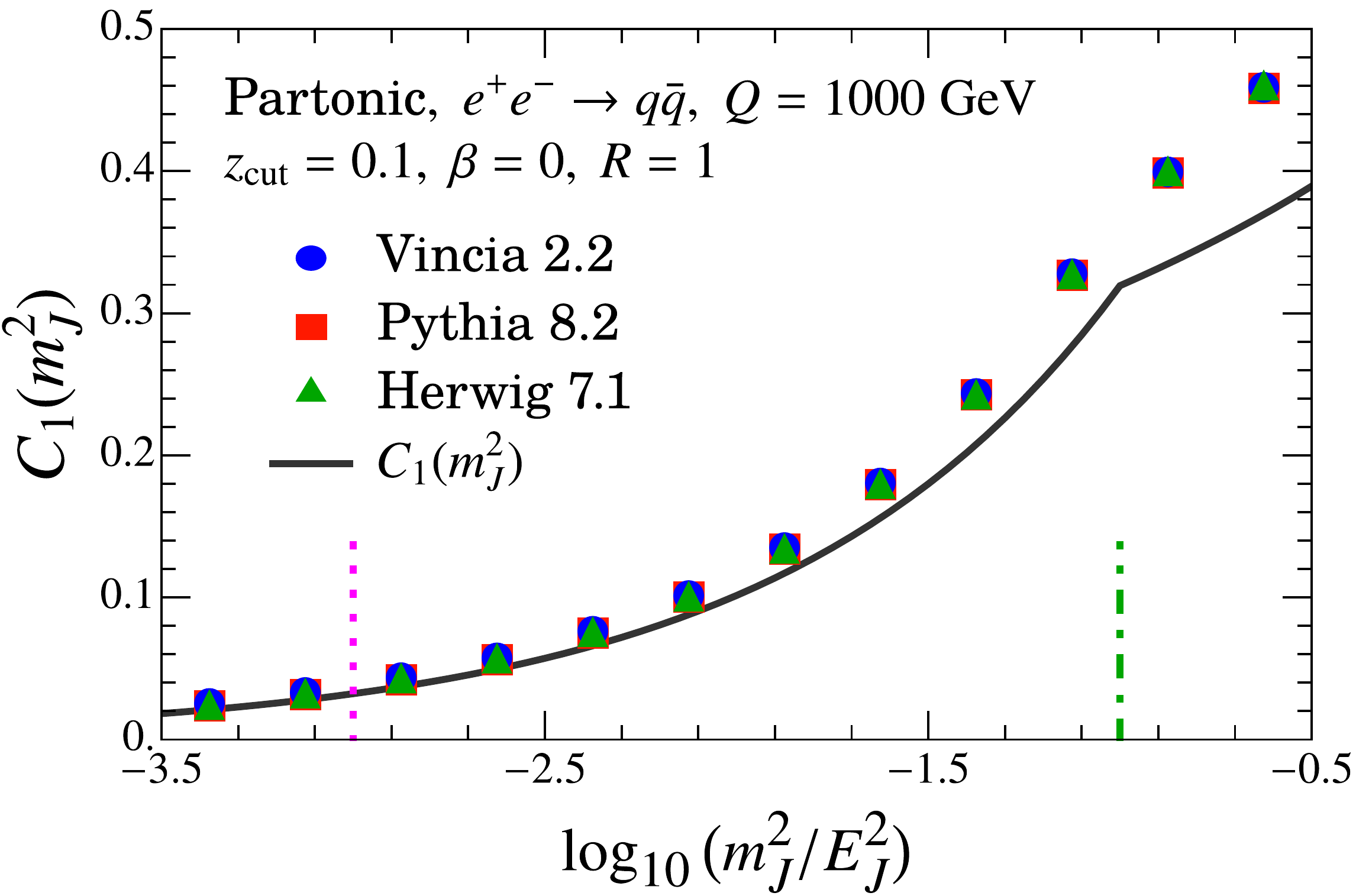}
\includegraphics[width=0.33\textwidth]{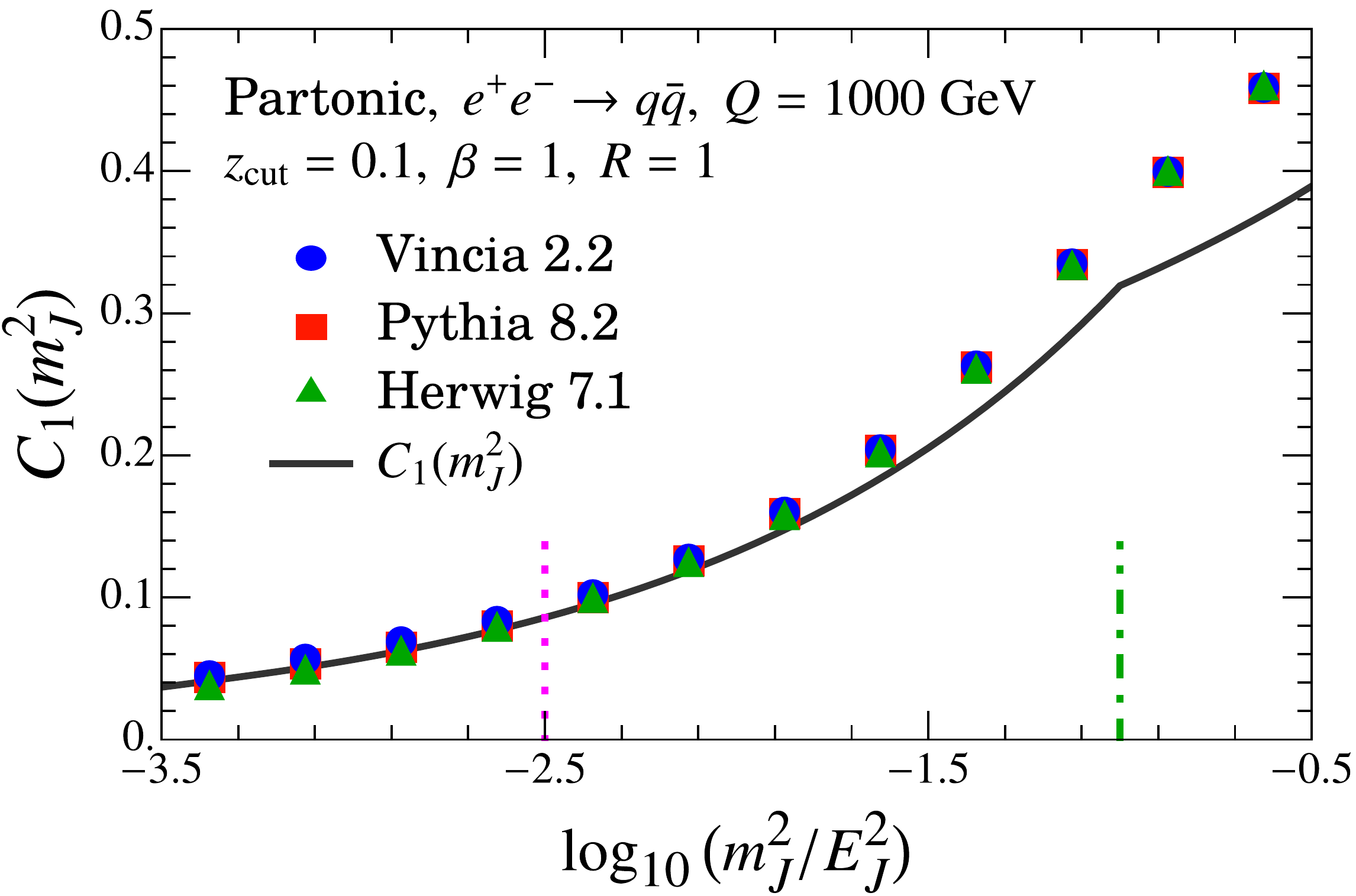}
\includegraphics[width=0.33\textwidth]{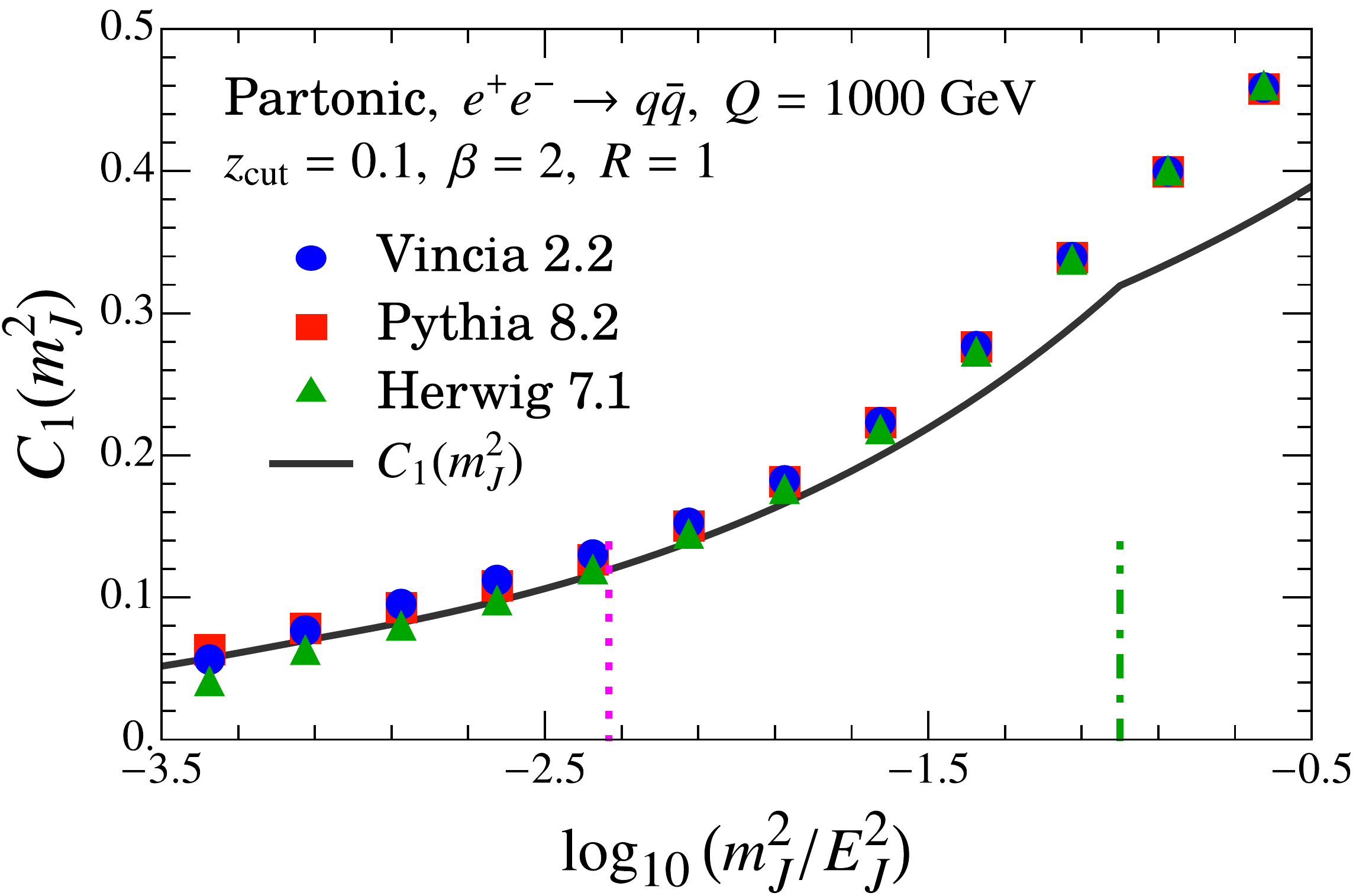}
\\[5pt]
\includegraphics[width=0.33\textwidth]{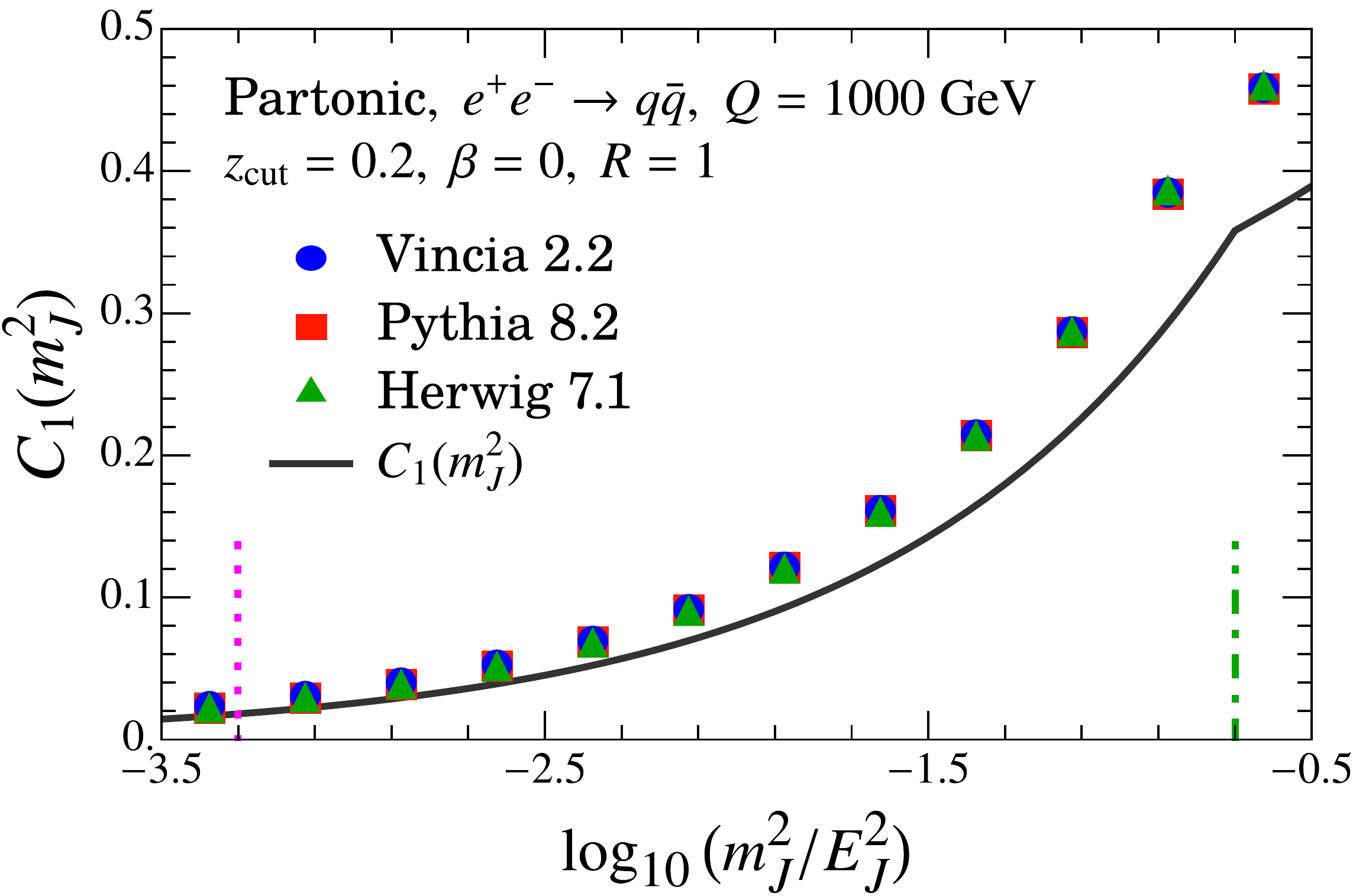}
\includegraphics[width=0.33\textwidth]{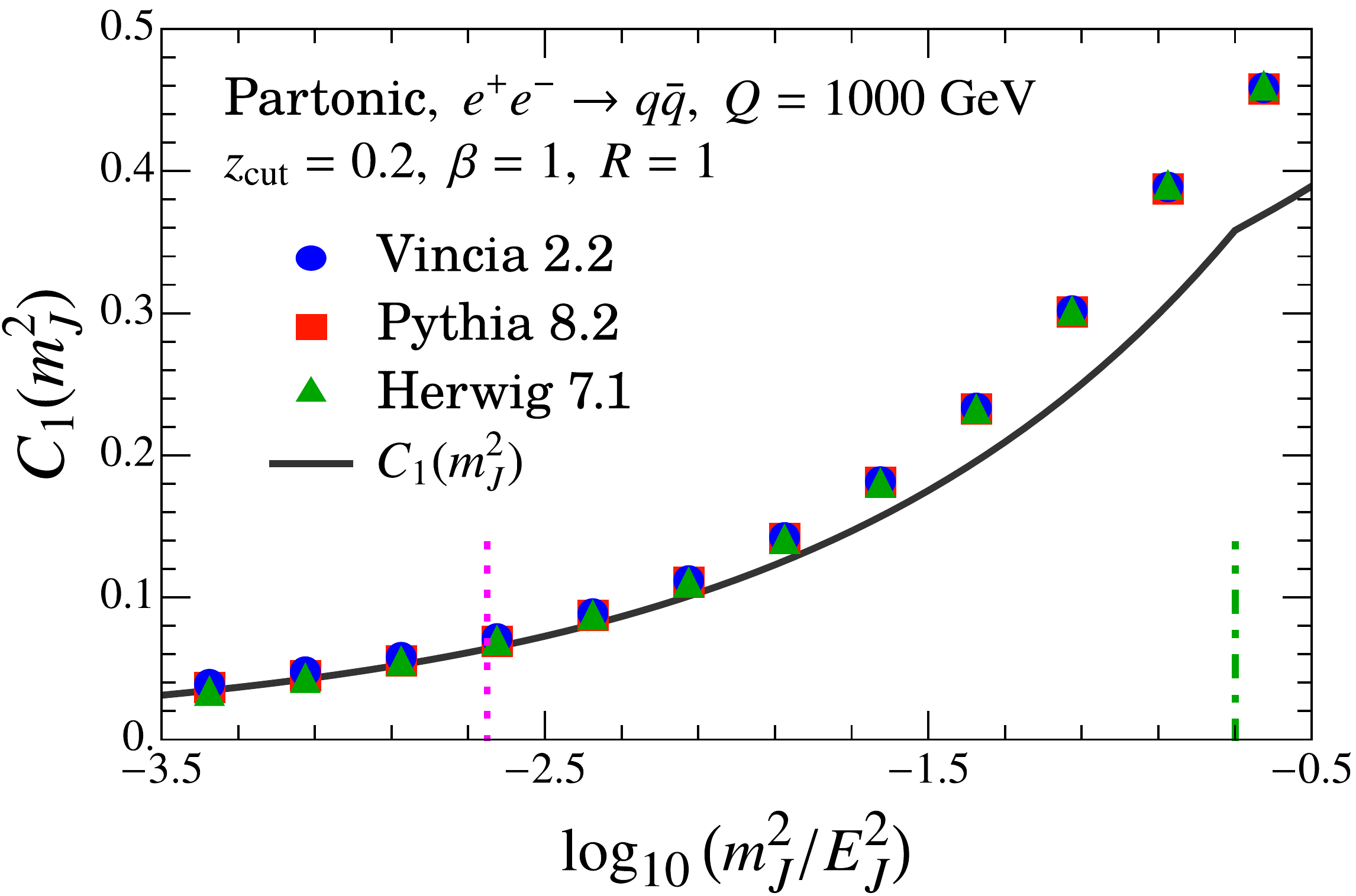}
\includegraphics[width=0.33\textwidth]{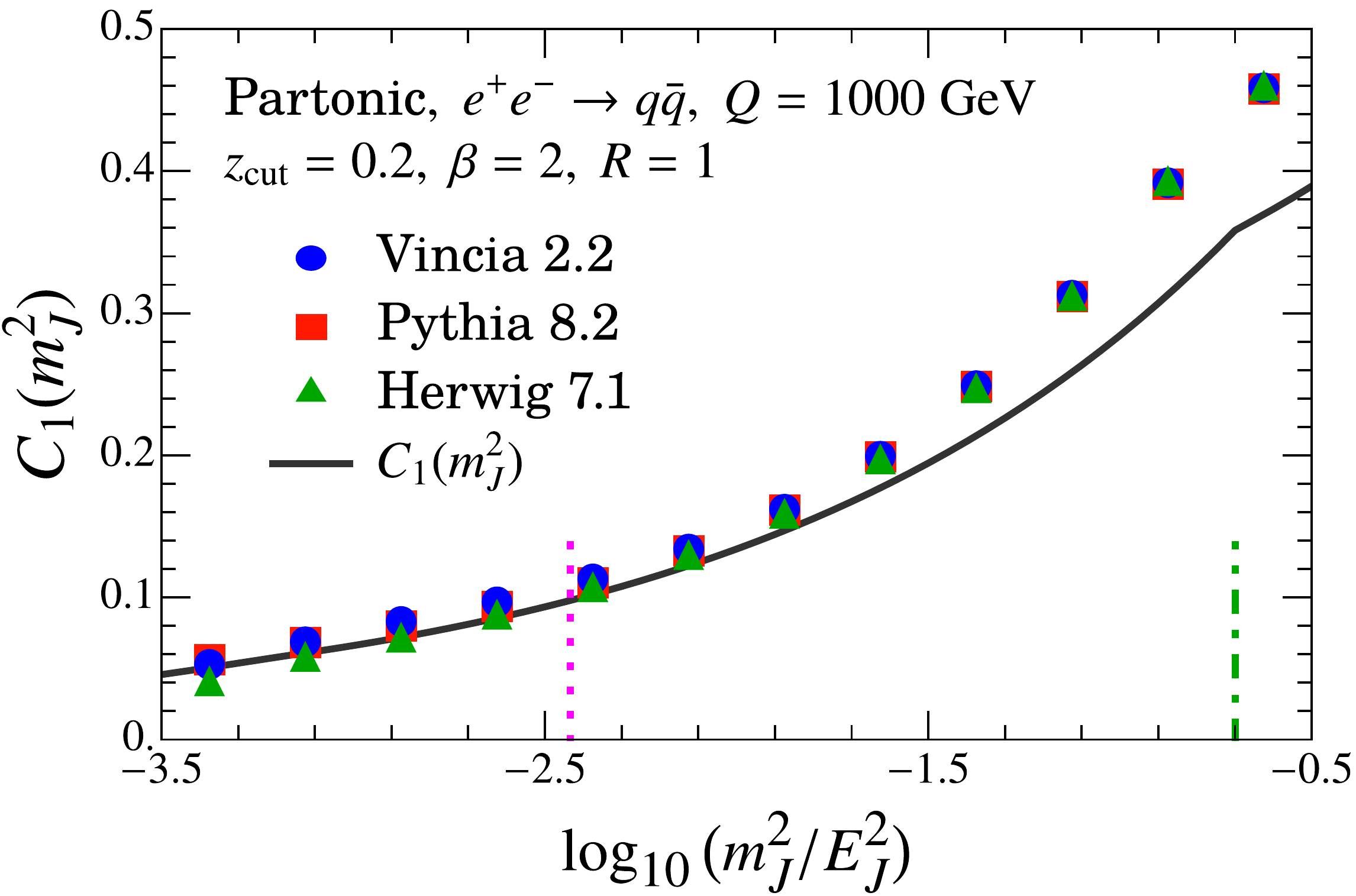}
\caption{Comparing our computation of $C_1(m_J^2)$ with partonic Monte Carlo for $\zcut = 0.1, 0.2$ (rows), $\beta = 0,1,2$ (columns). \label{fig:C1vsMC}}
\vspace{0.2cm}
\includegraphics[width=0.33\textwidth]{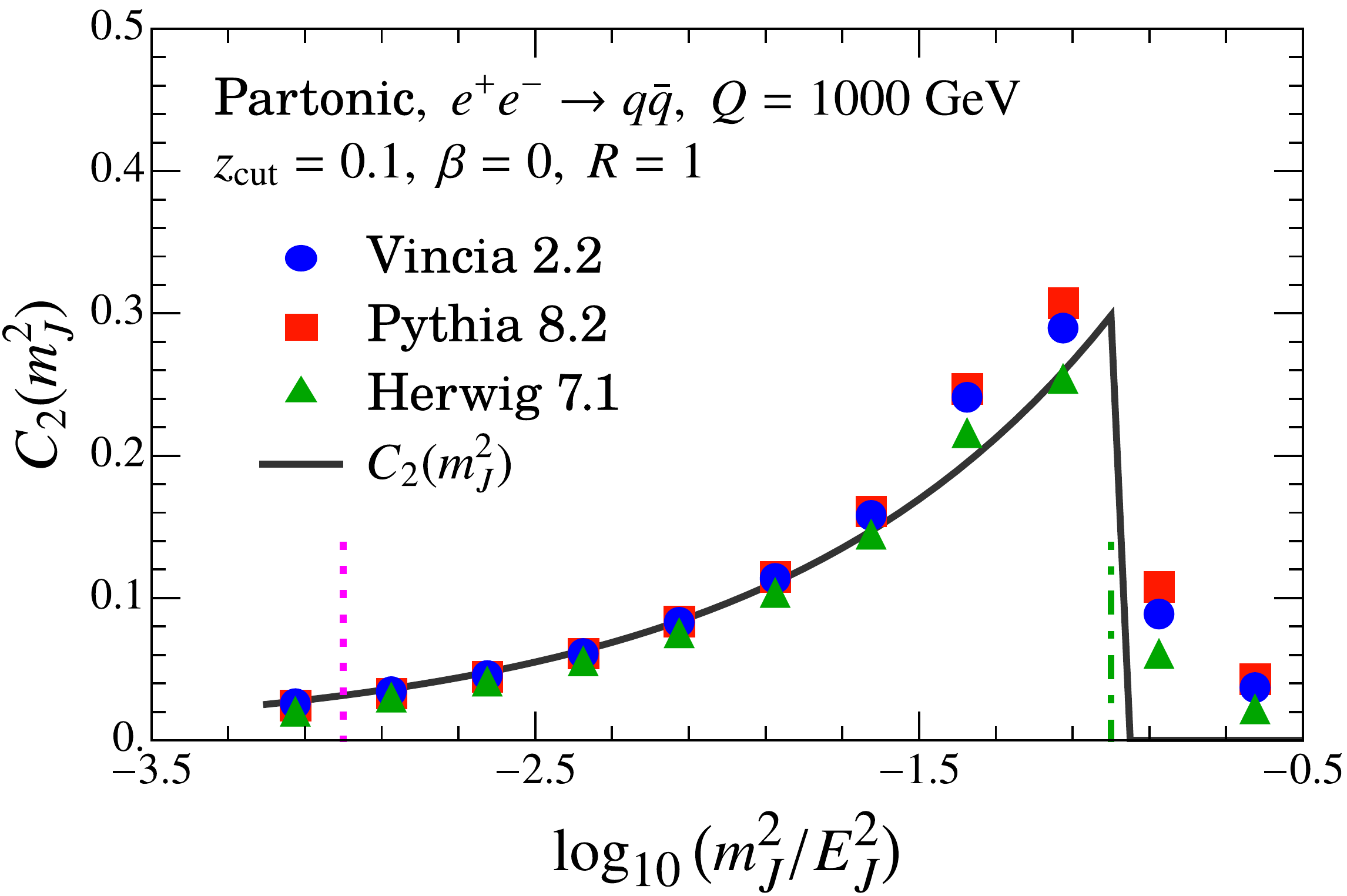}
\includegraphics[width=0.33\textwidth]{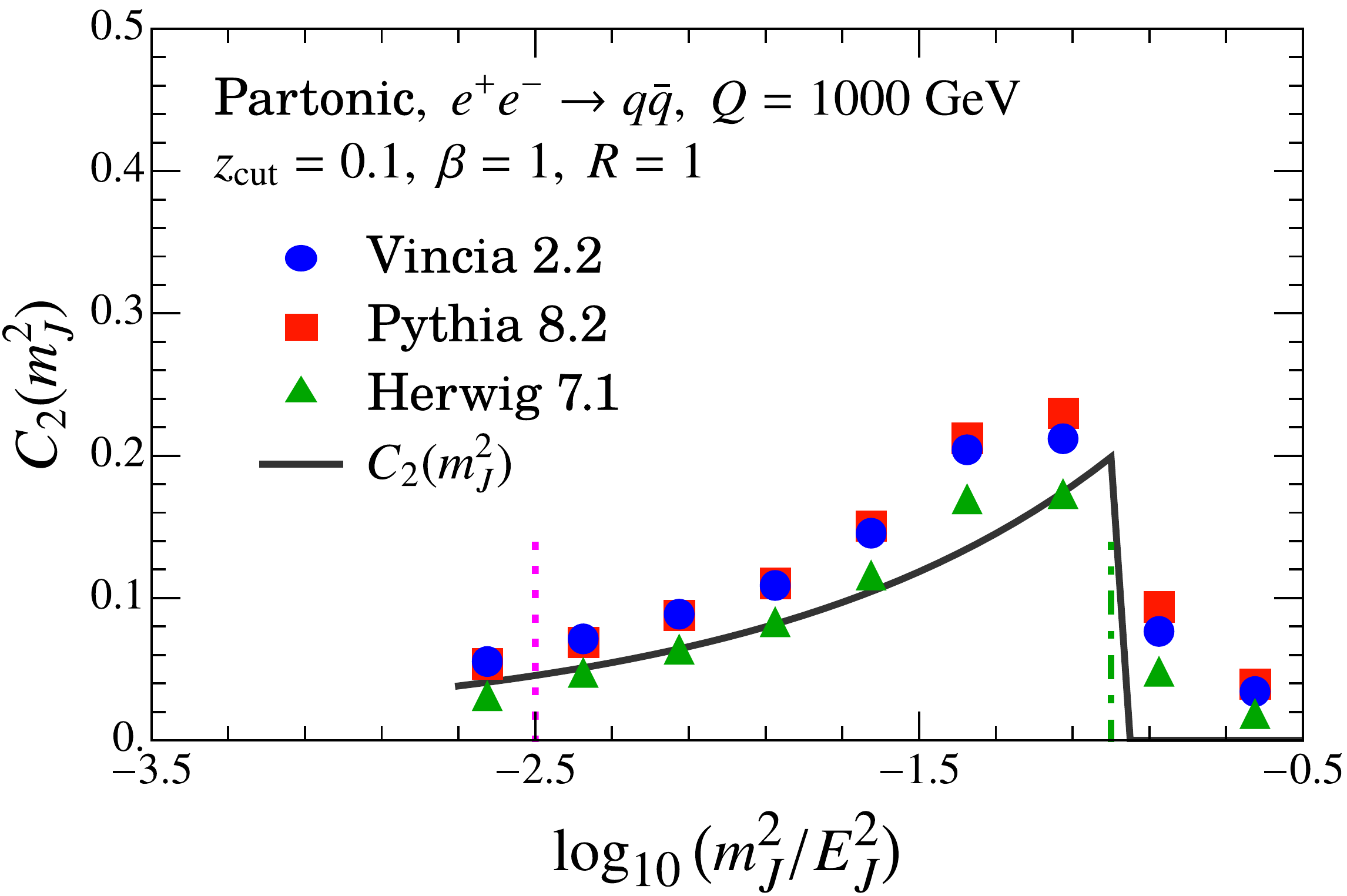}
\includegraphics[width=0.33\textwidth]{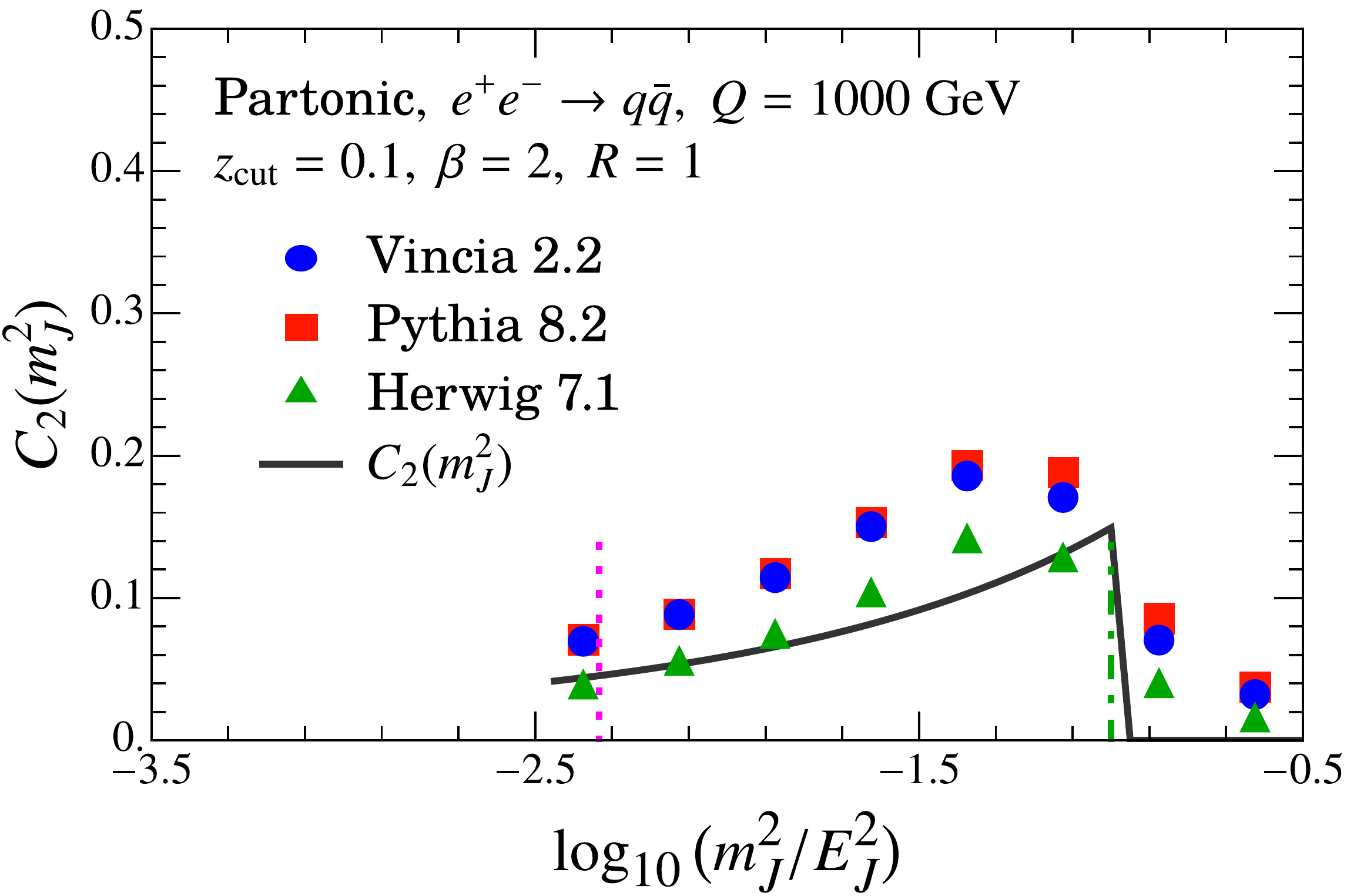}
\\[5pt]
\includegraphics[width=0.33\textwidth]{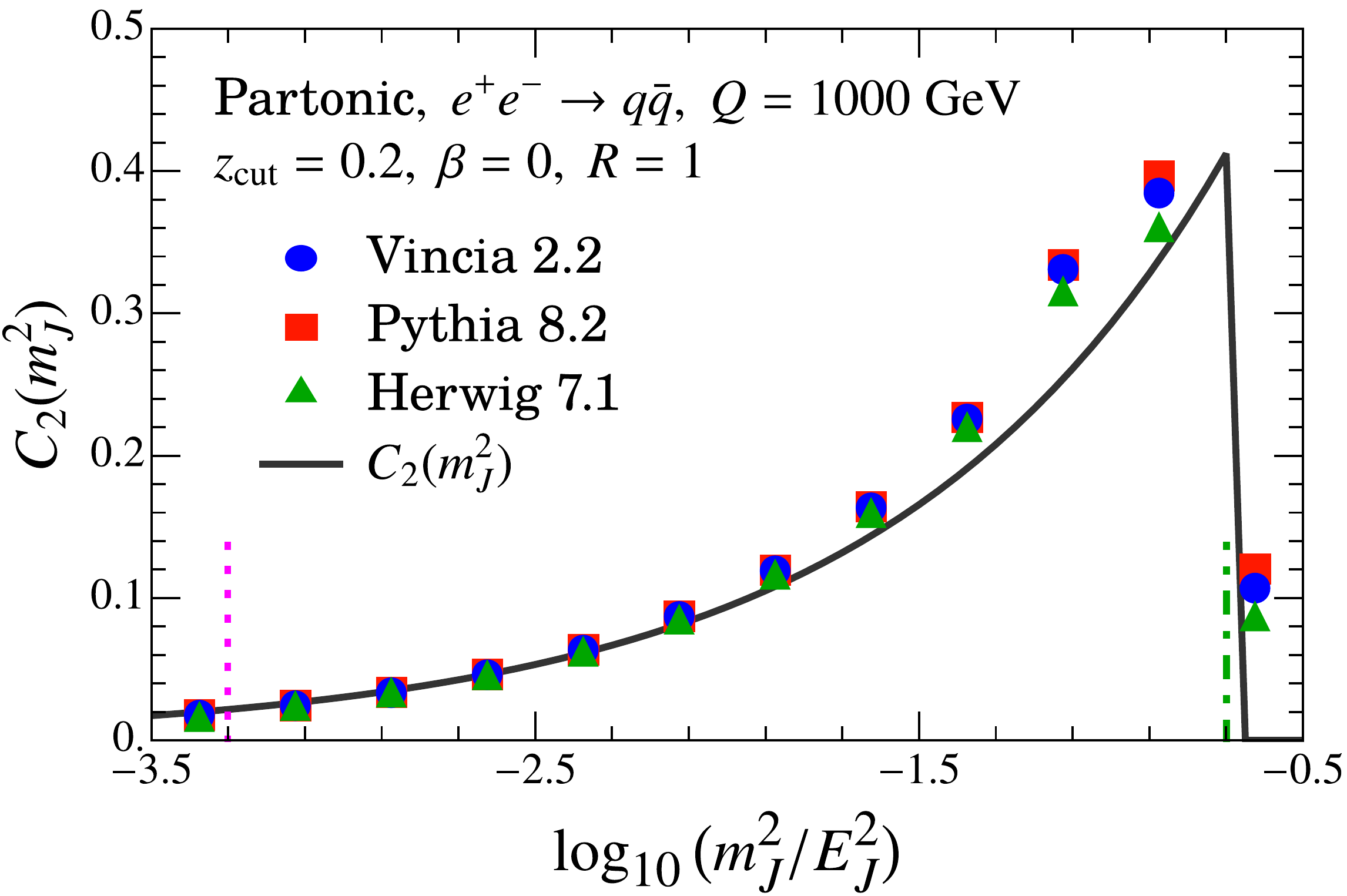}
\includegraphics[width=0.33\textwidth]{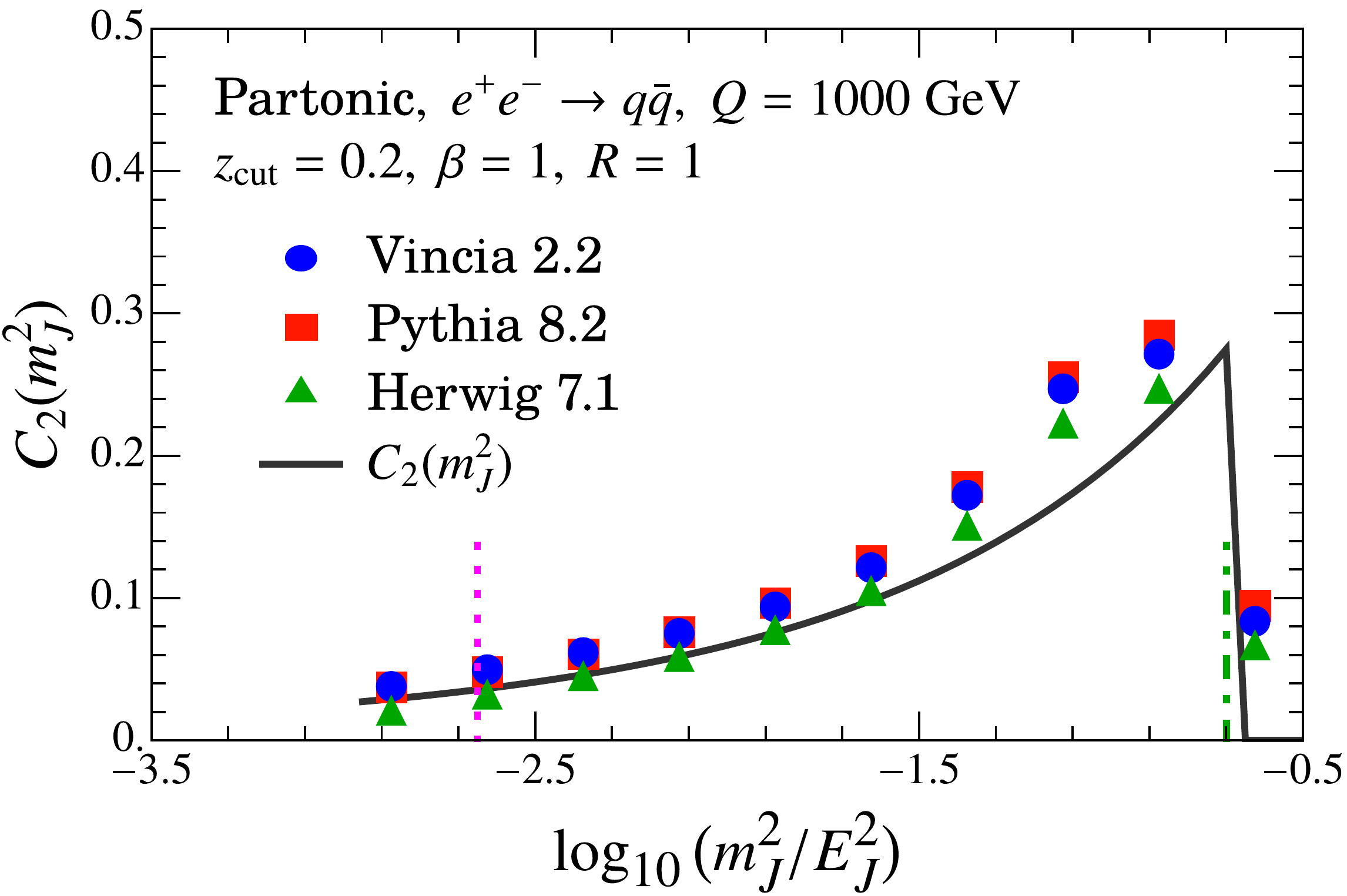}
\includegraphics[width=0.33\textwidth]{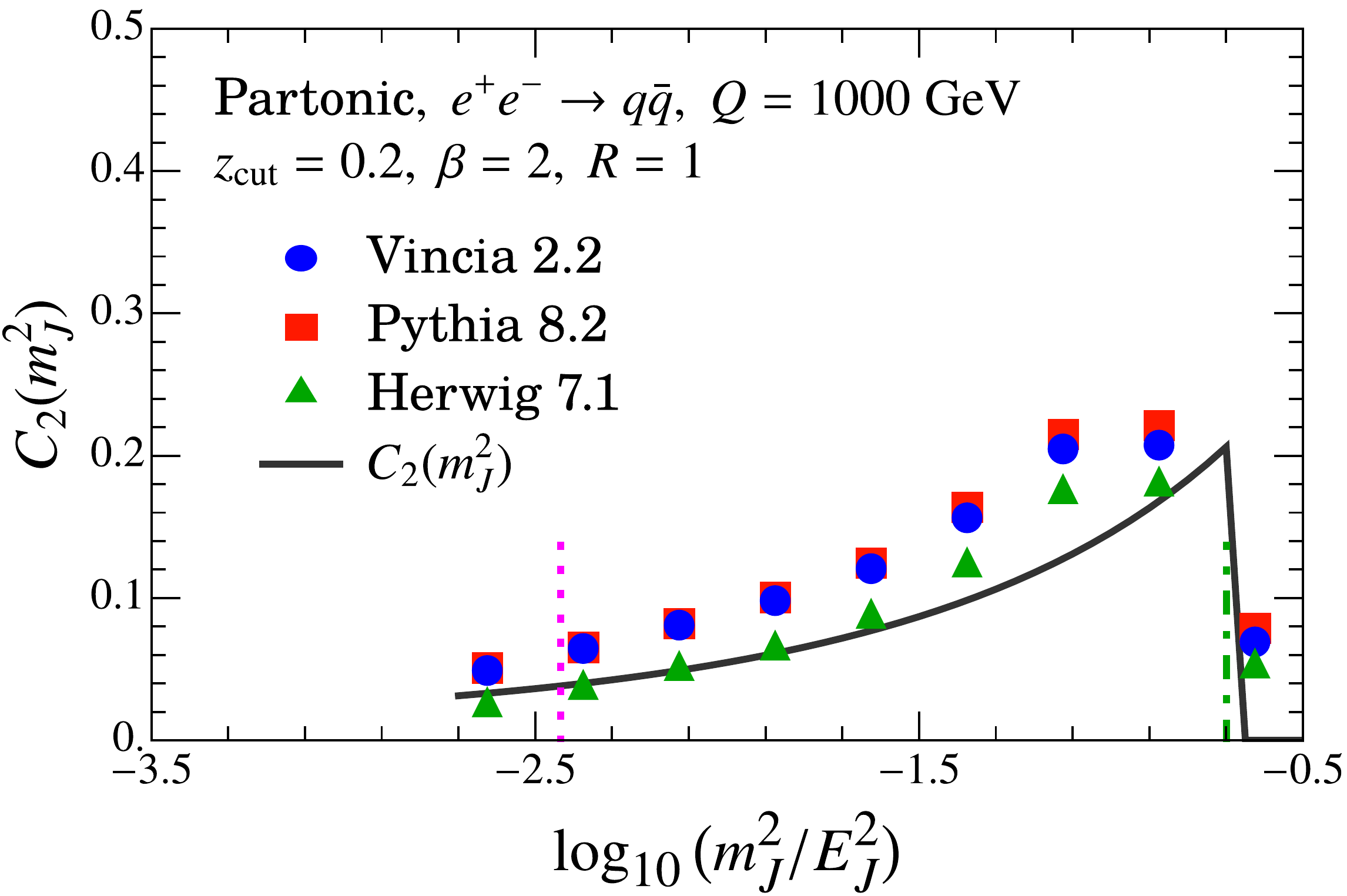}
\caption{Comparing our computation of $C_2(m_J^2)$ with partonic Monte Carlo for $\zcut = 0.1, 0.2$ (rows), $\beta = 0,1,2$ (columns). \label{fig:C2vsMC}}
\vspace{-0.2cm}
\end{figure}
As shown in \fig{modesLambdaprime}, for the SDOE region these $\Lambda'$ modes differ from the $\Lambda$ modes in \eq{pL} which represent the true dominant NP modes in the SDOE region. The $\Lambda'$ modes do no account for the existence of soft drop stopping angle, which is represented by the vertical orange line and which effectively reduces the jet radius to $R_g < R$.
In our SDOE analysis we have shown that the $\Lambda$ modes are the dominant nonperturbative modes. Their effects exceed those of the $\Lambda'$ modes since they have parametrically larger $p^+$ momenta, and lead to only an $\alpha_s$ suppression rather than extra Sudakov suppression. In the SDNP region, $m_J$ has decreased to a region where the CS, $\Lambda$, and $\Lambda'$ modes of \fig{modesLambdaprime} all have the same scaling, so that \eq{Lpmode} and \eq{LamCS} describe the same mode. We note, however, that the SDNP region does not have a simple connection to the SDOE region, so that the latter does not yield a simple normalization constraint for the SDNP the shape function, $F^\otimes_\kappa(k_{\rm NP},\beta)$ of \eq{SDNPfact}.

In \Refcite{Kang:2018vgn} a model equivalent to that of \Refcite{Frye:2016aiz} in \eq{FryeNP} was employed to describe soft drop angularity distributions~\cite{Berger:2003iw}. Our analysis of nonperturbative modes and conclusions can also be applied for angularities as long as one considers angularities that are sufficiently jet mass like and away from the broadening limit, i.e. $a < 1$. We leave a detailed analysis of the power corrections to groomed angularities to future work.
In \Refcite{Lee:2019lge} the angularity $a = 1.5$ was considered where the collinear mode becomes nonperturbative before the collinear-soft modes do. In this case our conclusions from the jet mass analysis do not apply.

\section{Monte Carlo Studies}

\label{sec:montecarlo}
In this section we present a Monte Carlo study to test our predictions for power corrections in the SDOE region. We consider three Monte Carlos: \Pythia 8.235~\cite{Sjostrand:2006za,Sjostrand:2014zea}, \Vincia 2.2~\cite{Fischer:2016vfv} and \Herwig 7.1~\cite{Bahr:2008pv,Bellm:2015jjp} with their respective default tunes for hadronization. The jet finding and soft drop grooming is implemented using the SoftDrop plugin in the \Fastjet 3.3 package~\cite{Cacciari:2011ma,Larkoski:2014wba}. We choose to simulate the $e^+e^- \rightarrow q \bar{q}$ dijet process at $Q = 500$ and 1000 GeV, take $R_0^{ee}=\pi/2$ and reconstruct two leading $R= 1$ jets with the anti-kT jet-finding algorithm~\cite{Cacciari:2008gp}.

\subsection{Comparing Wilson Coefficient Results with Monte Carlo}

We show in \fig{C1vsMC} a direct comparison of our calculation of $C_1(m_J^2)$ with partonic level Monte Carlo events. According to \eqs{C1}{C1R} $C_1(m_J^2)$ corresponds to the average of half the opening angle of the stopping pair for a given jet mass $m_J$. So to obtain the Monte Carlo result for a given $m_J$ value we can simply sample the opening angle of the stopping pair of the groomed jet in different jet mass bins. We show the comparison for $\zcut = \{0.1, 0.2\}$ and $\beta = \{0,1,2\}$ and $Q = 1000$ GeV with \Pythia, \Vincia and \Herwig. The purple and green vertical lines in \fig{C1vsMC} delineate the SDOE region as shown in \fig{logrhoplot}. We observe good agreement of all the Monte Carlos with the calculation in the SDOE region.

In order to extract $C_2(m_J^2)$ from Monte Carlo we implement the $\delta$ function in \eq{bndry03} by modifying the soft drop condition such that
\begin{align}
\label{eq:SoftDropEps}
\Theta \bigg(z - \tzcut \theta^\beta\bigg) \rightarrow
\Theta \bigg(z - \tzcut \theta^\beta + \frac{2}{\theta} \, \eps\bigg) \, ,
\end{align}
and take a numerical derivative with respect to $\eps$ of the jet mass cross section in a given jet mass bin. This yields $\tilde C_2(m_J^2)$ in \eq{C2def} times the partonic cross section in that bin. In \fig{C2vsMC} we compare the $C_2(m_J^2)$ obtained in this way from the Monte Carlo using the modification of the soft drop test in \eq{SoftDropEps} with our analytical result from \eqs{C2def}{C2}. Note that we do not show results for $C_2(m_J^2)$ from the Monte Carlo further to the left of the pink vertical line, i.e., beyond the SDOE region where the formalism with $C_2$ no longer applies.
We observe that unlike the case of $C_1(m_J^2)$ discussed in \fig{C1vsMC}, \Pythia and \Vincia results for $C_2(m_J^2)$ differ from \Herwig, with \Herwig being in a better agreement with our calculation of $C_2$.

\subsection{Catchment Area Geometry versus Jet Mass}
\begin{figure}[t!]
\centering
\includegraphics[width=0.45\textwidth]{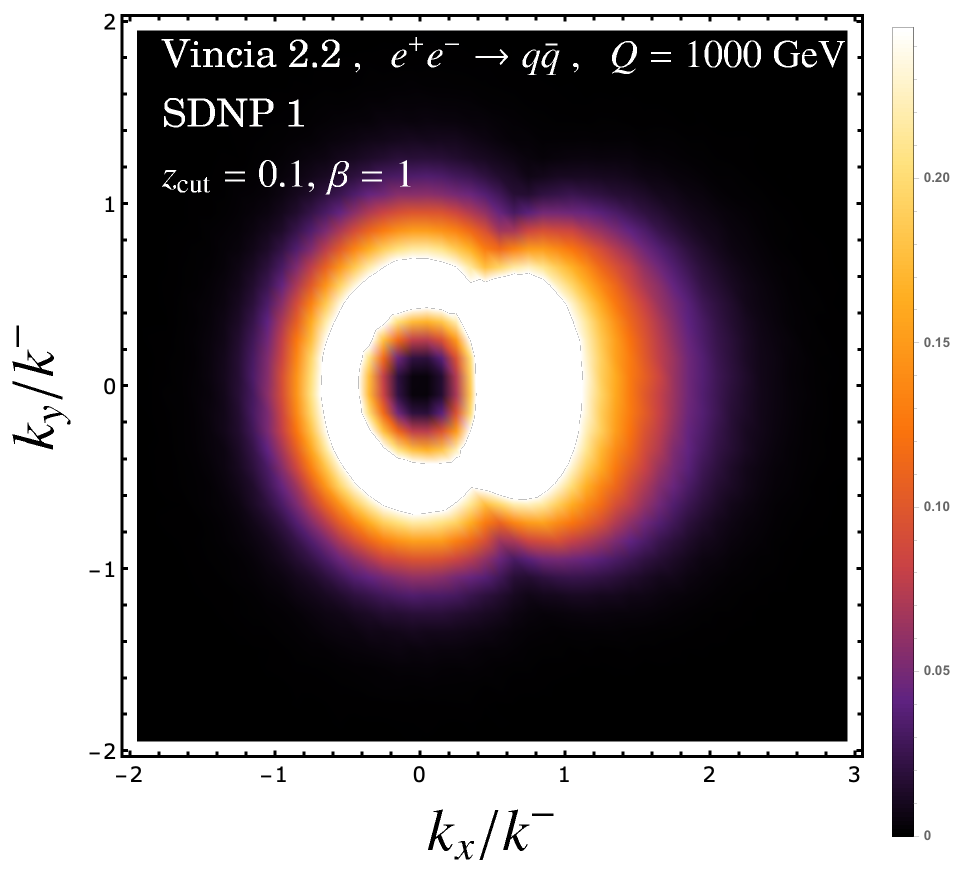}
\includegraphics[width=0.45\textwidth]{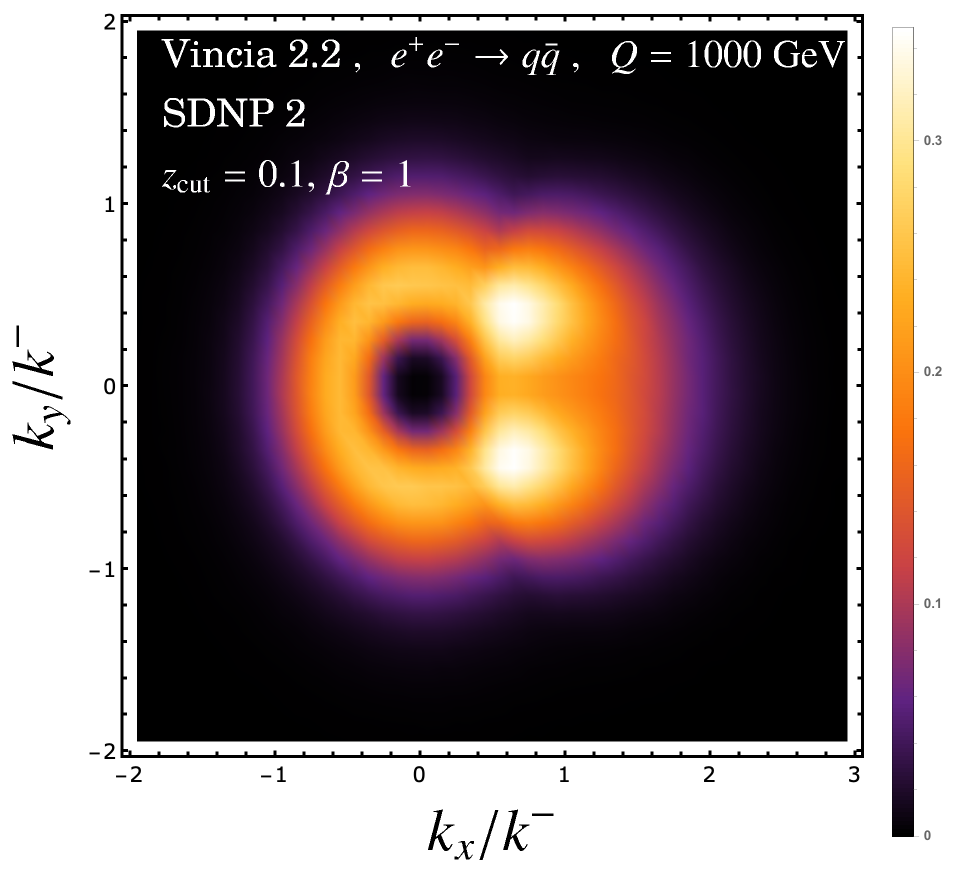}
\\
\includegraphics[width=0.45\textwidth]{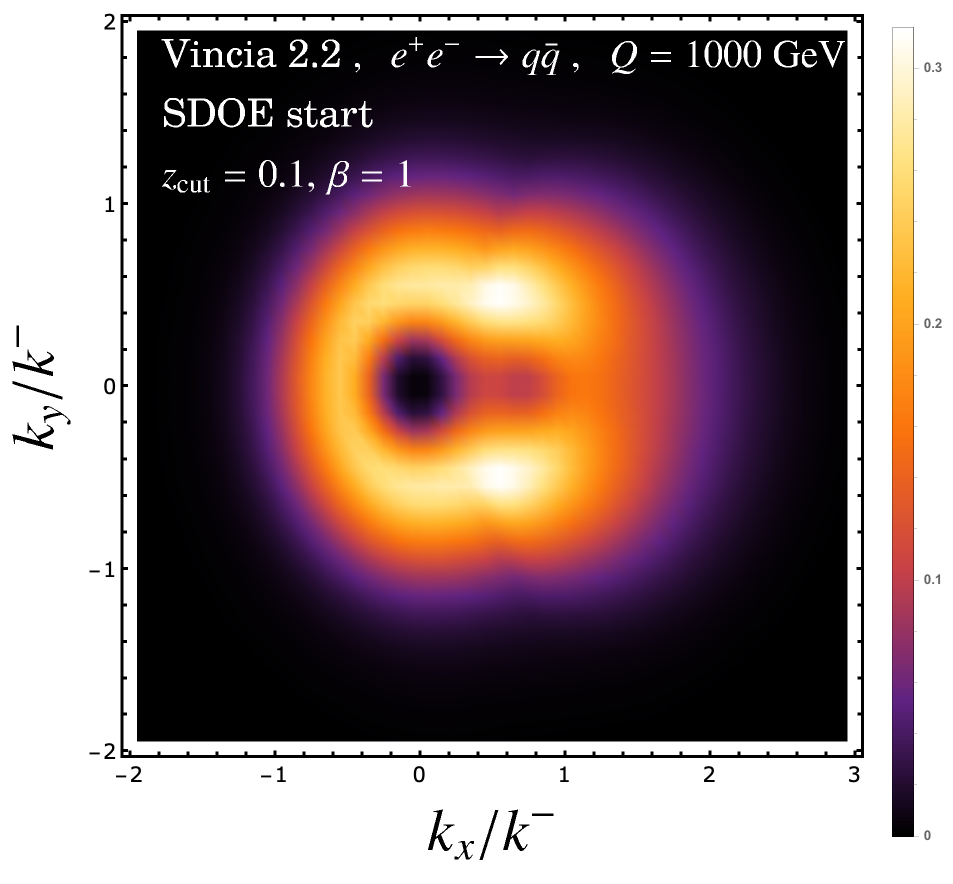}
\includegraphics[width=0.45\textwidth]{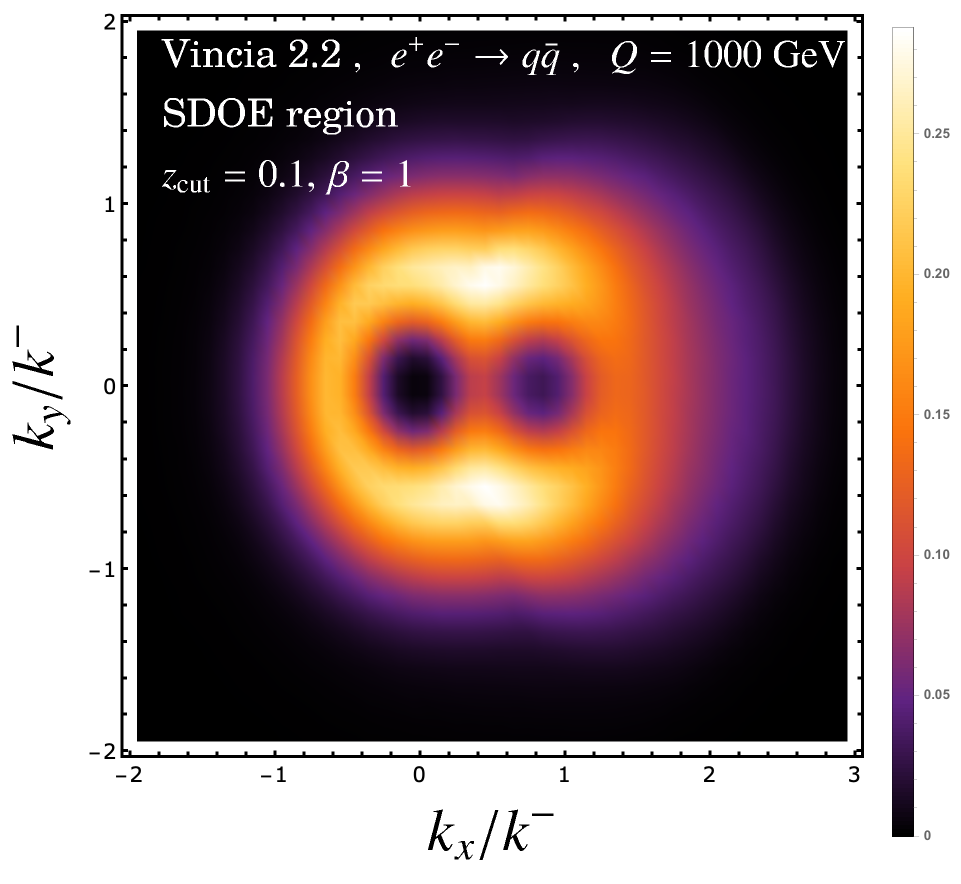}
\\
\includegraphics[width=0.45\textwidth]{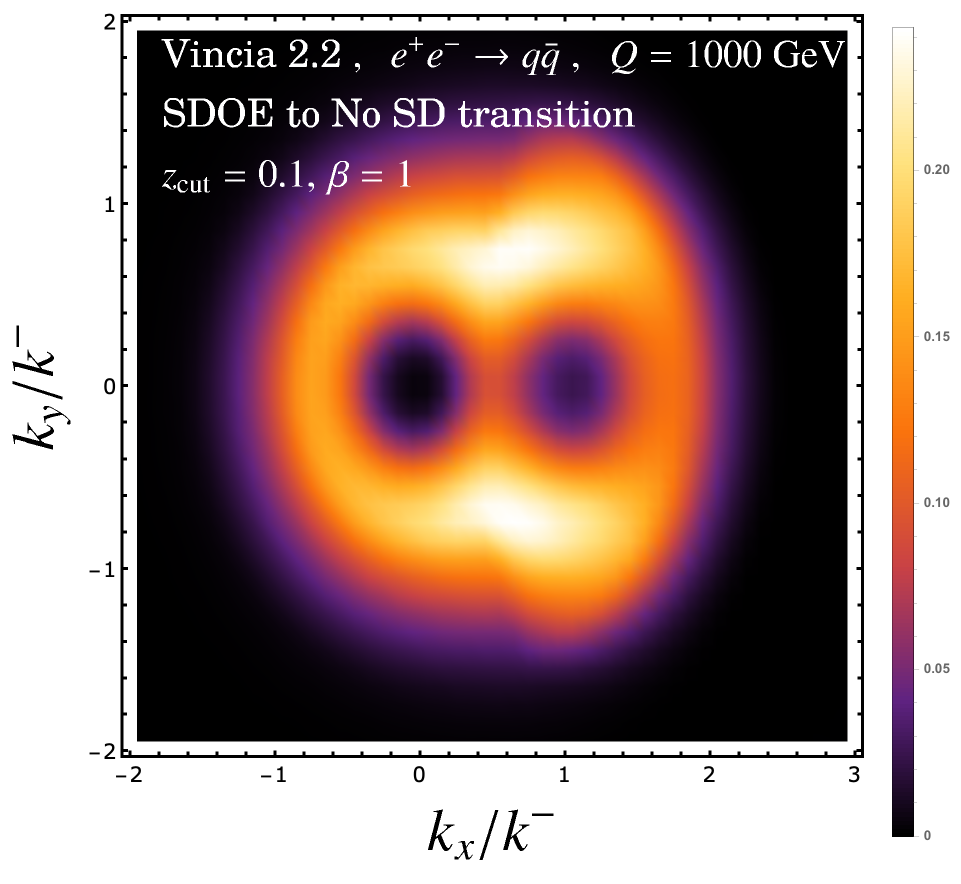}
\includegraphics[width=0.45\textwidth]{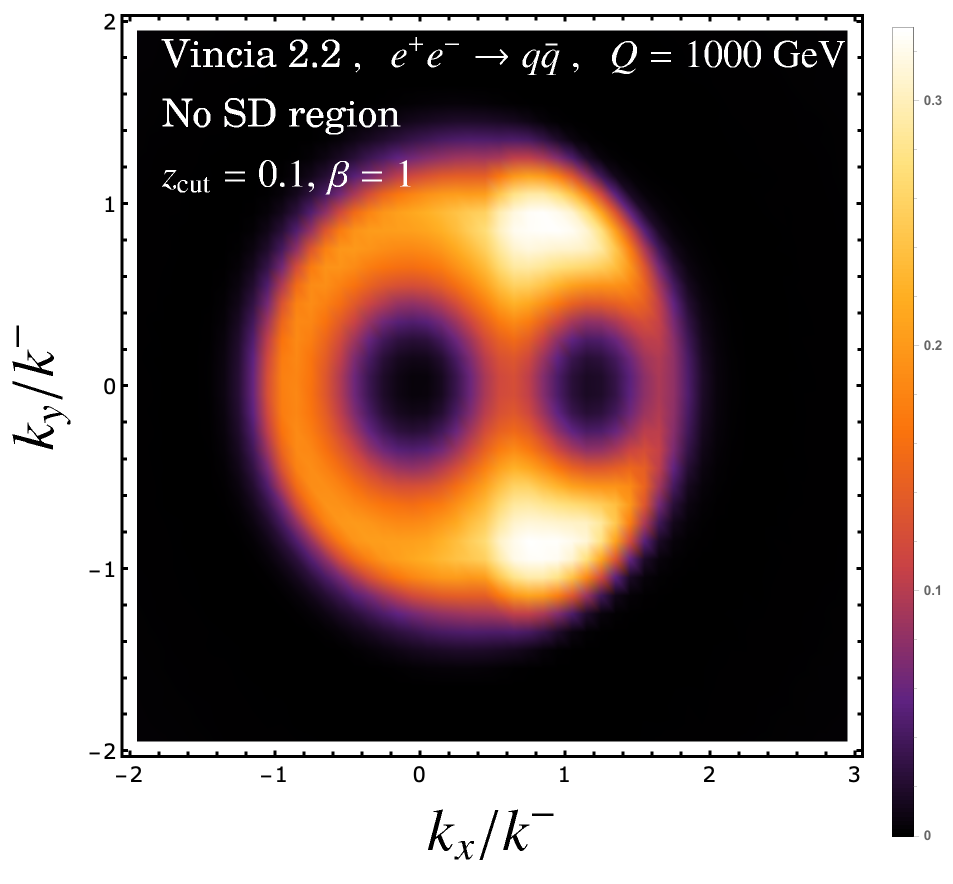}
\caption{``Heat maps'' showing probability distribution of the angular location of the NP subjet in the plane perpendicular to the jet axis in rescaled coordinates for different regions of jet mass spectrum. \label{fig:heatmaps}}
\vspace{-0.2cm}
\end{figure}
\begin{figure}[t!]
\centering
\includegraphics[width=0.75\textwidth]{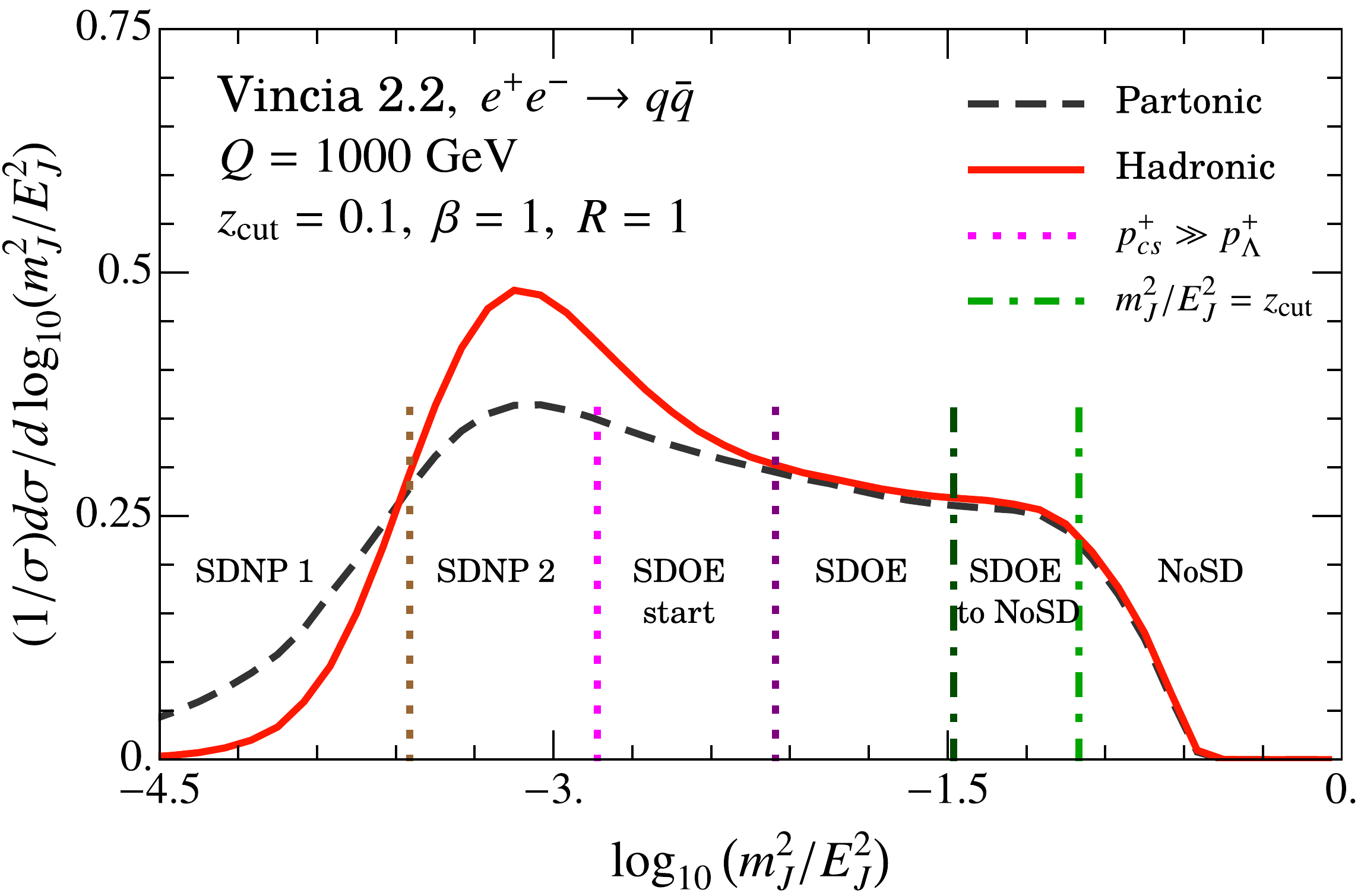}\\
\vspace{-0.2cm}
\caption{Vincia prediction for $m_J^2/E_J^2$ distribution indicating the bins of jet masses that correspond to the different panels shown in \fig{heatmaps}.
Here the inequality $\gg$ is replaced by a factor of 3.}
\label{fig:logrhoplotNPxy}
\vspace{-0.2cm}
\end{figure}
The key ideas that allowed us to describe the NP corrections in the SDOE and SDNP regions were our assumptions about the geometry of the catchment area of the NP subjets around the collinear and collinear-soft jet axes. Upon rescaling the NP subjet momenta according to \eq{rescaling} we were able to factorize the perturbative contributions and the nonperturbative matrix elements. To test the underlying kinematic approximations we show in \fig{heatmaps} heat maps illustrating the distribution of the rescaled angular location of the NP subjets in the plane perpendicular to the jet axis for different bins of the jet masses from hadron level \Vincia with at $Q = 1000$ GeV with $\zcut = 0.1$ and $\beta = 1$.
Here hotter colors correspond to a higher density of NP subjets.
In accordance with our definition for NP subjets in \sec{CAforNP} we uncluster the groomed jet until we find the first subjet that has energy $E \leq 1.0$ GeV. We then rescale the angle of this subjet with respect to the jet axis using \eq{rescaling} and our calculation for the opening angle $\theta_{cs} = 2\,C_1(m_J^2)$ for the corresponding jet mass value of the jet.
We then rotate the subjet in the azimuthal plane by $\phi_{cs}$, where the more energetic (collinear) subjet lies at the origin, and the softer subjet of the stopping pair is rotated onto the positive $x$-axis (according to \fig{NPfig8rescaled}). These transformation allow us to visualize the 2 dimensional distribution of the angular locations of the NP subjet in the rescaled coordinates shown in \figs{NPfig8rescaled}{NPfig8SDNP}b.

We divide the distributions in several jet mass bins in order to visualize how the catchment area geometry changes when going from the SDNP to the SDOE region, and from the SDOE to the ungroomed resummation region. These distinct regions of the jet mass spectrum have been defined as in \eqs{NPregions}{sdlimit} and are illustrated in \fig{logrhoplotNPxy}. We can see from \fig{heatmaps} that there are prominent voids around the collinear subjet and (once present) the perturbative collinear-soft subjet. These are primarily due to the fact that both hadronization and the CA algorithm cluster non-perturbative particles/hadrons with the energetic subjets close to the collinear or collinear-soft subjet axes, and these clusters always lie deeper than the first NP subjet in the CA clustering tree.
The top left panel corresponds to the events in the SDNP region. Here we see that the NP subjets are clustered along with the collinear subjet at angles $\sim \theta_{cs}$. The region extends slightly to the right of the collinear subjet catchment area since the stopping subjets are themselves nonperturbative and are included in the distribution. In the subsequent three panels we enter the SDOE region and see the expected geometry from \fig{NPfig8rescaled}a emerging. Since we rescaled the angles of the NP subjets by our perturbative prediction, $\theta_{cs}=2C_1$, it is a nontrivial check of our discussions in \sec{EFTinOPEregion} that the rightmost ``black hole'' at the collinear-soft subjet location in the center right panel is indeed centered at ${k_x}/{k^-}\simeq 1$ and ${k_y}/{k^-} \simeq 0$. The resulting regions of high density also confirm our expectation that the dominant NP modes are determined by the $\theta_{cs}$ angle, leading to the halos of radius $\simeq 1$ around the stopping subjets. In the bottom two panels we exit the SDOE region and start to enter the ungroomed resummation region. Here we see a distortion in the distributions due to the effect of the ungroomed jet boundary, which is a circle of radius $R=1$.

\subsection{Testing Universality with Fits for Hadronic Parameters}
\label{sec:MCfits}
\begin{figure}[t!]
\centering
\includegraphics[width=0.75\textwidth]{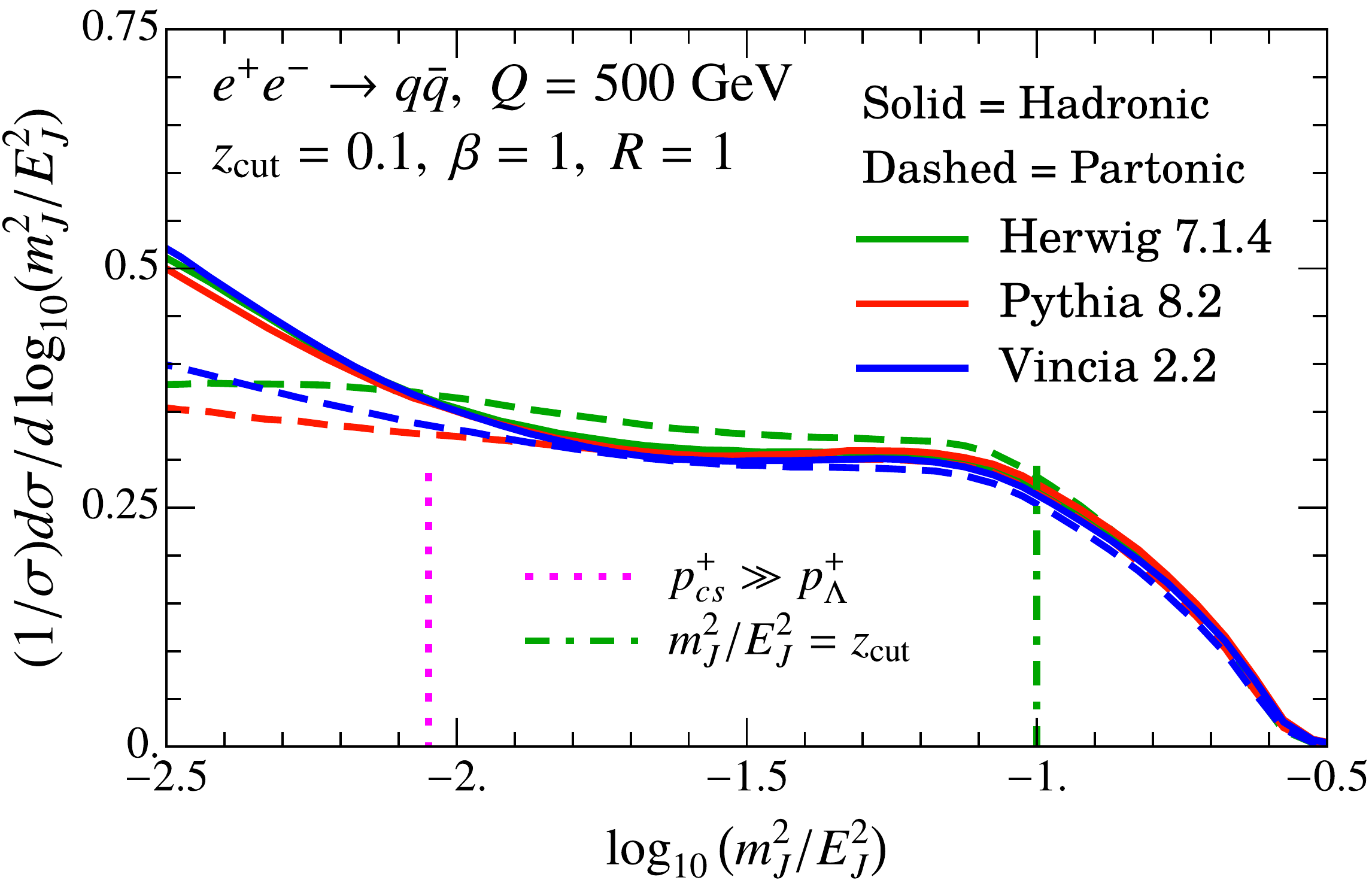}\\
\vspace{-0.2cm}
\caption{Differing hadronization models and partonic cross sections in Monte Carlos in the soft drop operator expansion region.}
\label{fig:partHadall}
\vspace{-0.2cm}
\end{figure}
\begin{table}[t!]
\begin{center}
\scalebox{1}{\begin{tabular}{c || c | c | c | c}
\hline \hline
Event Generator & $\Omega_{1q}^{\figeight}$ (GeV) & $\Upsilon_{1,0}^q$ (GeV) & $\Upsilon_{1,1}^q$ (GeV) & $\chi_{\rm min}^2$/dof
\\[0.5ex] \hline\hline
\Pythia 8.235 & 1.63 & -1.21 & 0.33 & 0.96
\\[0.5ex] \hline
\Vincia 2.2& 1.22 & -1.04 & 0.50 & 0.84
\\[0.5ex] \hline
\Herwig 7.1.4 (default) &1.14 & -1.73 & -0.15 & 2.53
\\[0.5ex] \hline
\Herwig 7.1 ($p_T$B) &1.14 & -1.32 & -0.11 & 0.77
\\ \hline\hline
\end{tabular}}
\end{center}
\caption{Summary of central fits for hadronic parameters in the SDOE region for different Monte Carlos. We also include a goodness of fit measure with $\chi_{\rm min}^2$/dof following the method described in the text.}
\label{tab:fitresults}
\end{table}

\begin{figure}[t!]
\centering
\includegraphics[width=0.33\textwidth]{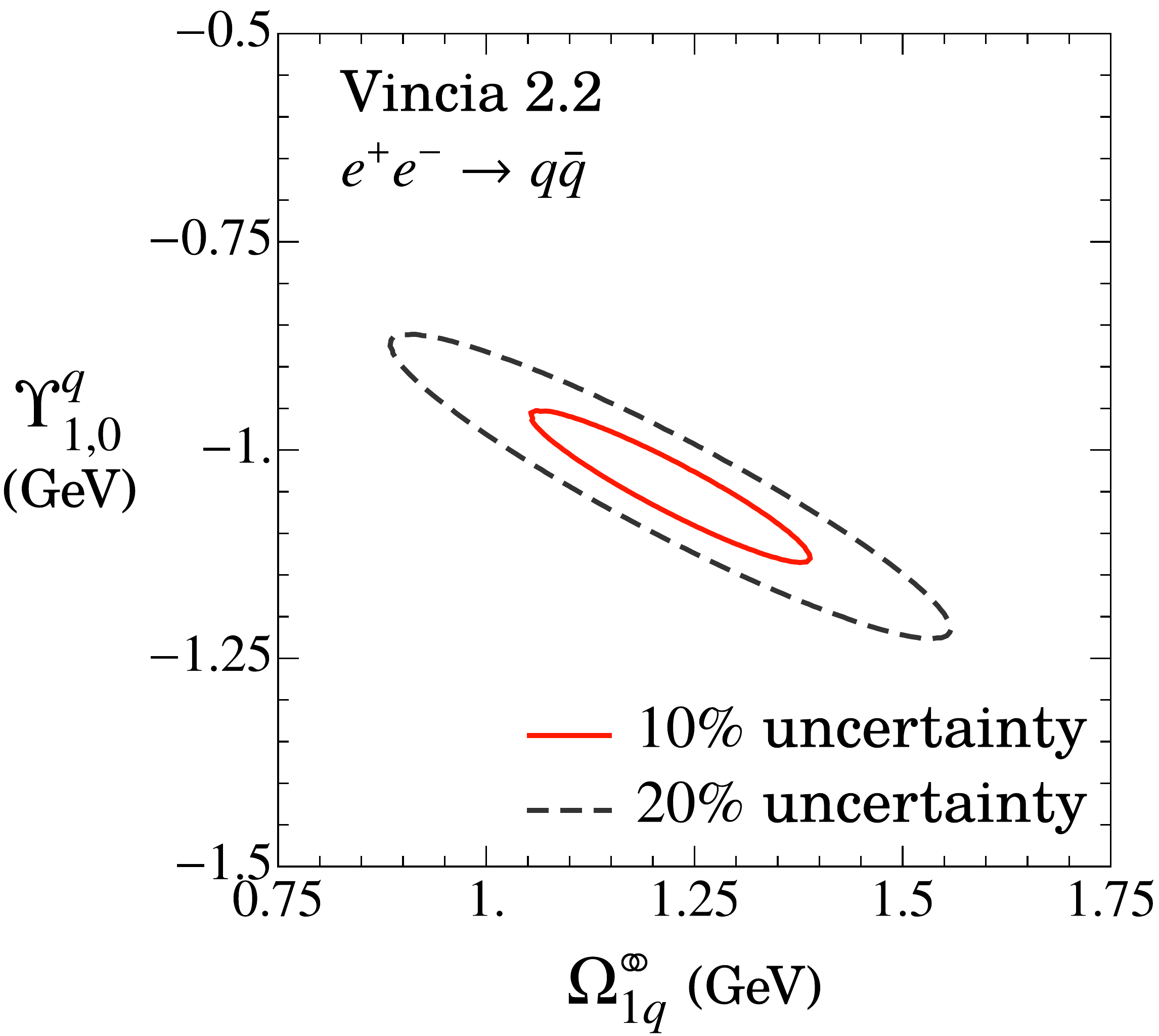}
\includegraphics[width=0.32\textwidth]{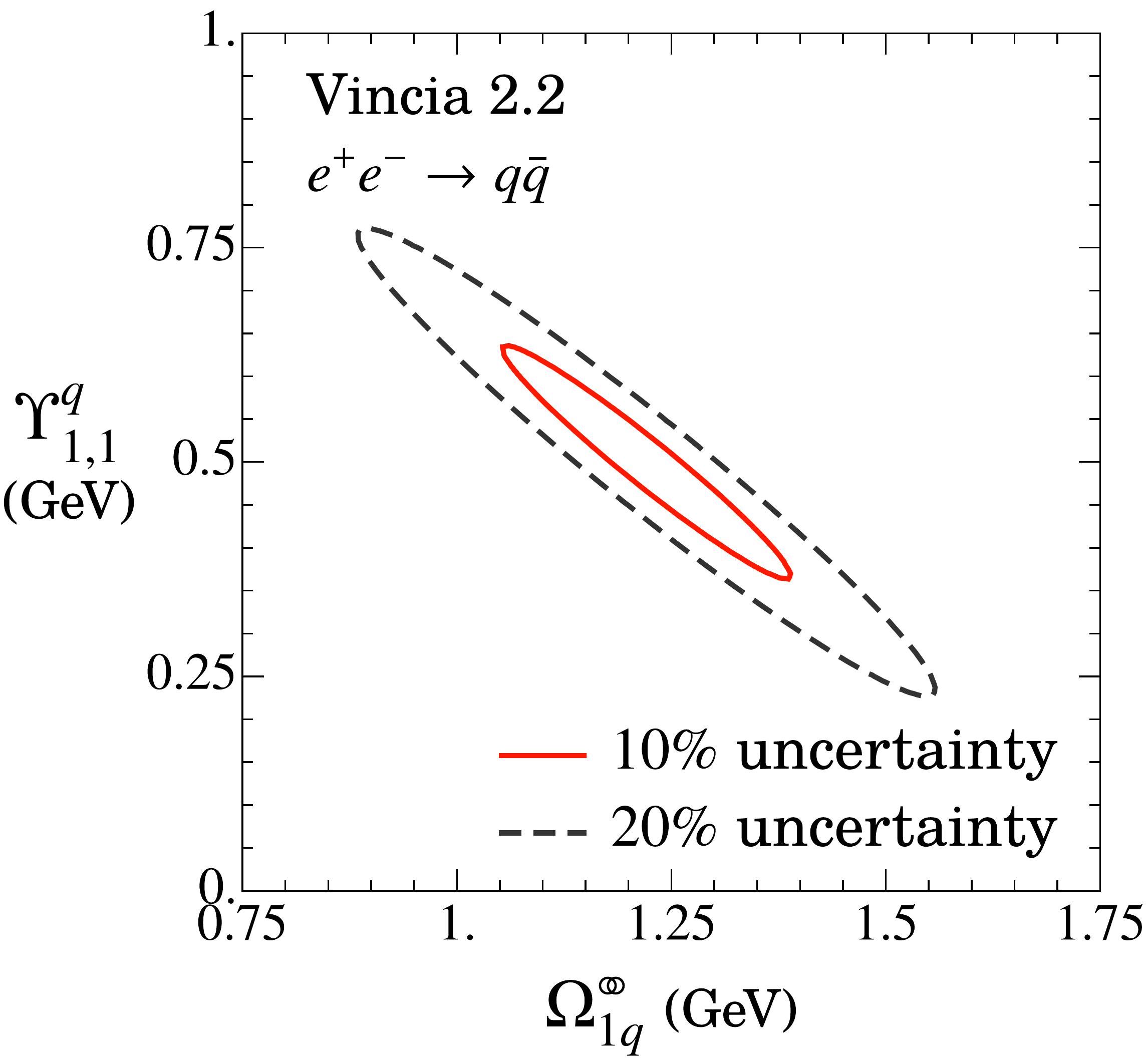}
\includegraphics[width=0.32\textwidth]{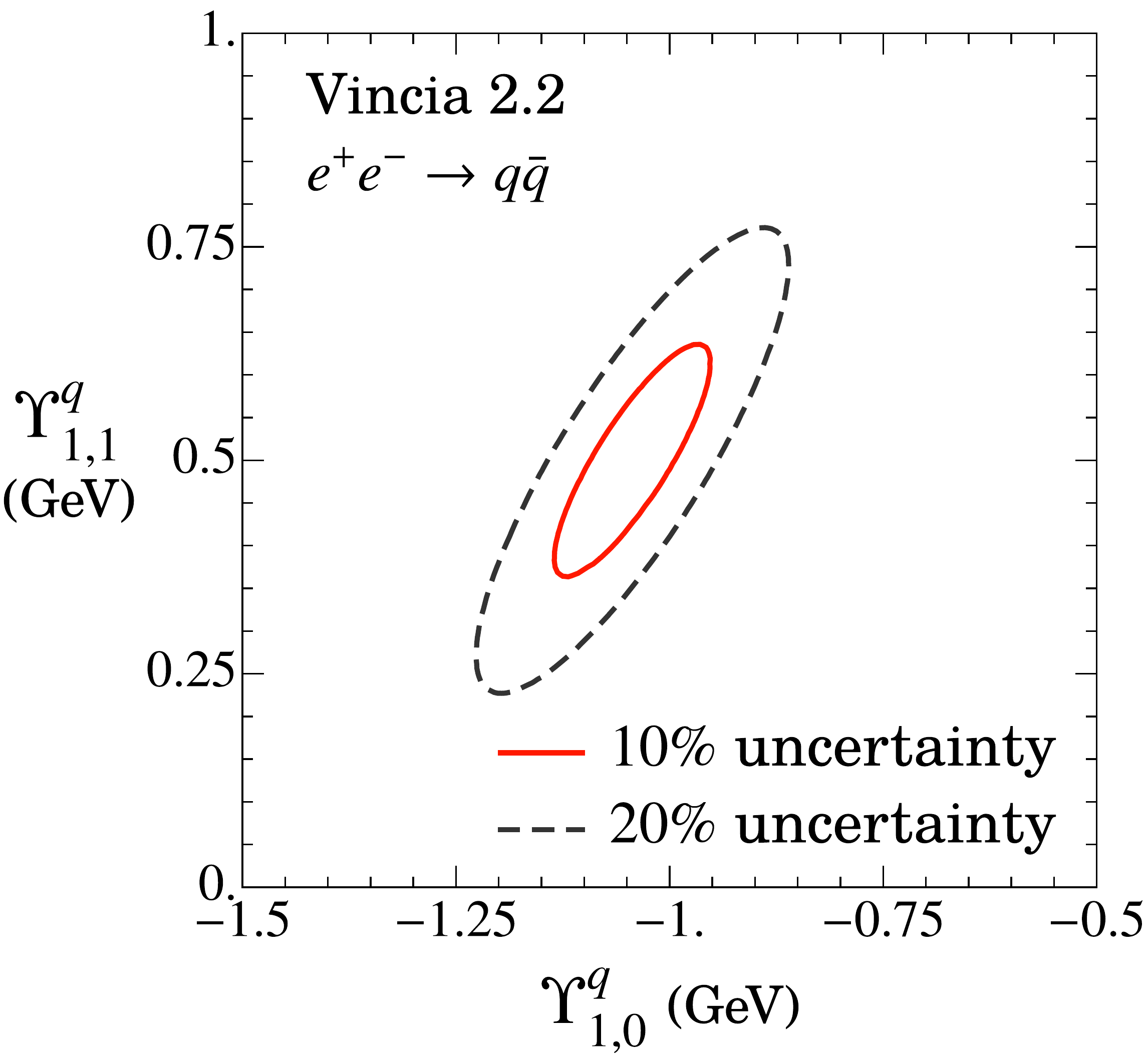}
\caption{70\% Confidence interval ellipses showing correlations between the hadronic parameters from \Vincia when assigning 10\% and 20\% hadronization uncertainties. Other Monte Carlos yield ellipses of similar size.\label{fig:ellipsesVincia}}
\vspace{-0.2cm}
\end{figure}

\begin{figure}[t!]
\centering
\includegraphics[width=0.48\textwidth]{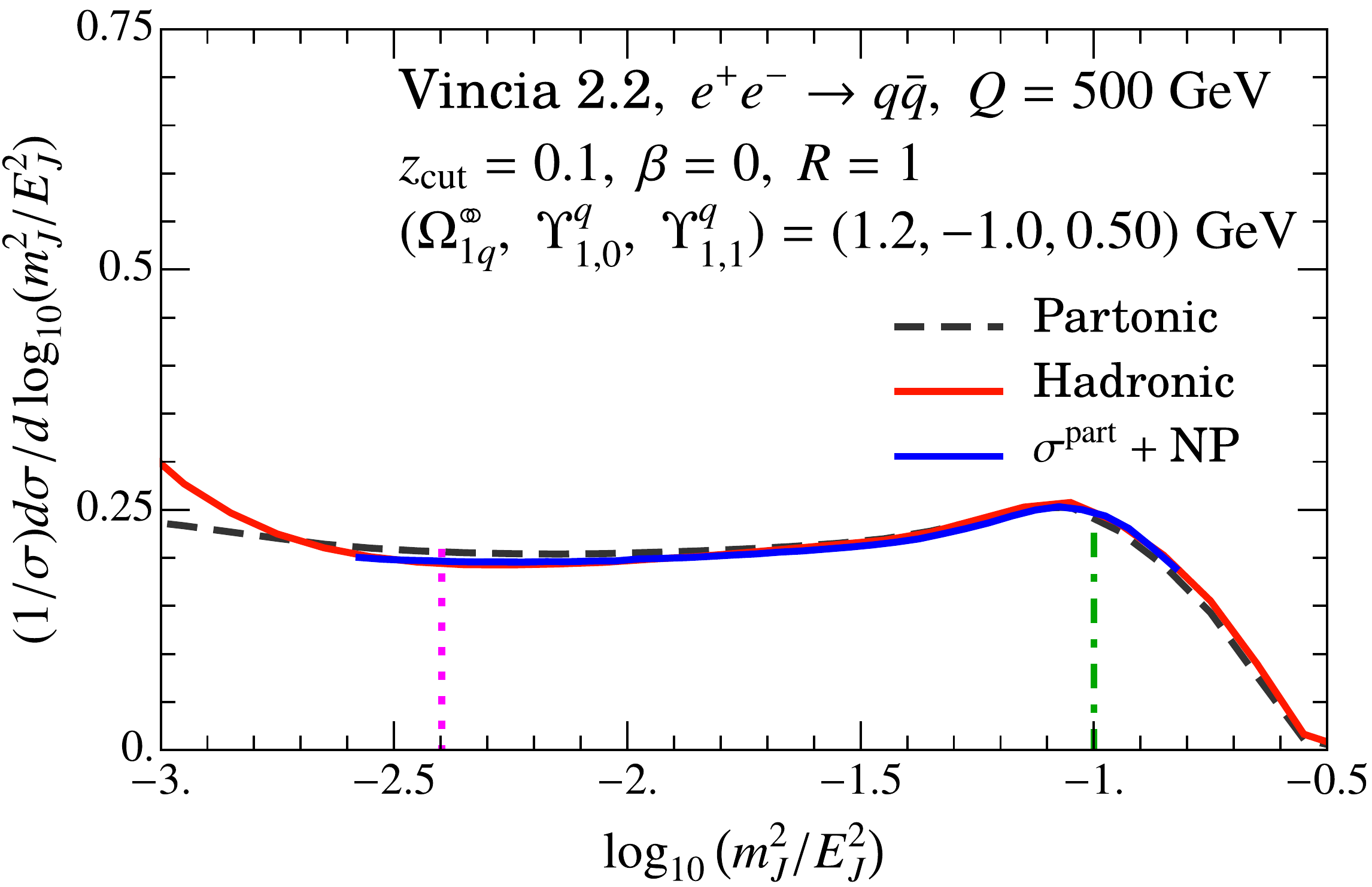}
\includegraphics[width=0.48\textwidth]{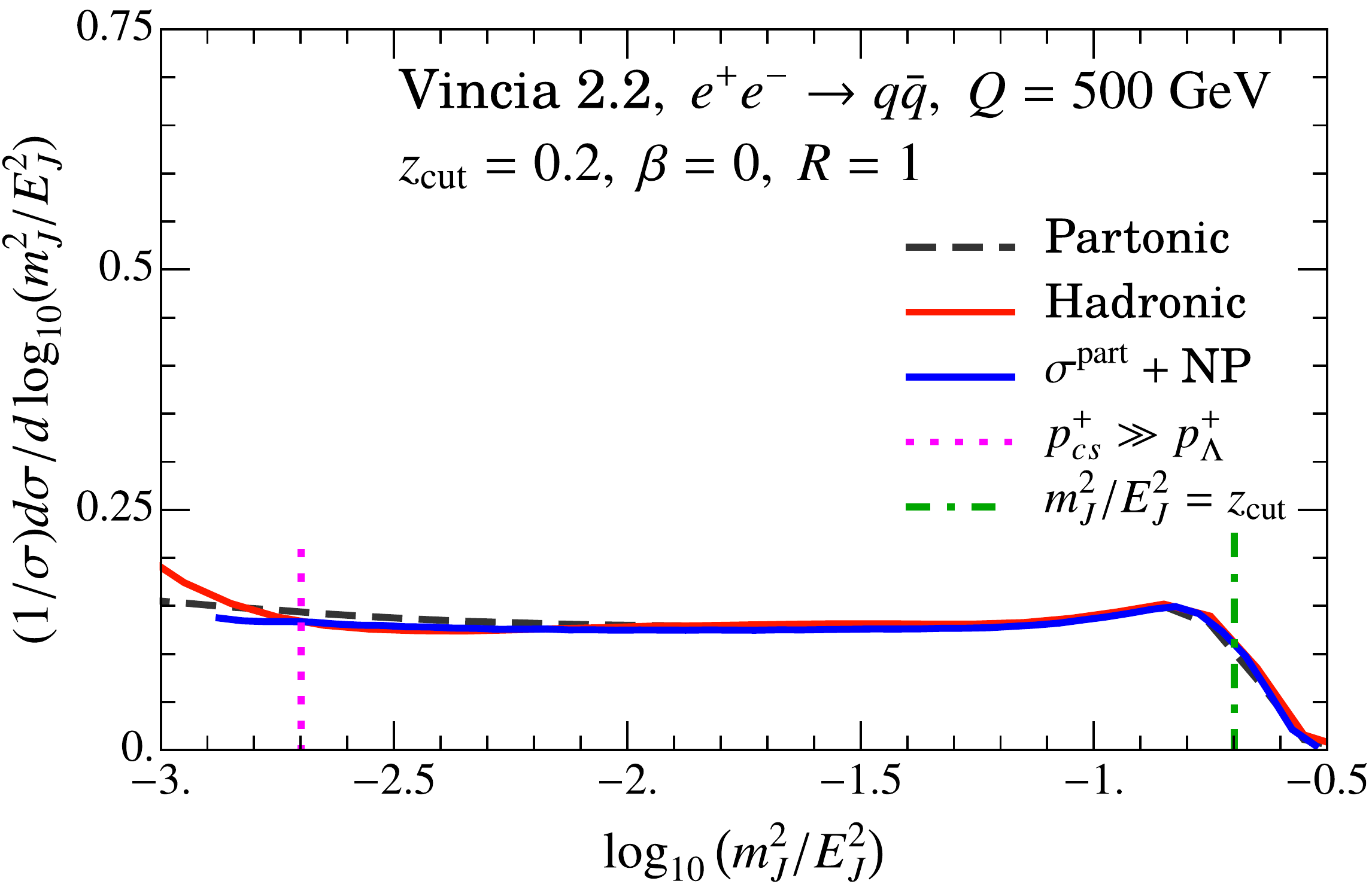}
\\
\includegraphics[width=0.48\textwidth]{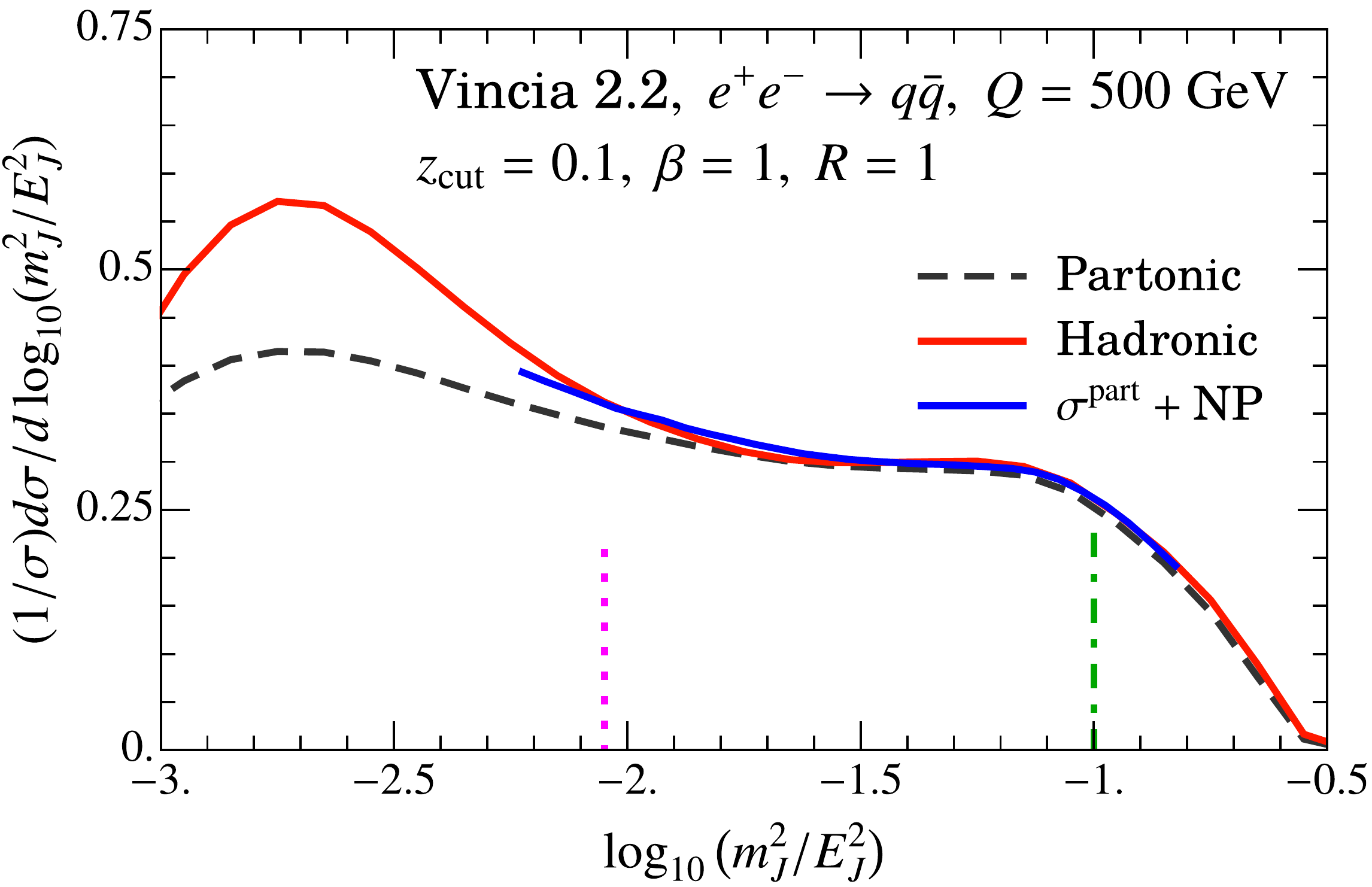}
\includegraphics[width=0.48\textwidth]{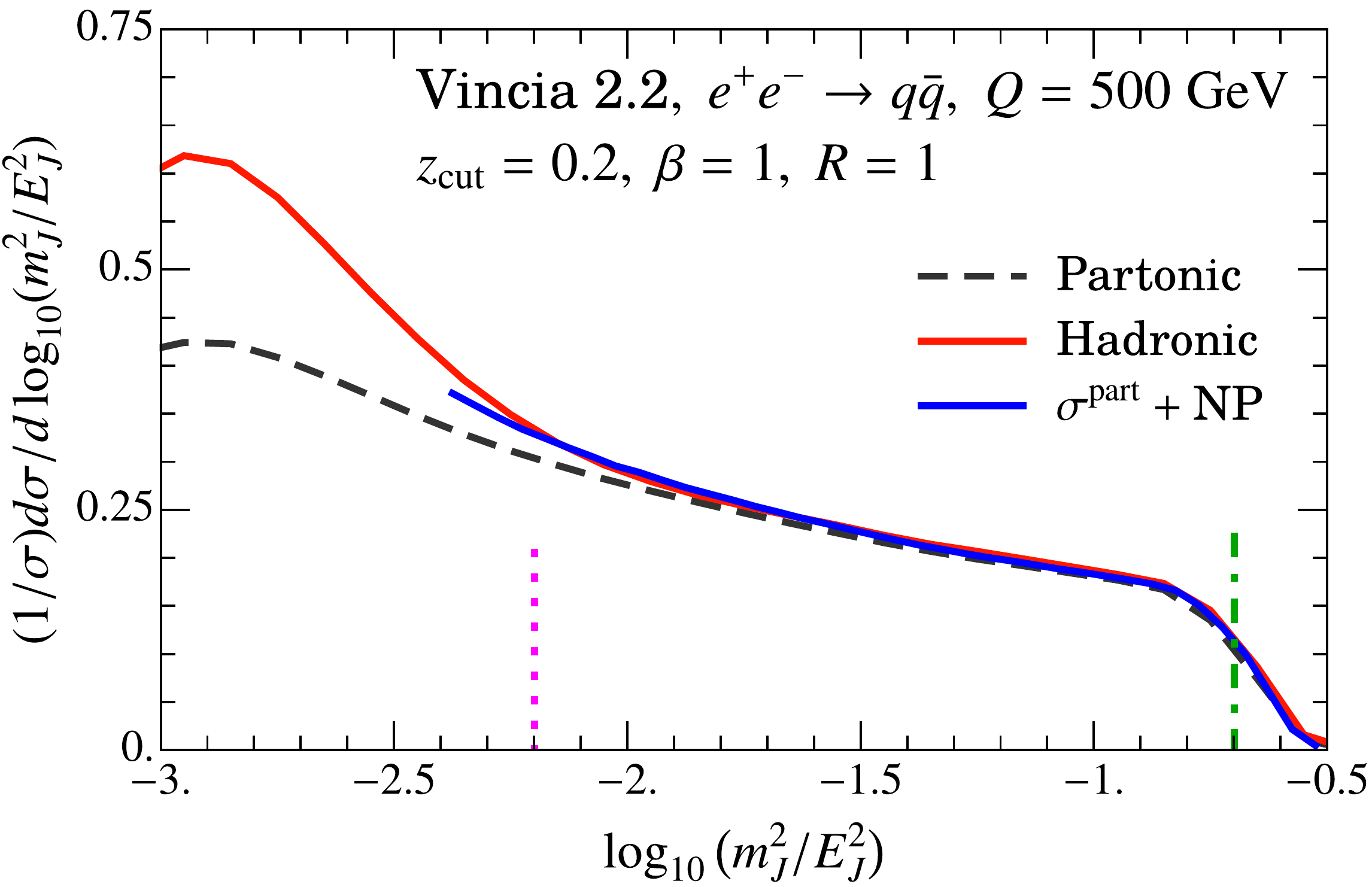}
\\
\includegraphics[width=0.48\textwidth]{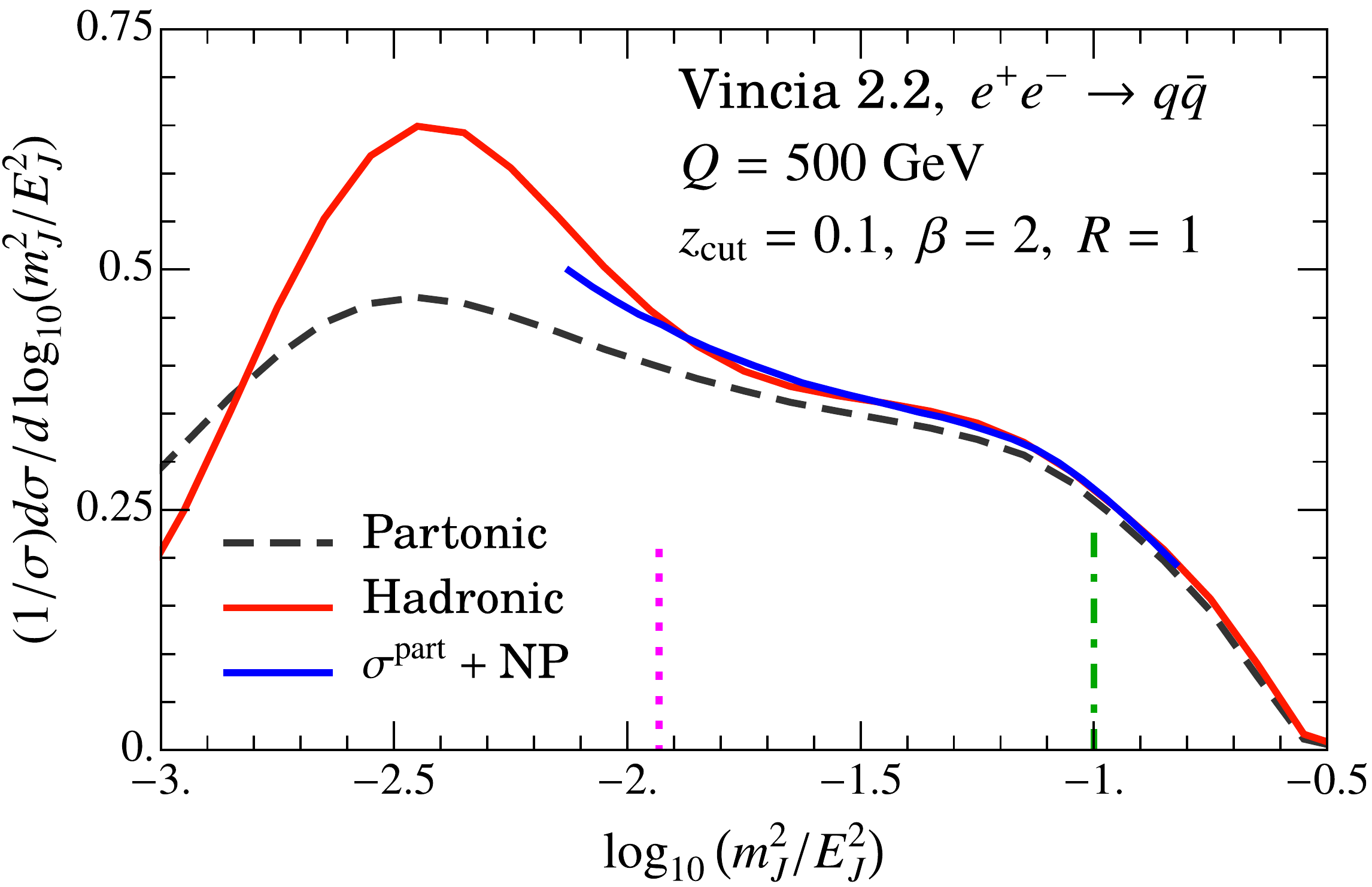}
\includegraphics[width=0.48\textwidth]{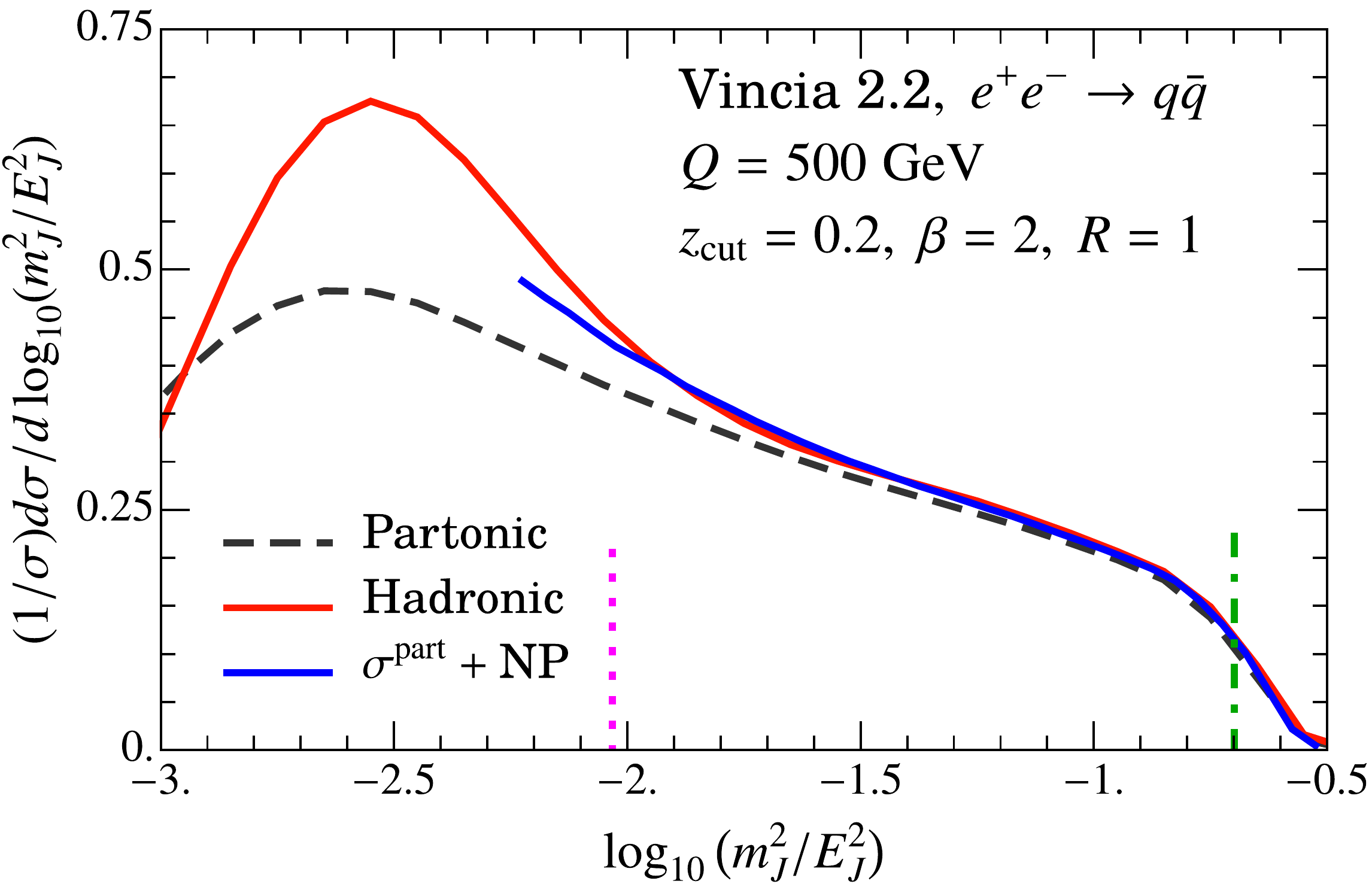}
\caption{Fit to hadronic \Vincia Monte Carlo in the SDOE region (indicated by the dotted vertical lines) for $\zcut = 0.1, 0.2$ (columns), $\beta = 0,1,2$ (rows) and $Q$ = 500 GeV. The same values of hadronic parameters displayed in the top left panel are used everywhere else. The vertical dashed and dot-dashed lines delineate the SDOE region as indicated in the top right panel.\label{fig:VinciaSDOEfit}}
\vspace{-0.2cm}
\end{figure}

\begin{figure}[t!]
\centering
\hspace{-2cm}
\includegraphics[width=0.55\textwidth]{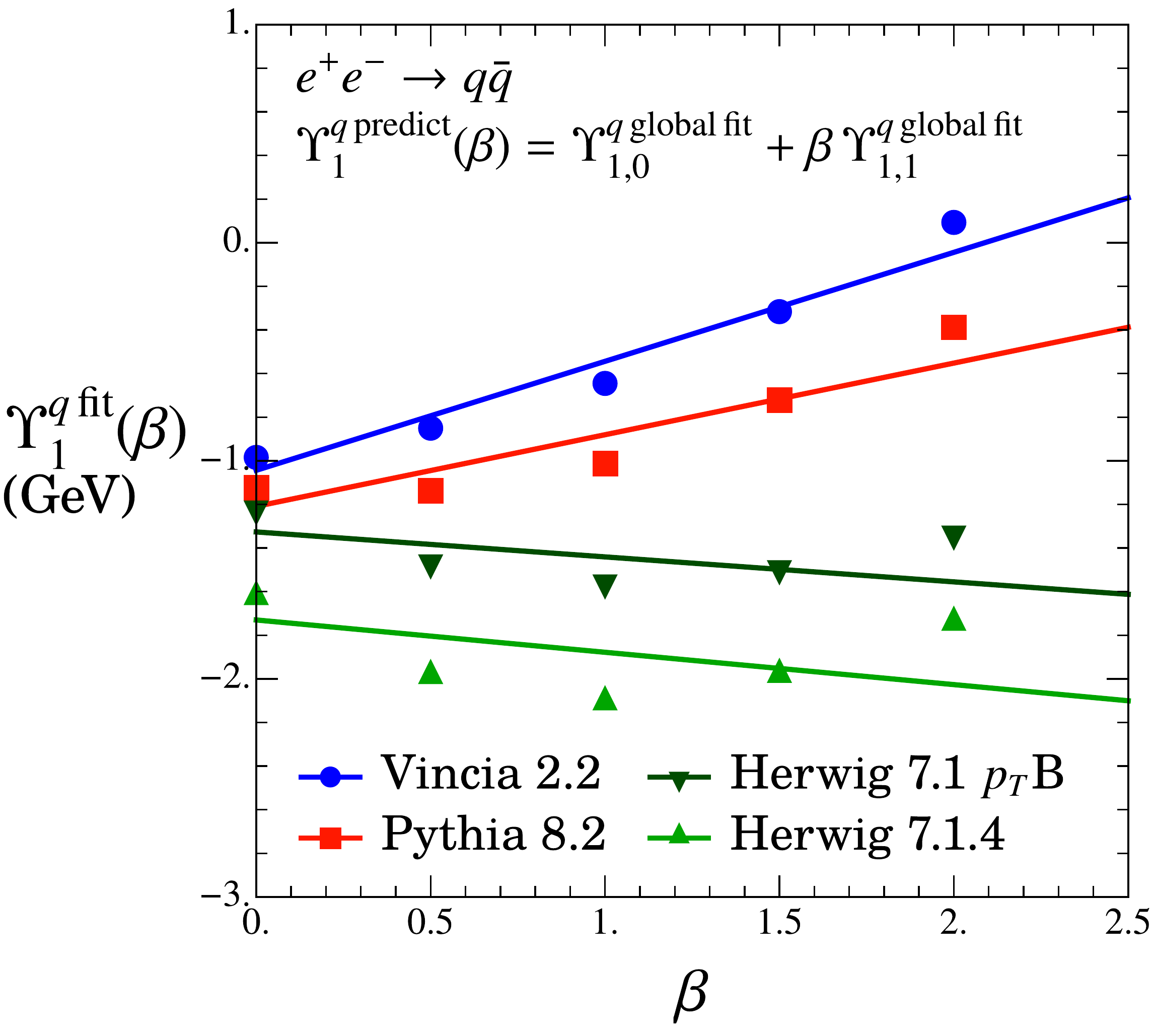}
\vspace{-0.4cm}
\caption{Linear behavior of the boundary power correction in $\beta$ is verified via comparison of $\Upsilon^q_1(\beta)$ for individual $\beta$ values and the linear prediction using global fit values of $\Upsilon^q_{1,0}$ and $\Upsilon^q_{1,1}$.}
\label{fig:upslinearity}
\end{figure}

\begin{figure}[t!]
\centering
\includegraphics[width=0.43\textwidth]{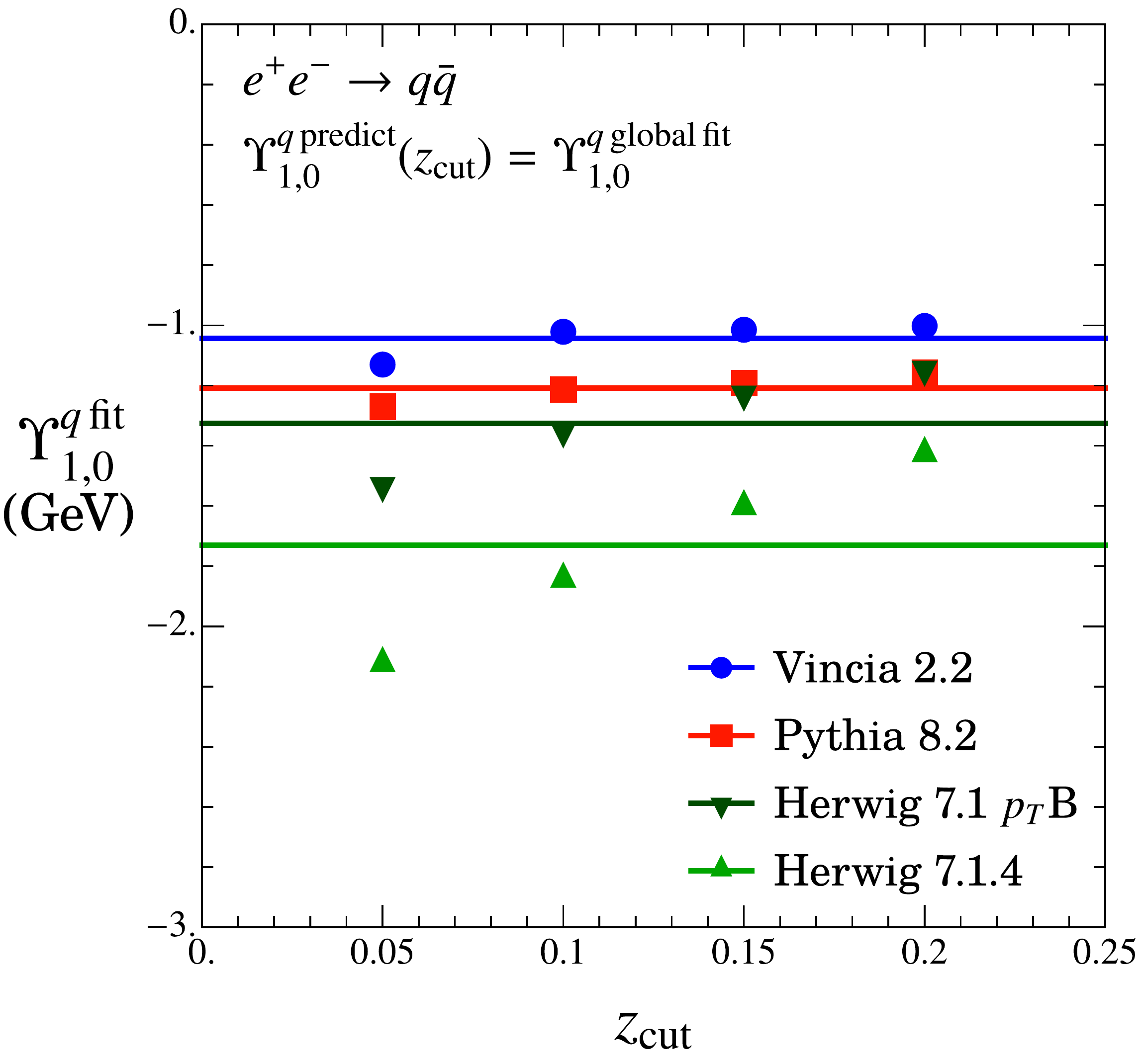}
\includegraphics[width=0.43\textwidth]{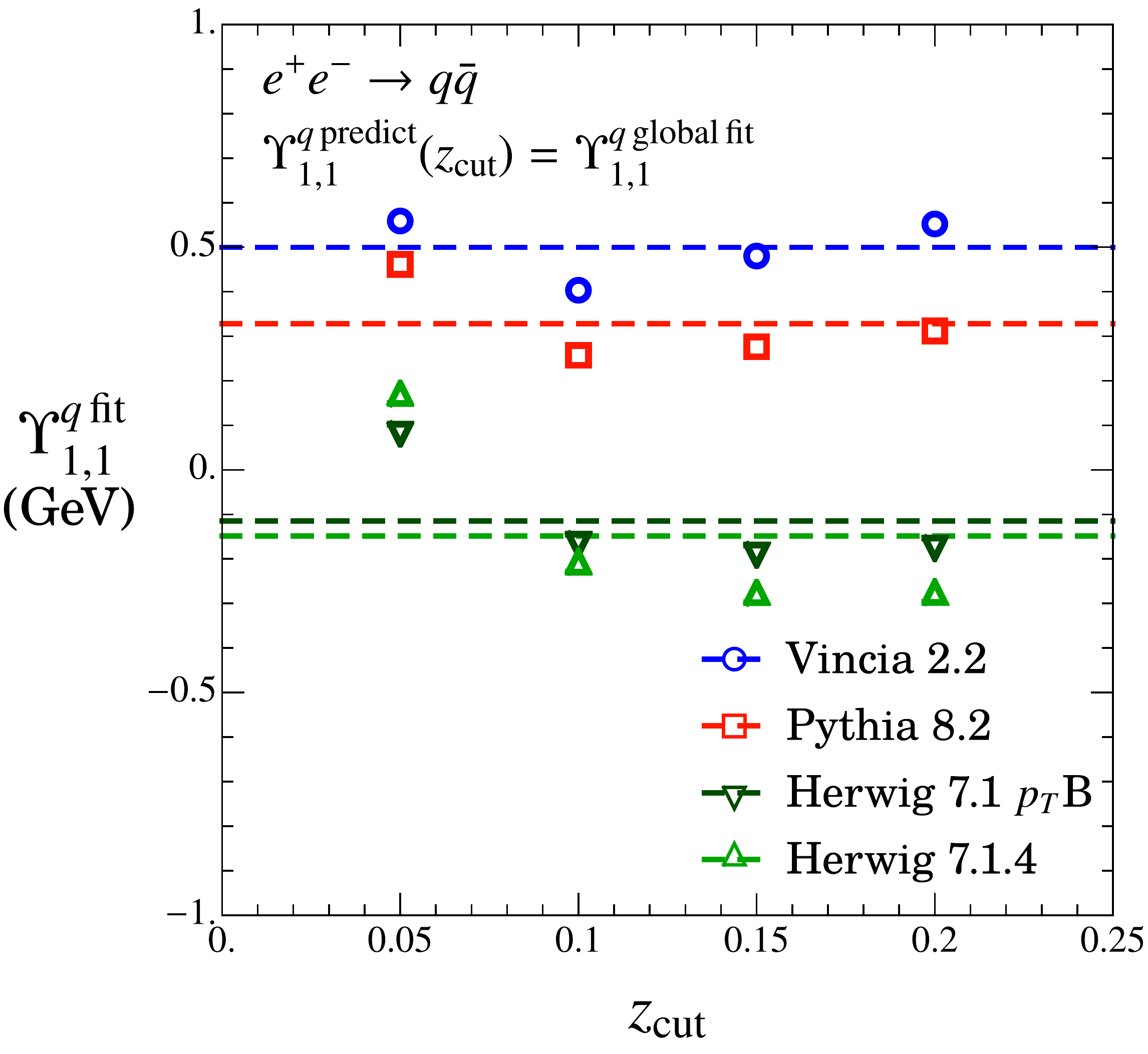}
\vspace{-0.2cm}
\caption{Testing $\zcut$ independence of $\Upsilon^q_{1,0}$ and $\Upsilon^q_{1,1}$ by fitting for individual $\zcut$ values and comparing against the global fit. \label{fig:zcutindep}}
\end{figure}

We now test the compatibility of the hadronization models of the three Monte Carlos with our description of power corrections in the SDOE region. In \fig{partHadall} the solid curves show the hadron level jet mass spectrum for $\zcut = 0.1$, $\beta = 1$, and $Q = 500$ GeV for the three Monte Carlos in the SDOE region. We note that while the hadron level predictions of the three Monte Carlos are quite close, they differ significantly in their parton level output, indicating in turn significant differences concerning their hadronization models.

To carry out a fit to determine the compatibility of the MC hadronization models with our description of hadronization, we consider a grid of jet mass spectra for the following sets of parameters:
\begin{align}
\label{eq:fitgrid}
Q = \{500,\, 1000\} \, \textrm{GeV} \, , \qquad \zcut = \{0.05, \,0.1, \,0.15, \,0.2\} \, , \qquad \beta = \{0, \,0.5, \,1.0,\, 1.5, \,2.0\} \, .
\end{align}
Following \eq{sigfullk} we then fit for the hadronic parameters $\Omega_{1q}^{\figeight}$, $\Upsilon^q_{1,0}$ and $\Upsilon^q_{1,1}$ in the SDOE region, taking the corresponding partonic MC cross section with hadronization turned off as the perturbative cross section $d \hat \sigma/dm_J^2$, and use our analytical results for $C_1^q(m_J^2)$ and $C_2^q(m_J^2)$ in \eqs{C1}{C2def}. We define the $\chi^2$-function as the sum of squared difference between the MC hadron level and the MC parton level cross sections including the power corrections of \eq{sigfullk}. We adopt $m_J$ independent uncertainties in the SDOE region. For the size of uncertainty, we first take 5\% as the approximate size of the hadronization corrections to the normalized cross section in this region, as is suggestive from \fig{partHadall}, and then define the uncertainty to be employed in the $\chi^2$-function to be 10\% of that.
We then fit in the SDOE region with approximately 20-30 bins, for all combinations of $Q$, $\zcut$ and $\beta$ values shown in \eq{fitgrid}. The result of the minimization is summarized in \tab{fitresults}.

We first note that the results of the fits all yield ${\cal O}$(1 GeV) values of the three universal hadronic parameters as expected. Although the extracted results from \Pythia and \Vincia differ, in both cases our three universal hadronic parameters provide an excellent description of the MC hadronization model. To further probe the reason for a rather poor fit results from default \Herwig 7.1.4 we carried out a fit for the $p_T$B tune defined in \Refcite{Reichelt:2017hts}. The default \Herwig 7.1.4 parton shower uses the jet mass preserving kinematic reconstruction, which was pointed out to be problematic in \Refcite{Hoang:2018zrp} where the shower-cut dependence of the parton level thrust distributions was analyzed. In \Refcite{Hoang:2018zrp} a better agreement with their analytical QCD results was observed with the $p_T$ preserving reconstruction originally proposed in \Refcite{Gieseke:2003rz} (see also~\cite{Bahr:2008pv}) which is the basis of the $p_T$B tune. This observation is further supported by studies in \Refscite{Dasgupta:2018nvj,Bewick:2019rbu}. We find here that the $p_T$B tune, with\footnote{Note that the $p_T$B tune of \Refcite{Reichelt:2017hts} also prescribes a change in the reference strong coupling value from $\alpha_s^{\rm CMW}(m_Z) = 0.127$ (corresponding to $\alpha_s(m_Z)=0.118$ in the $\overline{\rm MS}$ scheme) to $\alpha_s^{\rm CMW}(m_Z) = 0.1087$ as a part of the tune. However, we found that upon using the $\alpha_s^{\rm CMW}(m_Z) = 0.1087$ the fitting procedure failed to produce stable, and we therefore used the $p_T$B tune with $\alpha_s^{\rm CMW}(m_Z) = 0.127$.}
$\alpha_s(m_Z)=0.118$, yields similar values of the hadronic parameters as the \Herwig 7.1.4 but with a much improved $\chi^2$.
The correlations between the fit results for $\Omega_{1q}^{\figeight}$, $\Upsilon_{1,0}^q$ and $\Upsilon_{1,1}^q$ for our fits for \Vincia are displayed in \fig{ellipsesVincia}. Although there is some visible correlation between the three parameters, they are all still well constrained at the central values of the fits.

\begin{figure}[t!]
\centering
\includegraphics[width=0.43\textwidth]{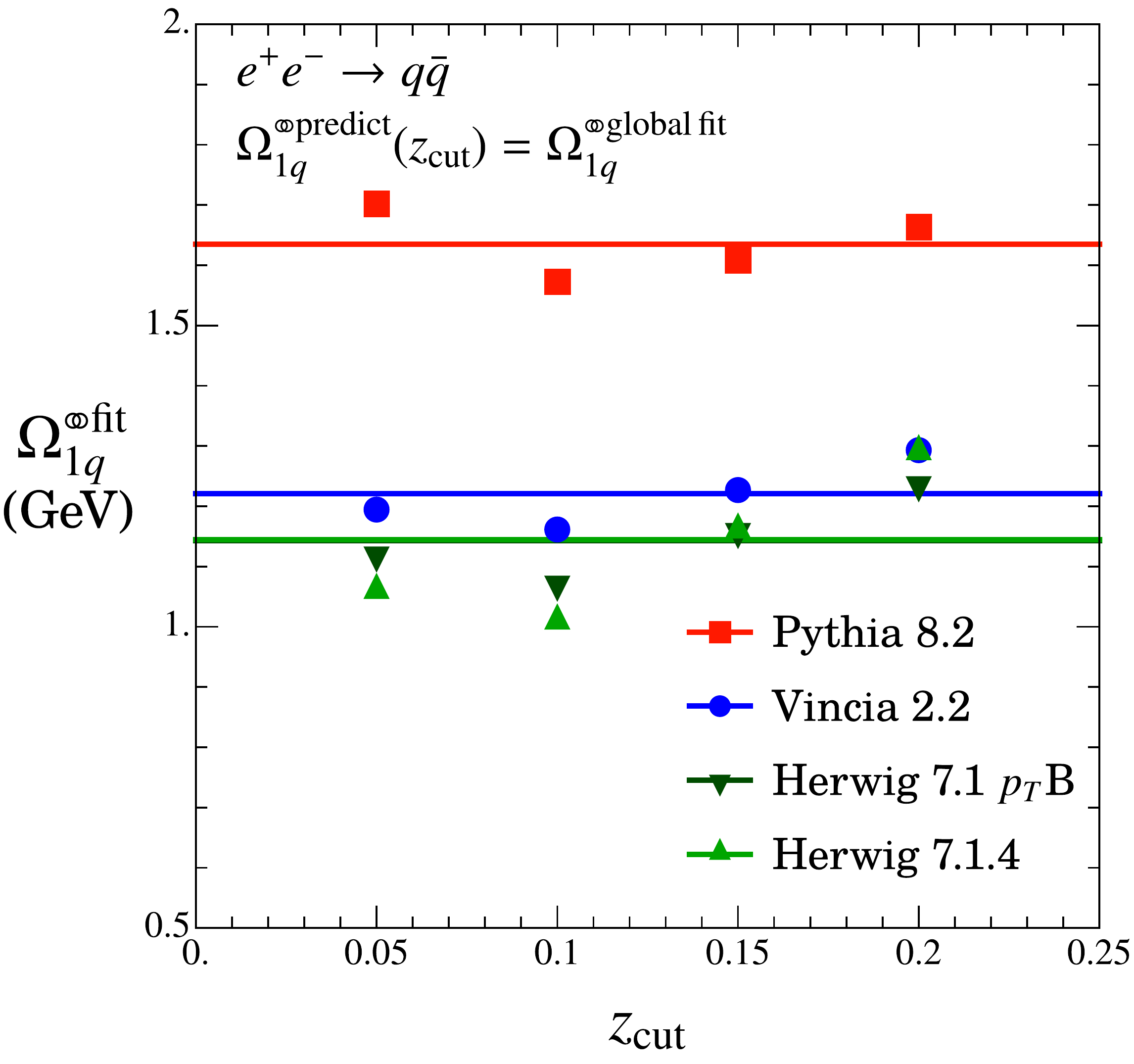}
\includegraphics[width=0.43\textwidth]{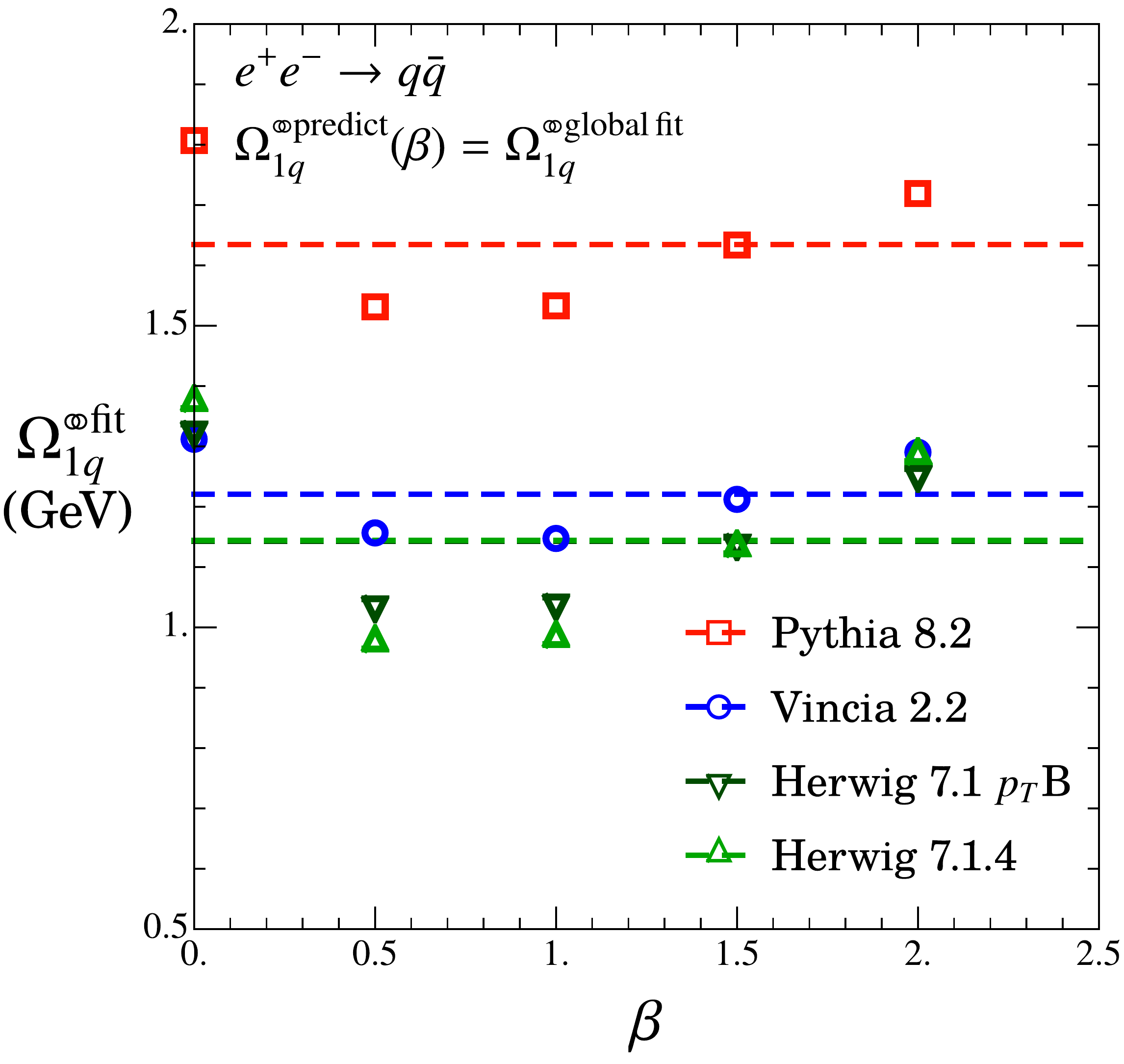}
\vspace{-0.2cm}
\caption{Testing $\zcut$ and $\beta$ independence of $\Omega_{1q}^{\figeight}$ by fitting for individual $\zcut$ and $\beta$ values and comparing against the global fit. The global fit values of $\Omega_{1q}^{\figeight}$ for \Herwig are the same for both tunes and hence the green lines are on top of each other.
\label{fig:O1zcutbetaindep}}
\end{figure}

In \fig{VinciaSDOEfit} we show a selection of the fit results (blue curves) for \Vincia hadron level results (red curves) and parton level results (black dashed curves) for a subset of the distributions included in the fits, using representative values of $Q$ = 500 GeV, $\zcut = \{0.1,0.2\}$ and $\beta = \{0, 1, 2\}$. We see that the three fitted values of the hadronic parameters $\Omega_{1q}^{\figeight}$, $\Upsilon_{1,0}^q$ and $\Upsilon_{1,1}^q$ are sufficient to describe the hadronization corrections for a range of $Q$, $\zcut$ and $\beta$ values. We obtain similar good quality results for $Q = 1000$ GeV and other values of $\zcut$ and $\beta$ not shown in \fig{VinciaSDOEfit}.

A prediction of \eq{O1Ups1} is that the boundary correction depends linearly on $\beta$. To check this within the context of our analysis we also fitted for $\Upsilon^q_1(\beta)$ for each value of $\beta$ separately, while fixing $\Omega_{1q}^{\figeight}$ to the global fit value shown in \tab{fitresults}. This allows us to compare the fit results for individual $\beta$ values against our prediction for $\Upsilon_{1}^q(\beta)$ using the global fit values of $\Upsilon^q_{1,0}$ and $\Upsilon^q_{1,1}$. The result of this exercise is displayed in \fig{upslinearity}. We observe a good agreement of the linear prediction and the explicit fit results for \Pythia, \Vincia and \Herwig $p_T$B, well within the $\pm 20\%$ estimated uncertainty from our predictions for $C_1^q$ and $C_2^q$. In \fig{zcutindep} we also test the $\zcut$-independence of the boundary power correction parameters $\Upsilon^q_{1,0}$ and $\Upsilon^q_{1,1}$. While the fits for these parameters obtained from \Pythia and \Vincia are roughly $\zcut$ independent, as expected from our theoretical considerations, we find a residual $\zcut$ dependence for the two versions of \Herwig. We see, however, for \Herwig some reduction of the $\zcut$ dependence when using the $p_T$B tune since the individual fits are significantly closer to the global fit. We conclude that the MC results support the universality of the $\Upsilon_{1,i}^q$ parameters at the expected level of precision of $\pm 20\%$ induced from the uncertainty in $C_1^q$ and $C_2^q$.

Finally, in \fig{O1zcutbetaindep} we test the universality of the shift correction with respect to variations in $\zcut$ and $\beta$. Here we fit for $\Omega_{1q}^{\figeight}$ for individual $\zcut$ and $\beta$ values (colored symbols) while fixing $\Upsilon^q_{1,0}$ and $\Upsilon^q_{1,1}$ to their global fit values. The results from these individual fits are then compared with those from the global fit values shown in \tab{fitresults} (horizontal solid and dashed lines). We find an independence to the value of $\zcut$ at the 10\% level for \Pythia and \Vincia, and up to 15\% for \Herwig 7.1 $p_T$B.
Again the MC results support the expected level of universality for $\Omega_{1q}^{\figeight}$.

\section{Conclusion}
\label{sec:conclusion}

We have presented a study of the dominant nonperturbative corrections to the groomed jet mass spectrum based on quantum field theory calculations in the framework of soft collinear effective theory and the coherent branching formalism. We considered the operator expansion region (SDOE), where the hadronization effects are power corrections, and the nonperturbative region (SDNP) of the jet mass spectrum, where they become leading order effects. In the SDOE region we identified two leading nonperturbative power corrections, called the ``shift'' and ``boundary'' corrections, which are related to the contribution of the nonperturbative radiation to the jet mass and to modifications in the soft drop test due to nonperturbative radiation respectively. The two power corrections have a perturbative dependence on the angle of the soft drop stopping collinear-soft subjet that sets the catchment area for the nonperturbative particles. We showed that this stopping angle dependence can be factored into the Wilson coefficient by an appropriate rescaling of the momenta of the nonperturbative modes. This allowed us to derive factorization at the level of both the measurement function and the nonperturbative matrix element. It resulted in jet mass dependent perturbative Wilson coefficients for the two effects, which involve the perturbative functions $C_1(m_J^2)$ and $C_2(m_J^2)$, that encode non-trivial dependence of the leading power corrections on the kinematic parameters $p_T$, $\eta_J$, the jet radius $R$, and grooming parameters $\zcut$ and $\beta$. We have explicitly calculated the coefficients $C_1(m_J^2)$ and $C_2(m_J^2)$ using the coherent branching formalism at LL order with a running coupling. As a result the leading power corrections in the SDOE region can be described by three universal hadronic parameters $\Omega_{1\kappa}^{\figeight}$, $\Upsilon^\kappa_{1,0}$ and $\Upsilon^\kappa_{1,1}$ that are independent of these parameters and have dimension of mass. Here $\kappa=q$ for a quark initiated jet, while $\kappa=g$ for a gluon initiated jet, and the parameters can differ for these two cases. In the SDNP region we showed that the leading order nonperturbative corrections can be described by a shape function that depends on $\beta$ which has a normalization less than one.

We tested our conceptual analysis on the structure of hadronization corrections in the SDOE region by carrying out Monte Carlo studies with \Pythia 8.235, \Vincia 2.2, and \Herwig 7.1. We extracted $C_1(m_J^2)$ and $C_2(m_J^2)$ from the Monte Carlos and found that they agree well with our LL calculations. We confirmed the kinematic approximations of our conceptual studies by analyzing the angular distributions of nonperturbative subjets produced by MC generators, and we found the expected geometries for the SDNP and SDOE jet mass regions. We then tested the hadronization models of the MC generators in detail by fitting for the three universal parameters $\Omega_{1q}^{\figeight}$, $\Upsilon_{1,0}^q$ and $\Upsilon_{1,1}^q$ for quark jets in $e^+e^-$ annihilation in the SDOE region. We found that the fitted values of the three parameters are ${\cal O}(1)$ GeV, and consistently describe the MC hadronization corrections for a range of $Q$, $\zcut$ and $\beta$ values. The parameters also showed the predicted universality for each of \Pythia, \Vincia and \Herwig, though with different values for the hadronic parameters.
An improvement of \Herwig 7.1.4 results was observed upon considering a different parton shower kinematic reconstruction from Ref.~\cite{Gieseke:2003rz}.
Thus our predictions for hadronization corrections were successfully confirmed by this comparison to MCs.

The results of this work enable a more precise description of hadronization effects for the groomed jet mass distribution, and make it possible to consider obtaining improved precision measurements of QCD parameters such as $\alpha_s$ and top mass $m_t$ from the LHC data using the groomed observables. We note that some of our results are also used in our theoretical analysis of soft drop groomed top quark jets in Ref.~\cite{Hoang:2017kmk}, where it is pointed out that these hadronization corrections are also universal between massive and massless quarks. While we focused on the case of jet mass in this work, we emphasize that the same techniques can also be applied to other groomed observables such as angularities~\cite{Berger:2003iw}. We also envisage, that our study may contribute to better calibrate and tune hadronization models of Monte Carlos, which have so far relied only on ungroomed observables. The universal hadronic parameters we obtained in our analysis involve combinations of $+$, $-$, and $\perp$ components of nonperturbative momenta and are hence more sophisticated than those obtained from classic event shapes.
Thus, our results for the power corrections for the groomed jet mass have interesting implications for dedicated tuning studies of MC hadronization models.

\begin{acknowledgments}
This work was supported by FWF Austrian Science Fund under the Project No. P28535-N27.
We thank FWF Austrian Science Fund under the Doctoral Program “Particles and Interactions” No.W1252-N27 for partial support.
This work was supported in part by the Office of Nuclear Physics of the U.S. Department of Energy under the Grant No. DE-SCD011090. I.S. was also supported by the Simons Foundation through the Investigator grant 327942.
We thank A.~Larkoski, S.~Pl\"atzer, and M.~Schwartz for helpful discussions.
We also thank the ESI workshop on ``Challenges and Concepts for Field Theory and Applications in the Era of LHC Run-2'' for hospitality and support while portions of this work were initiated.
AP thanks the Center for Theoretical Physics, MIT, for hospitality and support while part of this work was completed.

\end{acknowledgments}

\appendix

\section{Measurement Operator for the Boundary Term}
\label{app:bndry}

To illustrate the key physical ideas for understanding the operator for the boundary term in \eq{hatpbndry} we consider a simple example. Consider two perturbative subjets with energies $E_1$ and $E_2$ forming a jet with energy $E_J$ such that
\begin{align}
\label{eq:pertE1E2}
E_1 + E_2 = E_J \, .
\end{align}
We will consider the first subjet to be collinear-soft and the second to be collinear, so
\begin{align}
E_1 \ll E_2 \, .
\end{align}
For simplicity we consider the case with $\beta = 0$, such that the soft drop criteria is purely a comparison of energies. Then the soft drop comparison is
\begin{align}
\label{eq:SDpart}
\overline \Theta_{\rm sd}^{\rm part} = \Theta \bigg (\frac{E_1}{E_J} - \zcut \bigg) = \Theta \big ( z_1 - \zcut \big ) \, ,
\end{align}
and we are interested in the regime where $\zcut \ll 1$.

Turning on hadronization, involves including NP radiation with energy $E_\Lambda \sim \Lambda_{\rm QCD}/\zeta_{cs} \ll E_1$. This radiation encodes both the non-perturbative rearrangement of momenta into and out of the subjets as well as the binding of partons to hadrons. Since $E_1\ll E_2$, the leading hadronization effects will occur when hadronization modifies the energy of the $E_1$ subjet. There will be a reduction in energy from the regrouping done by hadronization causing a loss of particles from the $1$ subjet to other subjets (here subjet $2$). Similarly there can be an increase to $E_1$ from hadronization if the nonperturbative regrouping adds particles to the $1$ subjet.

We can probe these non-perturbative effects by using a single NP subjet of energy $E_\Lambda$ that participates in the CA clustering. At the stage where this NP subjet is not yet clustered, hadronization has changed the energy of subjet $1$ to $E_1' = E_1 - E_{\Lambda_1}$. We also note that $E_2' = E_2 - E_{\Lambda_2}$ is the energy of subjet $2$ after hadronization. Case by case the $E_{\Lambda_i}$ could be positive or negative, but on aggregate we expect a positive value. Here $E_{\Lambda_2}$ is always negligible relative to $E_2$, but can appear in comparisons with $E_1$. By energy conservation the total jet energy is $E_J\equiv E_1+E_2 = E_1' + E_2' + E_\Lambda$, which implies $E_{\Lambda_2} = E_\Lambda - E_{\Lambda_1}$.

We consider the effect on the soft drop condition in \eq{SDpart} on the hadronized subjets $1'$, $2'$ and $\Lambda$, and expand back in terms of the variables of the perturbative subjets $1$ and $2$. From the point of view of CA clustering, we have three situations:
\begin{enumerate}[(i)]
\item the CS subjet $1'$ and NP subjet are clustered first, followed by the collinear subjet $2'$,
\item collinear subjet $2'$ and NP subjet are clustered first, followed by the CS subjet $1'$,
\item the subjets $1'$ and $2'$ are clustered first, followed by the NP subjet.
\end{enumerate}
The case that occurs is entirely determined by a geometric test of the angles of the NP subjet, which divides up the phase space as
$1=\overline \Theta_{\rm NP}^{\rm (i)}+\overline\Theta_{\rm NP}^{\rm (ii)}+\overline\Theta_{\rm NP}^{\rm (iii)}$.
From the beginning we drop corrections $E_{\Lambda}$ and $E_{\Lambda_i}$ relative to $E_J$ and $E_2$ and only retain those appearing with $E_1$ or $E_1'$.
For case (i), we have
\begin{align} \label{eq:bndryCase1}
\overline \Theta_{\rm sd}^{\rm (i)} \overline \Theta_{\rm NP}^{\rm (i)}
&= \Theta \bigg(\frac{E_1' + E_{\Lambda}}{E_J} - \zcut \bigg)
\overline \Theta_{\rm NP}^{\rm (i)}
= \Theta \bigg(\frac{E_1 + E_{\Lambda_2}}{E_J} - \zcut \bigg)
\overline \Theta_{\rm NP}^{\rm (i)}
= \Theta \Big((z_1 + \,z_{\Lambda_2}) - \zcut \Big)
\overline \Theta_{\rm NP}^{\rm (i)}
\nn \\
& = \overline \Theta_{\rm sd}^{\rm part}\overline \Theta_{\rm NP}^{\rm (i)}
+ z_{\Lambda_2} \, \delta (z_1 - \zcut) \overline \Theta_{\rm NP}^{\rm (i)} + \ldots \, ,
\end{align}
where $z_1=E_1/E_J$, $z_{\Lambda_2}=E_{\Lambda_2}/E_J$, etc., and the ellipses denote higher order power corrections.
For case (ii) there is no correction from the CA clustering since the NP subjet is grouped with the collinear subjet. This is also the situation for case (iii) where the NP subjet is simply groomed away by soft drop in the SDOE region, since it is the comparison of $1'$ and $2'$ that stops soft drop. Expanding to the first non-trivial order, these cases give
\begin{align}
\label{eq:bndryCase2}
\overline \Theta_{\rm sd}^{\rm (ii)} \overline \Theta_{\rm NP}^{\rm (ii)}
& =\Theta \bigg(\frac{E_1' }{E_J} - \zcut \bigg)
\overline \Theta_{\rm NP}^{\rm (ii)}
= \overline \Theta_{\rm sd}^{\rm part}
\overline \Theta_{\rm NP}^{\rm (ii)}
- z_{\Lambda_1} \, \delta (z_1 - \zcut)
\overline \Theta_{\rm NP}^{\rm (ii)}
\,, \\
\label{eq:bndryCase3}
\overline \Theta_{\rm sd}^{\rm (iii)} \overline \Theta_{\rm NP}^{\rm (iii)}
&= \Theta \bigg(\frac{E_1' }{E_J} - \zcut \bigg)
\overline \Theta_{\rm NP}^{\rm (iii)}
= \overline \Theta_{\rm sd}^{\rm part}
\overline \Theta_{\rm NP}^{\rm (iii)}
- \, z_{\Lambda_1} \,\delta (z_1 - \zcut)
\overline \Theta_{\rm NP}^{\rm (iii)} \, .
\end{align}
Noting that $\overline \Theta_{\rm NP}^{\, \bndry} = \overline \Theta_{\rm NP}^{\rm (i)}$ and $\Theta_{\rm NP}^{\, \bndry} = \overline \Theta_{\rm NP}^{\rm (ii)}+\overline \Theta_{\rm NP}^{\rm (iii)}= 1 - \overline \Theta_{\rm NP}^{\, \bndry}$, the sum of results in Eqs.~(\ref{eq:bndryCase1}-\ref{eq:bndryCase3}) give
\begin{align}
\label{eq:ThetaSDhad}
\overline \Theta_{\rm sd}^{\rm had} &= \overline \Theta_{\rm sd}^{\rm part} - \Theta_{\rm NP}^{\, \bndry} (\theta_{\Lambda}, \, \theta_1,\,\Delta \phi_1) \, z_{\Lambda_1} \, \delta (z_1 - \zcut) +\overline \Theta_{\rm NP}^{\, \bndry} (\theta_{\Lambda}, \, \theta_1,\,\Delta \phi_1) \, z_{\Lambda_2} \, \delta (z_1 - \zcut)\, , \nn \\
& = \overline \Theta_{\rm sd}^{\rm part} + \delta (z_1 - \zcut) \, \Big[ - \Theta_{\rm NP}^{\, \bndry} (\theta_{\Lambda}, \, \theta_1,\,\Delta \phi_1)
\frac{q_1^-}{Q} \, +\overline \Theta_{\rm NP}^{\, \bndry} (\theta_{\Lambda}, \, \theta_1,\,\Delta \phi_1) \frac{q_2^-}{Q} \Big]
\,.
\end{align}
The terms in \eq{ThetaSDhad} track flow of energy in and out of the subjets. Generalization of this example to $\beta > 0$ is straightforward.

The final step is to determine an analog for \eq{ThetaSDhad} from the point of view of the operator expansion with NP fields. In this context the NP fields enable us to describe effects where a simple correspondence between partonic and hadronic expressions breaks down without double counting the hadronization effects related to particles that remain in the same region.
In the OPE with a single non-perturbative $\Lambda$ field, there is only a single $q^-$ momentum,
so that
at this order the OPE gives $q_1^- = q_2^-=q^-$. Thus the
NP source fields satisfying $\overline \Theta_{\rm NP}^{\, \bndry} (\theta_{\Lambda}, \, \theta_1,\,\Delta \phi_1) =1$ correspond to the term for region (i), and those with $\Theta_{\rm NP}^{\, \bndry} (\theta_{\Lambda}, \, \theta_1,\,\Delta \phi_1) =1$ correspond to the terms for regions (ii) and (iii).
This leads to the result quoted in \eq{measurement6}.

\section{Collinear-Soft Function with a Probe Nonperturbative Gluon}
\label{app:Eikonal}
Here we provide some details of the results presented in \sec{npsource} required to demonstrate the factorization of nonperturbative corrections, and to show that the boundary correction from soft drop failing subjets is subleading at LL accuracy.

\subsection{Analysis with One Perturbative Gluon} \label{app:oneloop}
\addtocontents{toc}{\protect\setcounter{tocdepth}{1}}

\subsubsection{Measurement Operator in SDOE region} \label{app:measurement}
\addtocontents{toc}{\protect\setcounter{tocdepth}{2}}
We first start by considering the measurement operator for the diagrams in \figs{np-abelian}{np-nonabelian}.
The momentum of the collinear-soft perturbative gluon is taken to be $p$, which can be real or virtual, and that of the NP source gluon to be $q$, which is always real. Here the total plus momentum $\ell^+$ of the real radiation kept by soft drop is measured.
For the real graphs with $p$ passing the cut the measurement operator is given by
\begin{align}
\label{eq:Mpq}
&{\cal M}^{p+q} =
\overline \Theta_{\rm NP}^{\figeight}(\theta_{q}, \theta_{p},\Delta \phi)\, \overline \Theta_{\rm NP}^{\, \bndry} (\theta_{q}, \theta_p,\Delta \phi) \, \bigg\{ \overline \Theta_{\rm sd}^{\, p+q}\, \delta(\ell^+ - p^+ - q^+) + \Theta_{\rm sd}^{\, p+q}\, \delta(\ell^+) \bigg\}
\\
&+ \overline \Theta_{\rm NP}^{\figeight}(\theta_{q}, \theta_{p},\Delta \phi)\,\Theta_{\rm NP}^{\bndry} (\theta_{q}, \theta_p,\Delta \phi)\, \bigg\{ \overline \Theta_{\rm sd}^{\, p - q}\delta(\ell^+ - p^+ - q^+) + \Theta_{\rm sd}^{\, p-q} \Big[ \overline \Theta_{\rm sd}^{\, q} \delta(\ell^+ - q^+) + \Theta_{\rm sd}^q \delta(\ell^+)\Big] \bigg\}\nn \\
& +\Theta_{\rm NP}^{\figeight}(\theta_{q}, \theta_{p},\Delta \phi) \, \bigg\{\overline \Theta_{\rm sd}^{\, q} \delta(\ell^+ - p^+ - q^+) + \Theta_{\rm sd}^{\, q} \Big[\overline \Theta_{\rm sd}^{\,p - q} \delta(\ell^+ - p^+) + \Theta_{\rm sd}^{\, p-q} \delta(\ell^+) \Big] \bigg\}
\nn \, ,
\end{align}
whereas for the virtual graphs for $p$ this is simply a soft drop test on the NP gluon $q$:
\begin{align}
\label{eq:Mq}
{\cal M}^q = \overline \Theta_{\rm sd}^{\, q} \delta(\ell^+ - q^+ )+ \Theta_{\rm sd}^{\, q} \delta(\ell^+ ) \, .
\end{align}
Here the unbarred $\Theta$'s are complements of the $\overline \Theta$'s, so $\Theta = 1 - \overline \Theta$. The three lines in \eq{Mpq} correspond to the three CA clustering situations we discussed above in \app{bndry}. In the first line in \eq{Mpq}, $p$ and $q$ are first combined together into a subjet and the combined momentum $p+q$ is tested for soft drop with the collinear jet.
In the second line, the first step of clustering combines $q$ with the collinear parton, with $q$ representing momentum lost by the CS subjet, and hence the soft drop test is first applied on $p-q$. The third line with $\Theta_{\rm NP}^{\figeight}$ corresponds to the case when $p$ and the collinear jet are clustered first, and thus the first soft drop test compares $q$ and the collinear$+p$ subjet. In this case the NP particle lies outside the shaded region in \fig{NPfig8}a.

The superscripts on the $\overline \Theta_{\rm sd}$ and $ \Theta_{\rm sd}$ operators correspond to the momentum that gets soft drop tested with the collinear parton. They represent soft drop passing and failing conditions respectively. In accordance with \eq{hatpbndry} and the discussion in \app{bndry}, in the first line where $q$ is clustered with the softer subjet, the momentum tested for soft drop is $p+q$. The next two cases in the second and the third line reduce the momentum of $p^\mu$ from the point of view of the soft drop condition. On the other hand, the NP particle contributes to the total momentum as long as it is part of either the collinear or the collinear-soft subjet, as shown by the arguments of the $\delta$-functions.

In the SDOE region, the probability of a NP emission to pass soft drop in this region alone is exponentially Sudakov suppressed. Hence, we set $\overline \Theta_{\rm sd}^{\,q} = 0$ and $ \Theta_{\rm sd}^q = 1$ in \eqs{Mpq}{Mq}. Thus, expanding to first order in the SDOE region, the measurement operators simplify to
\begin{align}
\label{eq:MpqSDOE}
{\cal M}^{p+q} & \stackrel{\rm SDOE}{=} \> \delta(\ell^+) + \overline \Theta_{\rm sd}^{\, p}\, \Big[ \delta(\ell^+ - p^+) - \delta(\ell^+) \Big ] \\
&+ \overline \Theta_{\rm NP}^{\figeight}(\theta_{q}, \, \theta_{p},\, \Delta \phi)\, \overline \Theta_{\rm sd}^{\, p}\,\Big [ \delta(\ell^+ - p^+ - q^+)- \delta(\ell^+ - p^+) \Big]\, \nn \\
&+ \Big (\overline \Theta_{\rm NP}^{\bndry} (\theta_{q}, \, \theta_p,\, \Delta \phi) - \Theta_{\rm NP}^{\bndry} (\theta_{q}, \, \theta_p,\, \Delta \phi) \Big)\, \Delta \overline \Theta_{\rm sd} \, \Big[ \delta(\ell^+ - p^+) - \delta(\ell^+) \Big ] \nn \\
\label{eq:MqSDOE}
{\cal M}^q & \stackrel{\rm SDOE}{=} \> \delta(\ell^+) \, ,
\end{align}
where $\Delta \overline \Theta_{\rm sd}$ is given by
\begin{align}
\label{eq:DeltaSDdef}
\Delta \overline \Theta_{\rm sd} = \delta \bigg ( \frac{p^-}{Q} - \zcut \Big(\frac{\theta_{p}}{R_0}\Big)^\beta \bigg)\, \frac{q^- }{Q} \, \bigg [\, 1 +\beta \,\Big(1- \frac{\theta_{q}}{\theta_p} \, \cos( \Delta \phi) \Big)\bigg]
\,.
\end{align}
In simplifying to \eq{MpqSDOE} we made use of the following relations related to the geometric regions:
\begin{align}
\overline \Theta_{\rm NP}^{\figeight} \, \overline \Theta_{\rm NP}^{\bndry} = \overline \Theta_{\rm NP}^{\bndry} \, , \\
\Theta_{\rm NP}^{\figeight} + \overline \Theta_{\rm NP}^{\figeight} \, \Theta_{\rm NP}^{\bndry} = \Theta_{\rm NP}^{\bndry} \,. \nn
\end{align}
We have written \eq{MpqSDOE} such that the second and third lines correspond to the shift correction in the $+$ momentum and the boundary correction, respectively, as we saw above in \eqs{sdoefull}{sdoefull2}. Note that we dropped the subleading $q^+$ shift in the boundary correction term. The term $-\delta (\ell^+ - p^+)$ in the shift correction in the second line in \eq{MpqSDOE} results from separating out the partonic measurement.

\addtocontents{toc}{\protect\setcounter{tocdepth}{1}}
\subsubsection{Abelian Graphs}
\addtocontents{toc}{\protect\setcounter{tocdepth}{2}}
\label{app:abelian}

In the following we give details of the computation of the abelian graphs shown in \fig{np-abelian} and discussed in \sec{abelian}. The sum of all the diagrams with the real perturbative gluon carrying momentum $p^\mu$ (with examples shown as the second and third diagrams in \fig{np-abelian}) reads:
\begin{align}
\label{eq:abelianreal}
\sum \textrm{(abelian, real)}&= C_\kappa(4C_\kappa-2C_A)(g^2 \tilde \mu^{2\epsilon})^2 \!\!\int\!\! \frac{d^d p}{(2\pi)^d}\, {\cal C}(p)\frac{n\cdot \bar{n}}{n\cdot p \, \bar{n} \cdot p}\,\int\!\! \frac{d^d q}{(2\pi)^d}\, \tilde {\cal C}(q)\,\frac{n \cdot \bar{n}}{n\cdot q \, \bar{n} \cdot q} \, {\cal M}^{p+q} \, ,
\end{align}
where $q^\mu$ is the momentum of the nonperturbative gluon and $C_\kappa = C_F$ or $C_A$ is the color factor for quark or gluon being the parent parton, respectively. ${\cal C}(p)$ and $\tilde{\cal C}(q)$ implement the onshell cut conditions:
\begin{align}
\label{eq:cut}
{\cal C}(p) = 2\pi \,\delta (p^2) \Theta(p^0)\, , \qquad\tilde {\cal C}(q) =2\pi \, \delta (q^2 - m^2) \Theta(q^0)\, ,
\end{align}
and the $\overline{\rm MS}$ factor $\tilde \mu^{2\epsilon} = (\mu^2 e^{\gamma_E}/(4\pi))^\eps$. For simplicity in \eq{abelianreal}, and also in the following, we include an implicit overall factor of $\qcut^{\frac{1}{1+\beta}}$ on the left hand side in the meaning of sum over graphs. In \eq{cut} $m^2 \sim \Lambda_{\rm QCD}^2$ is a mass term for the nonperturbative source gluon,
which for this calculation can be thought of as a proxy for the mass of the NP subjet.
We get two Eikonal factors for both the gluons emitted from the parent parton. The virtual diagrams (examples are the last two diagrams in \fig{np-abelian}) add to give
\begin{align}
\label{eq:abelianvirtual}
\sum \textrm{(abelian, virtual)}&= - C_\kappa(4C_\kappa-2C_A)(g^2 \tilde \mu^{2\epsilon})^2\!\! \int\!\! \frac{d^d p}{(2\pi)^d} {\cal C}(p)\frac{n\cdot \bar{n}}{n\cdot p \, \bar{n} \cdot p}\int\!\! \frac{d^d q}{(2\pi)^d}\, \tilde {\cal C}(q)\,\frac{n \cdot \bar{n}}{n\cdot q \, \bar{n} \cdot q} \,{\cal M}^q \, .
\end{align}
The sum of all the abelian graphs is then given by
\begin{align}
\label{eq:abelian1}
\sum &\textrm{(abelian)}= C_\kappa(4C_\kappa-2C_A) (g^2 \tilde \mu^{2\epsilon})^2 \!\!\int\!\! \frac{d^d p}{(2\pi)^d}\, {\cal C}(p)\frac{n\cdot \bar{n}}{n\cdot p \, \bar{n} \cdot p}\,\int \!\!\frac{d^d q}{(2\pi)^d}\, \tilde {\cal C}(q)\,\frac{n \cdot \bar{n}}{n\cdot q \, \bar{n} \cdot q} \, \big[ {\cal M}^{p+q}-{\cal M}^q \big] \nn \\
& =\int \frac{d^d q}{(2\pi)^d}\,\bigg[ \frac{ g^2\,(4C_\kappa-2C_A)\,\tilde \mu^{2\epsilon}\, \tilde{ \cal C}(q) }{ q^+ \,q^-} \bigg] \, \frac{\alpha_s C_\kappa}{\pi}\frac{(\mu^2 e^{\gamma_E})^\eps}{\Gamma(1-\eps)} \int_0^\infty \frac{d p^+ \,dp^-}{(p^+\, p^-)^{1+\eps}} \, \big[ {\cal M}^{p+q}-{\cal M}^q \big]\,.
\end{align}
Note that the $\delta(\ell^+)$ term in \eq{MqSDOE} in the virtual graphs cancels the first $\delta (\ell^+)$ term in \eq{MpqSDOE} from the real graphs. This yields the result stated in \eqs{scabelian}{scfig8}.

\addtocontents{toc}{\protect\setcounter{tocdepth}{1}}
\subsubsection{Non-Abelian Graphs}
\addtocontents{toc}{\protect\setcounter{tocdepth}{2}}
\label{app:nonabelian}

In this section we give details of the computation of the non-abelian diagrams shown in \fig{np-nonabelian} and discussed in \sec{nonabelian}. The sum of real radiation graphs (with examples given by the first two diagrams in \fig{np-nonabelian}) gives:
\begin{align}
\label{eq:nonabelianreal}
&\sum \textrm{(non-abelian, real)}= 4\, C_\kappa C_A (g^2 \tilde \mu^{2\epsilon})^2 \int \frac{d^d p}{(2\pi)^d}\, \int \frac{d^d q}{(2\pi)^d}\, {\cal C}(p)\, \tilde {\cal C}(q) \, {\cal M}^{p+q} \, \frac{1}{(p+q)^2}\\
&\times\! \bigg [ \frac{1}{p^+ (p^-\! +\! q^- )}\! +\! \frac{1}{q^+ (p^-\! +\! q^- )} \! +\! \ \frac{1}{(p^+\! +\! q^+ ) p^- }\! +\! \ \frac{1}{(p^+\! +\! q^+ ) q^- }\!-\! \frac{4}{(p^+\! +\! q^+ )(p^-\! +\! q^- )} \!+\! \frac{1}{q^+ p^-} \!+\! \frac{1}{p^+q^- } \bigg] \nn
\end{align}
Expanding the Eikonal propagators in the limit $p^\mu \gg q^\mu$ for all the components gives
\begin{align}
&\sum \textrm{(non-abelian, real)}\nn \\
&\qquad= 4\, C_\kappa C_A (g^2 \tilde \mu^{2\epsilon})^2 \int \frac{d^d p}{(2\pi)^d}\, \int \frac{d^d q}{(2\pi)^d}\, {\cal C}(p)\, \tilde {\cal C}(q) \, {\cal M}^{p+q} \, \frac{1}{(p+q)^2} \, \bigg [ \frac{2}{q^+ p^-} + \frac{2}{p^+ q^-} \bigg] \nn \, , \\
\label{eq:NAexpanded}
&\qquad = \frac{\alpha_s C_\kappa}{\pi}\frac{(\mu^2 e^{\gamma_E})^\eps}{\Gamma(1-\eps)} \int_0^\infty \frac{d p^+ \,dp^-}{(p^+\, p^-)^{\eps}} \, \int \frac{d \phi_p}{2 \pi} \,\int \frac{d^d q}{(2\pi)^d}\, \frac{ g^2\,C_A\,\tilde \mu^{2\epsilon}\, \tilde{ \cal C}(q) }{ q^+ \,q^-}\, {\cal M}^{p+q} \\
&\qquad \qquad \times \frac{2}{p^+ q^- + q^+ p^- - 2\sqrt{p^+p^-} | \vec q_\perp| \cos(\Delta \phi) } \, \bigg [ \frac{q^-}{p^-} + \frac{q^+}{p^+ }\bigg] \nn
\end{align}
where we made use of the onshell cut conditions in \eq{cut} to simplify the propagator. This yields
\begin{align}
\sum \textrm{(non-abelian, real)}&= \frac{\alpha_s C_\kappa}{\pi}\frac{(\mu^2 e^{\gamma_E})^\eps}{\Gamma(1-\eps)} \int_0^\infty \frac{d p^+ \,dp^-}{(p^+\, p^-)^{1+\eps}} \, \int \frac{d^d q}{(2\pi)^d}\, \frac{2\, g^2\,C_A\,\tilde \mu^{2\epsilon}\, \tilde{ \cal C}(q) }{ q^+ \,q^-}\, {\cal M}^{p+q} \\
&\qquad \times \frac{q^+ p^- + p^+ q^-}{p^+ q^- + q^+ p^- - 2 \sqrt{p^+ p^-} | \vec q_\perp| \cos (\Delta \phi)} \, .\nn
\end{align}

Next consider the virtual non-abelian graphs (examples are the last two diagrams in \fig{np-nonabelian}).\footnote{Unlike for the abelian graphs, here it is known that there can be additional $1/\eps$ UV poles which generate anomalous dimensions for the power correction parameters~\cite{Mateu:2012nk}, which here would sum single logarithms generated between $\Lambda_{\rm QCD}$ and the collinear-soft scale. Consideration of these effects is beyond the scope of this work.} The sum of all the non-abelian graphs reads
\begin{align}
\label{eq:nonabelian}
\sum \textrm{non-abelian}&= \frac{\alpha_s C_\kappa}{\pi}\frac{(\mu^2 e^{\gamma_E})^\eps}{\Gamma(1-\eps)} \int_0^\infty \frac{d p^+ \,dp^-}{(p^+\, p^-)^{1+\eps}} \, \int \frac{d^d q}{(2\pi)^d}\, \frac{2\, g^2\,C_A\,\tilde \mu^{2\epsilon}\, \tilde{ \cal C}(q) }{ q^+ \,q^-} \\
&\qquad \times\, \big[ {\cal M}^{p+q} - {\cal M}^{q} \big] \frac{q^+ p^- + p^+ q^-}{p^+ q^- + q^+ p^- - 2 \sqrt{p^+ p^-} | \vec q_\perp| \cos (\Delta \phi)} \, .\nn
\end{align}
Note that the $\delta(\ell^+)$ term again cancels in the difference ${\cal M}^{p+q} - {\cal M}^{q}$.
This yields the result stated in \eq{scnonabelian}.

\subsection{Analysis with Two Perturbative Emissions}

\subsubsection{Confirmation of the NP Factorization at ${\cal O}(\alpha_s^2)$}
\label{app:2pert}

\begin{figure}[t!]
\centering
\includegraphics[width=.8\columnwidth]{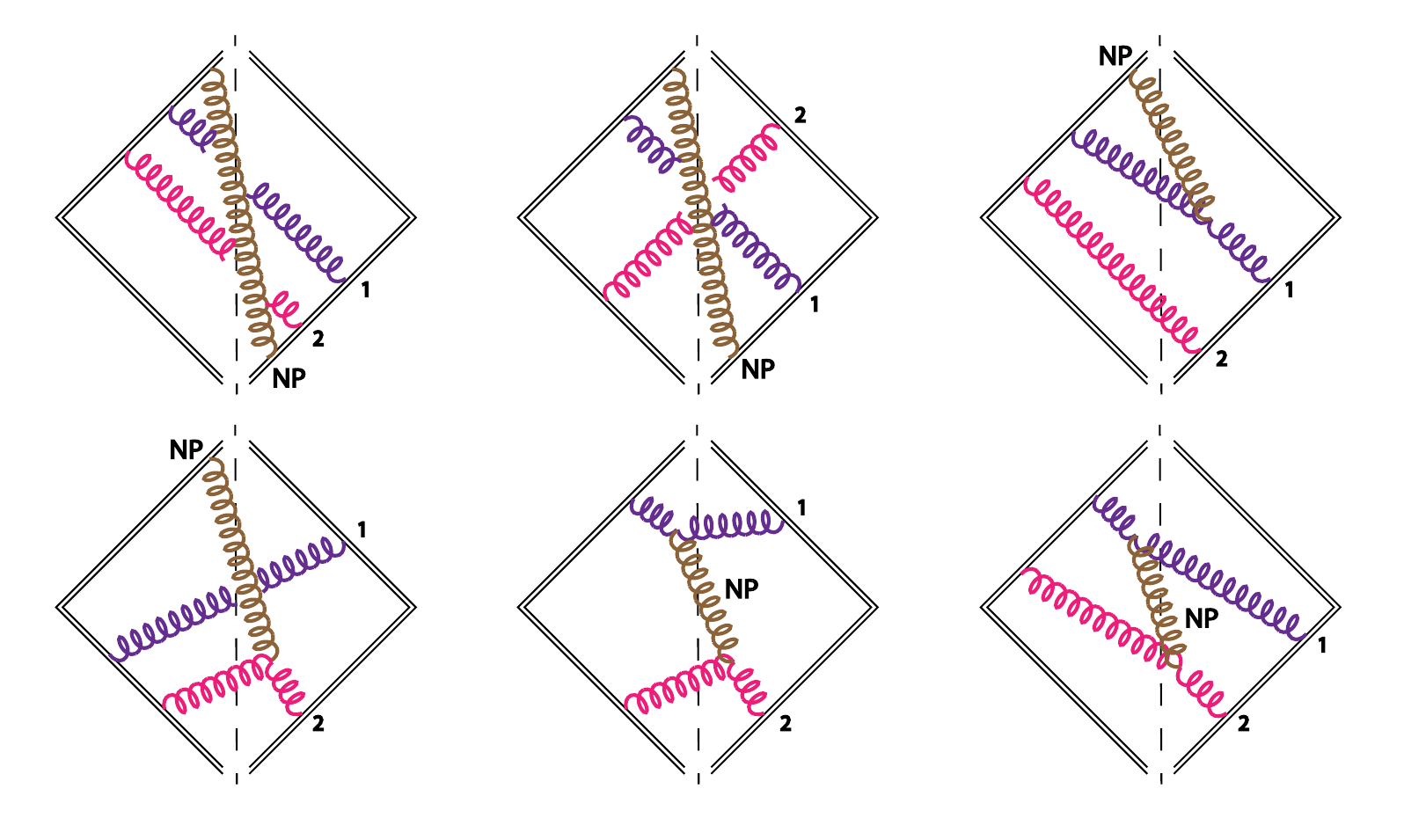}
\vspace{-10pt}
\caption{Examples of diagrams with one perturbative soft drop failing or unresolved gluon (purple, labeled 1), one perturbative soft drop passing gluon (magenta, labeled 2), and a nonperturbative gluon (brown, labeled NP). Dashed line represents measurement of the plus momentum of the passing gluon along with the soft drop operator. Eikonal emissions of perturbative gluons are considered. Diagrams with virtual perturbative gluons are not shown. Graphs with a virtual NP gluon are not considered.}
\label{fig:np-2-pert}
\end{figure}
In this appendix we perform a cross check on the factorization of the nonperturbative matrix element by using two perturbative emissions, one of which stops the soft drop and the other being either a failing or an unresolved emission. We show examples of such graphs in \fig{np-2-pert}. For simplicity, in this exercise, we will only evaluate the real radiation graphs, where two perturbative gluons and one nonperturbative source gluon run across the cut. The graphs with one or both the perturbative gluons being virtual simply serve to cancel certain terms and correspond to the cases already considered above in \app{oneloop}. At LL it suffices to consider the diagrams with both the perturbative gluons having eikonal couplings to the energetic collinear parton. The soft drop stopping gluon has the momentum $p^\mu_2$ and the additional gluon has $p^\mu_1$. At LL we can limit ourselves to the case where the angles $\theta_{p_1}$ and $\theta_{p_2}$ are strongly ordered, such that we have the following two scenarios:
\begin{align}
\label{eq:2pert-scenarios}
\text{$p_1^\mu$ fails soft drop} &: && R \gg \theta_{p_1} \gg \theta_{p_2} \, , \nn\\
\text{$p_1^\mu$ is an unresolved emission} &: && \theta_{p_1} \ll \theta_{p_2} \ll R \, .
\end{align}
Concerning attachments of the NP gluon, at leading order, we again only need to consider diagrams where the NP gluon is either emitted last from the collinear parton (described by a Wilson line), or is attached to one or both of the perturbative gluons. From \eqs{Fabelian}{Fnonabelian} we see that the combined NP source function from the abelian and non-abelian graphs derived with a single perturbative emission is given by
\begin{align}
\label{eq:Fcombined}
\tilde F (k^\mu) =4 g^2 \,\tilde \mu^{2\epsilon}\, \tilde{ \cal C}(k) \bigg [ \Big(C_\kappa - \frac{C_A}{2}\Big)\frac{1}{ k^+ \,k^-} + \frac{C_A}{2}\frac{1}{ k^+ \,k^-} \, \frac{k^+ + k^-}{k^+ + k^- - 2 \,| \vec k_\perp| \cos (\phi_k)}\bigg]
\,.
\end{align}
Here the variables $(k^+,k^-,k_\perp)$ are related to the NP gluon momentum $q^\mu$ by the rescaling in \eq{rescaling} with $\theta_{cs} = \theta_{p_2}$ being the angle of the stopping gluon with respect to the jet axis. The goal is to check that also in the presence of an additional perturbative emission, the same source function of \eq{Fcombined} emerges at LL accuracy. We anticipate that in presence of additional perturbative emissions at LL, we will obtain an additional $C_\kappa$ for each emission. The contributions from all the real emission diagrams of the kind shown in \fig{np-2-pert}, can be then decomposed into the following color basis:
\begin{align}
\label{eq:colbasis}
C_\kappa^2\Big(C_\kappa - \frac{C_A}{2}\Big) \,, \qquad C_\kappa^2\frac{C_A}{2} \, , \qquad C_\kappa\Big(\frac{C_A}{2}\Big)^2 \, .
\end{align}
The graphs involving two triple gluon vertices, such as the last two diagrams in \fig{np-2-pert}, and many more, contribute with a $C_\kappa (C_A/2)^2$ color factor. For simplicity, we will not include these diagrams in this analysis and hence will only consider contributions involving the first two color structures in \eq{colbasis}, which are generated by the abelian graphs and the non-abelian graphs involving a single triple gluon vertex.

We first consider the measurement operator ${\cal M}(\ell^+, p_1^\mu, p_2^\mu,q^\mu) $ for the case of two perturbative emissions. It is a higher order generalization of the two emission case in \eq{MpqSDOE} and can be expressed as follows
\begin{align}
\label{eq:2pert-measurement}
{\cal M}(\ell^+, p_1^\mu, p_2^\mu,q^\mu) &= {\cal M}(\ell^+, p_1^\mu, p_2^\mu ) + \Theta(\theta_{p_1} - \theta_{p_2}) \Theta_{\rm sd}^{p_1} \Delta {\cal M}(\ell^+, p_2^\mu,q^\mu) \\
& + \Theta(\theta_{p_2} - \theta_{p_1}) \Delta {\cal M}(\ell^+, p_1^\mu, p_2^\mu,q^\mu) \, . \nn
\end{align}
The first term on the right hand side is the measurement for the leading power term that only involves the two perturbative emissions $p_1^\mu$ and $p_2^\mu$. Since we can ignore clustering of $p_1^\mu$ and $p_2^\mu$ due to the strong ordering limits, it reads:
\begin{align}
\label{eq:2pert-M}
{\cal M}(\ell^+, p_1^\mu, p_2^\mu) &=\Theta(\theta_{p_1} - \theta_{p_2}) \Theta_{\rm sd}^{p_1}\Big[\overline \Theta_{\rm sd}^{p_2} \delta (\ell^+ - p_2^+) + \Theta_{\rm sd}^{p_2} \delta (\ell^+ ) \Big] \\
&+ \Theta(\theta_{p_2} - \theta_{p_1}) \overline \Theta_{\rm sd}^{p_2} \delta (\ell^+ - p_1^+ -p_2^+)
\nn \,.
\end{align}
The second term and the third term in \eq{2pert-measurement} account for the power corrections in the two cases in \eq{2pert-scenarios}, namely, when $p_1$ fails soft drop first and when $p_1$ lies at a smaller angle, respectively. They are given by
\begin{align}
\label{eq:2pert-deltaM}
\Delta {\cal M}(\ell^+, p_2^\mu,q^\mu) &= \overline \Theta_{\rm NP}^{\figeight}(\theta_{q}, \, \theta_{p_2},\, \Delta \phi)\, \overline \Theta_{\rm sd}^{\, p_2}\,\Big [ \delta(\ell^+ - p_2^+ - q^+)- \delta(\ell^+ - p_2^+) \Big] \\
&+ \Delta \overline \Theta_{\rm sd}^{p_2} \, \Big (\overline \Theta_{\rm NP}^{\bndry} (\theta_{q}, \, \theta_{p_2},\, \Delta \phi) - \Theta_{\rm NP}^{\bndry} (\theta_{q}, \, \theta_{p_2},\, \Delta \phi) \Big)\, \Big[ \delta(\ell^+ - p_2^+) - \delta(\ell^+) \Big ] \nn \, ,
\\
\Delta {\cal M}(\ell^+, p_1^\mu, p_2^\mu,q^\mu) &= \overline \Theta_{\rm NP}^{\figeight}(\theta_{q}, \, \theta_{p_2},\, \Delta \phi)\, \overline \Theta_{\rm sd}^{\, p_2}\,\Big [ \delta(\ell^+ - p_1^+- p_2^+ - q^+)- \delta(\ell^+ - p_1^+-p_2^+) \Big]\nn \\
&+ \Delta \overline \Theta_{\rm sd}^{p_2} \,\Big (\overline \Theta_{\rm NP}^{\bndry} (\theta_{q}, \, \theta_{p_2},\, \Delta \phi) - \Theta_{\rm NP}^{\bndry} (\theta_{q}, \, \theta_{p_2},\, \Delta \phi) \Big)\, \Big[ \delta(\ell^+ - p_2^+) - \delta (\ell^+) \Big ]\nn \\
&- \Delta \overline \Theta_{\rm sd}^{p_2} \, \Big (\overline \Theta_{\rm NP}^{\bndry} (\theta_{q}, \, \theta_{p_2},\, \Delta \phi) - \Theta_{\rm NP}^{\bndry} (\theta_{q}, \, \theta_{p_2},\, \Delta \phi) \Big) \, \overline \Theta_{\rm sd}^{p_1} \, \Big[ \delta(\ell^+ - p_1^+) - \delta (\ell^+) \Big ] \, .\nn
\end{align}
Here the final term involving $\overline \Theta_{\rm sd}^{p_1}$ in the boundary correction in $\Delta {\cal M}(\ell^+, p_1^\mu, p_2^\mu,q^\mu)$ results from the case where $p_2^\mu$ fails soft drop due to hadronization and the groomer then recurses to test the inner $p_1^\mu$ emission. We will consider this term in detail in \app{scfail} where we show it to be subleading at LL. We can thus ignore this term for now. Note that in the limit where $p_1^\mu$ is an unresolved emission with $p_1^+ \ll p_2^+ $, the two $\Delta {\cal M}$ formulas in \eq{2pert-deltaM} become equal and \eqs{2pert-measurement}{2pert-M} then simplify to the following expression:
\begin{align}
\label{eq:2pert-Munresolved}
{\cal M}(\ell^+, p_1^\mu, p_2^\mu,q^\mu) &= {\cal M}(\ell^+, p_1^\mu, p_2^\mu ) +\Big[1-\Theta(\theta_{p_1} - \theta_{p_2})\,\overline \Theta_{\rm sd}^{p_1} \Big] \,\Delta {\cal M}(\ell^+, p_2^\mu,q^\mu)
\,,
\end{align}
where the leading power term in this limit is given by
\begin{align}
{\cal M}(\ell^+, p_1^\mu, p_2^\mu) &\stackrel{p_1^+ \ll p_2^+}{=} \Big[1-\Theta(\theta_{p_1} - \theta_{p_2})\,\overline \Theta_{\rm sd}^{p_1} \Big] \, \overline \Theta_{\rm sd}^{p_2} \delta(\ell^+ - p_2^+) + \Theta(\theta_{p_1} - \theta_{p_2})\,\Theta_{\rm sd}^{p_1}\,\Theta_{\rm sd}^{p_2} \, \delta (\ell^+)
\,.
\end{align}

We now consider the abelian diagrams, such as the first two graphs in the top row in \fig{np-2-pert}. With the gluon carrying $q^\mu$ always being the last radiated, the sum of these diagrams is given by
\begin{align}
\label{eq:2pert-abelian}
\sum &\text{(abelian, real, LL)} =\Big( \frac{\alpha_s }{\pi}\frac{(\mu^2 e^{\gamma_E})^\eps}{\Gamma(1-\eps)}\Big)^2 \int_0^\infty \frac{d p_1^+ \,dp_1^-}{(p_1^+\, p_1^-)^{\eps}} \int_0^\infty \frac{d p_2^+ \,dp_2^-}{(p_2^+\, p_2^-)^{\eps}} \, \int\! \frac{d^d q}{(2\pi)^d} \\
& \times{\cal M}(\ell^+, p_1^\mu, p_2^\mu,q^\mu)\, \Bigg\{4g^2\,\tilde \mu^{2\epsilon}\, \tilde{ \cal C}(q)\,C_\kappa\Big(C_\kappa - \frac{C_A}{2}\Big)^2 \frac{1}{p_1^+p_1^-}\frac{1}{p_2^+p_2^-} \frac{1}{q^+q^-} + \ldots \Bigg \}\, ,\nn
\end{align}
where we only show the sum of terms from all the abelian diagrams that add up to give the independent emission result, as this will give the largest logarithmic contribution. The terms denoted by ellipsis with other color structures give subleading logarithms. In the following we will suppress such terms.

As an aside, we note that upon including the virtual terms, the leading power measurement in the unresolved limit of $p_1^\mu$ emission can be manipulated into the same form as in the coherent branching formalism in \eq{partonic},
\begin{align}
\label{eq:2pertfull}
{\cal M}(\ell^+, p_1^\mu, p_2^\mu) + {\cal M}_{\rm virtual}(\ell^+, p_2^\mu) &= \Theta(\theta_{p_1} - \theta_{p_2})\,\big[-\overline \Theta_{\rm sd}^{p_1}\big]\, \overline \Theta_{\rm sd}^{p_2}\big[\delta(\ell^+ - p_2^+) - \delta(\ell^+ ) \big]
\,.
\end{align}

Keeping only the LL terms, the sum of diagrams involving one triple-gluon vertex, with the NP gluon attached to either of the perturbative gluons, with examples being the third and the fourth diagrams in \fig{np-2-pert}, yield
\begin{align}
\label{eq:2pert-na1}
&\sum \text{(1 n.a., real, LL)} =\Big( \frac{\alpha_s }{\pi}\frac{(\mu^2 e^{\gamma_E})^\eps}{\Gamma(1-\eps)}\Big)^2 \int_0^\infty\! \frac{d p_1^+ \,dp_1^-}{(p_1^+\, p_1^-)^{1+\eps}} \int_0^\infty \! \frac{d p_2^+ \,dp_2^-}{(p_2^+\, p_2^-)^{1+\eps}} \int\! \frac{d^d q}{(2\pi)^d} \,{\cal M}(\ell^+, p_1^\mu, p_2^\mu,q^\mu)\nn\\
& \qquad \times 4g^2\,\tilde \mu^{2\epsilon}\, \tilde{ \cal C}(q)\,C_\kappa\frac{C_A}{2}\Big(C_\kappa - \frac{C_A}{2}\Big) \bigg[\Big(\frac{p_1^+}{q^+}+\frac{p_1^-}{q^-}\Big)\frac{1}{(p_1 + q)^2}+ \Big (\frac{p_2^+}{q^+}+\frac{p_2^-}{q^-}\Big)\frac{1}{(p_2 + q)^2}\bigg ] \, .
\end{align}
The two terms in the last line of \eq{2pert-na1} result from attachments of the NP gluon to $p_1$ and $p_2$ respectively.

In the next step, we remove the dependence of the non-perturbative part of the matrix element on $\theta_{p_2}$ and $\phi_{p_2}$ by performing a boost and a rotation to the NP gluon $q^\mu$ as in \eq{rescaling}. The result for the abelian graphs in \eq{2pert-abelian} becomes
\begin{align}
\label{eq:2pert-abelian2}
\sum \text{(abelian, real, LL)} &=\Big( \frac{\alpha_s }{\pi}\frac{(\mu^2 e^{\gamma_E})^\eps}{\Gamma(1-\eps)}\Big)^2C_\kappa\Big(C_\kappa - \frac{C_A}{2}\Big)^2 \int_0^\infty \frac{d p_1^+ \,dp_1^-}{(p_1^+\, p_1^-)^{1+\eps}} \int_0^\infty \frac{d p_2^+ \,dp_2^-}{(p_2^+\, p_2^-)^{1+\eps}} \,\nn \\
& \times \sum_a \tilde {\cal M}_a(\ell^+, p_1^\mu, p_2^\mu )\, \int\! \frac{d^d k}{(2\pi)^d}\,{\cal M}_a^{\rm NP}(k^\mu) \ 4g^2\, \tilde \mu^{2\epsilon}\, \tilde{ \cal C}(k)\, \frac{1}{k^+ k^-} \, ,
\end{align}
where the measurement function after rescaling has now factorized into a sum ${\cal M} \to \sum_a \tilde {\cal M}_a {\cal M}_a^{\rm NP}$, where the terms in the sum include the leading power result ($a = 0$), the shift ($a = \figeight$) and the boundary corrections ($a = \bndry$) in \eq{2pert-Munresolved}:
\begin{align}
\label{eq:2pert-Mfact}
\tilde {\cal M}_{0}(\ell^+, p_1^\mu, p_2^\mu ) &= {\cal M}(\ell^+, p_1^\mu, p_2^\mu) \,, \\
\tilde {\cal M}^{\rm NP}_0(k^\mu) &= 1 \, ,\nn \\
\tilde {\cal M}_{\figeight}(\ell^+, p_1^\mu, p_2^\mu ) &= - \frac{\theta_{p_2}}{2} \, \Big[1-\Theta(\theta_{p_1} - \theta_{p_2})\,\overline \Theta_{\rm sd}^{p_1} \Big] \, \overline \Theta_{\rm sd}^{p_2} \frac{d}{d \ell^+} \delta(\ell^+ - p_2^+)\nn \, , \\
\tilde {\cal M}^{\rm NP}_{\figeight}(k^\mu) &= k^+ \,\overline \Theta_{\rm NP}^{\, \figeight} \Big(\frac{k_\perp}{k^-}, \, 1, \, \phi_k \Big) \nn \, ,\\
\tilde {\cal M}_{\bndry} (\ell^+, p_1^\mu, p_2^\mu )&= \frac{2}{\theta_{p_2}}\, \Big[1-\Theta(\theta_{p_1} - \theta_{p_2})\,\overline \Theta_{\rm sd}^{p_1} \Big]\,\delta \Bigl( \frac{p_2^- }{Q} - \tilde z_{\rm cut} \, \theta_{p_2}^\beta \Bigr)\, \Big[ \delta(\ell^+ - p_2^+) - \delta (\ell^+) \Big ] \, , \nn \\
\tilde {\cal M}^{\rm NP}_{\bndry}(k^\mu) &=\Big (k^- + \beta\, \big( k^- - k_\perp \, \cos \phi_k\big) \Big) \, \bigg[\overline \Theta_{\rm NP}^{\, \bndry} \Big(\frac{k_\perp}{k^-}, \, 1, \, \phi_k \Big) - \Theta_{\rm NP}^{\, \bndry} \Big(\frac{k_\perp}{k^-}, \, 1, \, \phi_k \Big)\bigg]\nn \, .
\end{align}
We see that the factors $\tilde {\cal M}_a(\ell^+, p_1^\mu, p_2^\mu ) $ have no dependence on the NP momentum, and the restriction in the rescaled coordinates is now implemented via ${\cal M}_a^{\rm NP}(k^\mu)$. Furthermore, the operators ${\cal M}_{\figeight}^{\rm NP}(k^\mu)$ and ${\cal M}_{\bndry}^{\rm NP}(k^\mu)$ will implement the same moments of the source function as in \eq{O1Ups1} derived using one perturbative emission analysis.

While the NP matrix element for the abelian result in \eq{2pert-abelian2} remains the same, the terms appearing in the non-abelian results in parentheses in \eq{2pert-na1} yield the following expressions
\begin{align}
\label{eq:2pert-rescaling}
\Big(\frac{p_1^+}{q^+}+\frac{p_1^-}{q^-}\Big)\frac{1}{(p_1 + q)^2} &= \frac{1}{k^+k^-}\frac{ k^+\frac{\theta_{p_2}}{\theta_{p_1}} + k^-\frac{\theta_{p_1}}{\theta_{p_2}}}{k^+\frac{\theta_{p_2}}{\theta_{p_1}} + k^-\frac{\theta_{p_1}}{\theta_{p_2}} - 2|\vec k_\perp| \cos (\phi_k)} \, , \\
\Big(\frac{p_2^+}{q^+}+\frac{p_2^-}{q^-}\Big)\frac{1}{(p_2 + q)^2} &= \frac{1}{k^+k^-}\frac{ k^++ k^-}{k^+ + k^-- 2|\vec k_\perp| \cos (\phi_k)} \, , \nn
\end{align}
where we have used the small angle approximation and on-shell condition,
\begin{align}
\theta_{p_i} = 2\sqrt{\frac{p_i^+}{p_i^-}} \, , \qquad |\vec p_{i\perp} | = \sqrt{p_i^+ p_i^-}\, , \qquad i = 1,2 \, .
\end{align}
We see that upon rescaling the momentum $q^\mu$ with respect to $\theta_{p_2}$ leaves a residual $\theta_{p_1}/\theta_{p_2}$ dependence in the terms where the gluon is attached to $p_1$.
We can, however, simplify the expressions by making use of the fact that only a strong ordering of the angles as in \eq{2pert-scenarios} will contribute to the LL result. In these two limits we find
\begin{align}
\label{eq:2pert-lim}
\frac{ k^+\frac{\theta_{p_2}}{\theta_{p_1}} + k^-\frac{\theta_{p_1}}{\theta_{p_2}}}{k^+\frac{\theta_{p_2}}{\theta_{p_1}} + k^-\frac{\theta_{p_1}}{\theta_{p_2}} - 2|\vec k_\perp| \cos (\phi_k)} &\stackrel{\theta_{p_1} \gg \theta_{p_2}}{\longrightarrow} 1 \, , \\
\frac{ k^+\frac{\theta_{p_2}}{\theta_{p_1}} + k^-\frac{\theta_{p_1}}{\theta_{p_2}}}{k^+\frac{\theta_{p_2}}{\theta_{p_1}} + k^-\frac{\theta_{p_1}}{\theta_{p_2}} - 2|\vec k_\perp| \cos (\phi_k)} &\stackrel{\theta_{p_1} \ll \theta_{p_2}}{\longrightarrow} 1 \, . \nn
\end{align}
Thus, taking either of the limits in \eq{2pert-lim} we get the same result for the non-abelian graphs in \eq{2pert-na1}, which simplify to
\begin{align}
\label{eq:2pert-na12}
&\sum \text{(1 n.a., real, LL)} =\Big( \frac{\alpha_s }{\pi}\frac{(\mu^2 e^{\gamma_E})^\eps}{\Gamma(1-\eps)}\Big)^2\! C_\kappa\frac{C_A}{2}\Big(C_\kappa - \frac{C_A}{2}\Big)\! \int_0^\infty\! \frac{d p_1^+ \,dp_1^-}{(p_1^+\, p_1^-)^{1+\eps}} \! \int_0^\infty \! \frac{d p_2^+ \,dp_2^-}{(p_2^+\, p_2^-)^{1+\eps}} \\
& \times \sum_a \tilde {\cal M}_a(\ell^+, p_1^\mu, p_2^\mu ) \int\! \frac{d^d k}{(2\pi)^d}\,{\cal M}_a^{\rm NP}(k^\mu) 4g^2 \tilde \mu^{2\epsilon}\tilde{ \cal C}(k) \bigg[\frac{1}{k^+k^-} + \frac{1}{k^+k^-}\frac{ k^++ k^-}{k^+ + k^-- 2|\vec k_\perp| \cos (\phi_k)}\bigg ] \,. \nn
\end{align}
Thus, upon adding the results in \eqs{2pert-abelian2}{2pert-na12} we find that for the $C_\kappa^3$ and $C_\kappa^2 C_A$ terms we obtain the following expression:
\begin{align}
\label{eq:2pert-full}
&\sum \text{(real, LL, $C_\kappa^3$, $C_\kappa^2 C_A$)} =\! \sum_a\Big( \frac{\alpha_s C_\kappa }{\pi}\frac{(\mu^2 e^{\gamma_E})^\eps}{\Gamma(1-\eps)}\Big)^2 \!\! \int_0^\infty\! \frac{d p_1^+ \,dp_1^-}{(p_1^+\, p_1^-)^{1+\eps}} \! \int_0^\infty \!\frac{d p_2^+ \,dp_2^-}{(p_2^+\, p_2^-)^{1+\eps}} \, \tilde {\cal M}_a(\ell^+, p_1^\mu, p_2^\mu)\nn \\
& \times \int\! \frac{d^d k}{(2\pi)^d} \ 4\,g^2\, \tilde \mu^{2\epsilon}\, \tilde{ \cal C}(k)\,{\cal M}_a^{\rm NP}(k^\mu) \frac{1}{k^+k^-}\bigg [\Big(C_\kappa - \frac{C_A}{2}\Big) + \frac{C_A}{2} \, \frac{k^+ + k^-}{k^+ + k^- - 2 \,| \vec k_\perp| \cos (\phi_k)} \bigg ] \, .
\end{align}
We thus see that \eq{2pert-full} involves precisely the same source function as identified in \eq{Fcombined} from the analysis of one perturbative emission, and hence, following \eq{2pert-Mfact}, the same hadronic parameters. This confirms our expectation that the OPE involves precisely the same NP matrix element for any number of perturbative emissions at LL order.

\subsubsection{Boundary correction from failing subjets}
\label{app:scfail}

Here we evaluate the perturbative coefficient for the power correction to the soft drop tests for failing subjets that have collinear-soft scaling and show that their contribution is subleading at LL accuracy. This serves to demonstrate that at LL accuracy one does not need to consider boundary corrections from failing collinear-soft subjets. We consider the same setup as in \app{2pert} with two perturbative gluons that have the collinear-soft scaling, one of which lies at larger angle and fails soft drop, and the other one which passes, plus an additional nonperturbative probe gluon. More precisely, we focus on the last term involving $\overline \Theta_{\rm sd}^{p_1}$ in the boundary correction in $\Delta {\cal M}(\ell^+, p_1^\mu, p_2^\mu,q^\mu)$ in \eq{2pert-deltaM}, where the $p_2^\mu$ is now a soft drop failing emission. Having checked that the same source function and hadronic parameters appear in the case of two perturbative emissions, the contribution from this term is simply given by
\begin{align}
\label{eq:SChad2}
S_c^{{\rm had}(2)}(\ell^+,\qcut,\beta, \mu)&=\frac{ \Upsilon^{ \kappa}_1(\beta)}{Q} \,\Delta S_c^{\, \bndry, {\rm fail}} (\ell^+,\qcut,\beta, \mu) \, ,
\end{align}
where the Wilson Coefficient for failing subjets reads
\begin{align}
\label{eq:Scfail1}
\Delta S_c^{\, \bndry, {\rm fail}} (\ell^+,\qcut,\beta, \mu)&= \qcut^{\frac{-1}{1+\beta}}\bigg( \frac{\alpha_s C_\kappa}{\pi}\frac{(\mu^2 e^{\gamma_E})^\eps}{\Gamma(1-\eps)}\bigg)^2\!
\int_0^\infty \!\!\frac{d p_1^+ \,d p_1^-}{(p_1^+\, p_1^-)^{1+\eps}}
\int_0^\infty\!\! \frac{d p_2^+ \,d p_2^-}{(p_2^+\, p_2^-)^{1+\eps}}\, \tilde {\cal M}^\prime_{\bndry}(\ell^+, p_1^\mu,p_2^\mu)
\,.
\end{align}
The measurement in the perturbative sector $\tilde {\cal M}^\prime_{\bndry}(\ell^+, p_1^\mu,p_2^\mu)$ results from similar manipulations that led to \eq{2pert-Mfact}, and is given by
\begin{align}
\label{eq:calM2}
\tilde {\cal M}^\prime_{\bndry}(\ell^+, p_1^\mu,p_2^\mu) &=
\frac{2}{\theta_{p_2}}\Theta\big(\theta_{p_2} - \theta_{p_1}\big)
\bigg[-\delta \Bigl( \frac{p_2^- }{Q} - \tilde z_{\rm cut} \theta_{p_1}^\beta \Bigr)\bigg]
\Theta \Bigl( \frac{p_1^- }{Q} - \tilde z_{\rm cut} \, \theta_{p_1}^\beta \Bigr)
\Big[ \delta(\ell^+ - p_1^+) - \delta(\ell^+) \Big ] \, .
\end{align}
Since we are in the SDOE region we restrict ourselves to $\ell^+ > 0$ and discard the $\delta (\ell^+)$ term. On performing the remaining three integrations, we find that the integral is finite and can be evaluated with $\eps=0$, giving
\begin{align}
\label{eq:scfail}
\Delta S_c^{\, \bndry, {\rm fail}} (\ell^+,\qcut,\beta, \mu) &=
-\Big(\frac{\alpha_s C_\kappa}{\pi}\Big)^2 \,
\frac{2+\beta}{(1+\beta)^2}\, \frac{1}{\ell^+\qcut^{\frac{1}{1+\beta}}} \frac{Q}{\qcut^{\frac{1}{2+\beta}} \big(\ell^+\big)^{\frac{1+\beta}{2+\beta}}} \, .
\end{align}
Thus we see that the correction is ${\cal O}(\alpha_s^2)$ and does not involve a large logarithm, and is hence, subleading at LL accuracy.

\bibliography{../top3}

\providecommand{\href}[2]{#2}\begingroup\raggedright\begin{thebibliography}{10}

\bibitem{Gehrmann-DeRidder:2007nzq}
A.~Gehrmann-De~Ridder, T.~Gehrmann, E.~W.~N. Glover and G.~Heinrich,
  \emph{{Second-order QCD corrections to the thrust distribution}},
  {\emph{Phys. Rev. Lett.} {\bfseries 99} (2007) 132002},
  [\href{https://arxiv.org/abs/0707.1285}{{\ttfamily 0707.1285}}].

\bibitem{GehrmannDeRidder:2007hr}
A.~Gehrmann-De~Ridder, T.~Gehrmann, E.~W.~N. Glover and G.~Heinrich,
  \emph{{NNLO corrections to event shapes in e+ e- annihilation}}, {\emph{JHEP}
  {\bfseries 12} (2007) 094},
  [\href{https://arxiv.org/abs/0711.4711}{{\ttfamily 0711.4711}}].

\bibitem{Weinzierl:2008iv}
S.~Weinzierl, \emph{{NNLO corrections to 3-jet observables in electron-positron
  annihilation}},
  \href{http://dx.doi.org/10.1103/PhysRevLett.101.162001}{\emph{Phys. Rev.
  Lett.} {\bfseries 101} (2008) 162001},
  [\href{https://arxiv.org/abs/0807.3241}{{\ttfamily 0807.3241}}].

\bibitem{Weinzierl:2009ms}
S.~Weinzierl, \emph{{Event shapes and jet rates in electron-positron
  annihilation at NNLO}}, {\emph{JHEP} {\bfseries 06} (2009) 041},
  [\href{https://arxiv.org/abs/0904.1077}{{\ttfamily 0904.1077}}].

\bibitem{Becher:2008cf}
T.~Becher and M.~D. Schwartz, \emph{{A Precise determination of $\alpha_s$ from
  LEP thrust data using effective field theory}},
  \href{http://dx.doi.org/10.1088/1126-6708}{\emph{JHEP} {\bfseries 07} (2008)
  034}, [\href{https://arxiv.org/abs/0803.0342}{{\ttfamily 0803.0342}}].

\bibitem{Chien:2010kc}
Y.-T. Chien and M.~D. Schwartz, \emph{{Resummation of heavy jet mass and
  comparison to LEP data}}, {\emph{JHEP} {\bfseries 08} (2010) 058},
  [\href{https://arxiv.org/abs/1005.1644}{{\ttfamily 1005.1644}}].

\bibitem{Abbate:2010xh}
R.~Abbate, M.~Fickinger, A.~H. Hoang, V.~Mateu and I.~W. Stewart, \emph{{Thrust
  at N${}^3$LL with Power Corrections and a Precision Global Fit for
  $\alpha_s(m_Z)$}},
  \href{http://dx.doi.org/10.1103/PhysRevD.83.074021}{\emph{Phys. Rev.}
  {\bfseries D83} (2011) 074021},
  [\href{https://arxiv.org/abs/1006.3080}{{\ttfamily 1006.3080}}].

\bibitem{Hoang:2014wka}
A.~H. Hoang, D.~W. Kolodrubetz, V.~Mateu and I.~W. Stewart,
  \emph{{$C$-parameter distribution at N$^3$LL$^\prime$ including power
  corrections}},
  \href{http://dx.doi.org/10.1103/PhysRevD.91.094017}{\emph{Phys. Rev.}
  {\bfseries D91} (2015) 094017},
  [\href{https://arxiv.org/abs/1411.6633}{{\ttfamily 1411.6633}}].

\bibitem{Hoang:2015hka}
A.~H. Hoang, D.~W. Kolodrubetz, V.~Mateu and I.~W. Stewart, \emph{{Precise
  determination of $\alpha_s$ from the $C$-parameter distribution}},
  \href{http://dx.doi.org/10.1103/PhysRevD.91.094018}{\emph{Phys. Rev.}
  {\bfseries D91} (2015) 094018},
  [\href{https://arxiv.org/abs/1501.04111}{{\ttfamily 1501.04111}}].

\bibitem{Berger:2010xi}
C.~F. Berger, C.~Marcantonini, I.~W. Stewart, F.~J. Tackmann and W.~J.
  Waalewijn, \emph{{Higgs Production with a Central Jet Veto at NNLL+NNLO}},
  {\emph{JHEP} {\bfseries 04} (2011) 092},
  [\href{https://arxiv.org/abs/1012.4480}{{\ttfamily 1012.4480}}].

\bibitem{Tackmann:2012bt}
F.~J. Tackmann, J.~R. Walsh and S.~Zuberi, \emph{{Resummation Properties of Jet
  Vetoes at the LHC}},
  \href{http://dx.doi.org/10.1103/PhysRevD.86.053011}{\emph{Phys. Rev.}
  {\bfseries D86} (2012) 053011},
  [\href{https://arxiv.org/abs/1206.4312}{{\ttfamily 1206.4312}}].

\bibitem{Banfi:2012yh}
A.~Banfi, G.~P. Salam and G.~Zanderighi, \emph{{NLL+NNLO predictions for
  jet-veto efficiencies in Higgs-boson and Drell-Yan production}},
  \href{http://dx.doi.org/10.1007/JHEP06(2012)159}{\emph{JHEP} {\bfseries 06}
  (2012) 159}, [\href{https://arxiv.org/abs/1203.5773}{{\ttfamily 1203.5773}}].

\bibitem{Becher:2012qa}
T.~Becher and M.~Neubert, \emph{{Factorization and NNLL Resummation for Higgs
  Production with a Jet Veto}},
  \href{http://dx.doi.org/10.1007/JHEP07(2012)108}{\emph{JHEP} {\bfseries 07}
  (2012) 108}, [\href{https://arxiv.org/abs/1205.3806}{{\ttfamily 1205.3806}}].

\bibitem{Banfi:2012jm}
A.~Banfi, P.~F. Monni, G.~P. Salam and G.~Zanderighi, \emph{{Higgs and Z-boson
  production with a jet veto}},
  \href{http://dx.doi.org/10.1103/PhysRevLett.109.202001}{\emph{Phys. Rev.
  Lett.} {\bfseries 109} (2012) 202001},
  [\href{https://arxiv.org/abs/1206.4998}{{\ttfamily 1206.4998}}].

\bibitem{Liu:2012sz}
X.~Liu and F.~Petriello, \emph{{Resummation of jet-veto logarithms in hadronic
  processes containing jets}},
  \href{http://dx.doi.org/10.1103/PhysRevD.87.014018}{\emph{Phys. Rev.}
  {\bfseries D87} (2013) 014018},
  [\href{https://arxiv.org/abs/1210.1906}{{\ttfamily 1210.1906}}].

\bibitem{Stewart:2013faa}
I.~W. Stewart, F.~J. Tackmann, J.~R. Walsh and S.~Zuberi, \emph{{Jet $p_T$
  resummation in Higgs production at $NNLL'+NNLO$}},
  \href{http://dx.doi.org/10.1103/PhysRevD.89.054001}{\emph{Phys. Rev.}
  {\bfseries D89} (2014) 054001},
  [\href{https://arxiv.org/abs/1307.1808}{{\ttfamily 1307.1808}}].

\bibitem{Becher:2013xia}
T.~Becher, M.~Neubert and L.~Rothen, \emph{{Factorization and
  $N^{3}LL_{p}$+NNLO predictions for the Higgs cross section with a jet veto}},
  \href{http://dx.doi.org/10.1007/JHEP10(2013)125}{\emph{JHEP} {\bfseries 10}
  (2013) 125}, [\href{https://arxiv.org/abs/1307.0025}{{\ttfamily 1307.0025}}].

\bibitem{Dawson:2016ysj}
S.~Dawson, P.~Jaiswal, Y.~Li, H.~Ramani and M.~Zeng, \emph{{Resummation of jet
  veto logarithms at N$^3$LL$_a$ + NNLO for $W^+ W^-$ production at the LHC}},
  \href{http://dx.doi.org/10.1103/PhysRevD.94.114014}{\emph{Phys. Rev.}
  {\bfseries D94} (2016) 114014},
  [\href{https://arxiv.org/abs/1606.01034}{{\ttfamily 1606.01034}}].

\bibitem{Lee:2006fn}
C.~Lee and G.~Sterman, \emph{{Universality of nonperturbative effects in event
  shapes}},  \href{https://arxiv.org/abs/hep-ph/0603066}{{\ttfamily
  hep-ph/0603066}}.

\bibitem{Bauer:2000ew}
C.~W. Bauer, S.~Fleming and M.~E. Luke, \emph{{Summing Sudakov logarithms in $B
  \to X_s \gamma$ in effective field theory}}, {\emph{Phys. Rev.} {\bfseries
  D63} (2000) 014006}.

\bibitem{Bauer:2000yr}
C.~W. Bauer, S.~Fleming, D.~Pirjol and I.~W. Stewart, \emph{{An effective field
  theory for collinear and soft gluons: Heavy to light decays}}, {\emph{Phys.
  Rev. D} {\bfseries 63} (2001) 114020}.

\bibitem{Bauer:2001yt}
C.~W. Bauer, D.~Pirjol and I.~W. Stewart, \emph{{Soft-Collinear Factorization
  in Effective Field Theory}},
  \href{http://dx.doi.org/10.1103/PhysRevD.65.054022}{\emph{Phys. Rev.}
  {\bfseries D65} (2002) 054022}.

\bibitem{Bauer:2001ct}
C.~W. Bauer and I.~W. Stewart, \emph{{Invariant operators in collinear
  effective theory}}, {\emph{Phys. Lett. B} {\bfseries 516} (2001) 134--142}.

\bibitem{Bauer:2002nz}
C.~W. Bauer, S.~Fleming, D.~Pirjol, I.~Z. Rothstein and I.~W. Stewart,
  \emph{{Hard scattering factorization from effective field theory}},
  {\emph{Phys. Rev.} {\bfseries D66} (2002) 014017}.

\bibitem{Mateu:2012nk}
V.~Mateu, I.~W. Stewart and J.~Thaler, \emph{{Power Corrections to Event Shapes
  with Mass-Dependent Operators}},
  \href{http://dx.doi.org/10.1103/PhysRevD.87.014025}{\emph{Phys. Rev.}
  {\bfseries D87} (2013) 014025},
  [\href{https://arxiv.org/abs/1209.3781}{{\ttfamily 1209.3781}}].

\bibitem{Kang:2013nha}
D.~Kang, C.~Lee and I.~W. Stewart, \emph{{Using 1-Jettiness to Measure 2 Jets
  in DIS 3 Ways}}, {\emph{Phys. Rev.} {\bfseries D88} (2013) 054004},
  [\href{https://arxiv.org/abs/1303.6952}{{\ttfamily 1303.6952}}].

\bibitem{Jouttenus:2013hs}
T.~T. Jouttenus, I.~W. Stewart, F.~J. Tackmann and W.~J. Waalewijn, \emph{{Jet
  mass spectra in Higgs boson plus one jet at next-to-next-to-leading
  logarithmic order}},
  \href{http://dx.doi.org/10.1103/PhysRevD.88.054031}{\emph{Phys. Rev.}
  {\bfseries D88} (2013) 054031}.

\bibitem{Stewart:2014nna}
I.~W. Stewart, F.~J. Tackmann and W.~J. Waalewijn, \emph{{Dissecting Soft
  Radiation with Factorization}},
  \href{http://dx.doi.org/10.1103/PhysRevLett.114.092001}{\emph{Phys. Rev.
  Lett.} {\bfseries 114} (2015) 092001}.

\bibitem{Dokshitzer:1997iz}
Y.~L. Dokshitzer, A.~Lucenti, G.~Marchesini and G.~Salam, \emph{{Universality
  of 1 / Q corrections to jet shape observables rescued}},
  \href{http://dx.doi.org/10.1016/S0550-3213(97)00650-0;
  10.1016/S0550-3213(97)00650-0}{\emph{Nucl.Phys.} {\bfseries B511} (1998)
  396--418}, [\href{https://arxiv.org/abs/hep-ph/9707532}{{\ttfamily
  hep-ph/9707532}}].

\bibitem{Salam:2001bd}
G.~P. Salam and D.~Wicke, \emph{{Hadron masses and power corrections to event
  shapes}}, {\emph{JHEP} {\bfseries 05} (2001) 061},
  [\href{https://arxiv.org/abs/hep-ph/0102343}{{\ttfamily hep-ph/0102343}}].

\bibitem{Dasgupta:2003iq}
M.~Dasgupta and G.~P. Salam, \emph{{Event shapes in $e^+ e^-$ annihilation and
  deep inelastic scattering}},
  \href{http://dx.doi.org/10.1088/0954-3899/30/5/R01}{\emph{J.Phys.} {\bfseries
  G30} (2004) R143}, [\href{https://arxiv.org/abs/hep-ph/0312283}{{\ttfamily
  hep-ph/0312283}}].

\bibitem{Dasgupta:2007wa}
M.~Dasgupta, L.~Magnea and G.~P. Salam, \emph{{Non-perturbative QCD effects in
  jets at hadron colliders}},
  \href{http://dx.doi.org/10.1088/1126-6708/2008/02/055}{\emph{JHEP} {\bfseries
  02} (2008) 055}, [\href{https://arxiv.org/abs/0712.3014}{{\ttfamily
  0712.3014}}].

\bibitem{Sjostrand:2007gs}
T.~Sjostrand, S.~Mrenna and P.~Skands, \emph{{A Brief Introduction to PYTHIA
  8.1}},
  \href{http://dx.doi.org/10.1016/j.cpc.2008.01.036}{\emph{Comput.Phys.Commun.178:852-867,2008}
  (Oct., 2007) }, [\href{https://arxiv.org/abs/0710.3820v1}{{\ttfamily
  0710.3820v1}}].

\bibitem{Bahr:2008pv}
M.~Bahr, S.~Gieseke, M.~A. Gigg, D.~Grellscheid, K.~Hamilton, O.~Latunde-Dada
  et~al., \emph{{Herwig++ Physics and Manual}},
  \href{http://dx.doi.org/10.1140/epjc}{\emph{Eur.Phys.J.C58:639-707,2008}
  (Dec., 2008) }, [\href{https://arxiv.org/abs/0803.0883v3}{{\ttfamily
  0803.0883v3}}].

\bibitem{Dasgupta:2018nvj}
M.~Dasgupta, F.~A. Dreyer, K.~Hamilton, P.~F. Monni and G.~P. Salam,
  \emph{{Logarithmic accuracy of parton showers: a fixed-order study}},
  {\emph{JHEP} {\bfseries 09} (2018) 033},
  [\href{https://arxiv.org/abs/1805.09327}{{\ttfamily 1805.09327}}].

\bibitem{Hoang:2018zrp}
A.~H. Hoang, S.~Pl{\"a}tzer and D.~Samitz, \emph{{On the Cutoff Dependence of
  the Quark Mass Parameter in Angular Ordered Parton Showers}},
  \href{http://dx.doi.org/10.1007/JHEP10(2018)200}{\emph{JHEP} {\bfseries 10}
  (2018) 200}, [\href{https://arxiv.org/abs/1807.06617}{{\ttfamily
  1807.06617}}].

\bibitem{Alioli:2012fc}
S.~Alioli, C.~W. Bauer, C.~J. Berggren, A.~Hornig, F.~J. Tackmann et~al.,
  \emph{{Combining Higher-Order Resummation with Multiple NLO Calculations and
  Parton Showers in GENEVA}},
  \href{http://dx.doi.org/10.1007/JHEP09(2013)120}{\emph{JHEP} {\bfseries 1309}
  (2013) 120}, [\href{https://arxiv.org/abs/1211.7049}{{\ttfamily 1211.7049}}].

\bibitem{Hamilton:2013fea}
K.~Hamilton, P.~Nason, E.~Re and G.~Zanderighi, \emph{{NNLOPS simulation of
  Higgs boson production}}, {\emph{JHEP} {\bfseries 10} (2013) 222},
  [\href{https://arxiv.org/abs/1309.0017}{{\ttfamily 1309.0017}}].

\bibitem{Hoeche:2014aia}
S.~H{\"o}che, Y.~Li and S.~Prestel, \emph{{Drell-Yan lepton pair production at
  NNLO QCD with parton showers}}, {\emph{Phys. Rev.} {\bfseries D91} (2015)
  074015}, [\href{https://arxiv.org/abs/1405.3607}{{\ttfamily 1405.3607}}].

\bibitem{Karlberg:2014qua}
A.~Karlberg, E.~Re and G.~Zanderighi, \emph{{NNLOPS accurate Drell-Yan
  production}}, {\emph{JHEP} {\bfseries 09} (2014) 134},
  [\href{https://arxiv.org/abs/1407.2940}{{\ttfamily 1407.2940}}].

\bibitem{Hoche:2014dla}
S.~H{\"o}che, Y.~Li and S.~Prestel, \emph{{Higgs-boson production through gluon
  fusion at NNLO QCD with parton showers}}, {\emph{Phys. Rev.} {\bfseries D90}
  (2014) 054011}, [\href{https://arxiv.org/abs/1407.3773}{{\ttfamily
  1407.3773}}].

\bibitem{Hamilton:2015nsa}
K.~Hamilton, P.~Nason and G.~Zanderighi, \emph{{Finite quark-mass effects in
  the NNLOPS POWHEG+MiNLO Higgs generator}}, {\emph{JHEP} {\bfseries 05} (2015)
  140}, [\href{https://arxiv.org/abs/1501.04637}{{\ttfamily 1501.04637}}].

\bibitem{Alioli:2013hqa}
S.~Alioli, C.~W. Bauer, C.~Berggren, F.~J. Tackmann, J.~R. Walsh and S.~Zuberi,
  \emph{{Matching Fully Differential NNLO Calculations and Parton Showers}},
  {\emph{JHEP} {\bfseries 06} (2014) 089},
  [\href{https://arxiv.org/abs/1311.0286}{{\ttfamily 1311.0286}}].

\bibitem{Alioli:2015toa}
S.~Alioli, C.~W. Bauer, C.~Berggren, F.~J. Tackmann and J.~R. Walsh,
  \emph{{Drell-Yan production at NNLL'+NNLO matched to parton showers}},
  {\emph{Phys. Rev.} {\bfseries D92} (2015) 094020},
  [\href{https://arxiv.org/abs/1508.01475}{{\ttfamily 1508.01475}}].

\bibitem{Butterworth:2008iy}
J.~M. Butterworth, A.~R. Davison, M.~Rubin and G.~P. Salam, \emph{{Jet
  substructure as a new Higgs search channel at the LHC}},
  \href{http://dx.doi.org/10.1103/PhysRevLett.100.242001}{\emph{Phys. Rev.
  Lett.} {\bfseries 100} (2008) 242001},
  [\href{https://arxiv.org/abs/0802.2470}{{\ttfamily 0802.2470}}].

\bibitem{Ellis:2009me}
S.~D. Ellis, C.~K. Vermilion and J.~R. Walsh, \emph{{Recombination Algorithms
  and Jet Substructure: Pruning as a Tool for Heavy Particle Searches}},
  \href{http://dx.doi.org/10.1103/PhysRevD.81.094023}{\emph{Phys. Rev.}
  {\bfseries D81} (2010) 094023},
  [\href{https://arxiv.org/abs/0912.0033}{{\ttfamily 0912.0033}}].

\bibitem{Krohn:2009th}
D.~Krohn, J.~Thaler and L.-T. Wang, \emph{{Jet Trimming}},
  \href{http://dx.doi.org/10.1007/JHEP02(2010)084}{\emph{JHEP} {\bfseries 02}
  (2010) 084}, [\href{https://arxiv.org/abs/0912.1342}{{\ttfamily 0912.1342}}].

\bibitem{Larkoski:2014wba}
A.~J. Larkoski, S.~Marzani, G.~Soyez and J.~Thaler, \emph{{Soft Drop}},
  \href{http://dx.doi.org/10.1007/JHEP05(2014)146}{\emph{JHEP} {\bfseries 05}
  (2014) 146}, [\href{https://arxiv.org/abs/1402.2657}{{\ttfamily 1402.2657}}].

\bibitem{Larkoski:2017jix}
A.~J. Larkoski, I.~Moult and B.~Nachman, \emph{{Jet Substructure at the Large
  Hadron Collider: A Review of Recent Advances in Theory and Machine
  Learning}},  \href{https://arxiv.org/abs/1709.04464}{{\ttfamily 1709.04464}}.

\bibitem{Dasgupta:2013ihk}
M.~Dasgupta, A.~Fregoso, S.~Marzani and G.~P. Salam, \emph{{Towards an
  understanding of jet substructure}}, {\emph{JHEP} {\bfseries 09} (2013) 029},
  [\href{https://arxiv.org/abs/1307.0007}{{\ttfamily 1307.0007}}].

\bibitem{Aaboud:2017qwh}
{\scshape ATLAS} collaboration, M.~Aaboud et~al., \emph{{A measurement of the
  soft-drop jet mass in pp collisions at $\sqrt{s} = 13$ TeV with the ATLAS
  detector}},
  \href{http://dx.doi.org/10.1103/PhysRevLett.121.092001}{\emph{Phys. Rev.
  Lett.} {\bfseries 121} (2018) 092001},
  [\href{https://arxiv.org/abs/1711.08341}{{\ttfamily 1711.08341}}].

\bibitem{Sirunyan:2018xdh}
{\scshape CMS} collaboration, A.~M. Sirunyan et~al., \emph{{Measurements of the
  differential jet cross section as a function of the jet mass in dijet events
  from proton-proton collisions at $\sqrt{s} =$ 13 TeV}},
  \href{https://arxiv.org/abs/1807.05974}{{\ttfamily 1807.05974}}.

\bibitem{Walsh:2011fz}
J.~R. Walsh and S.~Zuberi, \emph{{Factorization Constraints on Jet
  Substructure}},  \href{https://arxiv.org/abs/1110.5333}{{\ttfamily
  1110.5333}}.

\bibitem{Dasgupta:2013via}
M.~Dasgupta, A.~Fregoso, S.~Marzani and A.~Powling, \emph{{Jet substructure
  with analytical methods}},
  \href{http://dx.doi.org/10.1140/epjc/s10052-013-2623-3}{\emph{Eur. Phys. J.}
  {\bfseries C73} (2013) 2623},
  [\href{https://arxiv.org/abs/1307.0013}{{\ttfamily 1307.0013}}].

\bibitem{Frye:2016aiz}
C.~Frye, A.~J. Larkoski, M.~D. Schwartz and K.~Yan, \emph{{Factorization for
  groomed jet substructure beyond the next-to-leading logarithm}},
  \href{http://dx.doi.org/10.1007/JHEP07(2016)064}{\emph{JHEP} {\bfseries 07}
  (2016) 064}, [\href{https://arxiv.org/abs/1603.09338}{{\ttfamily
  1603.09338}}].

\bibitem{Marzani:2017mva}
S.~Marzani, L.~Schunk and G.~Soyez, \emph{{A study of jet mass distributions
  with grooming}}, \href{http://dx.doi.org/10.1007/JHEP07(2017)132}{\emph{JHEP}
  {\bfseries 07} (2017) 132},
  [\href{https://arxiv.org/abs/1704.02210}{{\ttfamily 1704.02210}}].

\bibitem{Larkoski:2017iuy}
A.~J. Larkoski, I.~Moult and D.~Neill, \emph{{Analytic Boosted Boson
  Discrimination at the Large Hadron Collider}},
  \href{https://arxiv.org/abs/1708.06760}{{\ttfamily 1708.06760}}.

\bibitem{Kang:2018jwa}
Z.-B. Kang, K.~Lee, X.~Liu and F.~Ringer, \emph{{The groomed and ungroomed jet
  mass distribution for inclusive jet production at the LHC}},
  \href{http://dx.doi.org/10.1007/JHEP10(2018)137}{\emph{JHEP} {\bfseries 10}
  (2018) 137}, [\href{https://arxiv.org/abs/1803.03645}{{\ttfamily
  1803.03645}}].

\bibitem{Kang:2018vgn}
Z.-B. Kang, K.~Lee, X.~Liu and F.~Ringer, \emph{{Soft drop groomed jet
  angularities at the LHC}},
  \href{http://dx.doi.org/10.1016/j.physletb.2019.04.018}{\emph{Phys. Lett.}
  {\bfseries B793} (2019) 41--47},
  [\href{https://arxiv.org/abs/1811.06983}{{\ttfamily 1811.06983}}].

\bibitem{Baron:2018nfz}
J.~Baron, S.~Marzani and V.~Theeuwes, \emph{{Soft-Drop Thrust}},
  \href{http://dx.doi.org/10.1007/JHEP08(2018)105,
  10.1007/JHEP05(2019)056}{\emph{JHEP} {\bfseries 08} (2018) 105},
  [\href{https://arxiv.org/abs/1803.04719}{{\ttfamily 1803.04719}}].

\bibitem{Kardos:2018kth}
A.~Kardos, G.~Somogyi and Z.~Tr{\'o}cs{\'a}nyi, \emph{{Soft-drop event shapes
  in electron--positron annihilation at next-to-next-to-leading order
  accuracy}},
  \href{http://dx.doi.org/10.1016/j.physletb.2018.10.014}{\emph{Phys. Lett.}
  {\bfseries B786} (2018) 313--318},
  [\href{https://arxiv.org/abs/1807.11472}{{\ttfamily 1807.11472}}].

\bibitem{Hoang:2017kmk}
A.~H. Hoang, S.~Mantry, A.~Pathak and I.~W. Stewart, \emph{{Extracting a Short
  Distance Top Mass with Light Grooming}},
  \href{http://dx.doi.org/10.1103/PhysRevD.100.074021}{\emph{Phys. Rev.}
  {\bfseries D100} (2019) 074021},
  [\href{https://arxiv.org/abs/1708.02586}{{\ttfamily 1708.02586}}].

\bibitem{Lee:2019lge}
C.~Lee, P.~Shrivastava and V.~Vaidya, \emph{{Predictions for energy correlators
  probing substructure of groomed heavy quark jets}},
  \href{https://arxiv.org/abs/1901.09095}{{\ttfamily 1901.09095}}.

\bibitem{Marzani:2017kqd}
S.~Marzani, L.~Schunk and G.~Soyez, \emph{{The jet mass distribution after Soft
  Drop}}, \href{http://dx.doi.org/10.1140/epjc/s10052-018-5579-5}{\emph{Eur.
  Phys. J.} {\bfseries C78} (2018) 96},
  [\href{https://arxiv.org/abs/1712.05105}{{\ttfamily 1712.05105}}].

\bibitem{Larkoski:2017cqq}
A.~J. Larkoski, I.~Moult and D.~Neill, \emph{{Factorization and Resummation for
  Groomed Multi-Prong Jet Shapes}},
  \href{http://dx.doi.org/10.1007/JHEP02(2018)144}{\emph{JHEP} {\bfseries 02}
  (2018) 144}, [\href{https://arxiv.org/abs/1710.00014}{{\ttfamily
  1710.00014}}].

\bibitem{Dokshitzer:1997in}
Y.~L. Dokshitzer, G.~D. Leder, S.~Moretti and B.~R. Webber, \emph{{Better jet
  clustering algorithms}},
  \href{http://dx.doi.org/10.1088/1126-6708/1997/08/001}{\emph{JHEP} {\bfseries
  08} (1997) 001}, [\href{https://arxiv.org/abs/hep-ph/9707323}{{\ttfamily
  hep-ph/9707323}}].

\bibitem{Wobisch:1998wt}
M.~Wobisch and T.~Wengler, \emph{{Hadronization corrections to jet
  cross-sections in deep inelastic scattering}},  in \emph{{Monte Carlo
  generators for HERA physics. Proceedings, Workshop, Hamburg, Germany,
  1998-1999}}, pp.~270--279, 1998.
\newblock \href{https://arxiv.org/abs/hep-ph/9907280}{{\ttfamily
  hep-ph/9907280}}.

\bibitem{Frye:2016okc}
C.~Frye, A.~J. Larkoski, M.~D. Schwartz and K.~Yan, \emph{{Precision physics
  with pile-up insensitive observables}},
  \href{https://arxiv.org/abs/1603.06375}{{\ttfamily 1603.06375}}.

\bibitem{Bauer:2011uc}
C.~W. Bauer, F.~J. Tackmann, J.~R. Walsh and S.~Zuberi, \emph{{Factorization
  and Resummation for Dijet Invariant Mass Spectra}},
  \href{http://dx.doi.org/10.1103/PhysRevD.85.074006}{\emph{Phys. Rev.}
  {\bfseries D85} (2012) 074006},
  [\href{https://arxiv.org/abs/1106.6047}{{\ttfamily 1106.6047}}].

\bibitem{Procura:2014cba}
M.~Procura, W.~J. Waalewijn and L.~Zeune, \emph{{Resummation of
  Double-Differential Cross Sections and Fully-Unintegrated Parton Distribution
  Functions}}, \href{http://dx.doi.org/10.1007/JHEP02(2015)117}{\emph{JHEP}
  {\bfseries 02} (2015) 117},
  [\href{https://arxiv.org/abs/1410.6483}{{\ttfamily 1410.6483}}].

\bibitem{Larkoski:2015zka}
A.~J. Larkoski, I.~Moult and D.~Neill, \emph{{Non-Global Logarithms,
  Factorization, and the Soft Substructure of Jets}},
  \href{http://dx.doi.org/10.1007/JHEP09(2015)143}{\emph{JHEP} {\bfseries 09}
  (2015) 143}, [\href{https://arxiv.org/abs/1501.04596}{{\ttfamily
  1501.04596}}].

\bibitem{Pietrulewicz:2016nwo}
P.~Pietrulewicz, F.~J. Tackmann and W.~J. Waalewijn, \emph{{Factorization and
  Resummation for Generic Hierarchies between Jets}},
  \href{http://dx.doi.org/10.1007/JHEP08(2016)002}{\emph{JHEP} {\bfseries 08}
  (2016) 002}, [\href{https://arxiv.org/abs/1601.05088}{{\ttfamily
  1601.05088}}].

\bibitem{Hoang:2007vb}
A.~H. Hoang and I.~W. Stewart, \emph{{Designing Gapped Soft Functions for Jet
  Production}},
  \href{http://dx.doi.org/10.1016/j.physletb.2008.01.040}{\emph{Phys. Lett.}
  {\bfseries B660} (2008) 483--493},
  [\href{https://arxiv.org/abs/0709.3519}{{\ttfamily 0709.3519}}].

\bibitem{Ligeti:2008ac}
Z.~Ligeti, I.~W. Stewart and F.~J. Tackmann, \emph{{Treating the b quark
  distribution function with reliable uncertainties}},
  \href{http://dx.doi.org/10.1103/PhysRevD.78.114014}{\emph{Phys. Rev.}
  {\bfseries D78} (2008) 114014},
  [\href{https://arxiv.org/abs/0807.1926}{{\ttfamily 0807.1926}}].

\bibitem{Cacciari:2008gn}
M.~Cacciari, G.~P. Salam and G.~Soyez, \emph{{The Catchment Area of Jets}},
  {\emph{JHEP} {\bfseries 04} (2008) 005},
  [\href{https://arxiv.org/abs/0802.1188}{{\ttfamily 0802.1188}}].

\bibitem{Belitsky:2001ij}
A.~V. Belitsky, G.~P. Korchemsky and G.~Sterman, \emph{{Energy flow in QCD and
  event shape functions}},
  \href{http://dx.doi.org/10.1016/S0370-2693(01)00899-1}{\emph{Phys. Lett.}
  {\bfseries B515} (2001) 297--307},
  [\href{https://arxiv.org/abs/hep-ph/0106308}{{\ttfamily hep-ph/0106308}}].

\bibitem{Catani:1992ua}
S.~Catani, L.~Trentadue, G.~Turnock and B.~R. Webber, \emph{{Resummation of
  large logarithms in $e^+\, e^-$ event shape distributions}},
  \href{http://dx.doi.org/10.1016/0550-3213(93)90271-P}{\emph{Nucl. Phys.}
  {\bfseries B407} (1993) 3--42}.

\bibitem{Dokshitzer:1998kz}
Y.~L. Dokshitzer, A.~Lucenti, G.~Marchesini and G.~Salam, \emph{{On the QCD
  analysis of jet broadening}}, {\emph{JHEP} {\bfseries 9801} (1998) 011},
  [\href{https://arxiv.org/abs/hep-ph/9801324}{{\ttfamily hep-ph/9801324}}].

\bibitem{Banfi:2004yd}
A.~Banfi, G.~P. Salam and G.~Zanderighi, \emph{{Principles of general
  final-state resummation and automated implementation}}, {\emph{JHEP}
  {\bfseries 03} (2005) 073},
  [\href{https://arxiv.org/abs/hep-ph/0407286}{{\ttfamily hep-ph/0407286}}].

\bibitem{Banfi:2014sua}
A.~Banfi, H.~McAslan, P.~F. Monni and G.~Zanderighi, \emph{{A general method
  for the resummation of event shape distributions in $e^{+} e^{-}$
  annihilation}}, {\emph{JHEP} {\bfseries 05} (2015) 102},
  [\href{https://arxiv.org/abs/1412.2126}{{\ttfamily 1412.2126}}].

\bibitem{Larkoski:2015lea}
A.~J. Larkoski, S.~Marzani and J.~Thaler, \emph{{Sudakov Safety in Perturbative
  QCD}}, \href{http://dx.doi.org/10.1103/PhysRevD.91.111501}{\emph{Phys. Rev.}
  {\bfseries D91} (2015) 111501},
  [\href{https://arxiv.org/abs/1502.01719}{{\ttfamily 1502.01719}}].

\bibitem{Berger:2003iw}
C.~F. Berger, T.~K{\'u}cs and G.~Sterman, \emph{{Event shape / energy flow
  correlations}},
  \href{http://dx.doi.org/10.1103/PhysRevD.68.014012}{\emph{Phys. Rev. D}
  {\bfseries 68} (2003) 014012},
  [\href{https://arxiv.org/abs/hep-ph/0303051}{{\ttfamily hep-ph/0303051}}].

\bibitem{Sjostrand:2006za}
T.~Sjostrand, S.~Mrenna and P.~Z. Skands, \emph{{PYTHIA 6.4 Physics and
  Manual}}, \href{http://dx.doi.org/10.1088/1126-6708/2006/05/026}{\emph{JHEP}
  {\bfseries 05} (2006) 026},
  [\href{https://arxiv.org/abs/hep-ph/0603175}{{\ttfamily hep-ph/0603175}}].

\bibitem{Sjostrand:2014zea}
T.~Sj{\"o}strand, S.~Ask, J.~R. Christiansen, R.~Corke, N.~Desai, P.~Ilten
  et~al., \emph{{An Introduction to PYTHIA 8.2}},
  \href{http://dx.doi.org/10.1016/j.cpc.2015.01.024}{\emph{Comput. Phys.
  Commun.} {\bfseries 191} (2015) 159--177},
  [\href{https://arxiv.org/abs/1410.3012}{{\ttfamily 1410.3012}}].

\bibitem{Fischer:2016vfv}
N.~Fischer, S.~Prestel, M.~Ritzmann and P.~Skands, \emph{{Vincia for Hadron
  Colliders}},
  \href{http://dx.doi.org/10.1140/epjc/s10052-016-4429-6}{\emph{Eur. Phys. J.}
  {\bfseries C76} (2016) 589},
  [\href{https://arxiv.org/abs/1605.06142}{{\ttfamily 1605.06142}}].

\bibitem{Bellm:2015jjp}
J.~Bellm et~al., \emph{{Herwig 7.0/Herwig++ 3.0 release note}},
  \href{http://dx.doi.org/10.1140/epjc/s10052-016-4018-8}{\emph{Eur. Phys. J.}
  {\bfseries C76} (2016) 196},
  [\href{https://arxiv.org/abs/1512.01178}{{\ttfamily 1512.01178}}].

\bibitem{Cacciari:2011ma}
M.~Cacciari, G.~P. Salam and G.~Soyez, \emph{{FastJet User Manual}},
  \href{http://dx.doi.org/10.1140/epjc/s10052-012-1896-2}{\emph{Eur. Phys. J.}
  {\bfseries C72} (2012) 1896},
  [\href{https://arxiv.org/abs/1111.6097}{{\ttfamily 1111.6097}}].

\bibitem{Cacciari:2008gp}
M.~Cacciari, G.~P. Salam and G.~Soyez, \emph{{The Anti-k(t) jet clustering
  algorithm}},
  \href{http://dx.doi.org/10.1088/1126-6708/2008/04/063}{\emph{JHEP} {\bfseries
  04} (2008) 063}, [\href{https://arxiv.org/abs/0802.1189}{{\ttfamily
  0802.1189}}].

\bibitem{Reichelt:2017hts}
D.~Reichelt, P.~Richardson and A.~Siodmok, \emph{{Improving the Simulation of
  Quark and Gluon Jets with Herwig 7}},
  \href{http://dx.doi.org/10.1140/epjc/s10052-017-5374-8}{\emph{Eur. Phys. J.}
  {\bfseries C77} (2017) 876},
  [\href{https://arxiv.org/abs/1708.01491}{{\ttfamily 1708.01491}}].

\bibitem{Gieseke:2003rz}
S.~Gieseke, P.~Stephens and B.~Webber, \emph{{New formalism for QCD parton
  showers}}, \href{http://dx.doi.org/10.1088/1126-6708/2003/12/045}{\emph{JHEP}
  {\bfseries 12} (2003) 045},
  [\href{https://arxiv.org/abs/hep-ph/0310083}{{\ttfamily hep-ph/0310083}}].

\bibitem{Bewick:2019rbu}
G.~Bewick, S.~Ferrario~Ravasio, P.~Richardson and M.~H. Seymour,
  \emph{{Logarithmic Accuracy of Angular-Ordered Parton Showers}},
  \href{https://arxiv.org/abs/1904.11866}{{\ttfamily 1904.11866}}.

\end{thebibliography}\endgroup

\end{document}